\documentclass[aip,pof,reprint]{revtex4-1}

\usepackage{lipsum}
\usepackage{color}
\usepackage{tikz}
\usetikzlibrary{patterns,calc,shapes,arrows,decorations.text}
\usepackage[outline]{contour}
\usepackage{booktabs}
\usepackage{subcaption}
\usepackage{amsfonts}
\usepackage{amsmath}
\usepackage{amssymb}
\usepackage{amsthm}
\usepackage{multirow}

\newcommand{\rd}{\mathrm{d}}

\newcommand{\I}{{(i)}}
\newcommand{\J}{{(j)}}
\renewcommand{\IJ}{{(ij)}}
\renewcommand{\i}{{i}}
\renewcommand{\j}{{j}}

\newcommand{\one}{{(1)}}
\newcommand{\two}{{(2)}}

\newcommand{\Kn}{\mathrm{Kn}}
\newcommand{\Pa}{\mathrm{Pa}}
\newcommand{\isp}{{(N_2)}}
\newcommand{\jsp}{{(H_2O)}}

\DeclareSymbolFont{extraitalic}      {U}{zavm}{m}{it}
\DeclareMathSymbol{\Qoppa}{\mathord}{extraitalic}{161}
\DeclareMathSymbol{\qoppa}{\mathord}{extraitalic}{162}
\DeclareMathSymbol{\Stigma}{\mathord}{extraitalic}{167}
\DeclareMathSymbol{\Sampi}{\mathord}{extraitalic}{165}
\DeclareMathSymbol{\sampi}{\mathord}{extraitalic}{166}
\DeclareMathSymbol{\stigma}{\mathord}{extraitalic}{168}

\newcommand*{\shashank}[1]{\textcolor{blue}{#1}}

\usepackage[normalem]{ulem}

\begin{document}
\preprint{Purdue/AAE/}

\title{Quantification of thermally-driven flows in microsystems using Boltzmann equation in deterministic and stochastic contexts}

\author{Shashank Jaiswal}
\email{jaiswal0@purdue.edu}
\author{Aaron Pikus}
\email{pikus@purdue.edu}
\author{Andrew Strongrich}
\email{astrongr@purdue.edu}
\author{Israel B. Sebasti\~{a}o}
\email{sebastiao@purdue.edu}
\affiliation{ 
  School of Aeronautics and Astronautics, Purdue University, West Lafayette, Indiana 47907, USA
}
\author{Jingwei Hu}%
\email{jingweihu@purdue.edu}
\affiliation{Department of Mathematics, Purdue University, West Lafayette, Indiana 47907, USA}
\author{Alina A. Alexeenko}
\email{alexeenk@purdue.edu}
\affiliation{ 
  School of Aeronautics and Astronautics, Purdue University, West Lafayette, Indiana 47907, USA
}

\date{\today}

\begin{abstract}

When the flow is sufficiently rarefied, a temperature gradient, for example, between two walls separated by a few mean free paths, induces a gas flow---an observation attributed to the thermo-stress convection effects at microscale. The dynamics of the overall thermo-stress convection process is governed by the Boltzmann equation---an integro-differential equation describing the evolution of the molecular distribution function in six-dimensional phase space---which models dilute gas behavior at the molecular level to accurately describe a wide range of flow phenomena. Approaches for solving the full Boltzmann equation with general inter-molecular interactions rely on two perspectives: one stochastic in nature often delegated to the direct simulation Monte Carlo (DSMC) method; and the others deterministic by virtue. Among the deterministic approaches, the discontinuous Galerkin fast spectral (DGFS) method has been recently introduced for solving the full Boltzmann equation with general collision kernels, including the variable hard/soft sphere models---necessary for simulating flows involving diffusive transport. In this work, the deterministic DGFS method; Bhatnagar-Gross-Krook (BGK), Ellipsoidal statistical BGK, and Shakhov kinetic models; and the widely-used stochastic DSMC method, are utilized to assess the thermo-stress convection process in MIKRA---Micro In-Plane Knudsen Radiometric Actuator---a microscale compact low-power pressure sensor utilizing the Knudsen forces. BGK model under-predicts the heat-flux, shear-stress, and flow speed; S-model over-predicts; whereas ESBGK comes close to the DSMC results. On the other hand, both the statistical/DSMC and deterministic/DGFS methods, segregated in perspectives, yet, yield inextricable results, bespeaking the ingenuity of Graeme Bird who laid down the foundation of practical rarefied gas dynamics for microsystems.

\end{abstract}
\shashank{$\Sampi$ Special issue dedicated to Graeme Bird}

\maketitle 

\section{Introduction}
\label{sec_introduction}
In microscale flows, the length scale dictates the type of forces governing the physical phenomena. The surface to the volume ratio is high and hence the surface forces dominate. The Reynolds number is low and the viscous shear stresses are significantly increased\cite{ho1998micro}. Under sufficiently rarefied flow conditions, an application of temperature gradient, say, between two parallel plates separated by few mean free paths, induces a low velocity gas flow commonly identified as thermo-stress convection effects\cite{kogan1976stresses}. A necessary condition to induce a sufficiently \textit{useful} gaseous velocity requires the characteristic length scale of the thermal gradients $T/|\nabla_x T|$ to be comparable to the molecular mean free path $\lambda$. At macroscale, such magnitudes are prohibitive, necessitating thermal gradients on the order of $10^6\,K/m$. However, at microscale, such conditions are readily achieved allowing the thermo-stress effects to overcome the classically dominant viscous forces \cite{strongrich2017microscale}.

From a historical and experimental viewpoint, Knudsen, in 1910, explored the possibility of gas actuation under the influence of temperature gradients using evacuated glass bulbs separated by a long narrow tube, wherein heating one of the bulbs resulted in a pumping action creating a high pressure at the hot end and low pressure at the cold end\cite{knudsen1910thermischer,karniadakis2006microflows}. In 1950's\cite{knudsen1950kinetic}, Knudsen carried out various experiments using Crooke's radiometer\cite{crookes1874xv}, wherein a device consisting of a long thin and narrow platinum band with dark (hot) and bright (cold) sides, in a rarefied environment, exhibits a net force due to momentum imbalance of particles reflecting from the dark and bright sides. Without being exhaustive, we refer to (Ref.~\onlinecite{ketsdever2012radiometric}) for a comprehensive review of the radiometric phenomenon. From a theoretical viewpoint, Maxwell hypothesized that one of the possible causes of radiometric effects are temperature stresses. However, based on linearized kinetic theory and corresponding reduced macroscopic equations of motion (see section 15 in Ref.~\onlinecite{maxwell1879vii}), the author concluded that no motion can be produced by temperature stresses\cite{maxwell1879vii,kogan1976stresses}, which, in general, is incorrect. Later, Kogan, in 1976, introduced the theory of thermo-stress convection, wherein the bulk velocity is attributed to presence of higher order terms of temperature stresses (see eq. 2.6 in Ref.~\onlinecite{kogan1976stresses}), arrived in part by the second order Chapman-Enskog expansion commonly identified as the Burnett approximations. In the multi-species context, however, the phenomenon and the effect of thermo-stress convection on the flow concentration (and the subsequent induced velocity) is more apparent. 

Chapman \cite{chapman1970mathematical}, as early as 1953, developed the theory of diffusion processes (see eq.~(8.4, 7) in Ref.~\onlinecite{chapman1970mathematical} again derived using Chapman-Enskog expansion) wherein the difference in concentrations of two species is proportional to the thermal gradient term $k_T \nabla \ln{T}$, where $k_T$ is thermal diffusion factor. At normal conditions, this coefficient is very low, and is therefore not accounted in practice. For instance, as a classical example, Bird\cite{Bird} devised a self-diffusion test case (see section 12.6) where the diffusion coefficient was measured by ignoring the thermal gradient term $k_T \nabla \ln{T}$ of eq.~(8.4, 7) in Ref.~\onlinecite{chapman1970mathematical}. Note however that there is considerable thermal gradient in self-diffusion cases, see for instance Ref.~\onlinecite{jaiswal2019dgfsMulti}, where we presented the results for temperature variation for self diffusion cases. Although the temperature gradient is unaccounted for, the diffusion coefficient, which is measured by a self diffusion simulation, matches well with the experimentally\cite{chapman1970mathematical} observed diffusion coefficient. This suggests that $k_T$ is potentially low --- which is indeed the case, for instance, see Ref~\onlinecite{pavlov2019diffusion}, wherein the authors noted thermal diffusion coefficient on order of $10^{-3}$. In microscale flows where the per unit temperature drop can easily reach $10^6 K/m$, as noted earlier, $k_T \nabla \ln{T}$ can have appreciable contributions. This type of process has been interpreted in terms of thermo-stress convection due to concentration inhomogeneities by Kogan\cite{kogan1976stresses}. The overall thermo-stress convection phenomenon/effect is highly coupled and exhibits highly rich flow structures (as will be shown in section~\ref{subsec_mikra_results}), and an in-depth understanding can prove to be very useful for development of next generation of microsystems. 

To summarize,  Sone\cite{sone2012kinetic} identified three broad groups of the temperature driven flow based on its application in microsystems: a) thermal creep flow\cite{kennard1938kinetic,sone1966thermal,sharipov1998data} which is an induced flow around a body with non uniform temperature; b) thermal stress slip flow, which is induced by nonuniform temperature gradient over the boundary \cite{kogan1976stresses,aoki1998rarefied,sone1972flow,sone1997demonstration,selden2009area,fowee2016experimental,ibrayeva2017numerical}; c) and nonlinear thermal stress flow\cite{kogan1976stresses}, which is important only when the temperature gradient in the gas is high, and nonlinear terms of temperature variations in stress tensor should be taken into account. The present study is delegated to the third i.e., nonlinear thermal stress flow.

From a practical engineering viewpoint, thermo-stress convection has been applied for micro-structure actuation. Passian\cite{passian2002knudsen,passian2003thermal}, in 2003, demonstrated a micro-cantilever suspended over a substrate, which when heated via a pulsed laser generated deflections at the cantilever tip as a consequence of the Knudsen forces in the gap between the substrate and micro-cantilever. Foroutan\cite{foroutan2014levitation}, in 2014, demonstrated untethered levitation in concave micro-flying robots relying on Knudsen force. The phenomenon has been further explored in small satellite and spacecraft attitude control devices \cite{nallapu2017radiometric} and high-altitude propulsion systems \cite{cornella2012analysis}. 

The dynamics of the overall thermo-stress convection process is governed by the Boltzmann equation---an integro-differential equation describing the evolution of the distribution function in six-dimensional phase space---which models the dilute gas behavior at the molecular level to accurately describe a wide range of non-continuum flow phenomena. In the present work, we assess the thermo-stress convection process using the fundamental microscopic full Boltzmann equation. The approaches for numerical solution of the Boltzmann equation date back to as early as 1940s \cite{grad1949}. However, it was not until 1960s that the numerical simulations were feasible. In practice, the numerical simulations of Boltzmann equation was made possible by introduction of direct simulation Monte Carlo (DSMC) method \cite{bird1963approach,Bird}. Over sufficient small intervals, by decoupling the molecular motion and interaction processes, DSMC first advects the particles deterministically according to their velocities, also termed as \textit{free transport}, and then describes the collisions by statistical models with a specified interaction potential. 

\begin{figure}[!ht]
  \centering
  \includegraphics[width=85mm]{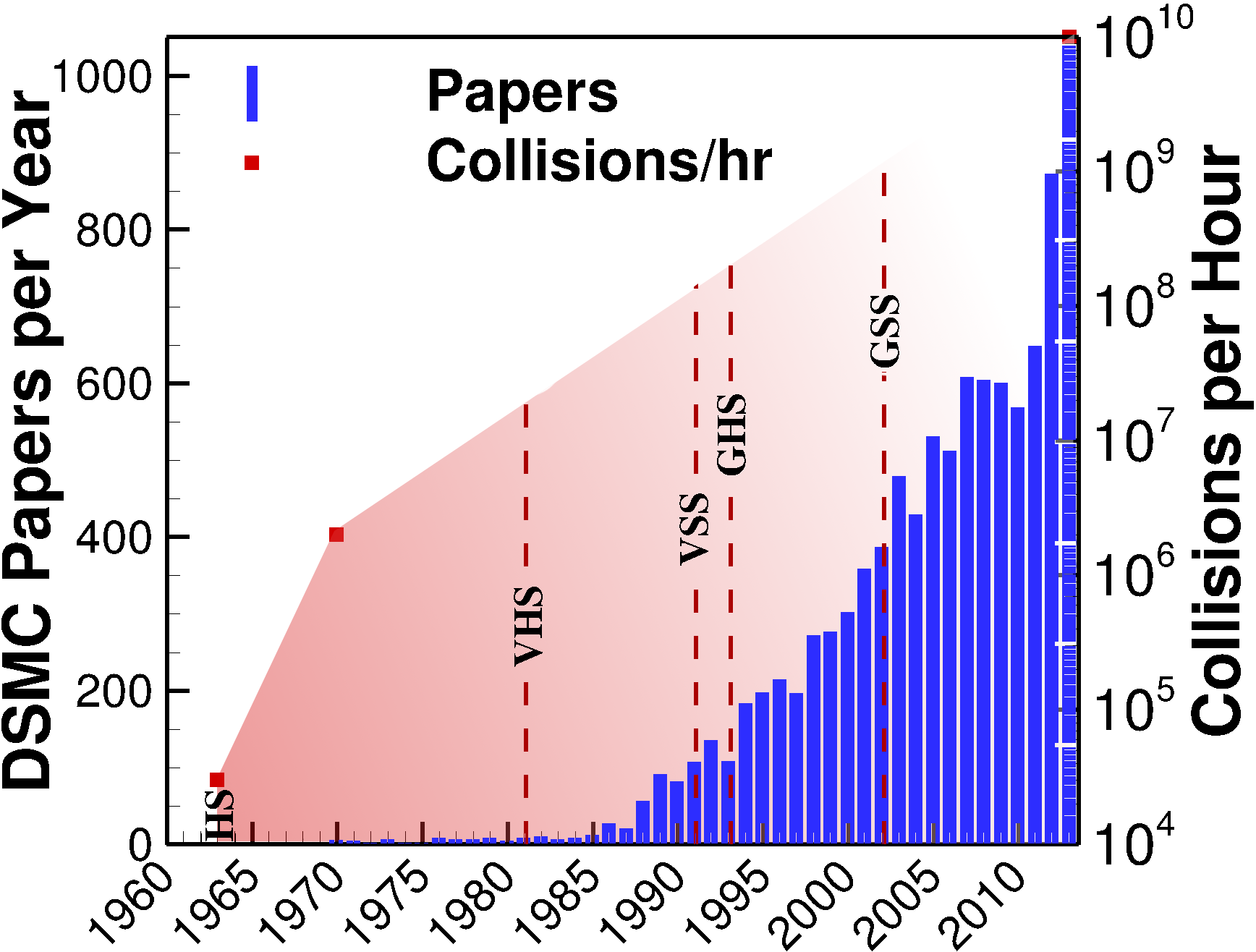}
  \caption{A holistic view of growth in DSMC publications over years; and the exponential increase in number of simulated collision events per hour.}
\label{fig_usageDSMC}
\end{figure}

The choice of interaction potential substantially affects the simulation fidelity and computational complexity. Early implementations of the DSMC method relied on purely repulsive hard sphere (HS) interaction model \cite{bird1963approach}. The HS model, however, deviates from experimental observations for common gases \cite{maitland1972critical} due to a square-root viscosity variation with temperature. The variable hard sphere (VHS) model proposed by Bird \cite{Bird} results in a more general power-law viscosity variation with temperature; and has been widely used for DSMC simulations of \textit{single-species} gas flows due to its computational efficiency and ease of implementation. The VHS model, however, deviates from experimental observations for common \textit{multi-species} flows \cite{koura1991test,koura1991variable} involving diffusive transport. Later, several variations of the VHS model were proposed, including, the variable soft sphere (VSS) \cite{koura1991variable}, M-1 \cite{kersch1994selfconsistent}, generalized soft sphere (GSS) \cite{fan2002generalized}, all of which belong to a class of repulsive interactions. The VSS model modifies the scattering law of the VHS model by using a scattering parameter ($\alpha$) that allows reproduction of measured diffusion coefficients in addition to the viscosity coefficient. M-1 model is a modification of VHS model to have a linear distribution of scattering angles in terms of the impact parameter. This modification allows M-1 to reproduce correct viscosity and diffusivity without the need of an additional parameter $(\alpha)$\cite{jaiswal2018dsmc}. The GSS model, although general, needs additional parameters for reproducing the viscosity and diffusion coefficients (see Ref.~\onlinecite{weaver2015assessment,Bird} for additional details/equations for these models). In particular, the complexity of DSMC algorithm is independent of number of species in the mixture, as well as the mass of the individual species. This makes the method highly useful and efficient for modeling sufficiently \textit{fast}\cite{kogan1976stresses} non-equilibrium flows. From a usage perspective, there is a growing number of applications requiring DSMC simulations. Statistics (see Fig.~\ref{fig_usageDSMC}) shows that over 1000 papers on DSMC are now published every year. Many of these papers are based on the codes made available to the research community in 1990s by Graeme Bird. Over the years---in accordance with the predictions of the Moore's law---the number of collisions performed per hour have increased exponentially. 


However, it is the stochastic nature of the DSMC that introduces high statistical noise in low-speed flows. In the present work, we study the thermo-stress convection process using the recently developed deterministic discontinuous Galerkin fast spectral (DGFS) method \cite{JAH19,jaiswal2019dgfsMulti} as well as DSMC: the primary tool for rarefied flow simulations. DGFS allows arbitrary unstructured geometries; high order accuracy in physical space time, and velocity space; arbitrary collision kernels, including, the well known VSS model \cite{jaiswal2019dgfsMulti}; and provides excellent nearly-linear scaling characteristics on massively parallel architectures \cite{jaiswal2018dgfsGPU,jaiswal2019dgfsMultiSpeciesGPU}. DGFS produces noise-free solutions and can simulate low-speed flows encountered in thermo-stress convection dominated devices. 

From a flow modelling viewpoint, Loyalka \cite{loyalka1977knudsen}, using a linearized Boltzmann equation, calculated the longitudinal and transversal Knudsen forces on the cylindrical surfaces of a hanging wire of a vacuum micro-balance. The authors noted Knudsen force maximum in the transitional regime for Helium--an observation attributed to the bimodal nature of radiometric forces\cite{selden2009origins}. Fierro\cite{fierro1981gas} studied the problem using a Bhatnagar-Gross-Krook (BGK) model for range of Knudsen numbers and different molecular species noting an inverted parabolic profile for variation of Knudsen force with pressure (which can be reinterpreted in terms of Knudsen number since a fixed size geometry was used for all cases). The authors observed a peak Knudsen force in $10-100\,N/m^2$ pressure range for Helium, Krypton, Hydrogen, Oxygen, and Carbon dioxide. Alexeenko\cite{alexeenko2006comparison} carried out numerical simulations around heated micro-beams using the conventional Navier-Stokes incorporating first order Maxwell slip and Smoluchowski temperature jump boundary conditions, DSMC, and primarily using a deterministic kinetic ellipsoidal statistical Bhatnagar-Gross-Krook (ESBGK) model employing a finite-difference-discrete-velocity scheme. The gas-damping coefficients on a moving micro-beam for quasi-static isothermal conditions were estimated by the three numerical methods for Knudsen numbers from 0.1 to 1.0. It was concluded that the Navier-stokes simulations overestimate the gas-damping force for Knudsen numbers larger than 0.1, while the ESBGK and DSMC methods are in good agreement for the slip and transitional flow regimes. Moreover, the Knudsen force peaks in the transitional regime at $\Kn \approx 2$, and the numerically predicted variation of the force is consistent with experimental observations of the displacement of a heated micro-beam. Zhu\cite{zhu2010origin} analyzed the problem specifically using DSMC in the slip, transition, and free molecular regimes noting \textit{qualitative} agreements between DSMC and experimental results of Passian\cite{passian2002knudsen,passian2003thermal}. Nabeth\cite{nabeth2011quantifying} analyzed the problem using the ESBGK model within a finite volume framework. Notably, the authors devised a semi-empirical relation between the force and the Knudsen number based on dynamic similarity. Anikin\cite{anikin2011numerical} studied the radiometric forces via a direct solution of Boltzmann equation on 2-D velocity grids via a discrete ordinate projection method\cite{tcheremissine1998conservative}. More recently, Lotfian\cite{lotfian2019radiometric} analyzed the various arrangements for radiometric pumps featuring vane and ratchet structures, including, zigzag triangular fins, using DSMC and finite volume based BGK-Shakhov model. 

In more complex scenarios, one can stack an array of micro-heaters to significantly enhance the Knudsen force output\cite{gimelshein2011impact,strongrich2014experimental,strongrich2017microscale}. Strongrich\cite{strongrich2014experimental} demonstrated the possibility of amplifying the Knudsen forces as well as reversing its direction by combining thermal gradients between several solid bodies. The idea was further explored, resulting in development of a Microscale In-Plane Knudsen Radiometric Actuator (MIKRA) sensor for flow actuation and measurement \cite{alexeenko2016microelectromechanical,strongrich2015microstructure,strongrich2017microscale}. MIKRA consists of array of hot and cold micro-beams termed as heater and shuttle arm. When the heater arm is heated under the application of electric current, the Knudsen force is generated in the gap between the shuttle and heater arm. The displacement of shuttle arm is then measured using a capacitor (specific details to follow in section~\ref{sec_mikra}). MIKRA presents an interesting problem for analyzing thermostress convection due to temperature gradients as well as concentration inhomogeneties, see Ref.~\onlinecite{pikus2019characterization} where authors observed species separation in MIKRA which \textit{might} be, in part, due to be the effect of $k_T \nabla \ln{T}$ term. We believe it's too early to make a definite conclusion on the topic.  

A key question, and a subject of ongoing research is the following: How well can the kinetic equations/methods/models, for instance, McCormack model \cite{mccormack1973construction}, Lattice Boltzmann method (LBM) \cite{luo2003theory}, Bhatnagar-Gross-Krook (BGK) \cite{bhatnagar1954model,sirovich1962kinetic,AAP02,haack2017conservative,bobylev2018general}, Ellipsoidal statistical Bhatnagar-Gross-Krook (ESBGK) \cite{holway1966new,brull2015ellipsoidal}, BGK-Shakhov (S-model)\cite{shakhov1968generalization}, Unified Gas Kinetic Scheme (UGKS) \cite{xu2010unified,guo2013discrete}, Discontinuous Galerkin Fast Spectral (DGFS) \cite{JAH19}, and direct simulation Monte Carlo (DSMC)\cite{Bird}, describe the thermo-stress convection process, including, their applicability regimes at wide range of rarefaction levels and temperature gradients, and required computational cost for reproducing the correct induced low speed velocity profile on a \textit{common standard} benchmark problems such as MIKRA where the experimental results are readily available. As noted by Kogan\cite{kogan1976stresses}, the overall thermo-stress convection process is \textit{complicated} function of concentration, of mass-ratio, molecule-collision cross section, etc. An in-depth understanding of the overall thermo-stress convection process at the microscale may potentially prove useful for development of a series of new MEMS devices without any moving parts (see, for instance,  Refs.~\onlinecite{alexeenko2016microelectromechanical,alexeenko2017microelectronic}). This paper, in part, focuses on quantifying the fidelity of results recovered from BGK, ESBGK, S-model, DGFS and DSMC for the Knudsen radiometric actuator MIKRA. 


The rest of this paper is organized as follows. In section~\ref{sec_theEquation}, we give an overview of the multi-species Boltzmann equation, the self/cross collision integrals, and the phenomenological VHS/VSS collision kernels used in practical engineering applications. Extensive numerical verification for BGK, ESBGK, S-model, and DGFS against DSMC are performed and discussed in Section~\ref{sec_verifications}. Section~\ref{sec_mikra} provides the description, problem statement, and results for the thermo-stress convection enabled MIKRA sensor. Section~\ref{sec_multispecies_mikra} presents the analysis of \textit{multi-species} thermo-stress convection in MIKRA sensor. Concluding remarks are given in section~\ref{sec_conclusions}. 



\section{Boltzmann equation}
\label{sec_theEquation}

In this section, we give a brief overview of the multi-species Boltzmann equation. Readers are referred to Ref.~\onlinecite{jaiswal2019dgfsMulti} for more details.

Suppose we consider a gas mixture of $s$ species ($s\geq 2$), each represented by a distribution function $f^\I(t,x,v)$, where $t\geq 0$ is the time, $x\in\Omega\subset \mathbb{R}^3$ is the position, and $v\in \mathbb{R}^3$ is the particle velocity ($f^\I\,\rd{x}\,\rd{v}$ gives the number of particles of species $i$ to be found in an infinitesimal volume $\rd{x}\,\rd{v}$ centered at the point $(x,v)$ of the phase space). The time evolution of $f^{(i)}$ is described by the multi-species Boltzmann equation written as \cite{Cercignani, Harris}
\begin{align} \label{MBE}
\partial_t f^\I &+ v\cdot \nabla_x f^\I = \sum_{j=1}^{s}\mathcal{Q}^\IJ(f^\I,f^\J), \quad i=1,\dots,s.
\end{align}
Here $\mathcal{Q}^\IJ$ is the collision operator that models the binary collisions between species $\i$ and $\j$, and acts only in the velocity space:
\begin{align} 
\mathcal{Q}^{(ij)}(f^{(i)},&f^{(j)})(v) = \int_{\mathbb{R}^3}\int_{S^2}\mathcal{B}_{ij}(v-v_*,\sigma)\nonumber \\ &\times \left[f^{(i)}(v')f^{(j)}(v_*')-f^{(i)}(v)f^{(j)}(v_*)\right]\rd{\sigma}\,\rd{v_*},
\end{align}
where $(v, v_*)$ and $(v', v'_*)$ denote the pre- and post- collision velocity pairs. During collisions, the momentum and energy are conserved:
\begin{align} 
m_iv+m_jv_* &= m_iv'+m_jv_*', \nonumber\\ 
m_i|v|^2+m_j|v_*|^2 &= m_i|v'|^2+m_j|v_*'|^2,
\end{align}
where $m_i$, $m_j$ denote the mass of particles of species $i$ and $j$ respectively. Hence one can parameterize $v'$ and $v_*'$ as follows
\begin{align}
v'=\frac{v+v_*}{2}+\frac{(m_i-m_j)}{2(m_i+m_j)}(v-v_*)+\frac{m_j}{(m_i+m_j)}|v-v_*|\sigma, \nonumber\\
\displaystyle v_*'=\frac{v+v_*}{2}+\frac{(m_i-m_j)}{2(m_i+m_j)}(v-v_*)-\frac{m_i}{(m_i+m_j)}|v-v_*|\sigma,
\end{align}
with $\sigma$ being a vector varying on the unit sphere $S^2$. $\mathcal{B}_{ij}=\mathcal{B}_{ji}(\geq 0)$ is the collision kernel characterizing the interaction mechanism between particles. It can be shown that
\begin{equation}
    \mathcal{B}_{ij} = B_{ij}(|v - v_*|, \cos\chi), \quad \cos \chi = \frac{\sigma \cdot (v - v_*)}{| v - v_*|},
\end{equation}
where $\chi$ is the deviation angle between $v-v_*$ and $v'-v'_*$.

Given the interaction potential between particles, the specific form of $B_{ij}$ can be determined using the classical scattering theory:
\begin{equation}
B_{ij}(|v - v_*|, \cos\chi)=|v-v_*|\,\Sigma_{ij}(|v-v_*|,\chi),
\label{eq_dim_B1}
\end{equation}
where $\Sigma_{ij}$ is the differential cross-section given by
\begin{equation}
\Sigma_{ij}(|v-v_*|,\chi)=\frac{b_{ij}}{\sin \chi}\left | \frac{\rd{b_{ij}}}{\rd{ \chi}} \right|,
\label{eq_differentialCrossSection}
\end{equation}
with $b_{ij}$ being the impact parameter.

With a few exceptions, the explicit form of $\Sigma_{ij}$ can be hard to obtain since $b_{ij}$ is related to $\chi$ implicitly. However, as stated in the introduction, the choice of interaction potential substantially affects the simulation fidelity and computational complexity. 
Proposed as a modification of Bird's VHS model, Koura et al. \cite{koura1991variable} introduced the so-called VSS model by assuming 
\begin{equation}
    \chi = 2 \cos^{-1} \{(b_{ij}/d_{ij})^{1/\alpha_{ij}}\},
\end{equation}
where $\alpha_{ij}$ is the scattering parameter, and $d_{ij}$ is the diameter borrowed from the VHS model (eqn.~(4.79) in Ref.~\onlinecite{Bird}):
\begin{equation}
    d_{ij} = d_{\mathrm{ref},ij} \Bigg[ \Bigg(\frac{2 k_B T_{\mathrm{ref},ij}}{\mu_{ij}|v-v_*|^2}\Bigg)^{\omega_{ij}-0.5} \frac{1}{\Gamma(2.5-\omega_{ij})} \Bigg]^{1/2},
    \label{eq_dVSS}
\end{equation}
with $\Gamma$ being the Gamma function, $\mu_{ij}=\frac{m_im_j}{m_i+m_j}$ the reduced mass, $d_{\mathrm{ref},ij}$, $T_{\mathrm{ref},ij}$, and $\omega_{ij}$, respectively, the reference diameter, reference temperature, and viscosity index. Substituting the eqns.~(\ref{eq_differentialCrossSection})-(\ref{eq_dVSS}) into (\ref{eq_dim_B1}), one can obtain $B_{ij}$ as
\begin{equation} \label{VSS}
    B_{ij} = b_{\omega_{ij},\,\alpha_{ij}} \, |v - v_*|^{2(1 - \omega_{ij})} \,(1 + \cos \chi)^{\alpha_{ij}-1},
\end{equation}
where $b_{\omega_{ij},\,\alpha_{ij}}$ is a constant given by
\begin{equation}
    b_{\omega_{ij},\,\alpha_{ij}} = \frac{d_{\mathrm{ref},ij}^2}{4} \Bigg(\frac{2 k_B T_{\mathrm{ref},ij}}{\mu_{ij}}\Bigg)^{\omega_{ij}-0.5} \frac{\alpha_{ij}}{\Gamma(2.5 - \omega_{ij})\,2^{\alpha_{ij}-1}}.
\end{equation}
In particular, the VHS kernel is obtained when $\alpha_{ij} = 1$ and $0.5\leq \omega_{ij} \leq 1$ ($\omega_{ij}=1$: Maxwell molecules; $\omega_{ij}=0.5$: HS); and the VSS kernel is obtained when $1< \alpha_{ij} \leq 2$ and $0.5\leq \omega_{ij} \leq 1$.


Given the distribution function $f^{(i)}$, the number density, mass density, velocity, and temperature of species $i$ are defined as
\begin{align}
&n^{(i)}=\int_{\mathbb{R}^3}f^{(i)}\,\rd{v}, \quad \rho^{(i)}=m_in^{(i)}, \quad \nonumber \\ &u^{(i)}=\frac{1}{n^{(i)}}\int_{\mathbb{R}^3}vf^{(i)}\,\rd{v}, \quad \nonumber \\ &T^{(i)}=\frac{m_i}{3n^{(i)}k_B}\int_{\mathbb{R}^3}(v-u^{(i)})^2f^{(i)}\,\rd{v}.
\end{align}
The total number density, mass density, and velocity are given by
\begin{equation}
n=\sum_{i=1}^s n^{(i)}, \quad \rho=\sum_{i=1}^s \rho^{(i)}, \quad u=\frac{1}{\rho}\sum_{i=1}^s\rho^{(i)}u^{(i)}.
\end{equation}
Further, the diffusion velocity, stress tensor, and heat flux vector of species $i$ are defined as
\begin{align}
&v^{(i)}_D=\frac{1}{n^{(i)}}\int_{\mathbb{R}^3}cf^{(i)}\,\rd{v}=u^{(i)}-u, \quad \nonumber \\ &\mathbb{P}^{(i)}=\int_{\mathbb{R}^3}m_ic\otimes c f^{(i)}\,\rd{v}, \nonumber \quad q^{(i)}=\int_{\mathbb{R}^3}\frac{1}{2}m_ic|c|^2 f^{(i)}\,\rd{v},
\end{align}
where $c=v-u$ is the peculiar velocity. Finally, the total stress, heat flux, pressure, and temperature are given by
\begin{equation}
\mathbb{P}=\sum_{i=1}^s\mathbb{P}^{(i)}, \quad q=\sum_{i=1}^s q^{(i)}, \quad p= n k_B T=\frac{1}{3}\text{tr}(\mathbb{P}).
\end{equation}

\subsection{Stochastic modelling}

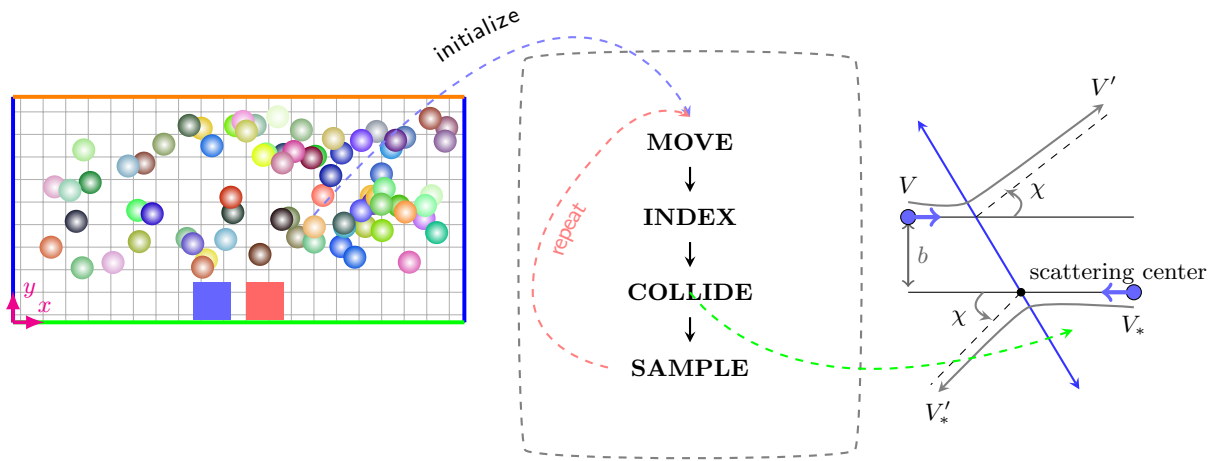
\begin{figure*}[!ht]
\definecolor{verdigris}{rgb}{0.26, 0.7, 0.68}
\newcommand\mycommfont[1]{{\large\bfseries\tt#1}}

\tikzstyle{block} = [rectangle, draw=blue!0, fill=blue!0,
    text width=6em, text centered, rounded corners, minimum height=2em]
\tikzstyle{line} = [thick, ->, >=stealth]
\tikzstyle{cloud} = [draw=none, ellipse,fill=red!10, node distance=2cm,
    minimum height=2em]
    
\begin{tikzpicture}[node distance = 1.cm, auto]
    \node [block] (move) {\bfseries MOVE};
    \node [block, below of=move] (index) {\bfseries INDEX};
    \node [block, below of=index] (collide) {\bfseries COLLIDE};
    \node [block, below of=collide] (sample) {\bfseries SAMPLE};
    \node [block, left of=sample, xshift=-4cm] (init) {};
    
    \node[inner sep=0pt, above of=init] (initimg) {}; 
    
    \begin{scope}[scale=0.2, above of=init, yshift=-13cm, xshift=-45cm]
        \def\Ho{20};

        \def\L{600 / \Ho};  
        \def\H{300 / \Ho};  
        \def\hG{10 / \Ho};  
        \def\hW{4 / \Ho};   
        
        \def\Lh{50 / \Ho};  
        \def\Hh{50 / \Ho};  
        
        \def \off{20};
        \def \dt{0.8};
        \def \pw{10};
        
        \coordinate (c1) at ({0}, {0});
        \coordinate (c2) at ({\L/2-\hG-\Lh}, {0});
        \coordinate (c3) at ({\L/2-\hG}, {0});
        \coordinate (c4) at ({\L/2+\hG}, {0});
        \coordinate (c5) at ({\L/2+\hG+\Lh}, {0});
        \coordinate (c6) at ({\L}, {0});
        \coordinate (c7) at ({\L}, {\hW});
        \coordinate (c8) at ({\L}, {\hW+\Hh});
        \coordinate (c9) at ({\L}, {\H});
        \coordinate (c10) at ({\L/2+\hG+\Lh}, {\H});
        \coordinate (c11) at ({\L/2+\hG}, {\H});
        \coordinate (c12) at ({\L/2-\hG}, {\H});
        \coordinate (c13) at ({\L/2-\hG-\Lh}, {\H});
        \coordinate (c14) at ({0}, {\H});
        \coordinate (c15) at ({0}, {\hW+\Hh});
        \coordinate (c16) at ({0}, {\hW});
        \coordinate (c17) at ({\L/2-\hG-\Lh}, {\hW});
        \coordinate (c18) at ({\L/2-\hG}, {\hW});
        \coordinate (c19) at ({\L/2-\hG}, {\hW+\Hh});
        \coordinate (c20) at ({\L/2-\hG-\Lh}, {\hW+\Hh});
        \coordinate (c21) at ({\L/2+\hG}, {\hW});
        \coordinate (c22) at ({\L/2+\hG+\Lh}, {\hW});
        \coordinate (c23) at ({\L/2+\hG+\Lh}, {\hW+\Hh});
        \coordinate (c24) at ({\L/2+\hG}, {\hW+\Hh});
        
        \draw (c1) -- (c2);
        \draw (c2) -- (c3);
        \draw (c3) -- (c4);
        \draw (c4) -- (c5);
        \draw (c5) -- (c6);
        \draw (c6) -- (c7);
        \draw (c7) -- (c8);
        \draw (c8) -- (c9);
        \draw (c9) -- (c10);
        \draw (c10) -- (c11);
        \draw (c11) -- (c12);
        \draw (c12) -- (c13);
        \draw (c13) -- (c14);
        \draw (c14) -- (c15);
        \draw (c15) -- (c16);
        \draw (c16) -- (c1);
        
        \draw[step=1.5,black!30,thin,xshift=0.5cm,yshift=0.5cm] (c1) grid (c9);
        
        \draw[pattern=north west lines, pattern color=black!50, line width = 0.1mm, thin, draw=none, fill=blue!60] (c17) -- (c18) -- (c19) -- (c20) -- (c17);
        
        \draw[pattern=north west lines, pattern color=black!50, line width = 0.1mm, thin, draw=none, fill=red!60] (c21) -- (c22) -- (c23) -- (c24) -- (c21);
        
        \draw[line width=1.5, blue] (c14) -- (c15) -- (c16) -- (c1);

        \draw[line width=1.5, blue] (c9) -- (c8) -- (c7) -- (c6);

        \draw[line width=1.5, orange] (c9) -- (c10) -- (c11) -- (c12) -- (c13) -- (c14);
        
        \draw[line width=1.5, green] (c1) -- (c2) -- (c3) -- (c4) -- (c5) -- (c6);
        
        \draw[-latex, line width=0.5mm, magenta] (c1) -- ([xshift=3*\off]c1) node[anchor=south] {$x$};
        
        \draw[-latex, line width=0.5mm, magenta] (c1) -- ([yshift=3*\off]c1) node[anchor=west] {$y$};
        
        \pgfmathsetseed{54321}
		\foreach \p in {1,...,75} { 
		    \pgfmathparse{0.5*(rand+1)}; \edef\a{\pgfmathresult};
		    \pgfmathparse{0.5*(rand+1)}; \edef\b{\pgfmathresult};
		    \pgfmathparse{0.5*(rand+1)}; \edef\c{\pgfmathresult};
		    \definecolor{MyColor}{rgb}{\a,\b,\c};
		    \node[circle, inner color=white, outer color=MyColor, shading=radial] at ([xshift={450+400*rand}, yshift={250+150*rand}]c1) {};
		    
		}
    \end{scope}

    \draw [line, xshift=4cm, draw=blue!50, dashed, thick, postaction={decorate,decoration={raise=2.5ex,text along path,text align=center,text={|\sffamily|initialize}}}] plot [smooth, tension=1] coordinates { ([yshift=1cm]initimg.north) ([xshift=-2cm,yshift=1cm]move.north) (move.north)};

    \draw [line] (move) -- (index);
    \draw [line] (index) -- (collide);
    \draw [line] (collide) -- (sample);
    \draw [line, draw=red!50, dashed, thick, postaction={decorate,decoration={text color=red!50,raise=-2.5ex,text along path,text align=center,text={|\sffamily|repeat}}}] plot [smooth, tension=1] coordinates { (sample.west) ([xshift=-1cm,yshift=1cm]sample.west) (move.west) (move.north)};
    \draw [line, draw=black!50, dashed, thick] plot [smooth cycle, tension=0.1] coordinates { ([xshift=-1.1cm,yshift=-1.1cm]sample.west) ([xshift=-1.1cm,yshift=1.1cm]move.west) ([xshift=1.1cm,yshift=1.1cm]move.east) ([xshift=1.1cm,yshift=-1.1cm]sample.east)};

    \begin{scope}[scale=0.1, left of=collide, yshift=-20cm, xshift=30cm]
        \def\Ho{20};

        \def\L{600 / \Ho};  
        \def\H{200 / \Ho};  

        \def \off{20};
        \def \dt{0.8};
        \def \pw{10};
        
        \coordinate (c1) at ({0}, {0});
        \coordinate (c2) at ({\L}, {0});
        \coordinate (c3) at ($(c1)!0.5!(c2)$);
        \coordinate (c4) at ({0}, {\H});
        \coordinate (c5) at ({\L}, {\H});
        \coordinate (c6) at ($(c4)!0.3!(c5)$);
        \coordinate (c7) at ([yshift=40*\H]$(c4)!0.9!(c5)$); 
        \coordinate (c8) at ([yshift=-35*\H]$(c1)!0.1!(c2)$); 

        \draw[] (c1) -- (c2);
        \draw[] (c4) -- (c5);
        \draw[blue!80, <->, thick, shorten >= -1.5cm, shorten <= -1.5cm, >=stealth] (c3) -- (c6);
        \draw[dashed] (c6) -- (c7);
        \draw[dashed] (c3) -- (c8);

        \draw[fill=black] (c3) circle (0.5cm) node[anchor=south west] {scattering center};
        
        \draw [line, draw=black!50, thick] plot [smooth, tension=1] coordinates { ([xshift=5cm]c6) ([xshift=6cm,yshift=2cm]c6) ([xshift=4cm,yshift=3.5cm]c6)} node[anchor=west, xshift=0.2cm] {$\chi$};

        \draw[line width=\dt, <->, shorten >=2pt, shorten <=2pt,  black!60] (c1) -- (c4) node[anchor=west,pos=0.5] {$b$};
        
        \draw[fill=blue!60] (c4) circle (1cm);
        \draw[line width=2*\dt, ->, shorten >=2pt, shorten <=2pt,  blue!60] (c4) -- ([xshift=5cm]c4);

        \draw[fill=blue!60] (c2) circle (1cm);
        \draw[line width=2*\dt, ->, shorten >=2pt, shorten <=2pt,  blue!60] (c2) -- ([xshift=-5cm]c2);
        
        \draw [line, draw=black!50, thick] plot [smooth, tension=0.3] coordinates { ([yshift=2cm]c4) ([xshift=-1cm,yshift=2cm]c6) ([xshift=-1cm,yshift=1cm]c7)}; 
        \draw[] ([yshift=2cm]c4) node[anchor=south] {$V$};
        \draw[] ([xshift=-1cm,yshift=1cm]c7) node[anchor=south] {$V'$};

        \draw [line, draw=black!50, thick] plot [smooth, tension=0.3] coordinates { ([yshift=-2cm]c2) ([xshift=1cm,yshift=-2cm]c3) ([xshift=1cm,yshift=-1cm]c8)}; 
        \draw[] ([yshift=-2cm]c2) node[anchor=north] {$V_*$};
        \draw[] ([xshift=1cm,yshift=-1cm]c8) node[anchor=north] {$V_*'$};
        
        \draw [line, draw=black!50, thick] plot [smooth, tension=1] coordinates { ([xshift=-5cm]c3) ([xshift=-6cm,yshift=-2cm]c3) ([xshift=-4cm,yshift=-3.5cm]c3)} node[anchor=east, xshift=-0.2cm] {$\chi$};
    \end{scope}
    
    \draw [line, xshift=4cm, draw=green, dashed, thick, postaction={decorate,decoration={raise=1ex,text along path,text align=center,text={|\sffamily|}}}] plot [smooth, tension=1] coordinates { (collide) ([xshift=2cm,yshift=-1cm]collide) ([xshift=4cm,yshift=-0.5cm]collide.east)};

\end{tikzpicture}
\caption{Flowchart of the DSMC process.\label{fig_dsmcAlgorithm}}
\end{figure*}

From a stochastic viewpoint, DSMC, as introduced by Bird\cite{bird1963approach,Bird}, incorporates four principal steps: a) index, b) move, c) collide, and d) sample. The flowchart of a standard DSMC algorithm is illustrated in Fig.~\ref{fig_dsmcAlgorithm}. DSMC uses a spatial grid to contain the simulated molecules and perform sampling. The algorithm starts with the distribution of molecules in the spatial domain according to the pre-specified initial condition: bulk velocity $u_{ini}$, temperature $T_{ini}$, and number densities $n_{ini}^\I$. The individual molecules must be tracked, and therefore an \text{indexing} mechanism is used to track which molecules are in which cell of the spatial domain. Repeated calling of the index subroutine is necessitated by molecular movement. 

The \textit{move} subroutine moves each molecule according to their velocity a distance appropriate for the specified time step. This velocity is assumed constant over each time step. The velocity of a molecule is changed either by external forces, such as electrostatic forces, or by the scattering resulting from a collision. In the absence of external forces, the velocity will only change as a result of a collision. In the present work, we consider elastic collisions i.e., collisions in which the total kinetic energy is unchanged. 

The \textit{collide} subroutine randomly selects a pair of molecules to collide using the acceptance-rejection method\cite{Bird,weaver2015assessment}. The collision is accepted with a probability
\begin{equation}
    P = \frac{|v-v_*|\,\Sigma_{ij}}{(|v-v_*|\,\Sigma_{ij})_\text{max}}
\end{equation}
where $(|v-v_*|\,\Sigma_{ij})_\text{max}$ is the maximum, effective volume swept out by a molecule. This maximum value is recorded in each cell such that each cell may have a separate collision frequency. Taking this into account, the goal of the \textit{collide} is to determine scattering angles and post-collision energies, as well as to obtain correct collision frequencies and microscopic properties. 

Finally, \textit{sample} performs sampling over all cells to determine macroscopic properties. The microscopic properties from each simulated molecule, for instance, molecular velocities and translational energy, are averaged in each cell to compute the macroscopic properties such as density, bulk velocity, pressure, translational temperature, etc. For example, in a given \textit{cell}, consider $M^\I_{cell}$ particles of species $i$, the mass density of the species is simply the total mass per unit cell volume i.e.,
\begin{equation}
    \rho^\I_{cell} = \frac{1}{V_{cell}} \sum_{j=1}^{M^\I_{cell}} m^\I_{j,\,cell}
\end{equation}
Details about sampling of other macroscopic properties, specifically in the stochastic context, can be found in Ref.~\onlinecite{Bird}.



\subsection{Deterministic modelling}

From a deterministic viewpoint, we use the recently introduced discontinuous Galerkin fast spectral (DGFS) method \cite{JAH19,jaiswal2019dgfsMulti}. DGFS directly approximates the Boltzmann equation (\ref{MBE}), where the transport term (spatial derivative) is discretized by the classical DG method and the collision term (integral in $v$) is discretized by the fast Fourier spectral method \cite{GHHH17,jaiswal2019dgfsMulti}. The discretized system is then advanced in time using the Runge-Kutta method. 

The coupling of two kinds of methods (DG in the physical space and spectral method in the velocity space) is possible due to the special structure of the Boltzmann equation -- the collision operator acts only in $v$ wherein $t$ and $x$ can be treated as parameters. Simply speaking, given the distribution functions $f^\I$ and $f^\J$ of species $\i$ and $\j$ at $N^3$ velocity grid, the fast Fourier spectral method produces $\mathcal{Q}^\IJ(f^\I,f^\J)$ at the same grid with $O(M N_\rho N^3\log N)$ complexity, where $M \ll N^2$ is the number of discretization points on the sphere and $N_\rho \sim O(N)$ is the number of Gauss-Legendre quadrature/discretization points in the radial direction needed for low-rank decomposition. Further details can be found in Refs.~\onlinecite{JAH19,jaiswal2019dgfsMulti}. 

The overall DGFS method is simple from mathematical and implementation perspective; highly accurate in both physical and velocity spaces as well as time; robust, i.e. applicable for general geometry and spatial mesh; exhibits nearly linear parallel scaling; and directly applies to general collision kernels needed for high fidelity modelling. Due to these features, we use DGFS for deterministic modelling of flows considered in this work. 

\section{BGK/ESBGK/S-model/DGFS: Verifications}
\label{sec_verifications}
Due to the non-linearity and complexity of Boltzmann collision term $\mathcal{Q}$, the collision operator is often simplified for practical reasons---a major motivation behind the development of kinetic models\cite{bhatnagar1954model,holway1966new,shakhov1968generalization,xu2010unified,AAP02,mieussens2004numerical}. In this section, we shall restrict our discussion to single-species system i.e., $s=1,\,i=\{1\}$. We will drop superscripts $^{(i)}$ and $^{(ij)}$ for simplicity.

While devising kinetic models for single species system, the collision term $\mathcal{S}$---in this work we denote kinetic models by symbol $\mathcal{S}$ to differentiate it from the full Boltzmann collision integral $\mathcal{Q}$---is expected to have the following four properties\cite{mieussens2000discrete,mieussens2004numerical,Cercignani}:
\begin{enumerate}
\item It guarantees the conservation of mass, momentum, and energy, i.e.,
\begin{align} \label{eq_kinetic_momCons}
\int_{\mathbb{R}^3} \mathcal{S}\,\rd{v} = \int_{\mathbb{R}^3} v\,\mathcal{S}\,\rd{v} =\int_{\mathbb{R}^3} v^2\, \mathcal{S}\,\rd{v} = 0.
\end{align}
\item The entropy production is always positive, i.e.,
\begin{align} \label{eq_kinetic_entropy}
-\int_{\mathbb{R}^3} \ln(f)\,\mathcal{S}\,\rd{v} \geq 0
\end{align}
\item Due to specific form of $\mathcal{S}$, the phase density in equilibrium is a Maxwellian i.e., 
\begin{align} \label{eq_kinetic_maxwellRelax}
\mathcal{S} = 0 \Leftrightarrow \int_{\mathbb{R}^3} \ln(f)\,\mathcal{S}\,\rd{v}=0 \Leftrightarrow f = \mathcal{M} 
\end{align}
where 
\begin{align} \label{eq_maxwellian}
\mathcal{M}=\frac{n}{(2 \pi R T)^{3/2}} \exp\Big( -\frac{(v-u)^2}{2 R T} \Big)
\end{align}
\item The Prandtl number is close to $2/3$ for monoatomic gases, i.e., 
\begin{align} \label{eq_kinetic_prandtl}
\mathrm{Pr} = \frac{5}{2}\frac{k_B}{m}\frac{\mu}{\kappa}, 
\end{align}
where $\mu$ and $\kappa$, respectively, refer to dynamic viscosity and thermal conductivity.
\end{enumerate}

Among popular kinetic models, BGK/ESBGK collision operators are relaxation type kernels given as:
\begin{align}
    \mathcal{S} = \nu \; (f_\gamma - f)
\end{align}
where $f_\gamma$ is the local equilibrium function, and $\nu=\mathrm{Pr}\;p/\mu$ is collision frequency. Here $p$ denotes pressure. For BGK, $f_\gamma$ is a local Maxwellian given as 
\begin{align}
    f_\gamma^\text{BGK} = \mathcal{M}
\end{align}
whereas for ESBGK, $f_\gamma$ is anisotropic Gaussian given as 
\begin{align}
    f_\gamma^\text{ESBGK} &= \frac{n}{\sqrt{\det(2\pi \mathbb{T})}} \exp\Big(-\frac{1}{2}\,(v-u)^T\,\mathbb{T}^{-1}\,(v-u)\Big) \nonumber\\
    \rho \mathbb{T} &= \frac{1}{\mathrm{Pr}}\rho R T \,\mathrm{Id} + \Big(1-\frac{1}{\mathrm{Pr}}\Big) \rho \ominus \nonumber \\
    \rho \ominus &= \int_{\mathbb{R}^3} c \otimes c \, f\, \rd{v}, \quad \rho R T = \int_{\mathbb{R}^3} c \otimes c \, f_\gamma^\text{BGK}\, \rd{v}
\end{align}
where $\mathrm{Id}$ is an identity matrix. For S-model, $f_\gamma$ is given as 
\begin{align}
    f_\gamma^\text{S-model} &= f_\gamma^\text{BGK} \Big[ 1 + \frac{1-\mathrm{Pr}}{5} \frac{S \,c}{pRT}\Big(\frac{c^2}{2RT} - \frac{5}{2}\Big) \Big] \nonumber\\
    S &= \int_{\mathbb{R}^3} c\, c^2 \, f\, \rd{v}
\end{align}
It can be easily shown that BGK, ESBGK, and S-model satisfy the  conditions (\ref{eq_kinetic_momCons}, \ref{eq_kinetic_entropy}, \ref{eq_kinetic_maxwellRelax}). S-model, and ESBGK satisfy (\ref{eq_kinetic_prandtl}), whereas BGK doesn't.

To put things more concretely, we consider four Fourier-Couette flow cases, and a flow over a micro-electronic chip to verify BGK/ESBGK/S-model/DGFS method against DSMC. 

\subsection{Verification: Fourier-Couette flows}
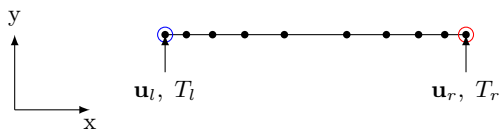
\begin{figure}[!ht]
	\centering
    \begin{tikzpicture}
       \def\pr{1.23};
       \foreach \x in {1,...,5}
         \fill ({-\pr + (\pr)^\x},0cm) circle (0.05cm);

       \foreach \x in {1,...,5}
         \fill ({ 4 + \pr - (\pr)^(\x)},0cm) circle (0.05cm);

       \draw(0,0) -- (4,0) ;
       \draw[blue] (0.0,0.0) circle (0.1cm);
       \draw[red] (4.0,0.0) circle (0.1cm);

       \draw[-latex] (0.0,-0.5) -- (0.0, 0.0) node[below, yshift=-0.5cm] {$\mathbf{u}_l,\;T_l$};
       \draw[-latex] (4.0,-0.5) -- (4.0, 0.0) node[below, yshift=-0.5cm] {$\mathbf{u}_r,\;T_r$};

       \draw[-latex] (-2.0,-1) -- (-1.0, -1) node[anchor=north] {x};
       \draw[-latex] (-2.0,-1) -- (-2.0, 0) node[anchor=south] {y};
    \end{tikzpicture}
	\caption{Numerical setup for 1D Fourier-Couette flow.}
	\label{fig_fourierCouetteFlowSchematic}
\end{figure}

\begin{figure}[!ht]

\begin{subfigure}[b]{.5\textwidth}
  \centering
  \includegraphics[width=68mm]{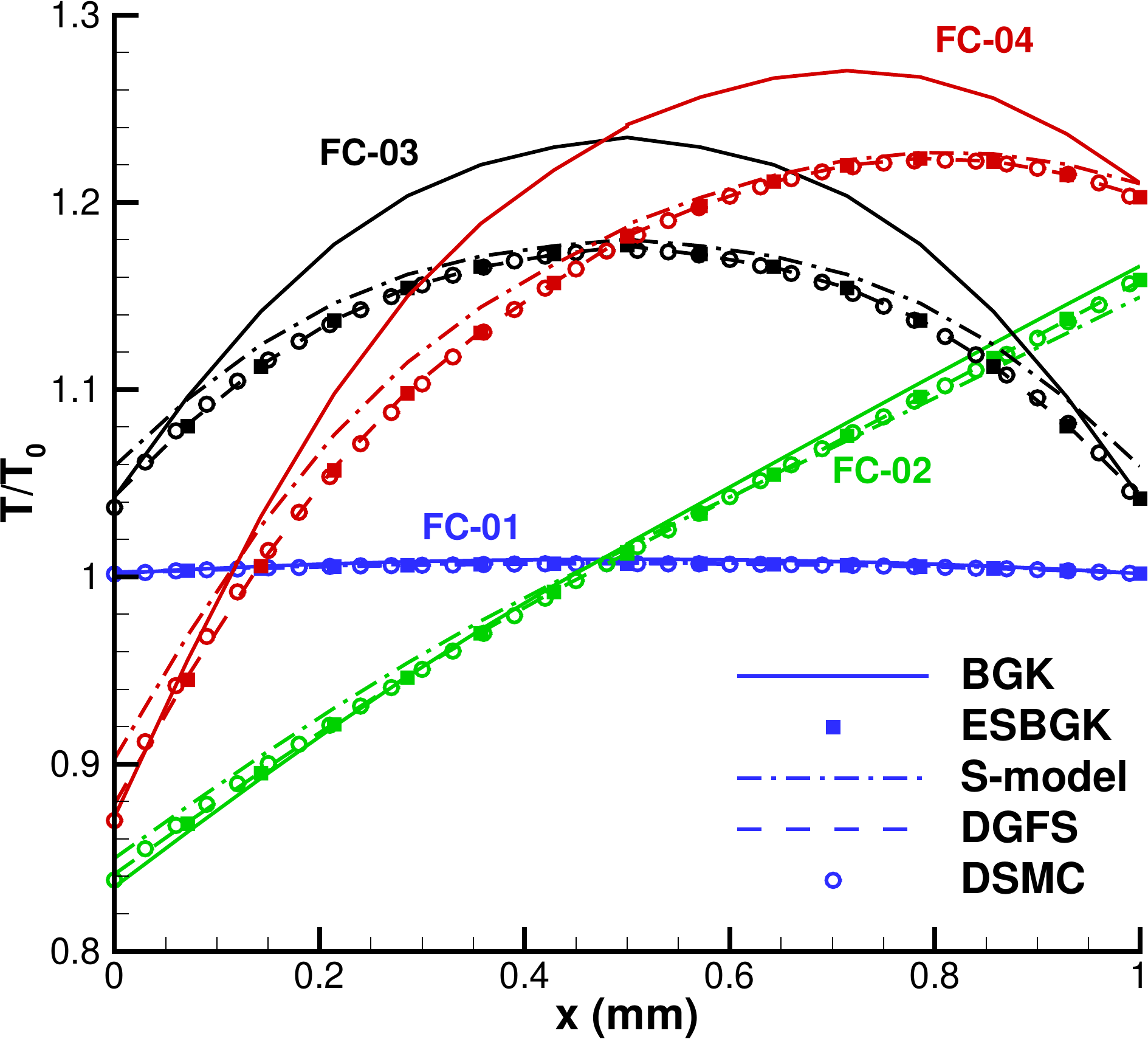}
  \caption{normalized temperature: Cases FC-$0\{1,2,3,4\}$}
\end{subfigure}%

\begin{subfigure}[b]{.5\textwidth}
  \centering
  \includegraphics[width=68mm]{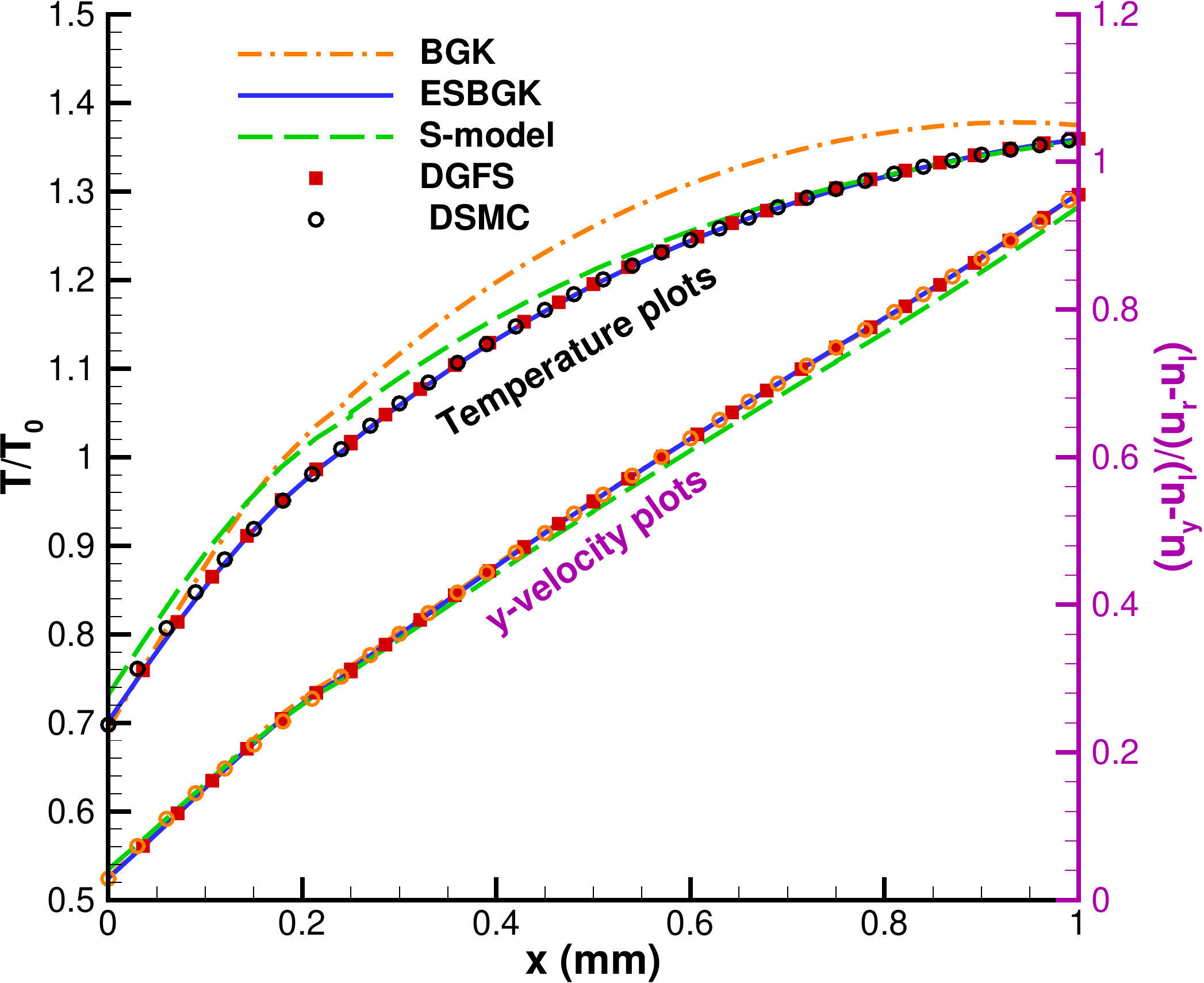}
  \caption{normalized $y$-velocity, and temperature for case FC-05}
\end{subfigure}
\caption{Variation of flow properties along the domain for Fourier-Couette flow cases obtained with BGK, DGFS, and DSMC for Argon.}
\label{fig_fourierCouetteVHS_U_T}
\end{figure}

In the current test case, we consider the effect of velocity and temperature gradients on the solution. The coordinates are chosen such that the walls are parallel to the $y$ direction and $x$ is the direction perpendicular to the walls. The geometry as well as boundary conditions are shown in Figure~\ref{fig_fourierCouetteFlowSchematic}. Specific case details have been provided in Tabs.~\ref{tab_fouriercouette_conditions} and \ref{tab_fouriercouette_cases}. Figure~\ref{fig_fourierCouetteVHS_U_T} illustrates the velocity and temperature along the domain length, wherein we note an excellent agreement between DGFS and DSMC. The velocity profiles from BGK/ESBGK are in good agreement with DGFS and DSMC, whereas the temperature profiles from ESBGK are in good agreement with DGFS and DSMC. The deviation in BGK temperature profiles is due to its Prandtl number defect. 

\begin{table}[!ht]
\centering
\begin{ruledtabular}
\begin{tabular}{@{}lcccc@{}}
\multicolumn{2}{l}{Common Parameters} \\ 
\hline
Molecular mass: $m_1$ ($\times 10^{27}\,kg$) & $66.3$ \\ 
Non-dim physical space & $[0,\,1]$ \\ 
Spatial elements & 2 \\
DG order & 3 \\
Time stepping & Euler \\
Viscosity index: $\omega$ & $0.81$ \\
Scattering parameter: $\alpha$ & $1.4$ \\
Ref. diameter: $d_{\text{ref}}$ ($\times 10^{10} m$) & $4.17$ \\
Ref. temperature: $T_{\text{ref}}$ ($K$) & $273$ \\
Ref. viscosity: $\mu_{\text{ref}}$ ($\Pa \cdot s$) & $2.117 \times 10^{-5}$ \\
Characteristic mass: $m_0$ ($\times 10^{27}\,kg$) & $66.3$ \\
Characteristic length: $H_0$ ($mm$) & 1 \\
Characteristic velocity: $u_0$ ($m/s$) & 337.2 \\
Characteristic temperature: $T_0$ ($K$) & 273 \\
Characteristic no. density: $n_0$ ($m^{-3}$) & $3.538 \times 10^{22}$ \\
\hline
\multicolumn{2}{l}{Initial conditions} \\
Velocity: $u$ ($m/s$) & 0 \\
Temperature: $T$ ($K$) & 273 \\
Number density: $n$ ($m^{-3}$) & $3.538\times10^{22}$ \\
Knudsen number\footnotemark[1]: $(\Kn)$ & $0.036$ \\
\end{tabular}
\end{ruledtabular}
\footnotetext[1]{Based on variable hard-sphere definition (see Ref.~\onlinecite{JAH19,Bird})}
\caption{Common Numerical parameters for Fourier-Couette flow.}
\label{tab_fouriercouette_conditions}
\end{table}

\begin{table*}[!ht]
\centering
\begin{ruledtabular}
\begin{tabular}{@{}lccccc@{}}
Parameter & Case FC-01 & Case FC-02 & Case FC-03 & Case FC-04 & Case FC-05 \\
\hline
Non-dim velocity space\footnotemark[1] & $[-5,\,5]^3$ & $[-5,\,5]^3$ & $[-5,\,5]^3$ & $[-5,\,5]^3$ & $[-8,\,8]^3$ \\
$\{N^3,\,N_\rho,\,M\}$\footnotemark[2] & $\{24^3,\,6,\,6\}$ & $\{24^3,\,6,\,6\}$ & $\{24^3,\,6,\,6\}$ & $\{24^3,\,6,\,6\}$ & $\{32^3,\,16,\,6\}$ \\
\hline
\multicolumn{2}{l}{Left wall (purely diffuse) conditions} \\
Velocity: $u_l$ ($m/s$) & $(0,\,-50,\,0)$ & $(0,\,-50,\,0)$ & $(0,\,-250,\,0)$ & $(0,\,-250,\,0)$ & $(0,\,-250,\,0)$ \\
Temperature: $T_l$ ($K$) & 273 & 223 & 273 & 223 & 173\\
\hline
\multicolumn{2}{l}{Right wall (purely diffuse) conditions} \\
Velocity: $u_r$ ($m/s$) & $(0,\,50,\,0)$ & $(0,\,50,\,0)$ & $(0,\,250,\,0)$ & $(0,\,250,\,0)$ & $(0,\,250,\,0)$ \\
Temperature: $T_r$ ($K$) & 273 & 323 & 273 & 323 & 373\\
\end{tabular}
\end{ruledtabular}
\footnotetext[1]{Non-dimensional (see Refs.~\onlinecite{JAH19,jaiswal2019dgfsMulti} for details on non-dimensionalization)}
\footnotetext[2]{Required only in the fast Fourier spectral low-rank decomposition for DGFS method (see Refs.~\onlinecite{GHHH17,JAH19})}
\caption{Numerical parameters for Fourier-Couette cases.}
\label{tab_fouriercouette_cases}
\end{table*}

\subsection{Verification: Flow around a micro-electronic chip}
\begin{figure*}[!ht]
\begin{subfigure}[b]{.5\textwidth}
  \definecolor{caribbeangreen}{rgb}{0.0, 0.8, 0.6}
\definecolor{verdigris}{rgb}{0.26, 0.7, 0.68}
\definecolor{turquoisegreen}{rgb}{0.63, 0.84, 0.71}
\begin{tikzpicture}[scale=0.5]		
\def\Ho{2};

\def\L{13 / \Ho};  
\def\H{30 / \Ho};  
\def\hG{3 / \Ho};  

\def\Lh{10 / \Ho};  
\def\Hh{20 / \Ho};  

\def \off{20};
\def \dt{0.8};
\def \pw{10};

\coordinate (c1) at ({0}, {0});
\coordinate (c2) at ({\Lh}, {0});
\coordinate (c3) at ({\L}, {0});
\coordinate (c4) at ({\L}, {\hG});
\coordinate (c5) at ({\L}, {\hG+\Hh});
\coordinate (c6) at ({\L}, {\H});
\coordinate (c7) at ({\Lh}, {\H});
\coordinate (c8) at ({0}, {\H});
\coordinate (c9) at ({0}, {\hG+\Hh});
\coordinate (c10) at ({\Lh}, {\hG+\Hh});
\coordinate (c11) at ({\Lh}, {\hG});
\coordinate (c12) at ({0}, {\hG});

\draw (c1) -- (c2);
\draw (c2) -- (c3);
\draw (c3) -- (c4);
\draw (c4) -- (c5);
\draw (c5) -- (c6);
\draw (c6) -- (c7);
\draw (c7) -- (c8);
\draw (c8) -- (c9);
\draw (c9) -- (c10);
\draw (c10) -- (c11);
\draw (c11) -- (c12);
\draw (c12) -- (c1);

\draw[dashed] (c2) -- (c11);
\draw[dashed] (c11) -- (c4);
\draw[dashed] (c10) -- (c5);
\draw[dashed] (c10) -- (c7);

\draw[line width=1.5, blue] (c8) node[anchor=north east, xshift=-0.5*\off] {symmetry} -- (c9);
\draw[line width=\dt, <->, shorten >=2pt, shorten <=2pt, blue] ([xshift=-\off]c8) -- ([xshift=-\off]c9) node[anchor=east, pos=0.5] {$7\,\mu m$};

\draw[line width=1.5, blue] (c12) node[anchor=north east] {symmetry} -- (c1);
\draw[line width=\dt, <->, shorten >=2pt, shorten <=2pt,  blue!60] ([xshift=\off]c1) -- ([xshift=\off]c12) node[anchor=south, pos=-0.0, yshift=0.1*\off, xshift=\off, fill=none, rounded corners=2pt] {$3\,\mu m$};

\draw[line width=1.5, blue] (c3) -- (c4) -- (c5) node[anchor=south west, pos=0.5, xshift=0.5*\off] {symmetry} -- (c6);
\draw[line width=\dt, <->, shorten >=2pt, shorten <=2pt, blue] ([xshift=\off]c3) -- ([xshift=\off]c6) node[anchor=west, pos=0.52] {$30\,\mu m$};

\draw[line width=1.5, orange] (c6) -- (c7) node[anchor=south east, pos=0.5, xshift=0.5*\off, yshift=\off] {diffuse wall at $T_h=296.5\,K$} -- (c8);
\draw[line width=\dt, <->, shorten >=2pt, shorten <=2pt, orange] ([yshift=\off]c6) -- ([yshift=\off]c8) node[anchor=south west, pos=0.5] {$13\,\mu m$};

\draw[line width=1.5, magenta] (c1) -- (c2) node[anchor=north east, pos=1.25, yshift=-\off] {diffuse wall at $T_c=295.5\,K$} -- (c3);
\draw[line width=\dt, <->, shorten >=2pt, shorten <=2pt, magenta] ([yshift=-\off]c1) -- ([yshift=-\off]c3) node[anchor=north west, pos=0.5] {$13\,\mu m$};

\draw[pattern=north west lines, pattern color=black!50, line width = 0.1mm, thin, draw=none] (c12) -- (c11) -- (c10) -- (c9) -- (c12);
\draw[line width=1.5, green] (c12) -- (c11) node[anchor=south east, fill=white, rounded corners=2pt, yshift=2*\off] {substrate, $T_s=296K$} -- (c10) -- (c9);

\draw[line width=\dt, <->, shorten >=2pt, shorten <=2pt,  black!60] ([yshift=4*\off]c10) -- ([yshift=4*\off]c5) node[anchor=south, pos=0.3, yshift=-0.8*\off, fill=white, rounded corners=2pt] {$3\,\mu m$};

\draw[-latex, line width=0.5mm, magenta] ([xshift=-6*\off,yshift=-2*\off]c1) -- ([xshift=-3*\off,yshift=-2*\off]c1) node[anchor=south] {$x$};

\draw[-latex, line width=0.5mm, magenta] ([xshift=-6*\off,yshift=-2*\off]c1) -- ([xshift=-6*\off,yshift=1*\off]c1) node[anchor=east] {$y$};

\end{tikzpicture}
  \caption{Schematic}
\end{subfigure}%
\begin{subfigure}[b]{.5\textwidth}
  \centering
  \includegraphics[width=38mm]{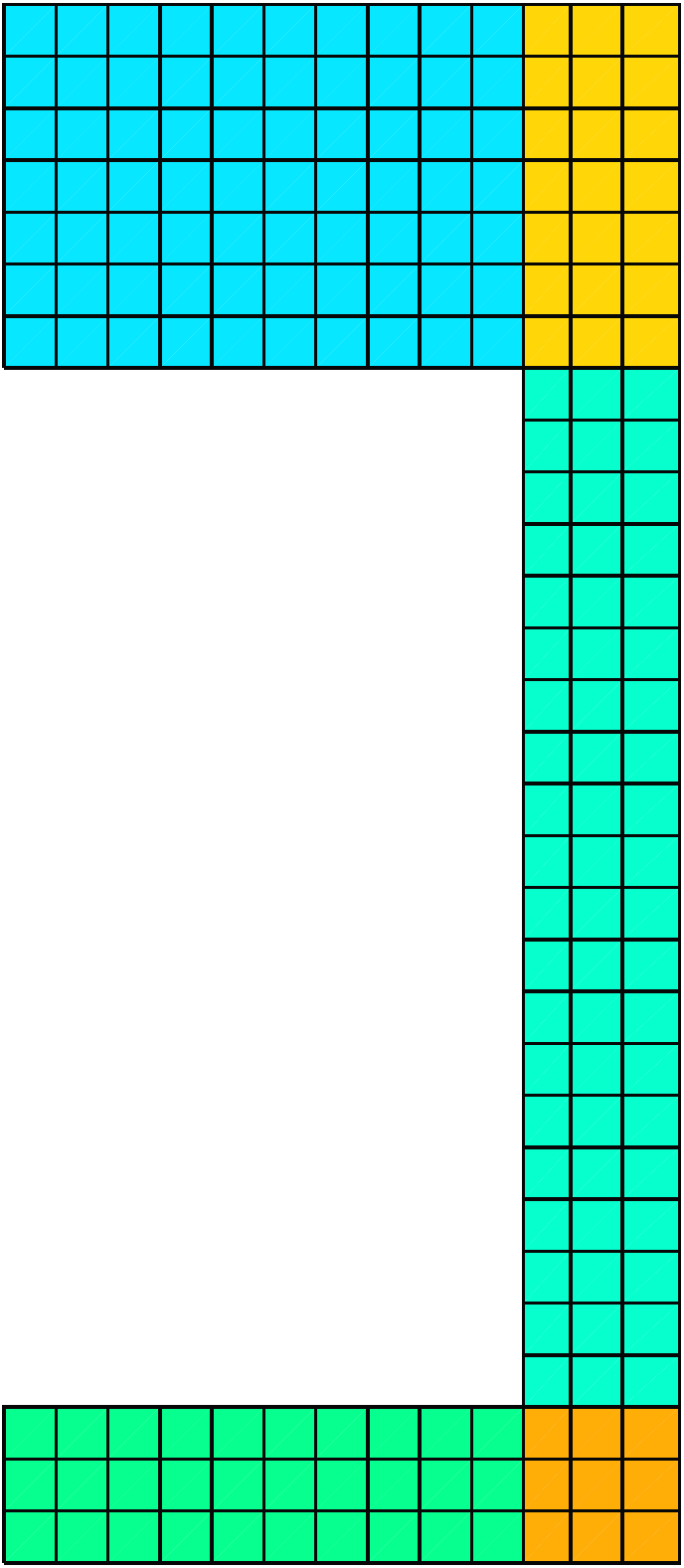}
  \caption{Mesh for DGFS simulations. For DSMC simulations, we subdivide each cell of the mesh above into $5 \times 5$ sub-cells.}
\end{subfigure}%
\caption{Numerical setup for the flow around a micro-electronic chip.}
\label{fig_electronicMicroDevicesSchematic}
\end{figure*}

In the current test case, we consider the effect of temperature gradients on a solid substrate placed in a rarefied environment. The problem schematic, geometry, as well as boundary conditions are shown in Figure~\ref{fig_electronicMicroDevicesSchematic}. Case details have been provided in Tab.~\ref{tab_emd_conditions}. 

\subsubsection{Numerical details}
We employ DSMC and DGFS to carry out simulation of flow around a micro-electronic chip. The simulation specific numerical parameters as well as differences between stochastic (DSMC) and deterministic (DGFS) modelling is described next.

\begin{itemize}
\item \textbf{DSMC}: SPARTA\cite{gallis2014direct} has been employed for carrying out DSMC simulations in the present work. It implements the DSMC method as proposed by Bird \cite{Bird}. The solver takes into account the translational/rotational/vibrational kinetic energies associated with the molecular motion. The solver has been benchmarked \cite{gallis2014direct} and widely used for studying hypersonic, subsonic and thermal \cite{gallis2017molecular, gallis2016direct, sebastiao2018direct,jaiswal2018femta,jaiswal2018dsmc,JAH19,jaiswal2019dgfsMulti,jaiswal2018dgfsGPU} gas flow problems. In this work, cell size less than $\lambda/3$ has been ensured in all the test cases. The no-time collision (NTC) algorithm is used in conjunction with Bird's VHS scattering model. The simulations are first run for 200,000 unsteady steps wherein the particles move, collide, and allowed to equilibrate. No sampling is performed at this stage. Next, the simulation is run for another 4,000,000 steady steps wherein the samples of flow properties namely number density, flow velocity, temperature, stress, and heat-flux, are taken for sufficiently long time so as to produce a meaningful bulk properties as well as minimize the statistical noise therein. In the present case, the DSMC domain is discretized with a uniform cell size of $0.2\,\mu m$, with 300 particles per cell on average during initialization. A time step of $10^{-9}$ sec is used during \textit{move} step of DSMC algorithm throughout the course of simulation. $N_2$ is used as the working gas in simulations. The properties of the working gas is given in Tab.~\ref{tab_props_N2}. We want to emphasize that for DSMC simulations, we take rotational/vibrational degrees of freedom into account i.e., $N_2$ is treated as a diaotomic species. DSMC simulations on 30 cores of Intel(R) Xeon(R) CPU E5-2670 v2 \@ 2.50GHz, took $\sim73$ hours. 

\item \textbf{DGFS}: We use the DGFS implementation described in Ref.~\onlinecite{JAH19}. The spatial domain consists of 281 uniform square cells of $1\,\mu m$ each. Since we are seeking a steady state solution, the time-step is selected based on the CFL constraints of the forward Euler scheme. Other case specific DGFS parameters have been provided in Tab.~\ref{tab_emd_conditions}. Note that, we employ $N_2$ as the working gas in simulations, since MIKRA experiments\cite{strongrich2017microscale} were performed in $N_2$ medium. $N_2$ is diatomic, however, DGFS, as of now, is applicable for monoatomic gases only. Since the working temperature range is low, we anticipate the effects of vibrational degrees of freedom to be negligible. DGFS simulations on 2 Nvidia-P100 GPUs took $\sim9$ hours.
\end{itemize}


\begin{table}[!ht]
\centering
\begin{ruledtabular}
\begin{tabular}{@{}lc@{}}
Mass: $m$ ($kg$) & $46.5 \times 10^{-27}$ \\
Viscosity index: $\omega$ ($-$) & $0.74$ \\ 
Scattering index: $\alpha$ ($-$) & $1.0$ \\
Ref. diameter: $d_{\mathrm{ref}}$ ($m$) & $4.17 \times 10^{-10}$ \\
Ref. temperature: $T_{\mathrm{ref}}$ ($K$) & 273 \\ 
Ref. viscosity: $\mu_{\mathrm{ref}}$ ($\mathrm{Pa}\cdot s$) & $1.656 \times 10^{-5}$ \\ 
\hline 
\multicolumn{2}{l}{DSMC specific parameters} \\
Rotational degrees of freedom: $\zeta_R$ ($-$) & 2 \\
Rotational relaxation: $Z_R$ ($-$) & 2 \\
Vibrational degrees of freedom: $\zeta_V$ ($-$) & 2 \\ 
Vibrational relaxation $Z_V$ ($-$) & $1.90114 \times 10^{-5}$ \\ 
Vibrational temperature $T_V$ ($K$) & 3371 
\end{tabular}
\end{ruledtabular}
\caption{$N_2$ gas VHS parameters used in 2-D single-species DSMC and DGFS simulations. Note that DGFS, being in very early stage of research, treats $N_2$ as a monoatomic species.}
\label{tab_props_N2}
\end{table}

\begin{table}[!ht]
\centering
\begin{ruledtabular}
\begin{tabular}{@{}lcccc@{}}
Parameters & MEC-01 \\ 
\hline
Spatial elements & 190 quadrilaterals \\
DG order & 3 \\
Time stepping & Euler \\
Points in velocity mesh: $N^3$ & $24^3$ \\
Points in radial direction\footnotemark[2]: $N_\rho$ & $6$ \\
Points on \textit{half} sphere\footnotemark[2]: $M$ & $6$ \\
Size of velocity mesh\footnotemark[3] & $[-5,\,5]^3$ \\
Characteristic length: $H_0$ ($\mu m$) & 3 \\
Characteristic velocity: $u_0$ ($m/s$) & 402.54 \\
Characteristic temperature: $T_0$ ($K$) & 273 \\
Characteristic no. density: $n_0$ ($m^{-3}$) & $4.894 \times 10^{23}$ \\
\hline
\multicolumn{2}{l}{Initial conditions} \\
Velocity: $u$ ($m/s$) & 0 \\
Temperature: $T$ ($K$) & 273 \\
Number density: $n$ ($m^{-3}$) & $4.894 \times 10^{23}$ \\
Knudsen number\footnotemark[1]: $(\Kn)$ & $0.88158$ \\
\end{tabular}
\end{ruledtabular}
\footnotetext[1]{Based on variable hard-sphere definition (see Ref.~\onlinecite{JAH19,Bird})}
\footnotetext[2]{Non-dimensional (see Refs.~\onlinecite{JAH19,jaiswal2019dgfsMulti} for details on non-dimensionalization)}
\footnotetext[3]{Required only in the fast Fourier spectral low-rank decomposition for DGFS method (see Refs.~\onlinecite{GHHH17,JAH19})}
\caption{Numerical parameters for flow around micro-electronic chip.}
\label{tab_emd_conditions}
\end{table}

\subsubsection{Results and discussion}
\begin{figure*}[!ht]
\centering
\begin{subfigure}{.5\textwidth}
  \centering
  \includegraphics[width=70mm]{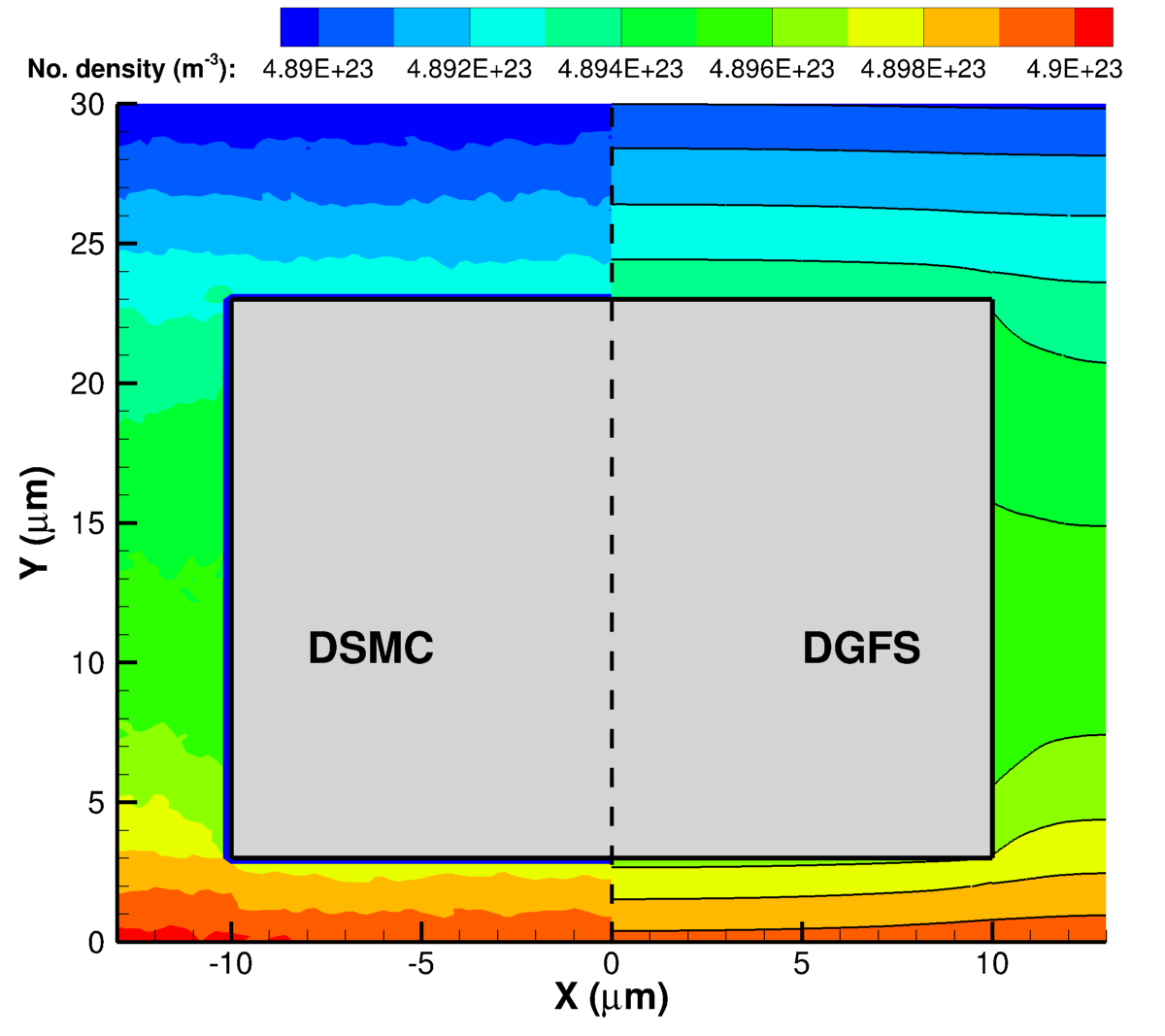}
  \caption{Number density ($m^{-3}$)}
\end{subfigure}%
\begin{subfigure}{.5\textwidth}
  \centering
  \includegraphics[width=70mm]{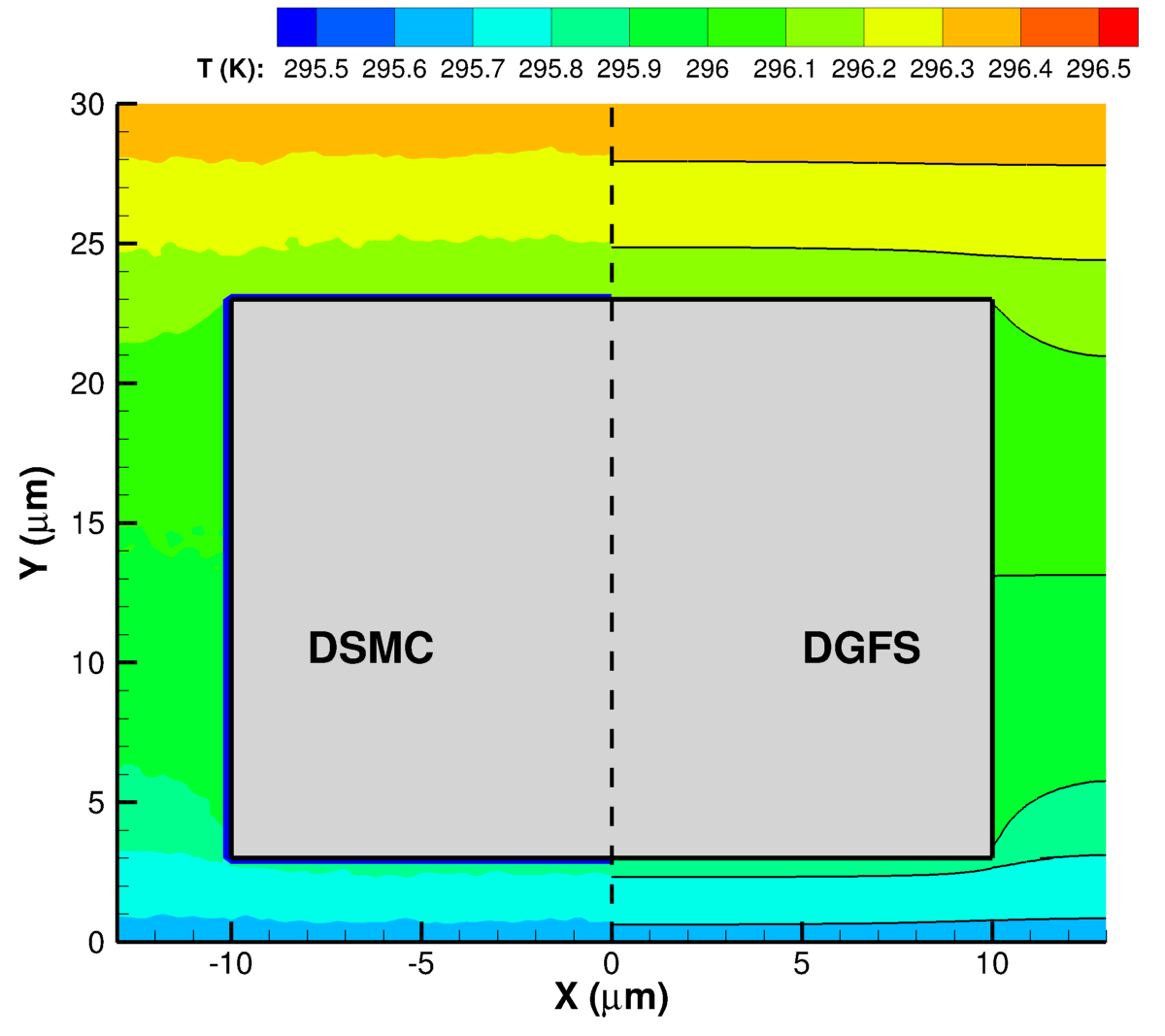}
  \caption{Temperature ($K$)}
\end{subfigure}
\begin{subfigure}{.5\textwidth}
  \centering
  \includegraphics[width=70mm]{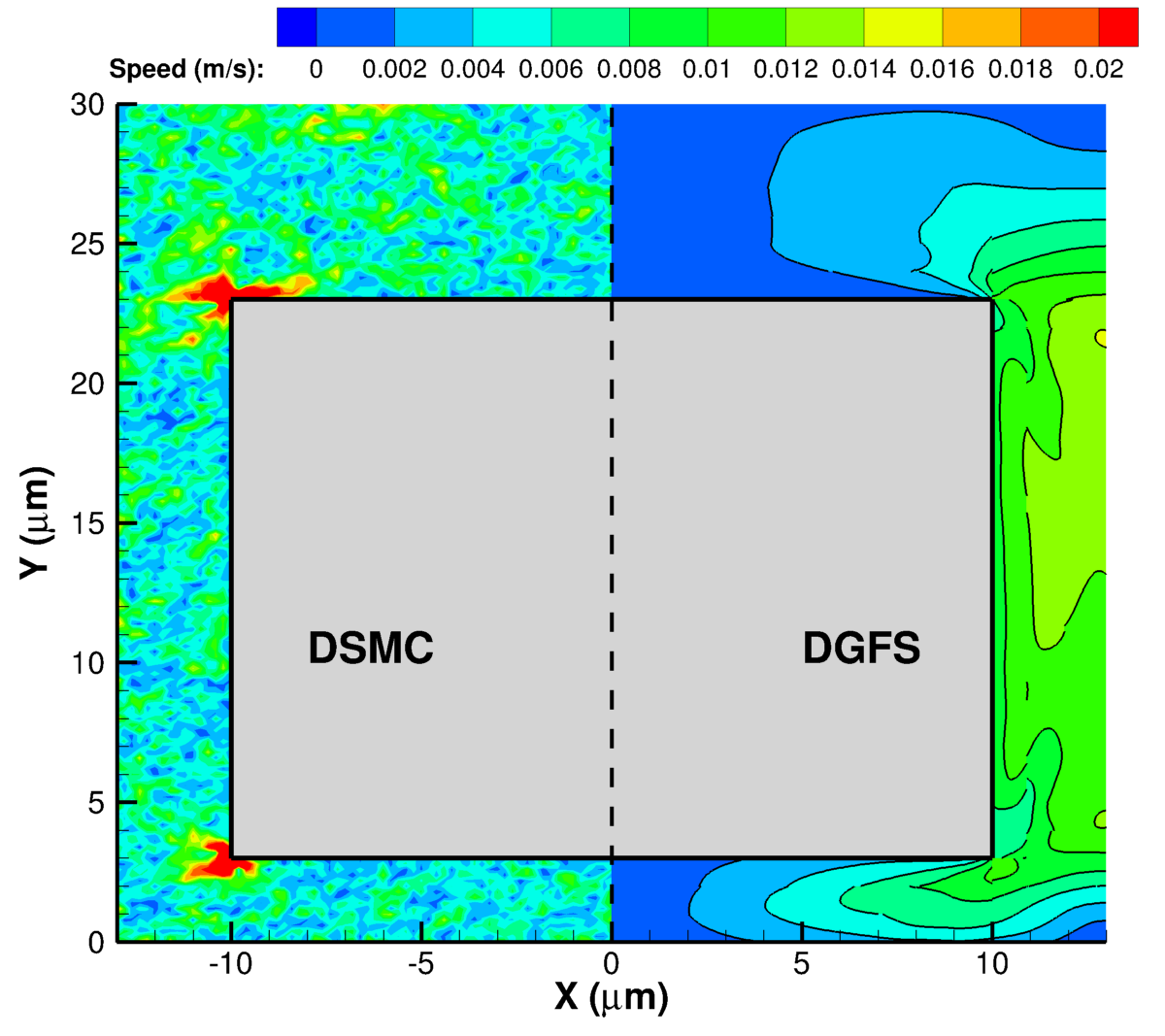}
  \caption{Speed ($m/s$)}
\end{subfigure}%
\begin{subfigure}{.5\textwidth}
  \centering
  \includegraphics[width=70mm]{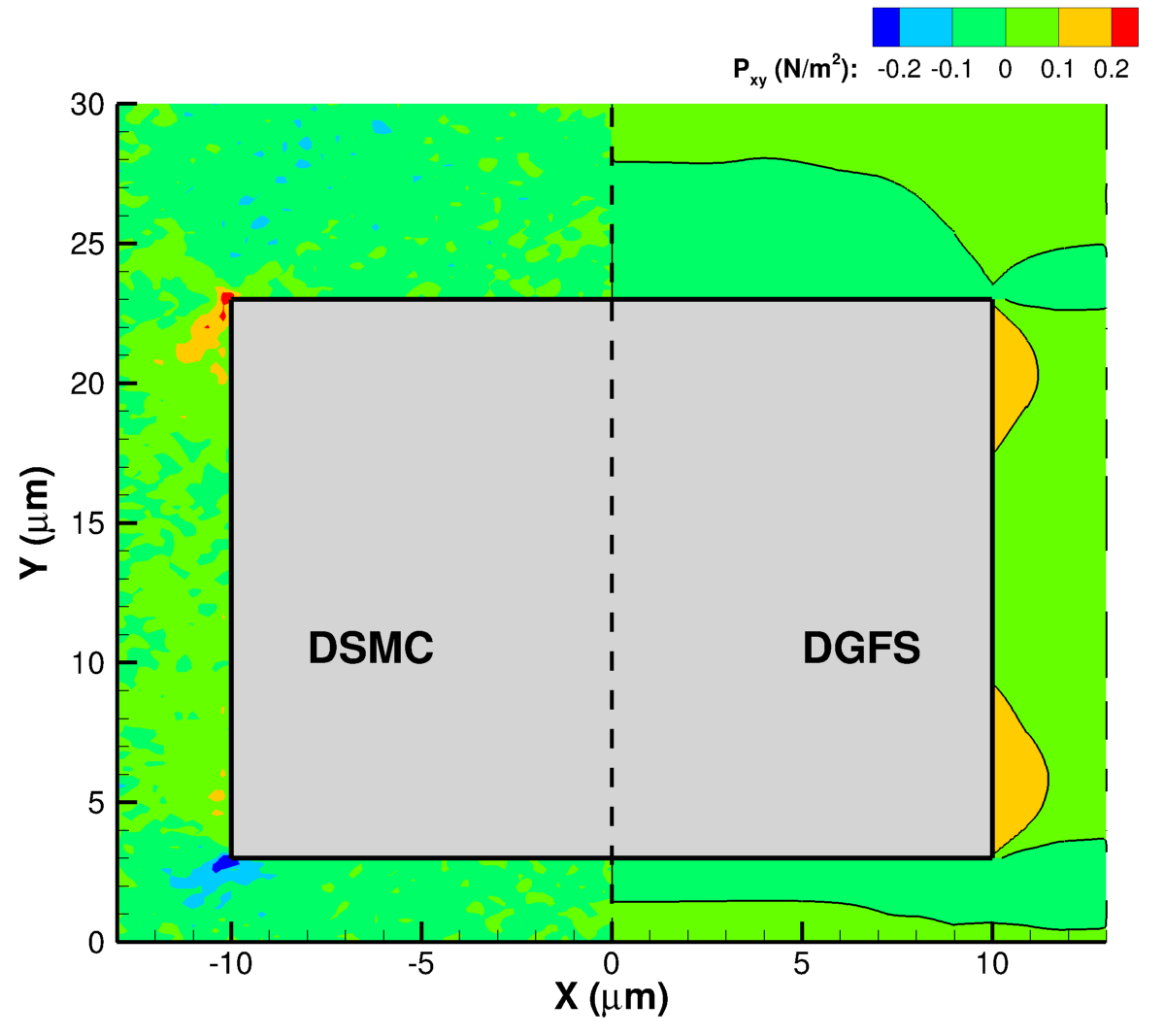}
  \caption{$xy$-component of stress ($N/m^2$)}
\end{subfigure}
\begin{subfigure}{.5\textwidth}
  \centering
  \includegraphics[width=70mm]{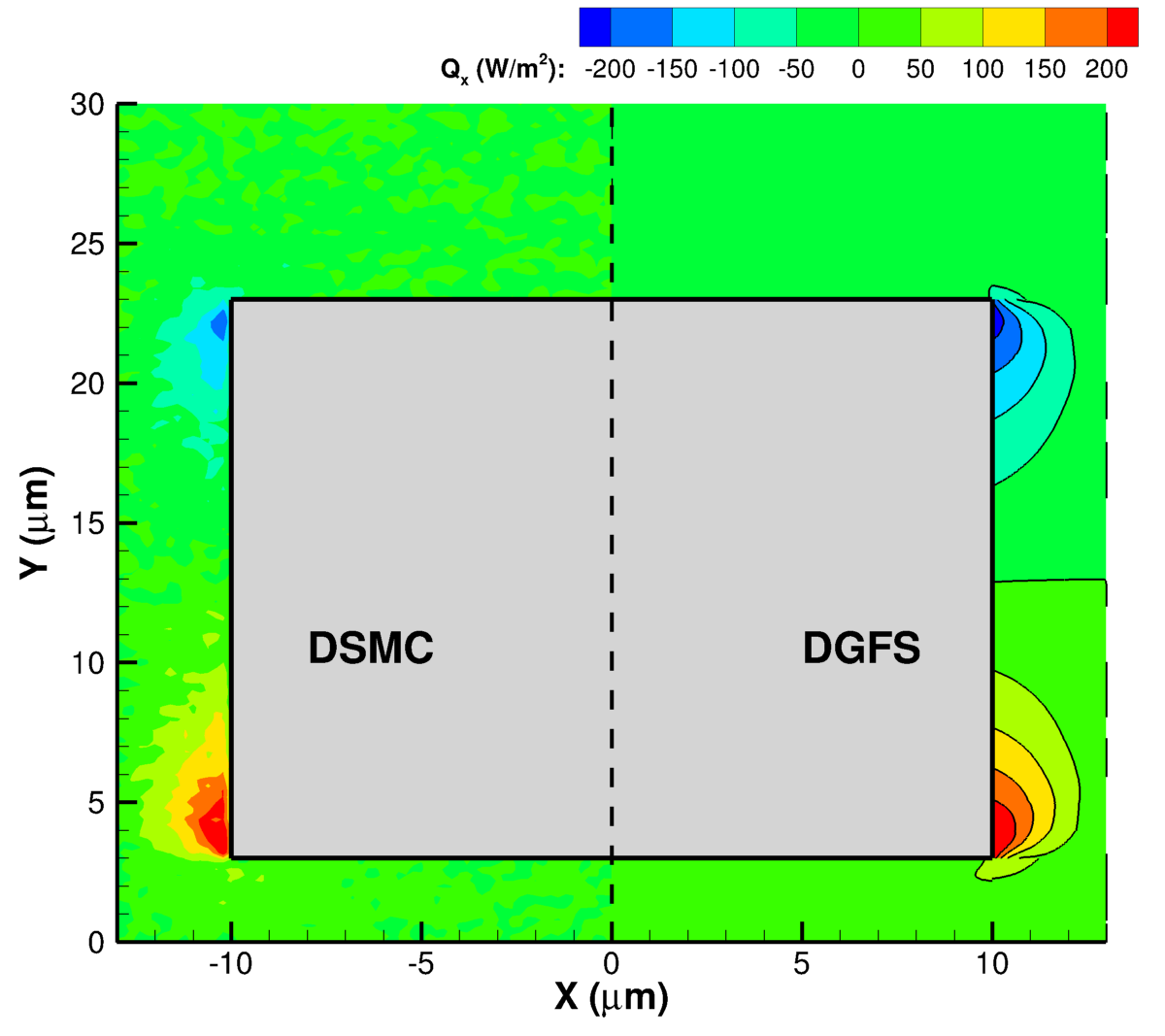}
  \caption{$x$-component of heat-flux ($W/m^2$)}
\end{subfigure}%
\begin{subfigure}{.5\textwidth}
  \centering
  \includegraphics[width=70mm]{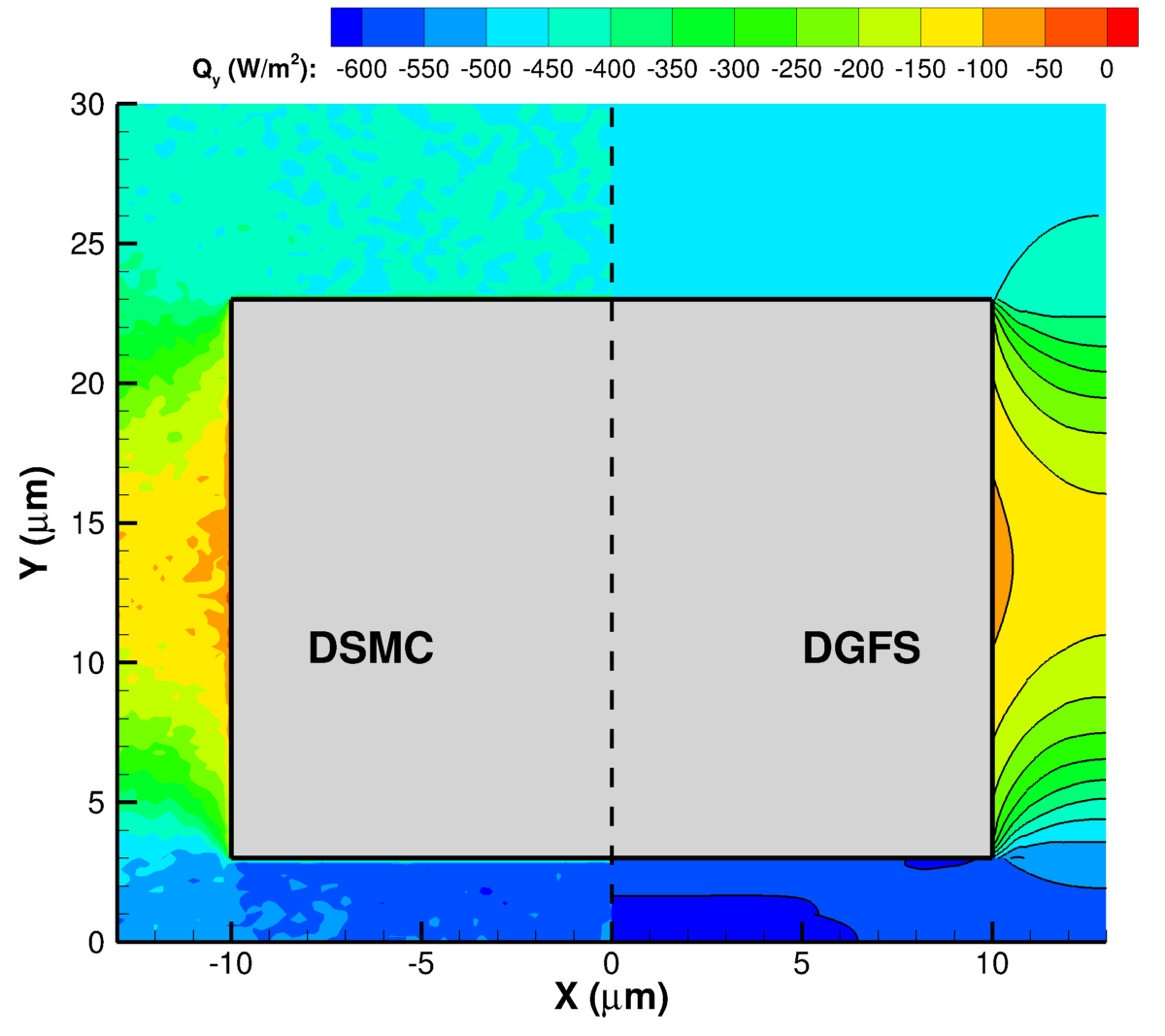}
  \caption{$y$-component of heat-flux ($W/m^2$)}
\end{subfigure}
\caption{Flow properties at steady state for micro-electronic chip obtained from DSMC and DGFS using VHS collision model. For each of these figures, DSMC results (mirrored along y-axis) have been shown in the second quadrant ($-17\,\mu m \leq x < 0\,\mu m$), whereas DGFS results have been illustrated in the first quadrant ($0\,\mu m \leq x < 17\,\mu m$). Observe the legend for number-density.}
\label{fig_emd_props}
\end{figure*}

Figures~\ref{fig_emd_props} illustrate the contours of various flow properties for the flow around the solid chip/substrate. Ignoring the statistical noise, we observe excellent agreement between DSMC and DGFS. In particular, DGFS reproduces noise-free smooth results. 

Next we compute the force acting on the substrate as a result of the temperature gradients initially present in the flow. In general, the pressure force on a surface is given by 
\begin{equation} \label{eq_pressureForce}
F = - \int_{\rd{A}} p\,n\,\rd{A}
\end{equation}
where $n$ is the unit surface normal, $p$ is the pressure on the surface, and $A$ is the area of the surface. 


\begin{table}
\begin{ruledtabular}
\begin{tabular}{@{}cccc@{}}
\multirow{2}{*}{Pressure (Pa)} & \multirow{2}{*}{$\Kn$} & \multicolumn{2}{c}{Force ($\mu N/\mu m$)} \\
\cline{3-4}
& & DSMC & DGFS \\
\hline
2000 &  0.88158 & -0.040008843 & -0.040010413 \\ 
\end{tabular}
\end{ruledtabular}
\caption{$x$-component of force on the substrate for MEC-01 case, obtained using DSMC and DGFS simulations.}
\label{tab_emd_surfaceprops_forces}
\end{table}
Table~\ref{tab_emd_surfaceprops_forces} presents the $x$-component of force on the substrate for the micro-electronic chip verification case. Again, we note reasonable agreement between the values recovered from DSMC and DGFS simulations. 

\subsection{Verification: Flow in short microchannels}
\begin{figure*}[!ht]
\begin{subfigure}{\textwidth}
  \begin{tikzpicture}[scale=1.4]		
\def\H{0.5};
\def\Lin{2};
\def\Lout{2};
\def\Hin{2};
\def\Hout{2};
\def\L{5};
\def \off{12};
\def \pt{3};
\def \dt{0.8};

\coordinate (c1) at ({0}, {0});
\coordinate (c2) at ({\Lin}, {0});
\coordinate (c3) at ({\Lin+\L}, {0});
\coordinate (c4) at ({\Lin+\L+\Lout}, {0});
\coordinate (c5) at ({\Lin +\L+\Lout}, {\H});
\coordinate (c6) at ({\Lin+\L+\Lout}, {\Hout});
\coordinate (c7) at ({\Lin+\L}, {\Hout});
\coordinate (c8) at ({\Lin+\L}, {\H});
\coordinate (c9) at ({\Lin}, {\H});
\coordinate (c10) at ({\Lin}, {\Hin});
\coordinate (c11) at ({0}, {\Hin});
\coordinate (c12) at ({0}, {\H});

\draw (c1) -- (c2);
\draw (c2) -- (c3);
\draw (c3) -- (c4);
\draw (c4) -- (c5);
\draw (c5) -- (c6);
\draw (c6) -- (c7);
\draw (c7) -- (c8);
\draw (c8) -- (c9);
\draw (c9) -- (c10);
\draw (c10) -- (c11);
\draw (c11) -- (c12);
\draw (c12) -- (c1);

\draw[line width=1.5, blue] (c10) -- (c11) node[anchor=south east] {inlet at $600K$} -- (c12) -- (c1);
\draw[line width=\dt, <->, shorten >=2pt, shorten <=2pt, blue] ([xshift=-\off]c11) -- ([xshift=-\off]c1) node[anchor=east, pos=0.5] {$H_{in}=2\,\mu m$};

\draw[line width=1.5, orange] (c1) -- (c2) -- (c3) -- (c4) node[anchor=south east] {symmetry};
\draw[line width=\dt, <->, shorten >=2pt, shorten <=2pt, orange] ([yshift=-\off]c1) -- ([yshift=-\off]c2) node[anchor=north, pos=0.5] {$L_{in}=2\,\mu m$};
\draw[line width=\dt, <->, shorten >=2pt, shorten <=2pt, orange] ([yshift=-\off]c2) -- ([yshift=-\off]c3) node[anchor=north, pos=0.5] {$L=5\,\mu m$};
\draw[line width=\dt, <->, shorten >=2pt, shorten <=2pt, orange] ([yshift=-\off]c3) -- ([yshift=-\off]c4) node[anchor=north, pos=0.5] {$L_{out}=2\,\mu m$};

\draw[line width=1.5, red] (c4) -- (c5) -- (c6) node[anchor=south west] {outlet at $300K$} -- (c7);
\draw[line width=\dt, <->, shorten >=2pt, shorten <=2pt, red] ([xshift=\off]c4) -- ([xshift=\off]c6) node[anchor=west, pos=0.5] {$H_{out}=2\,\mu m$};

\draw[line width=1.5, black] (c7) -- (c8) -- (c9) node[anchor=south east] {walls} -- (c10);
\draw[line width=\dt, <->, shorten >=1p t, shorten <=1pt, black] (c8) -- (c3) node[anchor=east, pos=0.5] {$H/2=0.5\,\mu m$};
\draw[pattern=north west lines, pattern color=black!50, line width = 0.1mm, thin, draw=none] (c7) -- (c8) -- ([xshift=-\pt]c8) -- ([xshift=-\pt]c7);
\draw[pattern=north west lines, pattern color=black!50, line width = 0.1mm, thin, draw=none] (c8) -- (c9) -- ([yshift=\pt]c9) -- ([yshift=\pt]c8);
\draw[pattern=north west lines, pattern color=black!50, line width = 0.1mm, thin, draw=none] (c9) -- (c10) -- ([xshift=\pt]c10) -- ([xshift=\pt]c9);

\draw[line width=1.2, green!50, dashed, postaction={decorate,decoration={text color=green!50,raise=2ex,text along path,text align=center,text={|\sffamily|linearly decreasing temperature along the wall}}}] ([yshift=0]c10) node[anchor=south east] {$600K$} -- ([yshift=0]c8) node[anchor=south west] {$300K$};

\draw[-latex, line width=0.5mm, magenta] (c1) -- ([yshift=3*\off]c1) node[anchor=west] {$x$};

\draw[-latex, line width=0.5mm, magenta] (c1) -- ([xshift=3*\off]c1) node[anchor=south] {$y$};

\end{tikzpicture}
  \caption{Schematic}
\end{subfigure}%

\begin{subfigure}{\textwidth}
  \centering
  \includegraphics[width=0.8\textwidth]{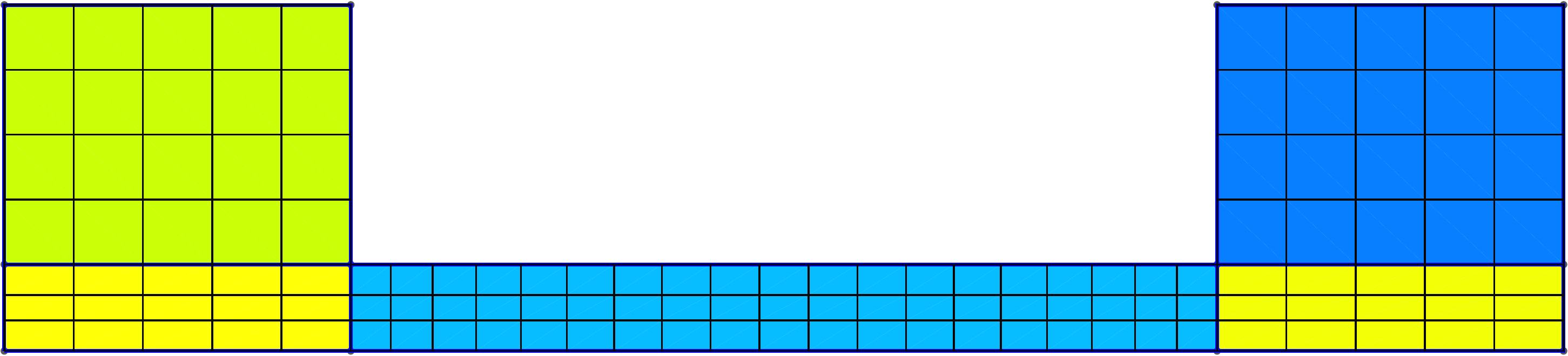}
  \caption{Mesh for DGFS simulations.}
  \label{fig_shortMicroChannelSchematic_mesh}
\end{subfigure}%
\caption{Numerical setup for the flow in short microchannels. On the horizontal channel walls, we impose a linearly decreasing temperature profile similar to case I(a) in Ref.~\onlinecite{alexeenko2006kinetic}.}
\label{fig_shortMicroChannelSchematic}
\end{figure*}

The present test case closely follows case-I(a) from Ref.~\onlinecite{alexeenko2006kinetic}. In the current test case, two reservoirs filled with $N_2$ gas, at different temperatures, are connected by a two-dimensional capillary tube, both with a finite length $L$ and height $H/2$, are considered. The problem schematic, geometry, as well as boundary conditions are shown in Figure~\ref{fig_shortMicroChannelSchematic}. Case details have been provided in Tab.~\ref{tab_shortMicroChannel_conditions}. Note in particular, we introduce a linearly decreasing temperature profile at the top wall. 

\subsubsection{Numerical details}
\begin{itemize}
\item \textbf{DSMC}: The no-time collision (NTC) algorithm is used in conjunction with Bird's VHS scattering model. The simulations are first run for 500,000 unsteady steps wherein the particles move, collide, and allowed to equilibrate. Next, the simulation is run for another 100,000 steady steps wherein the samples of flow properties are taken. In the present case, the DSMC domain is discretized with a uniform cell size of $0.01\,\mu m$, with 30 particles per cell on average during initialization (Note that SPARTA uses hierarchical Cartesian grid over the simulation domain: used to track particles and to co-locate particles in the same grid cell for performing collision and chemistry operations. At the junction, where the walls join the inlet and outlet regions, one can identify two boundary cells. We further refine, specifically, these two boundary cells into $10\times 10$ sub-cells. These two cells are unique i.e., for each of these cells, the top face is marked as inlet, and the left face is marked as solid wall. The cell-size has been made smaller to avoid any potential leakage). A time step of $10^{-10}$ sec is used. $N_2$ is (Tab.~\ref{tab_props_N2}) used as the working gas in simulations.

\item \textbf{DGFS}: The spatial domain consists of 127 non-uniform quadrilateral elements as shown in Fig.~(\ref{fig_shortMicroChannelSchematic_mesh}). Case specific DGFS parameters have been provided in Tab.~\ref{tab_shortMicroChannel_conditions}. 
\end{itemize}

\begin{table}[!ht]
\centering
\begin{ruledtabular}
\begin{tabular}{@{}lcccc@{}}
Parameters & SM-01 \\ 
\hline
Spatial elements & 127 quadrilaterals\\
DG order & 3 \\
Time stepping & Euler \\
Points in velocity mesh: $N^3$ & $32^3$ \\
Points in radial direction\footnotemark[2]: $N_\rho$ & $8$ \\
Points on \textit{half} sphere\footnotemark[2]: $M$ & $6$ \\
Size of velocity mesh\footnotemark[3] & $[-5.72,\,5.72]^3$ \\
Characteristic length: $H_0$ ($\mu m$) & 1 \\
Characteristic velocity: $u_0$ ($m/s$) & 421.98 \\
Characteristic temperature: $T_0$ ($K$) & 300 \\
Characteristic no. density: $n_0$ ($m^{-3}$) & $6.62 \times 10^{24}$ \\
\hline
\multicolumn{2}{l}{Initial conditions} \\
Velocity: $u$ ($m/s$) & 0 \\
Temperature: $T$ ($K$) & 300 \\
Number density: $n$ ($m^{-3}$) & $6.62 \times 10^{24}$ \\
Knudsen number\footnotemark[1]: $(\Kn)$ & $0.2$ \\
\hline
\multicolumn{2}{l}{Inlet condition} \\
Velocity: $u_{in}$ ($m/s$) & 0 \\
Temperature: $T_{in}$ ($K$) & 600 \\
Number density: $n_{in}$ ($m^{-3}$) & $3.31 \times 10^{24}$ \\
Pressure: $p_{in}$ ($N/m$) & $27420$ \\
\hline
\multicolumn{2}{l}{Outlet condition} \\
Velocity: $u_{out}$ ($m/s$) & 0 \\
Temperature: $T_{out}$ ($K$) & 300 \\
Number density: $n_{out}$ ($m^{-3}$) & $6.62 \times 10^{24}$ \\
Pressure: $p_{out}$ ($N/m$) & $27420$ \\
\end{tabular}
\end{ruledtabular}
\footnotetext[1]{Based on variable hard-sphere definition (see Ref.~\onlinecite{JAH19,Bird})}
\footnotetext[2]{Non-dimensional (see Refs.~\onlinecite{JAH19,jaiswal2019dgfsMulti} for details on non-dimensionalization)}
\footnotetext[3]{Required only in the fast Fourier spectral low-rank decomposition for DGFS method (see Refs.~\onlinecite{GHHH17,JAH19})}
\caption{Numerical parameters for flow in short microchannels.}
\label{tab_shortMicroChannel_conditions}
\end{table}

\subsubsection{Results and discussion}
\begin{figure*}[!ht]
\centering
\begin{subfigure}{.5\textwidth}
  \centering
  \includegraphics[width=80mm]{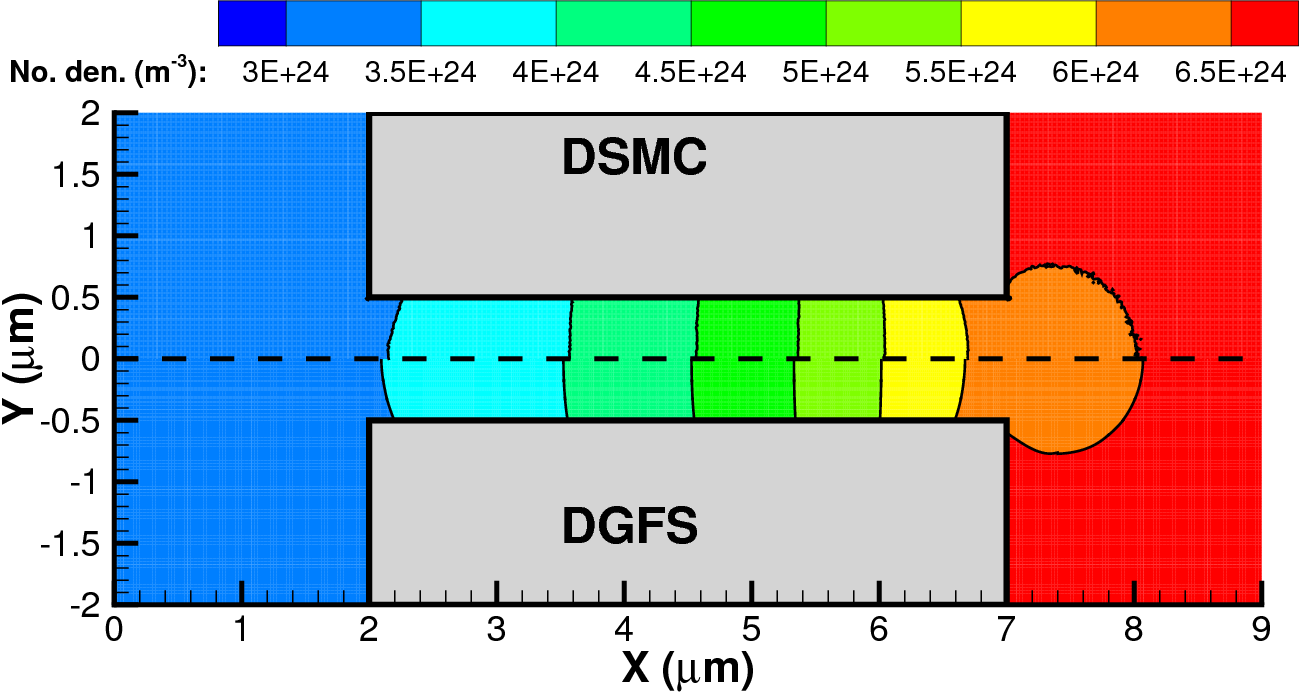}
  \caption{Number density ($m^{-3}$)}
\end{subfigure}%
\begin{subfigure}{.5\textwidth}
  \centering
  \includegraphics[width=80mm]{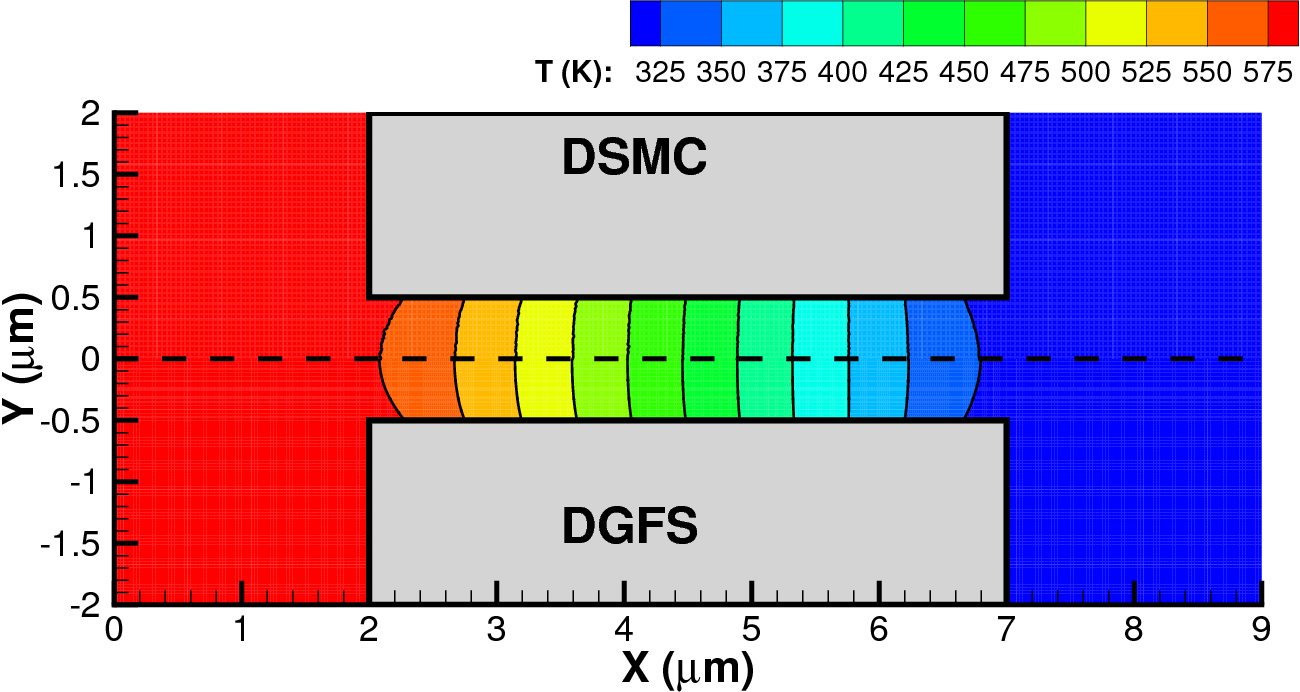}
  \caption{Temperature ($K$)}
\end{subfigure}
\begin{subfigure}{.5\textwidth}
  \centering
  \includegraphics[width=80mm]{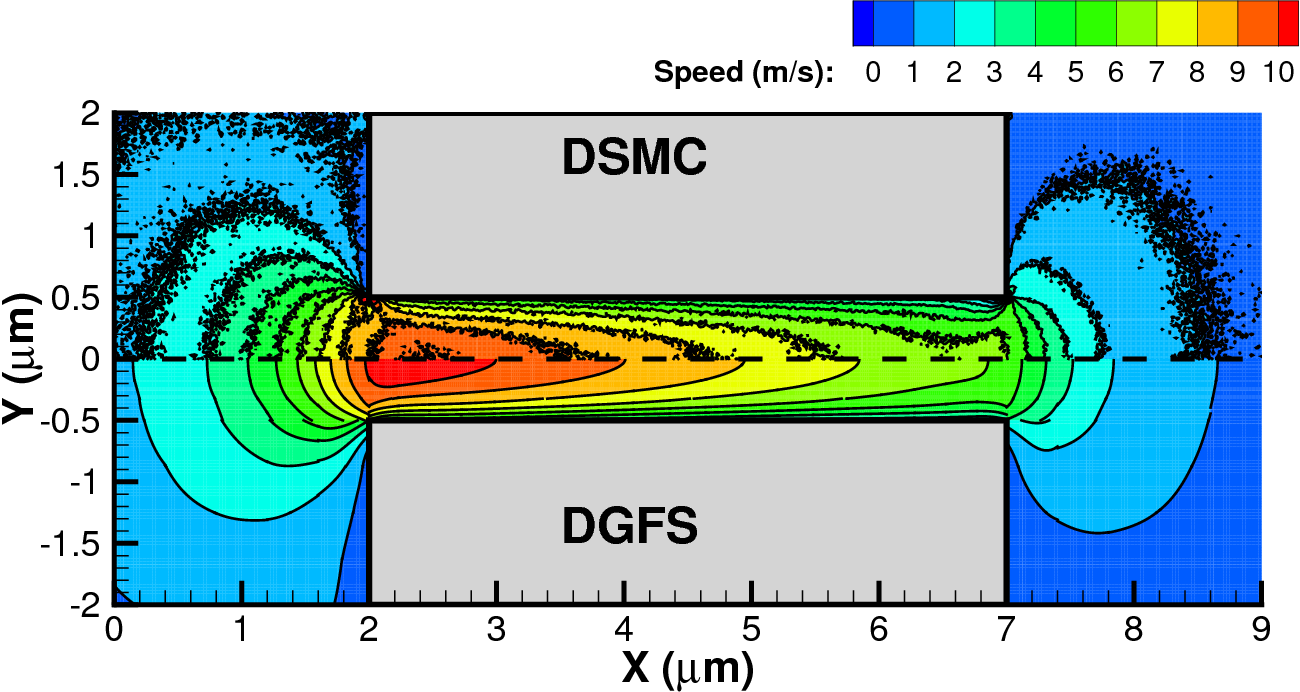}
  \caption{Speed ($m/s$)}
\end{subfigure}%
\begin{subfigure}{.5\textwidth}
  \centering
  \includegraphics[width=80mm]{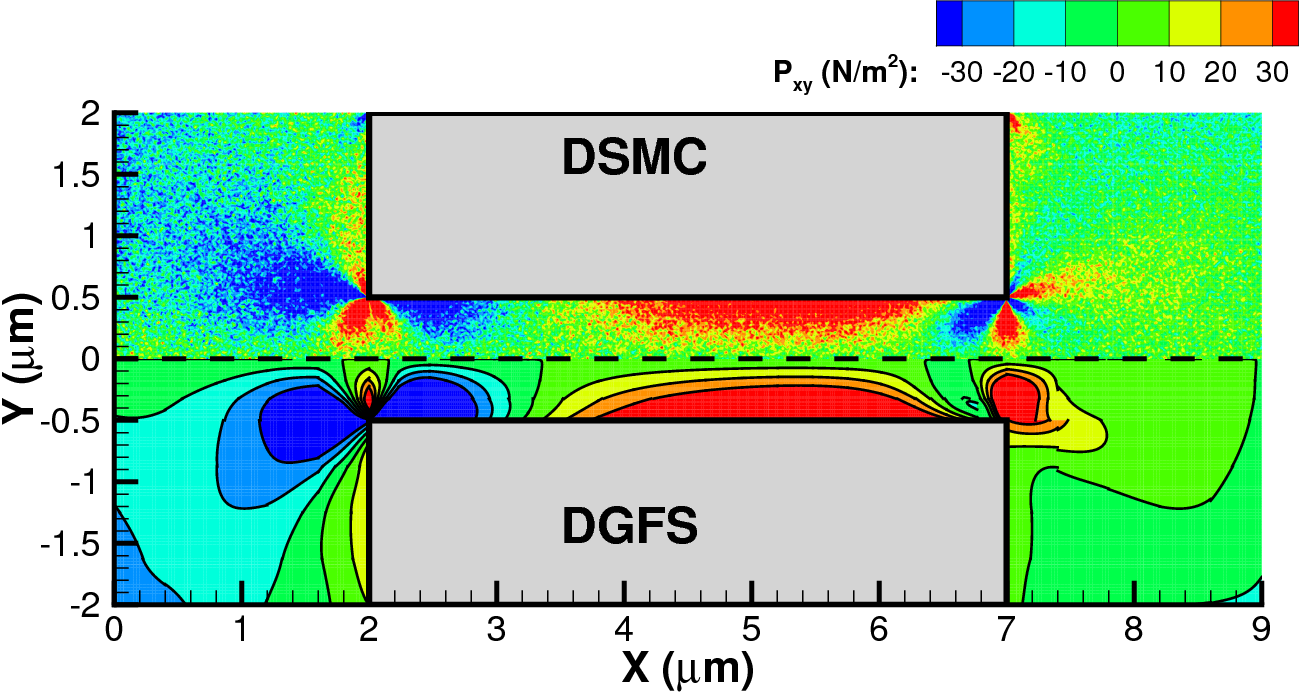}
  \caption{$xy$-component of stress ($N/m^2$)}
\end{subfigure}
\begin{subfigure}{.5\textwidth}
  \centering
  \includegraphics[width=80mm]{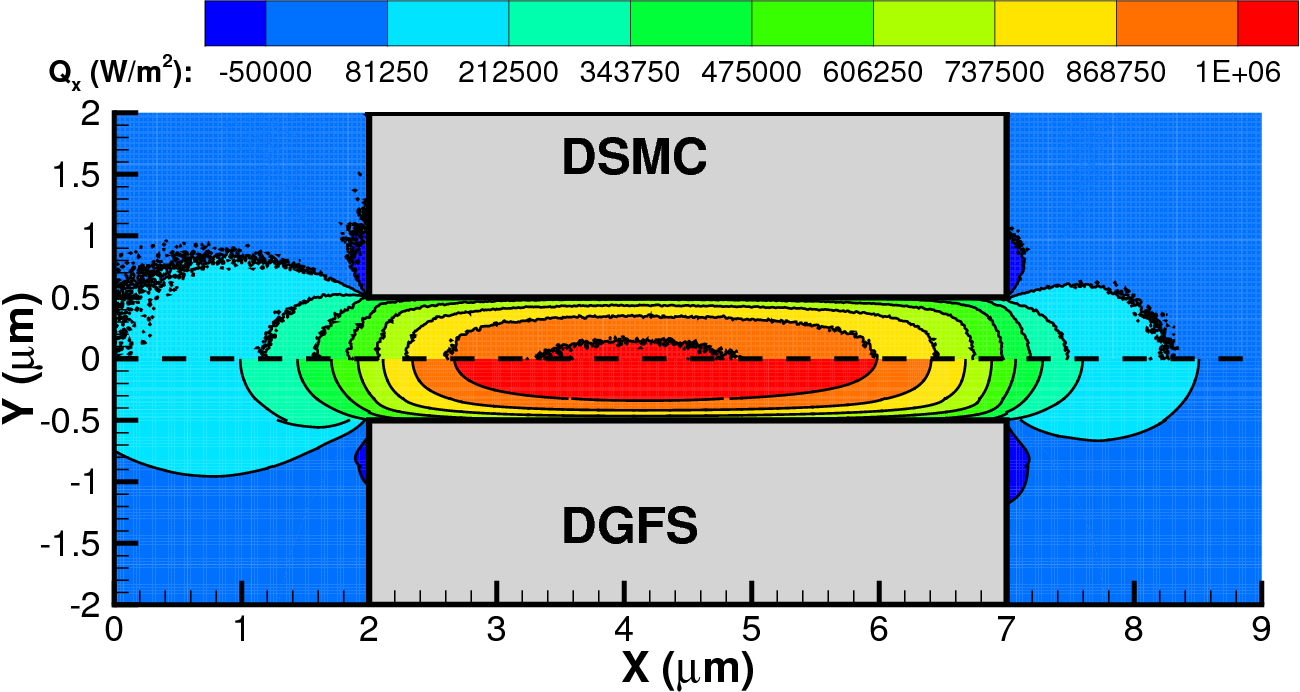}
  \caption{$x$-component of heat-flux ($W/m^2$)}
\end{subfigure}%
\begin{subfigure}{.5\textwidth}
  \centering
  \includegraphics[width=80mm]{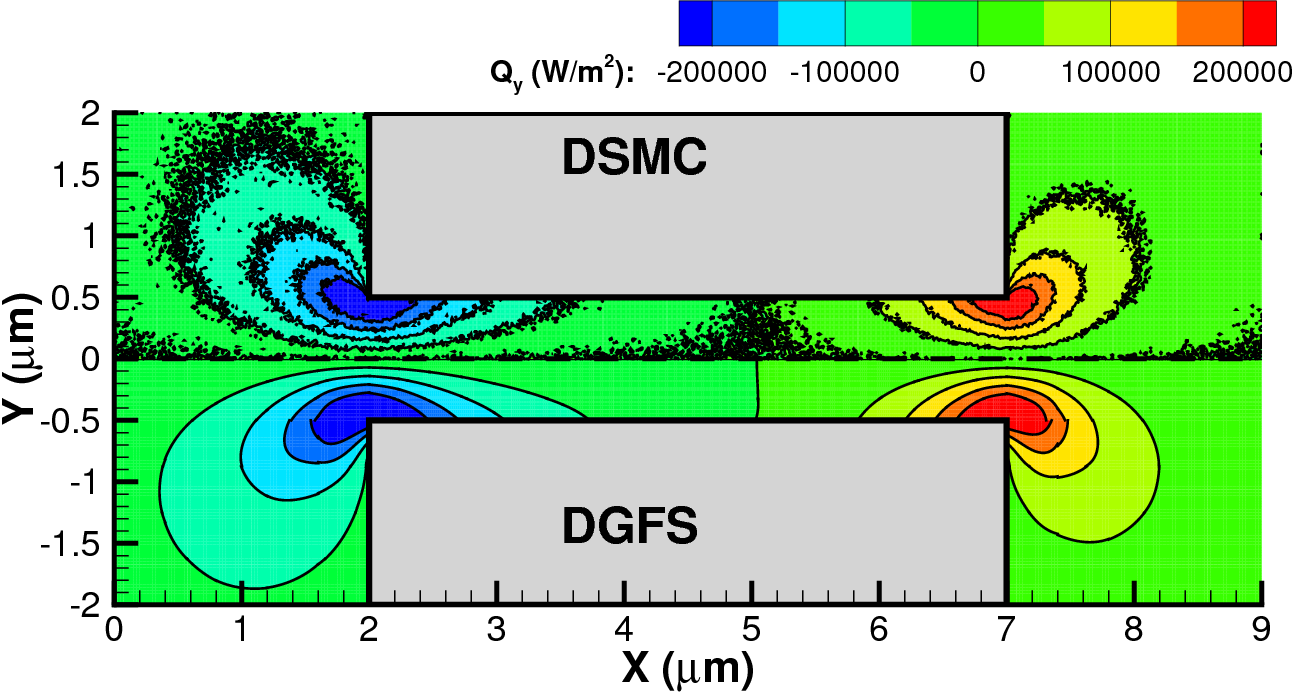}
  \caption{$y$-component of heat-flux ($W/m^2$)}
\end{subfigure}
\caption{Flow properties for short microchannel test-case obtained from DSMC and DGFS using VHS collision model. For each of these figures, DSMC results have been shown in the first quadrant ($0\,\mu m \leq y < 2\,\mu m$), whereas DGFS results (mirrored along x-axis) have been illustrated in the fourth quadrant ($-2\,\mu m \leq y < 0\,\mu m$). Differences in $Q_x$ can be attributed to the fact that DSMC simulations consider rotational degrees-of-freedom of $N_2$ into account, whereas DGFS doesn't.}
\label{fig_shortMicroChannel_props}
\end{figure*}

Figures~\ref{fig_shortMicroChannel_props} illustrate the contours of various flow properties for the flow around the solid chip/substrate. Ignoring the statistical noise, we gain note excellent agreement between DSMC and DGFS. In particular, minor differences in $x$-component of heat-flux i.e., $Q_x$ can be attributed to the fact that DSMC simulations consider rotational degrees-of-freedom of $N_2$ into account, whereas DGFS doesn't. 

\begin{figure*}[!ht]
\centering
\begin{subfigure}{.5\textwidth}
  \centering
  \includegraphics[width=80mm]{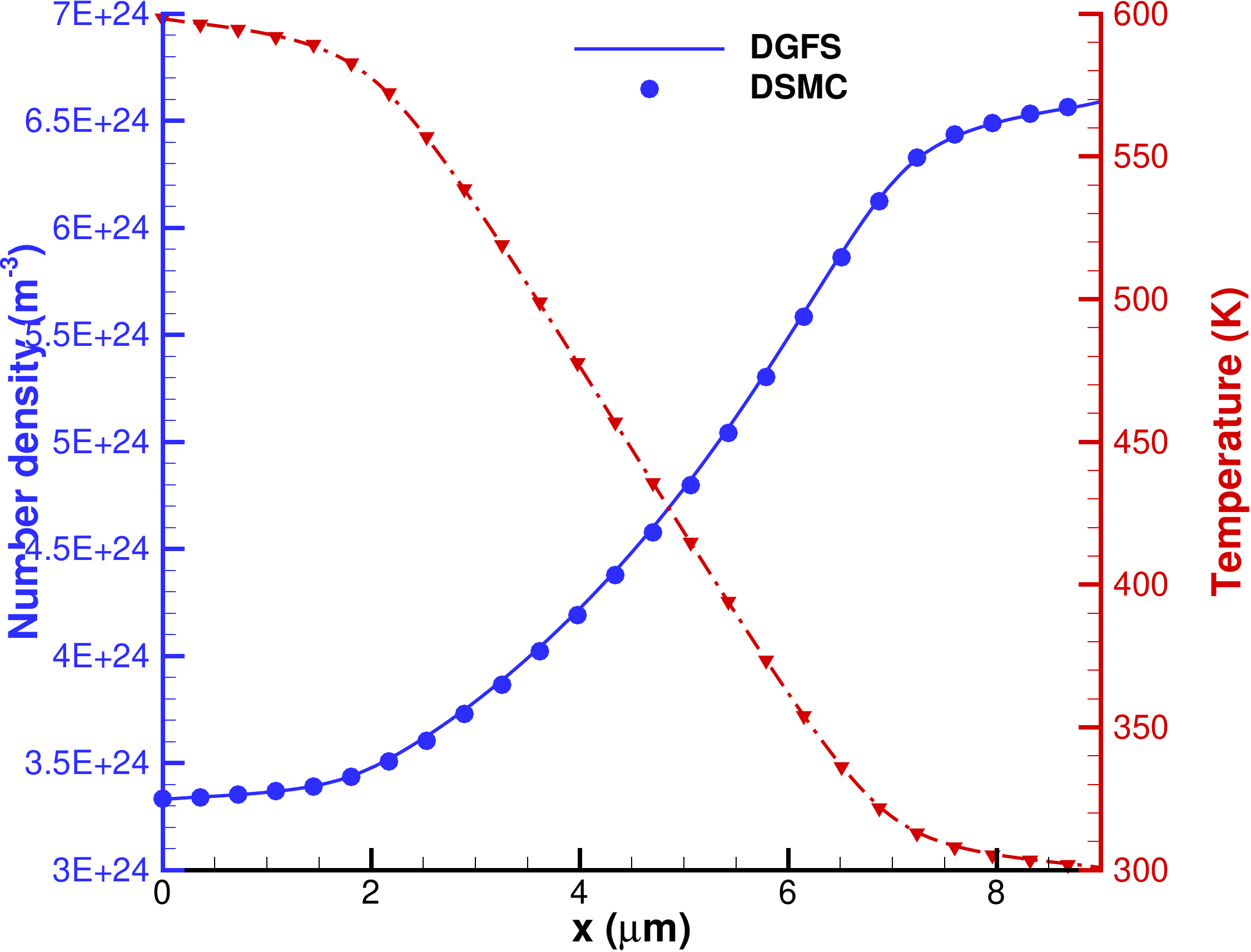}
  \caption{Number density ($m^{-3}$), and Temperature ($K$)}
\end{subfigure}%
\begin{subfigure}{.5\textwidth}
  \centering
  \includegraphics[width=80mm]{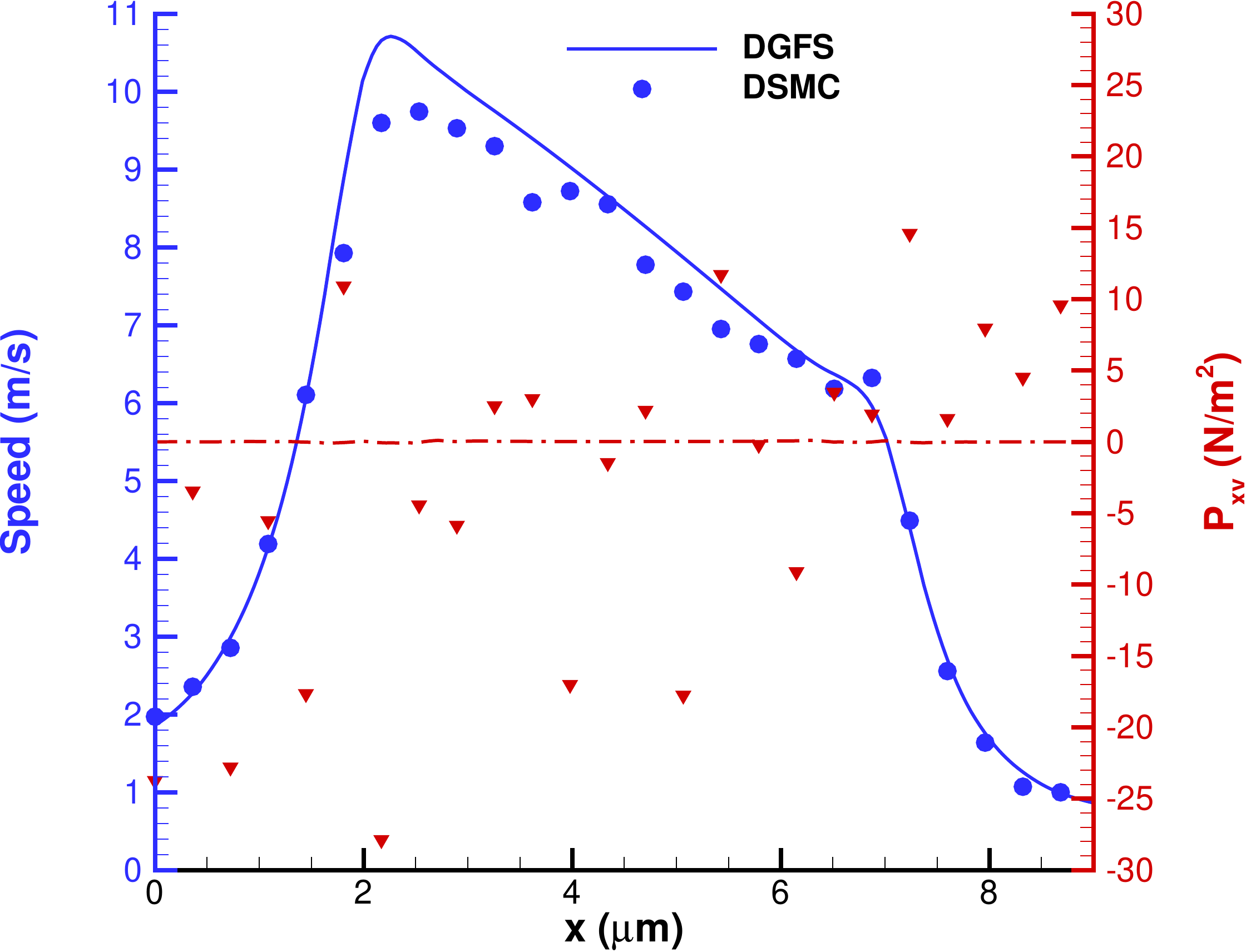}
  \caption{Speed ($m/s$), and xy-component of stress ($N/m^2$)}
\end{subfigure}
\caption{Flow properties on the horizontal centerline ($y=0\,\mu m$) for short microchannel test-case obtained from DSMC and DGFS using VHS collision model.}
\label{fig_shortMicroChannel_propsOverCenterline}
\end{figure*}

Figures~\ref{fig_shortMicroChannel_propsOverCenterline} shows the variation of flow properties over the vertical centerline, wherein we again observe an excellent agreement. 

\section{MIKRA: Micro In-Plane Knudsen Radiometric Actuator}
\label{sec_mikra}

\begin{figure*}
\includegraphics[width=0.8\textwidth]{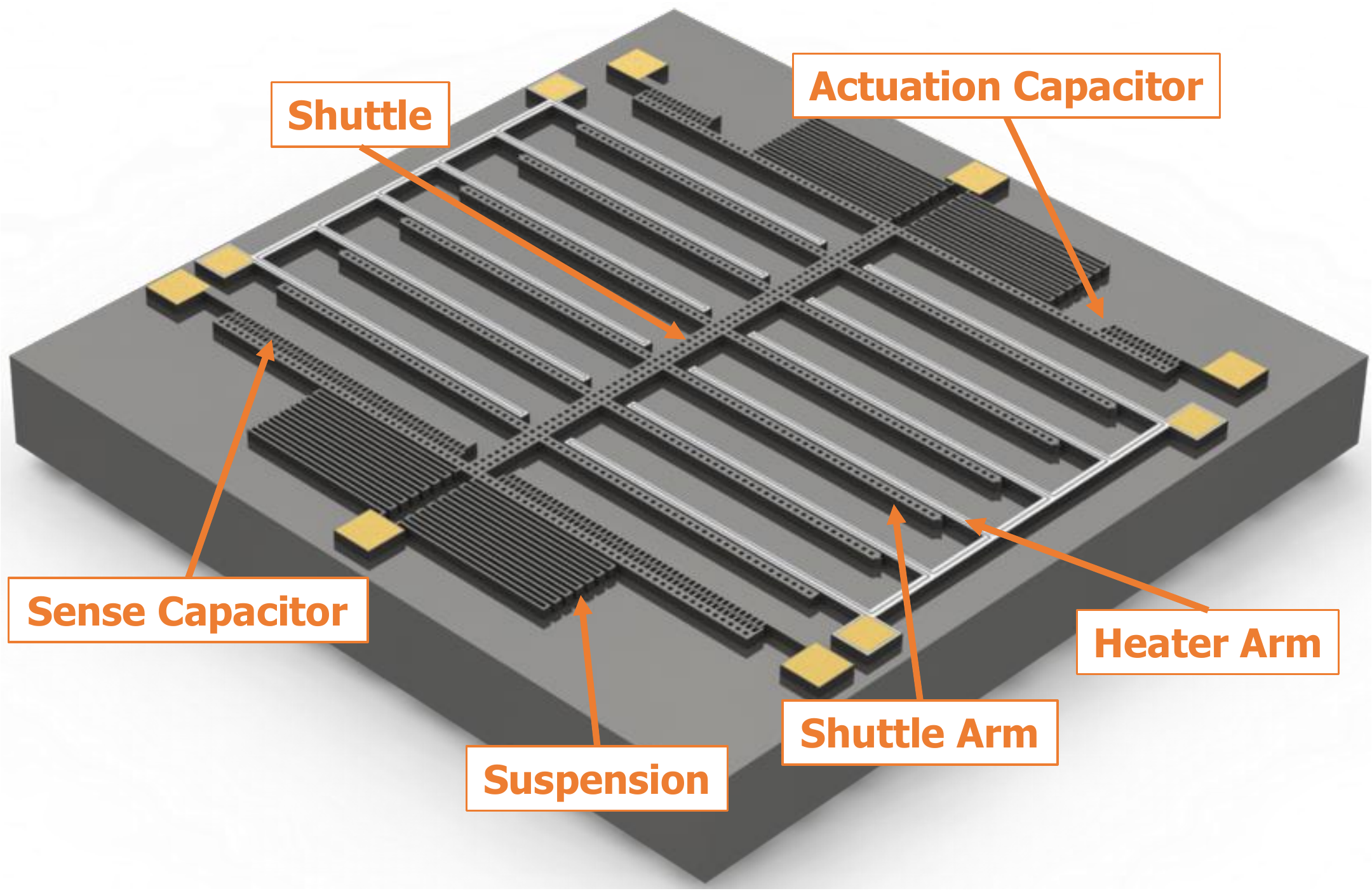}
\caption{The CAD model for Gen1 Micro In-Plane Knudsen Radiometric Actuator (MIKRA) \cite{strongrich2017microscale}.\label{fig_mikraCAD}}
\end{figure*}

MIKRA, acronym for Micro In-Plane Knudsen Radiometric Actuator, is a microscale compact low-power pressure sensor. A CAD representation of the device has been illustrated in Fig.~\ref{fig_mikraCAD}. Simply speaking, the device consists of an array of (tweleve) microbeams labelled as \textit{Shuttle Arm} and \textit{Heater Arm} in Fig.~\ref{fig_mikraCAD}. The heater arm is heated, and a thermal motion is induced in the gap between the heater and the shuttle. Subsequently, the shuttle arm experiences forces on order of few micro-newtons. This force is commonly identified as Knudsen force. Depending on the temperature of the heater, the shuttle gets displaced, and this displacement is measured capacitively. The magnitude of displacement is then used to estimate the ambient pressure. Specific details on MIKRA can be found in Refs.~\onlinecite{alexeenko2016microelectromechanical,strongrich2017microscale,pikus2019characterization}.


\subsection{Problem Statement}
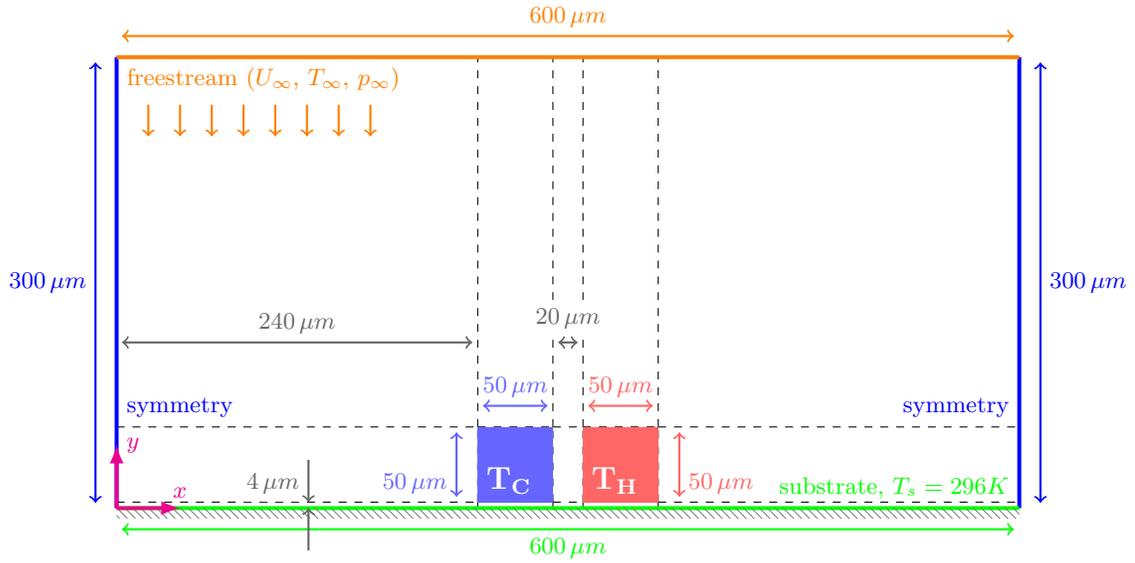
\begin{figure*}
\definecolor{caribbeangreen}{rgb}{0.0, 0.8, 0.6}
\definecolor{verdigris}{rgb}{0.26, 0.7, 0.68}
\definecolor{turquoisegreen}{rgb}{0.63, 0.84, 0.71}
\begin{tikzpicture}[scale=0.4]		
\def\Ho{20};

\def\L{600 / \Ho};  
\def\H{300 / \Ho};  
\def\hG{10 / \Ho};  
\def\hW{4 / \Ho};   

\def\Lh{50 / \Ho};  
\def\Hh{50 / \Ho};  

\def \off{20};
\def \dt{0.8};
\def \pw{10};

\coordinate (c1) at ({0}, {0});
\coordinate (c2) at ({\L/2-\hG-\Lh}, {0});
\coordinate (c3) at ({\L/2-\hG}, {0});
\coordinate (c4) at ({\L/2+\hG}, {0});
\coordinate (c5) at ({\L/2+\hG+\Lh}, {0});
\coordinate (c6) at ({\L}, {0});
\coordinate (c7) at ({\L}, {\hW});
\coordinate (c8) at ({\L}, {\hW+\Hh});
\coordinate (c9) at ({\L}, {\H});
\coordinate (c10) at ({\L/2+\hG+\Lh}, {\H});
\coordinate (c11) at ({\L/2+\hG}, {\H});
\coordinate (c12) at ({\L/2-\hG}, {\H});
\coordinate (c13) at ({\L/2-\hG-\Lh}, {\H});
\coordinate (c14) at ({0}, {\H});
\coordinate (c15) at ({0}, {\hW+\Hh});
\coordinate (c16) at ({0}, {\hW});
\coordinate (c17) at ({\L/2-\hG-\Lh}, {\hW});
\coordinate (c18) at ({\L/2-\hG}, {\hW});
\coordinate (c19) at ({\L/2-\hG}, {\hW+\Hh});
\coordinate (c20) at ({\L/2-\hG-\Lh}, {\hW+\Hh});
\coordinate (c21) at ({\L/2+\hG}, {\hW});
\coordinate (c22) at ({\L/2+\hG+\Lh}, {\hW});
\coordinate (c23) at ({\L/2+\hG+\Lh}, {\hW+\Hh});
\coordinate (c24) at ({\L/2+\hG}, {\hW+\Hh});

\draw (c1) -- (c2);
\draw (c2) -- (c3);
\draw (c3) -- (c4);
\draw (c4) -- (c5);
\draw (c5) -- (c6);
\draw (c6) -- (c7);
\draw (c7) -- (c8);
\draw (c8) -- (c9);
\draw (c9) -- (c10);
\draw (c10) -- (c11);
\draw (c11) -- (c12);
\draw (c12) -- (c13);
\draw (c13) -- (c14);
\draw (c14) -- (c15);
\draw (c15) -- (c16);
\draw (c16) -- (c1);

\draw[dashed] (c2) -- (c17);
\draw[dashed] (c3) -- (c18);
\draw[dashed] (c4) -- (c21);
\draw[dashed] (c5) -- (c22);

\draw[dashed] (c16) -- (c17);
\draw[dashed] (c17) -- (c18);
\draw[dashed] (c18) -- (c21);
\draw[dashed] (c21) -- (c22);
\draw[dashed] (c22) -- (c7);

\draw[dashed] (c17) -- (c20);
\draw[dashed] (c18) -- (c19);
\draw[dashed] (c21) -- (c24);
\draw[dashed] (c22) -- (c23);

\draw[dashed] (c15) -- (c20);
\draw[dashed] (c20) -- (c19);
\draw[dashed] (c19) -- (c24);
\draw[dashed] (c24) -- (c23);
\draw[dashed] (c23) -- (c8);

\draw[dashed] (c20) -- (c13);
\draw[dashed] (c19) -- (c12);
\draw[dashed] (c24) -- (c11);
\draw[dashed] (c23) -- (c10);

\draw[pattern=north west lines, pattern color=black!50, line width = 0.1mm, thin, draw=none, fill=blue!60] (c17) -- (c18) -- (c19) -- (c20) -- (c17) node[anchor=south west, color=white] {\large $\mathbf{T_C}$};
\draw[line width=\dt, <->, shorten >=2pt, shorten <=2pt,  blue!60] ([xshift=-\off]c17) -- ([xshift=-\off]c20) node[anchor=north east, pos=0.5] {$50\,\mu m$};
\draw[line width=\dt, <->, shorten >=2pt, shorten <=2pt,  blue!60] ([yshift=\off]c19) -- ([yshift=\off]c20) node[anchor=south, pos=0.5] {$50\,\mu m$};

\draw[pattern=north west lines, pattern color=black!50, line width = 0.1mm, thin, draw=none, fill=red!60] (c21) -- (c22) -- (c23) -- (c24) -- (c21) node[anchor=south west, color=white] {\large $\mathbf{T_H}$};
\draw[line width=\dt, <->, shorten >=2pt, shorten <=2pt,  red!60] ([xshift=\off]c22) -- ([xshift=\off]c23) node[anchor=north west, pos=0.5] {$50\,\mu m$};
\draw[line width=\dt, <->, shorten >=2pt, shorten <=2pt,  red!60] ([yshift=\off]c23) -- ([yshift=\off]c24) node[anchor=south, pos=0.5] {$50\,\mu m$};

\draw[line width=1.5, blue] (c14) -- (c15) node[anchor=south west] {symmetry} -- (c16) -- (c1);
\draw[line width=\dt, <->, shorten >=2pt, shorten <=2pt, blue] ([xshift=-\off]c14) -- ([xshift=-\off]c1) node[anchor=east, pos=0.5] {$300\,\mu m$};

\draw[line width=1.5, blue] (c9) -- (c8) node[anchor=south east] {symmetry} -- (c7) -- (c6);
\draw[line width=\dt, <->, shorten >=2pt, shorten <=2pt, blue] ([xshift=\off]c9) -- ([xshift=\off]c6) node[anchor=west, pos=0.5] {$300\,\mu m$};

\draw[line width=1.5, orange] (c9) -- (c10) -- (c11) -- (c12) -- (c13) -- (c14) node[anchor=north west] {freestream ($U_\infty$, $T_\infty$, $p_\infty$)};
\draw[line width=\dt, <->, shorten >=2pt, shorten <=2pt, orange] ([yshift=\off]c9) -- ([yshift=\off]c14) node[anchor=south, pos=0.5] {$600\,\mu m$};
\foreach \i in {1,...,8}
{
    \draw[line width=\dt, <-, shorten >=2pt, shorten <=2pt, orange] ([xshift=1.5*\i*\off,yshift=-4*\off]c14) -- ([xshift=1.5*\i*\off,yshift=-2*\off]c14);
}

\draw[line width=1.5, green] (c1) -- (c2) -- (c3) -- (c4) -- (c5) -- (c6) node[anchor=south east] {substrate, $T_s=296K$};
\draw[line width=\dt, <->, shorten >=2pt, shorten <=2pt,  green] ([yshift=-\off]c1) -- ([yshift=-\off]c6) node[anchor=north, pos=0.5] {$600\,\mu m$};
\draw[pattern=north west lines, pattern color=black!50, line width = 0.1mm, thin, draw=none] (c1) -- (c2) -- (c3) -- (c4) -- (c5) -- (c6) -- ([yshift=-\pw]c6) -- ([yshift=-\pw]c5) -- ([yshift=-\pw]c4) -- ([yshift=-\pw]c3) -- ([yshift=-\pw]c2) -- ([yshift=-\pw]c1);

\draw[line width=\dt, <->, shorten >=2pt, shorten <=2pt,  black!60] ([yshift=4*\off]c19) -- ([yshift=4*\off]c24) node[anchor=south, pos=0.5, yshift=0.1*\off, fill=white, rounded corners=2pt] {$20\,\mu m$};

\draw[line width=\dt, <->, shorten >=2pt, shorten <=2pt,  black!60] ([yshift=4*\off]c20) -- ([yshift=4*\off]c15) node[anchor=south, pos=0.5] {$240\,\mu m$};

\draw[line width=\dt, ->, black!60] ([xshift=-8*\off,yshift=-2*\off]c2) -- ([xshift=-8*\off]c2) node[anchor=south east, pos=0.5, yshift=0.5*\off, fill=white, rounded corners=2pt] {$4\,\mu m$};
\draw[line width=\dt, ->, black!60] ([xshift=-8*\off,yshift=2*\off]c17) -- ([xshift=-8*\off]c17);

\draw[-latex, line width=0.5mm, magenta] (c1) -- ([xshift=3*\off]c1) node[anchor=south] {$x$};

\draw[-latex, line width=0.5mm, magenta] (c1) -- ([yshift=3*\off]c1) node[anchor=west] {$y$};

\end{tikzpicture}
\caption{Schematic for numerical simulation of thermo-stress convection in MIKRA Gen1\cite{strongrich2017microscale}. The interior dashed thin black lines indicate the blocks used for structured mesh generation. Specifically for deterministic DGFS simulations, a linear gradient is applied within blocks such that the cells are finer in the near-vane region.\label{fig_mikraSchematic}}
\end{figure*}

The flow configuration is shown in Fig.~\ref{fig_mikraSchematic}. Consider the 2D uniform flow of $N_2$ with freestream velocity $U_\infty$, freestream temperature $T_\infty$, and freestream pressure $p_\infty$ over two two-dimensional square vanes, each with side lengths of $50\,\mu m$, separated by a gap of $20\,\mu m$ (also used as the nondimensionalizing length scale). The vanes are modeled as purely diffuse solid walls. The left vane, indicated in blue, is kept at a lower/cold temperature which we denote by $T_C$. The right vane, indicated in red, is kept at a higher/hot temperature which we denote by $T_H$. The substrate, indicated in green, forms the lower boundary of the domain, and is modelled as a purely diffuse solid wall. The end goal is to simulate the motion of gas flows in the gap between the two vanes, subject to different initial pressures $p_\infty$, hot ($T_H$) and cold ($T_C$) vane temperatures as listed in Tab.~\ref{tab_mikraCases}, in order to identify the correct circulation, induced low velocity, temperature gradient, and Knudsen forces from the vanes. The results are to be obtained from both stochastic (DSMC) and deterministic (DGFS) simulations. 

\begin{table}[!ht]
\centering
\begin{ruledtabular}
\begin{tabular}{@{}lccc@{}}
Parameter & \multicolumn{3}{c}{Cases} \\ 
 & M-01 & M-02 & M-03 \\ 
\hline
Pressure: $p$ (Torr) & $1.163$ & $2.903$ & $7.246$ \\
Number density: $n$ ($\times 10^{21}\;m^{-3}$) & $37.8609$ & $94.5058$ & $235.8901$ \\
Knudsen number\footnotemark[1]: $\Kn$ & $1.85$ & $0.74$ & $0.30$ \\
Cold vane temperature: $T_C$ ($K$) & $306$ & $306$ & $304$ \\
Hot vane temperature: $T_H$ ($K$) & $363$ & $356$ & $331$\\
\hline
\multicolumn{4}{l}{DGFS parameters} \\
Points in velocity mesh: $N^3$ & $24^3$ & $24^3$ & $24^3$ \\
Points in radial direction\footnotemark[2]: $N_\rho$ & $6$ & $6$ & $6$ \\
Points on \textit{half} sphere\footnotemark[2]: $M$ & $6$ & $6$ & $6$ \\
Size of velocity mesh\footnotemark[3] & $[-5,\,5]^3$ & $[-5,\,5]^3$ & $[-5,\,5]^3$ \\
\hline
\multicolumn{4}{l}{BGK/ESBGK/S-model parameters} \\
Points in velocity mesh: $N^3$ & $48^3$ & $24^3$ & $24^3$ \\
Size of velocity mesh\footnotemark[3] & $[-7,\,7]^3$ & $[-5,\,5]^3$ & $[-5,\,5]^3$ \\
\end{tabular}
\end{ruledtabular}
\footnotetext[1]{Based on hard-sphere definition (see Ref.~\onlinecite{Bird})}
\footnotetext[2]{Required only in the fast Fourier spectral low-rank decomposition (see Refs.~\onlinecite{GHHH17,JAH19})}
\footnotetext[3]{Non-dimensional (see Refs.~\onlinecite{JAH19,jaiswal2019dgfsMulti} for details on non-dimensionalization)}
\caption{Numerical parameters for thermo-stress convection in MIKRA Gen1 simulations for DSMC and DGFS using VHS collision model for $N_2$ molecules.}
\label{tab_mikraCases}
\end{table}

\subsection{Numerical details}
\label{subsec_numdetails}
The simulation is carried out at wide range of Knudsen number for flows in early slip to early free molecular regime. The simulation specific numerical parameters as well as differences between stochastic (DSMC) and deterministic (DGFS) modelling is described next.

\begin{figure*}
\includegraphics[width=0.8\textwidth]{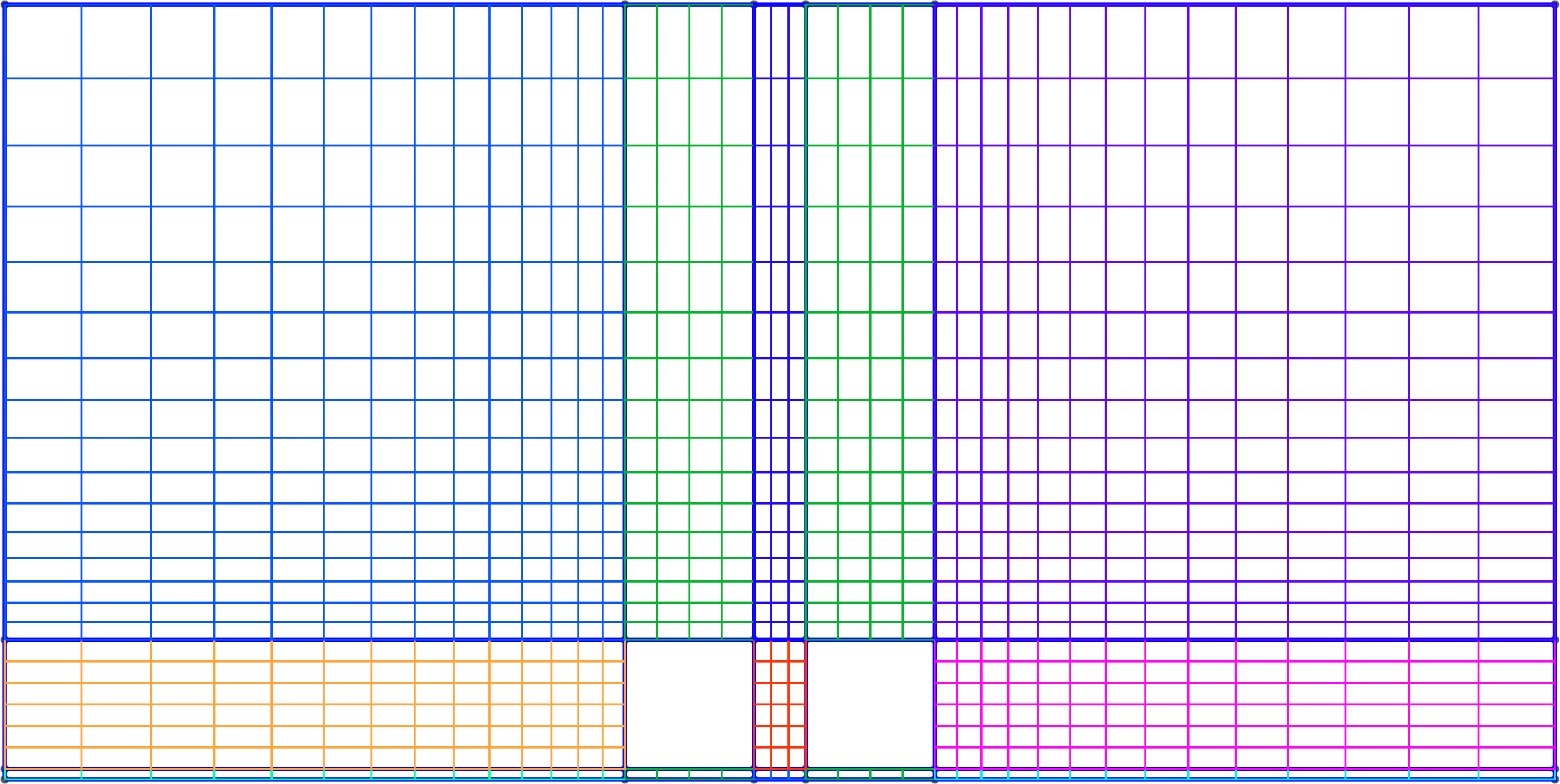}
\caption{Spatial mesh used for carrying out DGFS simulations for MIKRA Gen1 device. A linear gradient is applied within blocks such that the cells are finer in the near-vane region. A 3rd order nodal/sem DG scheme has been used. \label{fig_mikraMesh}}
\end{figure*}

\begin{itemize}
\item \textbf{DSMC}: SPARTA\cite{gallis2014direct} has been employed for carrying out DSMC simulations in the present work. 
The simulations are first run for 200,000 unsteady steps wherein the particles move, collide, and allowed to equilibrate. No sampling is performed at this stage. Next, the simulation is run for another 5,000,000 steady steps wherein the samples of flow properties namely number density, flow velocity, temperature, stress, and heat-flux, are taken for sufficiently long time so as to produce a meaningful bulk properties as well as minimize the statistical noise therein. In the present case, the DSMC domain is discretized into $300 \times 150$ cells, resulting in a uniform cell size of $2\,\mu m$, with 50 particles per cell on average during initialization. A time step of $10^{-9}$ sec is used during \textit{move} step of DSMC algorithm throughout the course of simulation. Note that these DSMC parameters have been taken from Ref.~\onlinecite{strongrich2017microscale} wherein the authors performed multiple verification cases with different time-steps, grid-size, domain length, particles per cell, etc. $N_2$ is used as the working gas in simulations, since MIKRA experiments\cite{strongrich2017microscale} were performed in $N_2$ medium. The properties of the working gas is given in Tab.~\ref{tab_props_N2}. DSMC simulations treat $N_2$ as diatomic species, and takes rotational degrees of freedom into account. 

\item \textbf{DGFS/BGK/ESBGK/S-model}: We use the DGFS implementation described in Ref.~\onlinecite{JAH19}. The spatial domain consists of 849 elements ($39 \times 23$ (total) - $2 \times 6 \times 4$ (remove the vane regions)). We use a linearly refined structured grid as illustrated in Fig.~\ref{fig_mikraMesh}. While structured grids might seem inflexible compared to unstructured grids, they are known to produce more stable scheme with superior convergence rates\cite{biswas1994parallel,cockburn2000development}, are amenable to highly efficient adaptive $h/p$ mesh refinement via recursive element splitting \cite{bakhtiari2016parallel} (Nevertheless, DGFS is more general, and test cases on general grids will be reported in future works). Since we are seeking a steady state solution, the time-step is selected based on the CFL constraints of the forward Euler scheme. Other case specific DGFS parameters have been provided in Tab.~\ref{tab_mikraCases}. 
\end{itemize}

It is worth noting that both the methods have different cell size requirements. In DSMC method, the contribution of particle collision to the transport properties is affected by strict spatial cell size requirements. In DGFS, however, the transport properties are strongly affected by local 3-D velocity space resolution rather than spatial resolution. As we show later, one can resolve the flow properties with fewer cells using DGFS.

\subsection{Results and Discussion}
\label{subsec_mikra_results}

\subsubsection{Flow pattern}
\begin{figure*}[!ht]
\centering
\begin{subfigure}{0.5\textwidth}
  \centering
  \includegraphics[width=0.95\textwidth]{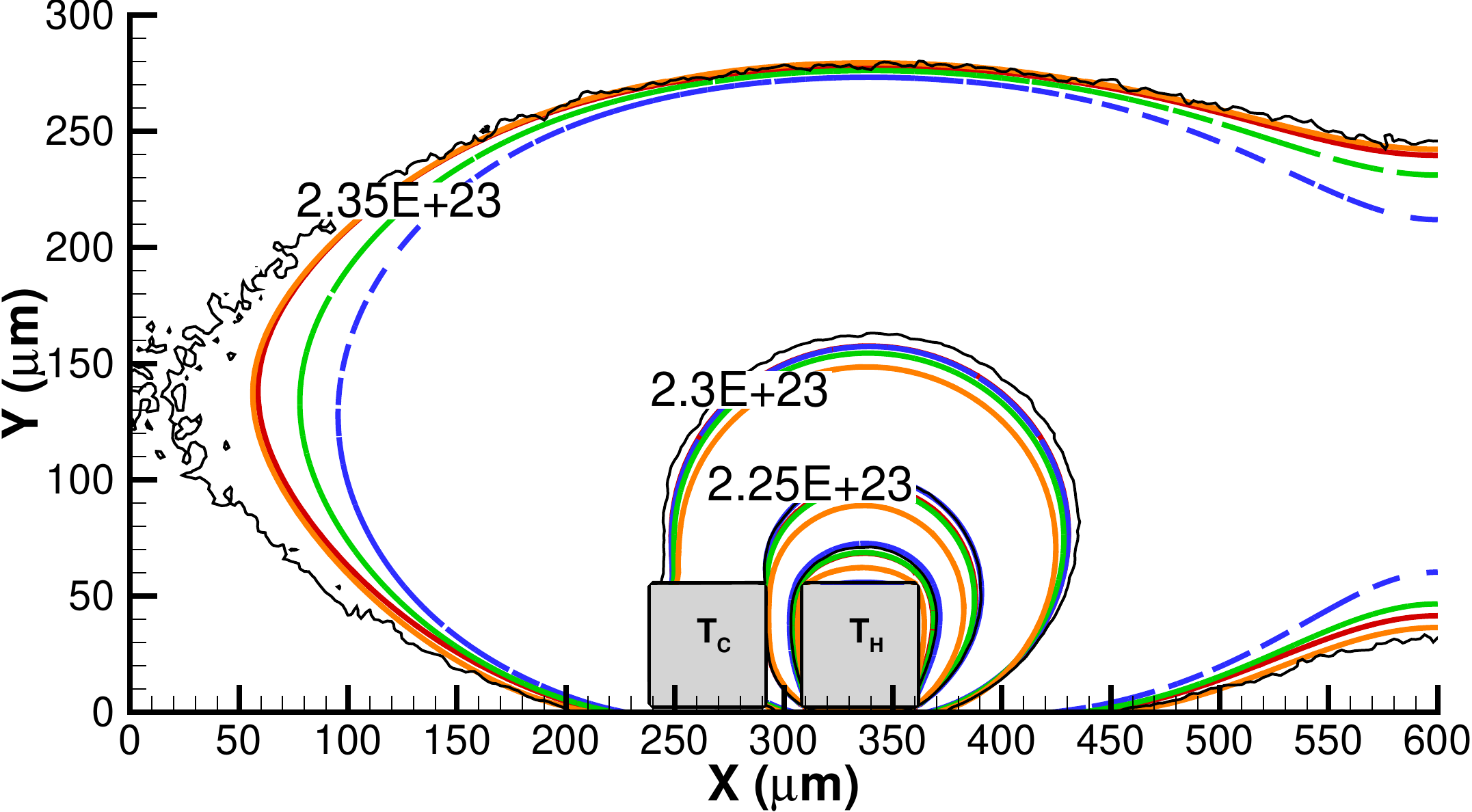}
  \caption{$\Kn=0.3$, Number density ($m^{-3}$)}
  \label{subfig_mikra_flowfield_966N2_nden}
\end{subfigure}%
\begin{subfigure}{0.5\textwidth}
  \centering
  \includegraphics[width=0.95\textwidth]{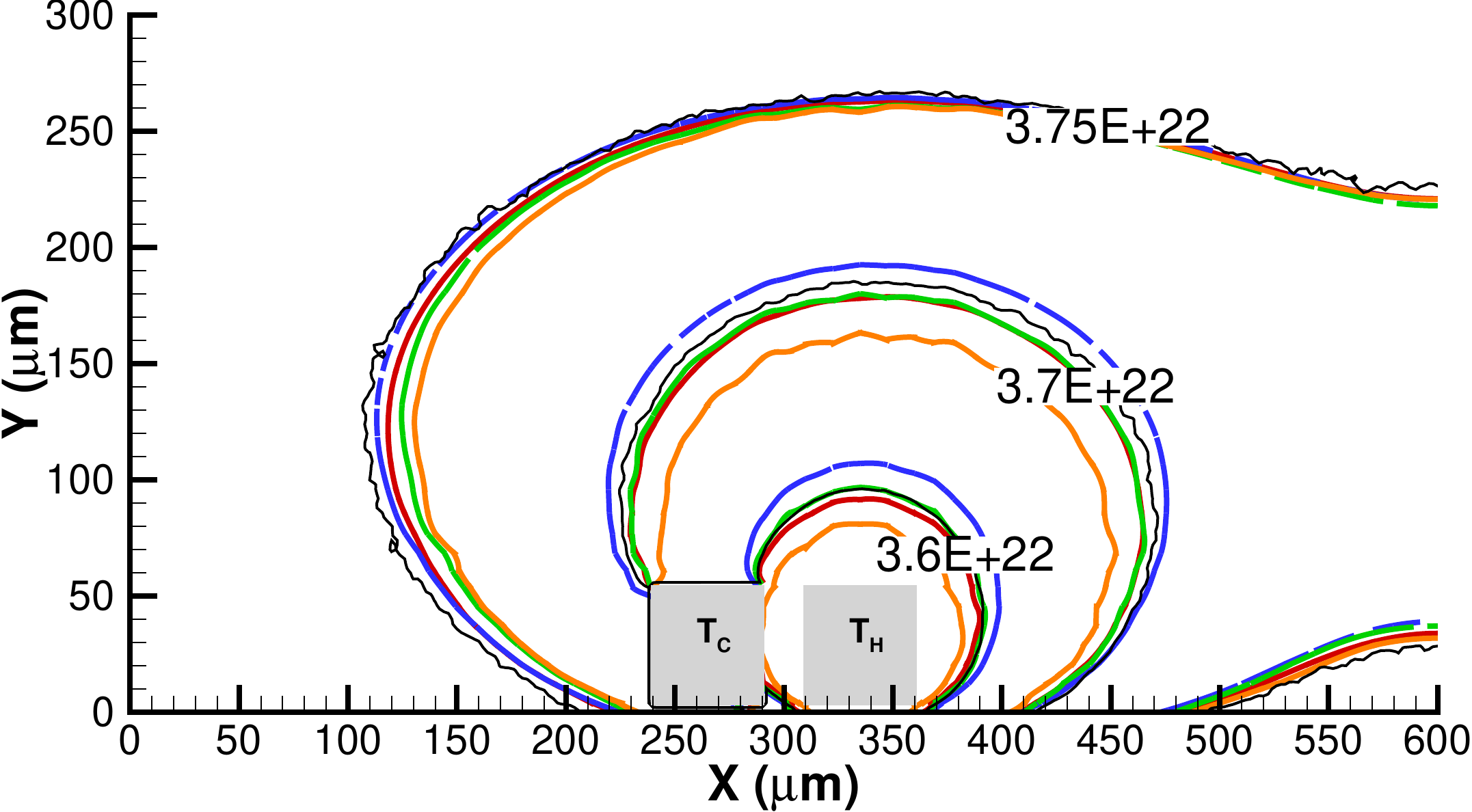}
  \caption{$\Kn=1.85$, Number density ($m^{-3}$)}
  \label{subfig_mikra_flowfield_155N2_nden}
\end{subfigure}
\begin{subfigure}{0.5\textwidth}
  \centering
  \includegraphics[width=0.95\textwidth]{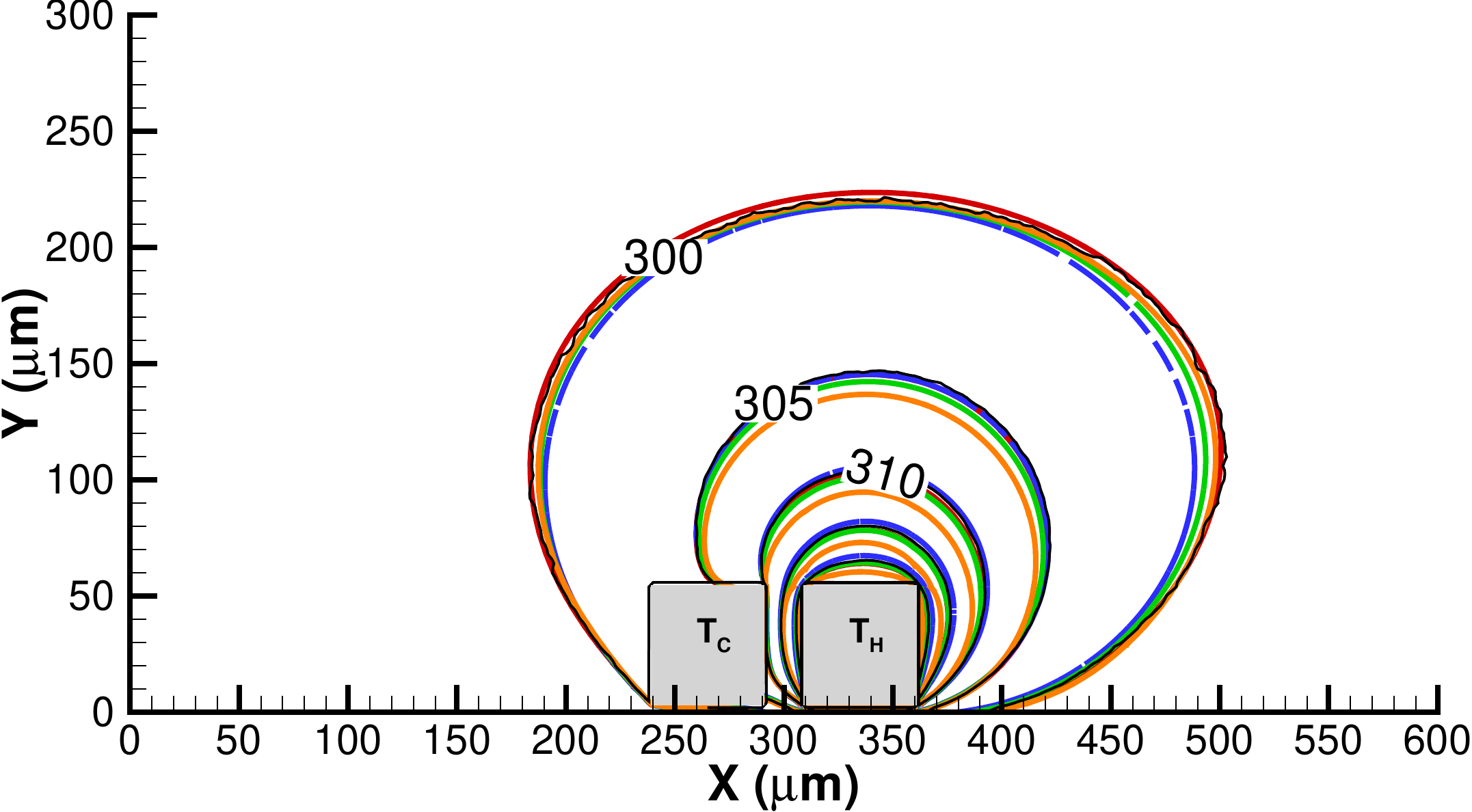}
  \caption{$\Kn=0.3$, Temperature ($K$)}
  \label{subfig_mikra_flowfield_966N2_T}
\end{subfigure}%
\begin{subfigure}{0.5\textwidth}
  \centering
  \includegraphics[width=0.95\textwidth]{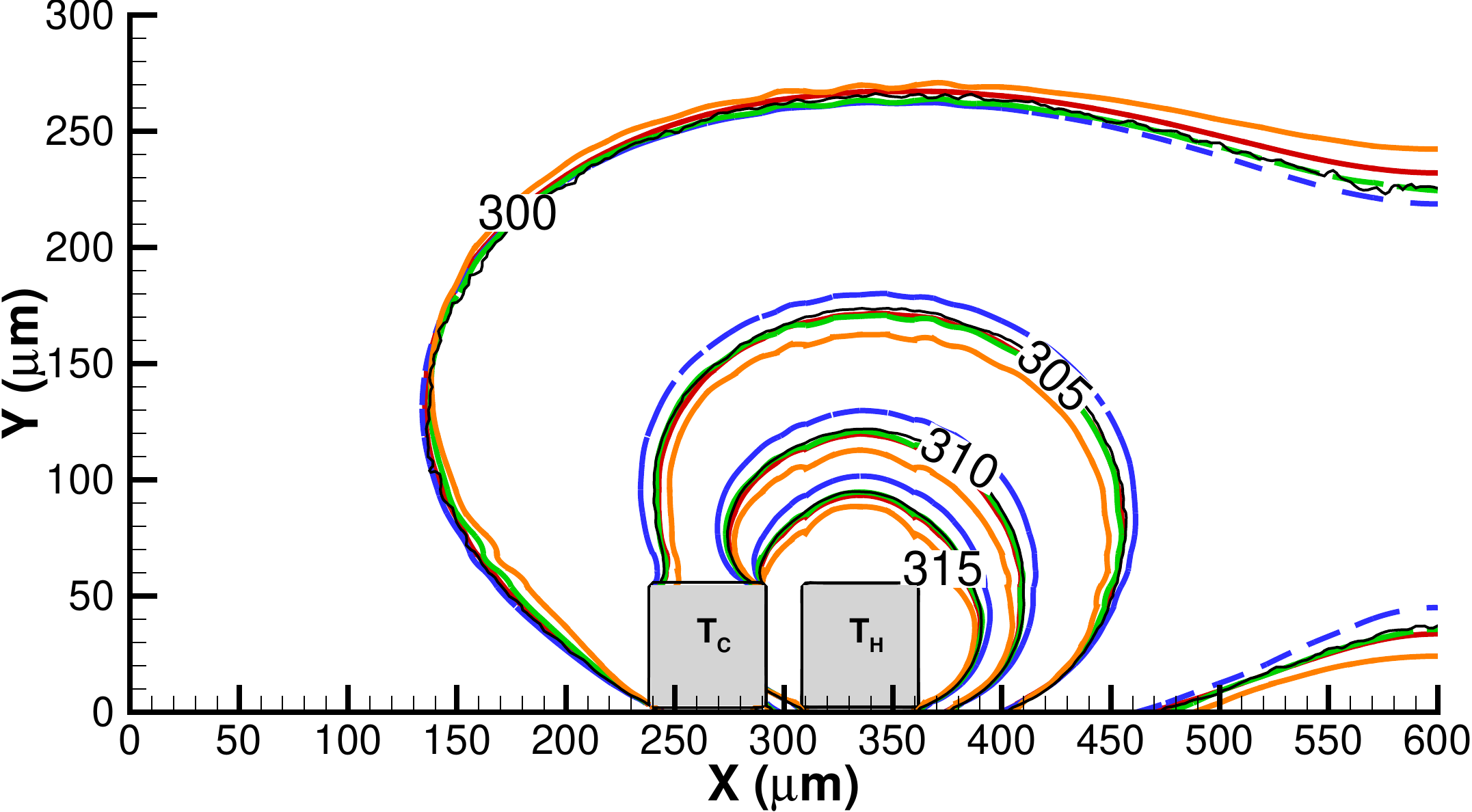}
  \caption{$\Kn=1.85$, Temperature ($K$)}
  \label{subfig_mikra_flowfield_155N2_T}
\end{subfigure}
\begin{subfigure}{0.5\textwidth}
  \centering
  \includegraphics[width=0.95\textwidth]{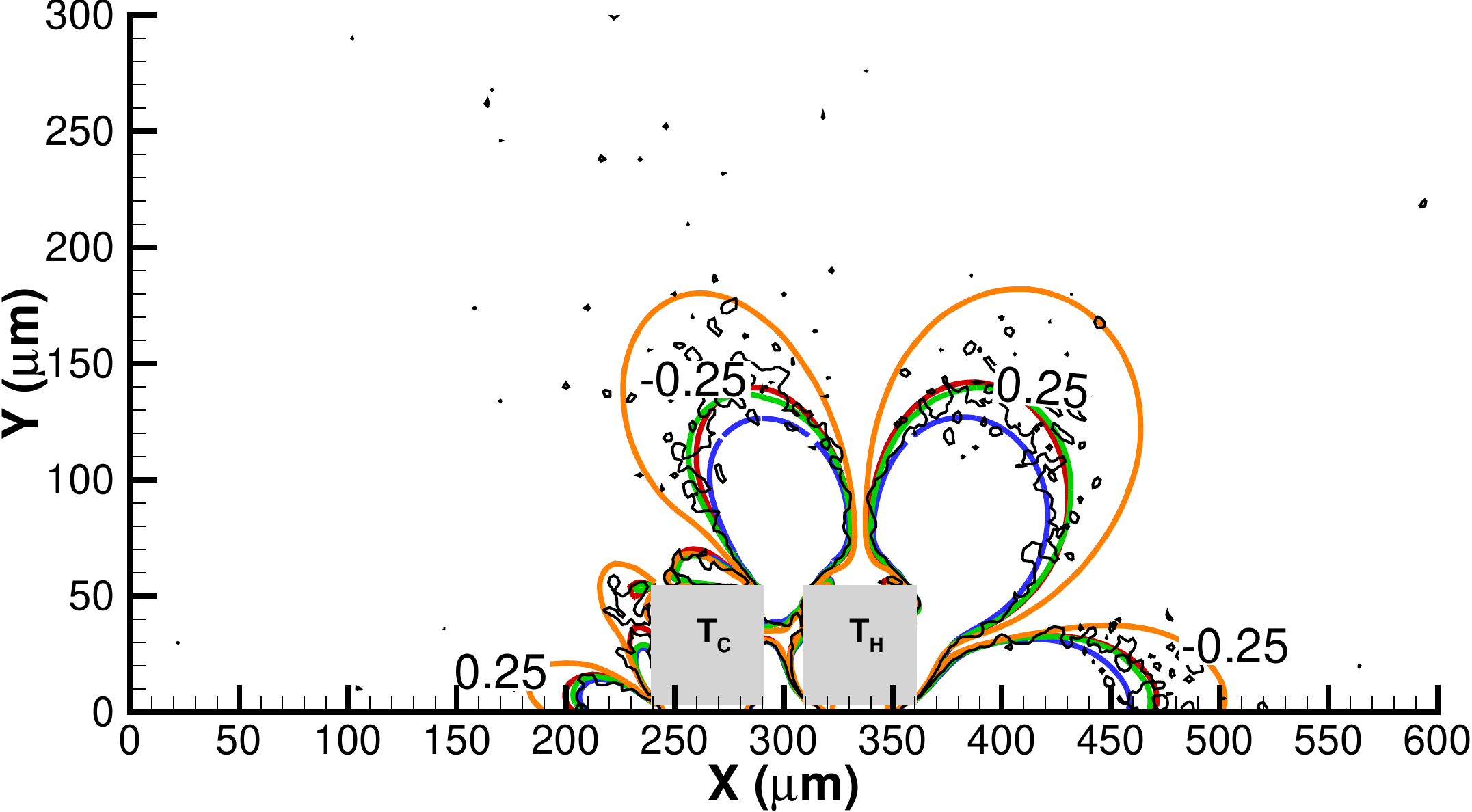}
  \caption{$\Kn=0.3$, $xy$-component of stress ($N/m^2$)}
  \label{subfig_mikra_flowfield_966N2_Pxy}
\end{subfigure}%
\begin{subfigure}{0.5\textwidth}
  \centering
  \includegraphics[width=0.95\textwidth]{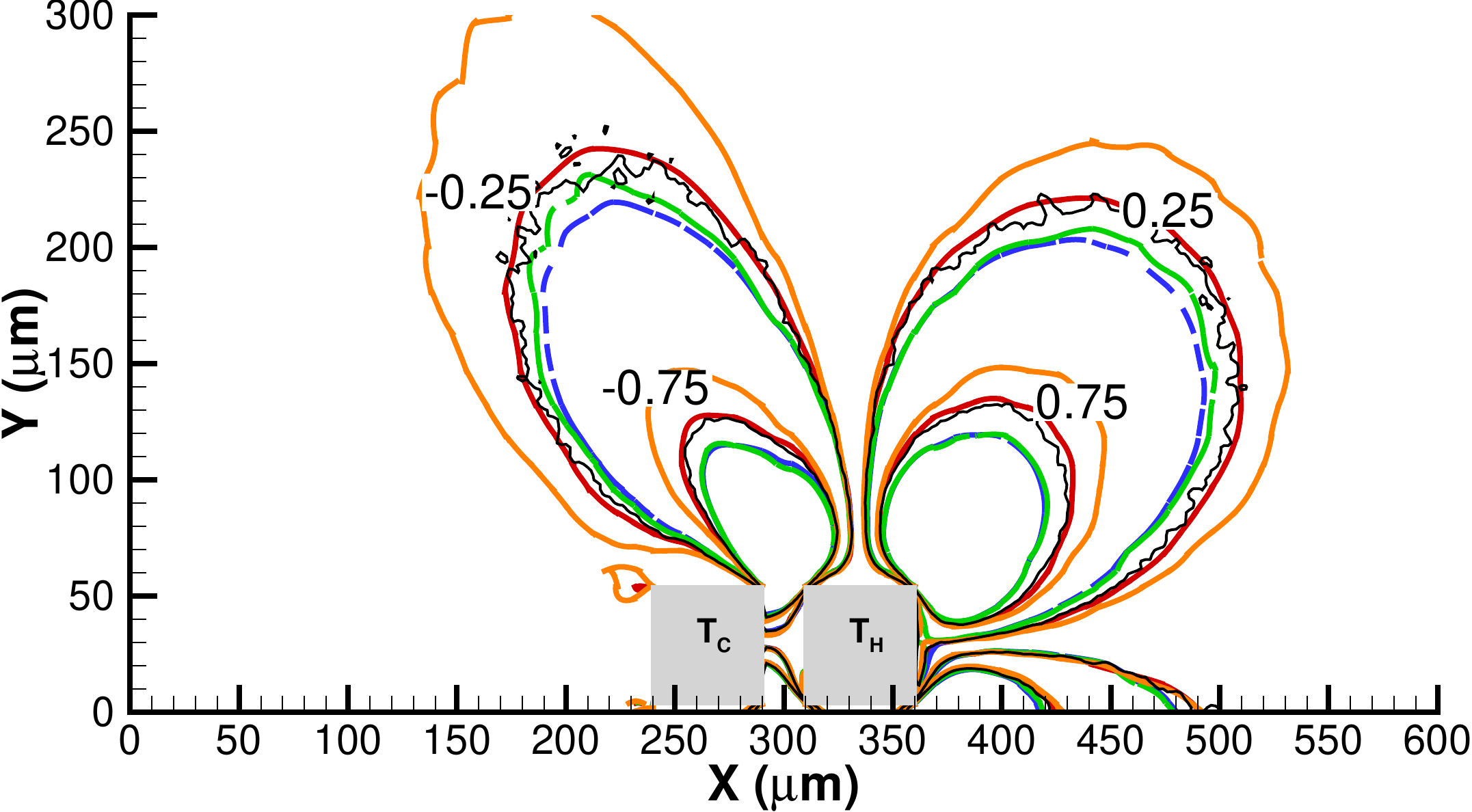}
  \caption{$\Kn=1.85$, $xy$-component of stress ($N/m^2$)}
  \label{subfig_mikra_flowfield_155N2_Pxy}
\end{subfigure}
\caption{Variation of flow properties along the domain for MIKRA Gen1 cases (M-01: $\Kn=1.85$, and M-03: $\Kn=0.3$) obtained from DSMC (thin black lines), DGFS (thick red lines), BGK (thick blue lines), ESBGK (thick green lines), and S-model (thick orange lines).}
\label{fig_mikra_flowfield_966N2_nden_T_Pxy}
\end{figure*}

\begin{figure*}[!ht]
\centering
\begin{subfigure}{0.5\textwidth}
  \centering
  \includegraphics[width=0.95\textwidth]{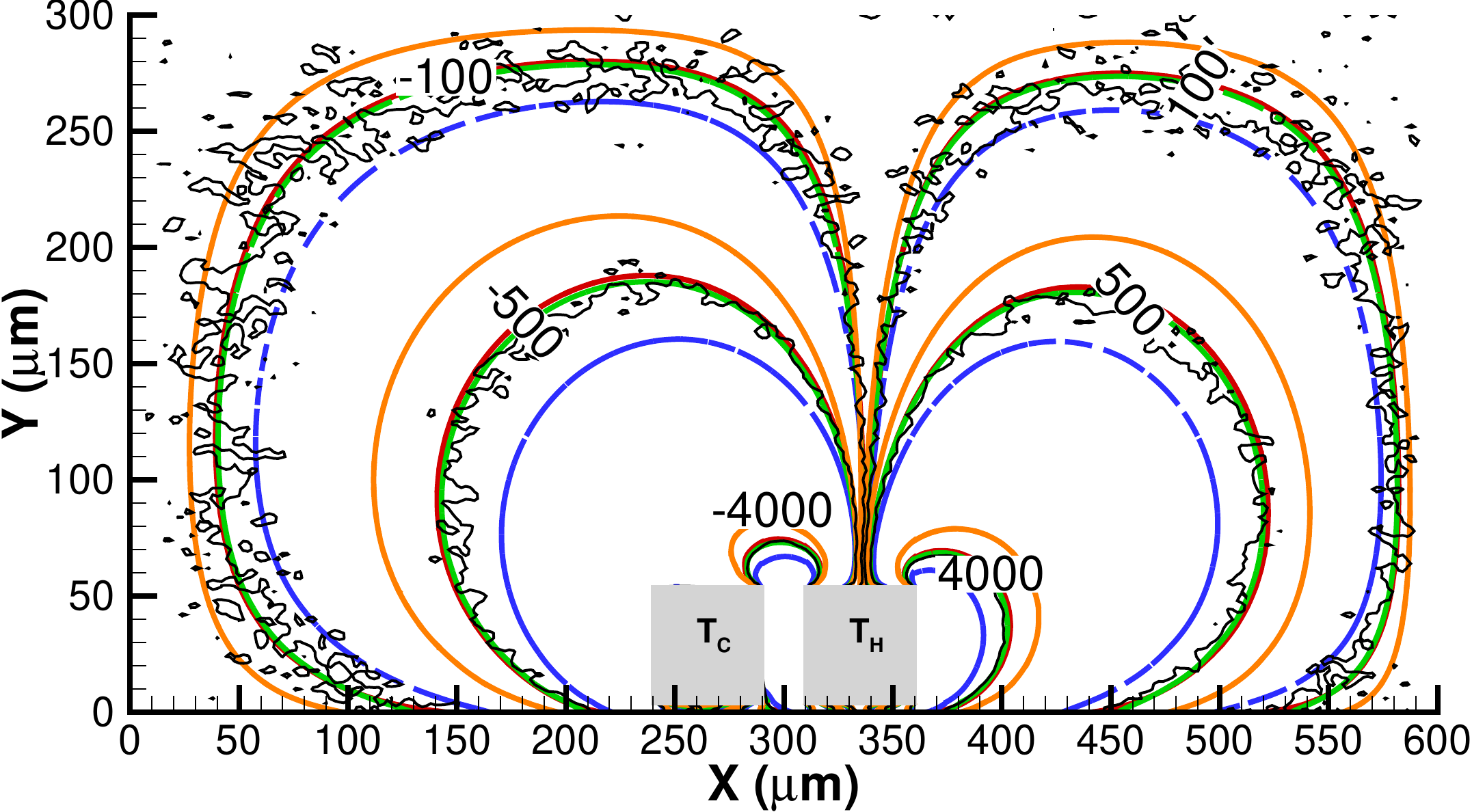}
  \caption{$\Kn=0.3$, $x$-component of heat-flux ($W/m^2$)}
  \label{subfig_mikra_flowfield_966N2_Qx}
\end{subfigure}%
\begin{subfigure}{0.5\textwidth}
  \centering
  \includegraphics[width=0.95\textwidth]{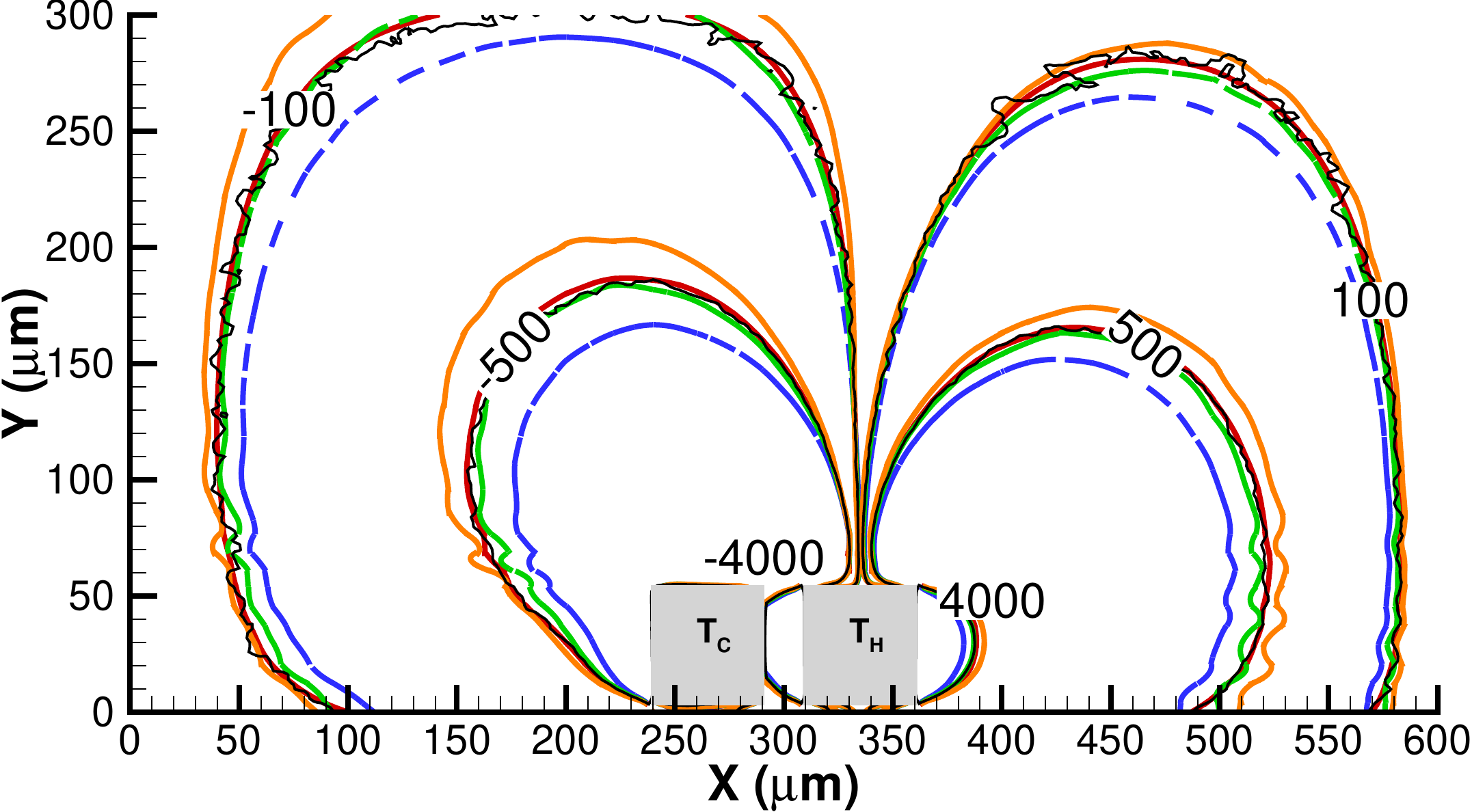}
  \caption{$\Kn=1.85$, $x$-component of heat-flux ($W/m^2$)}
  \label{subfig_mikra_flowfield_155N2_Qx}
\end{subfigure}
\begin{subfigure}{0.5\textwidth}
  \centering
  \includegraphics[width=0.95\textwidth]{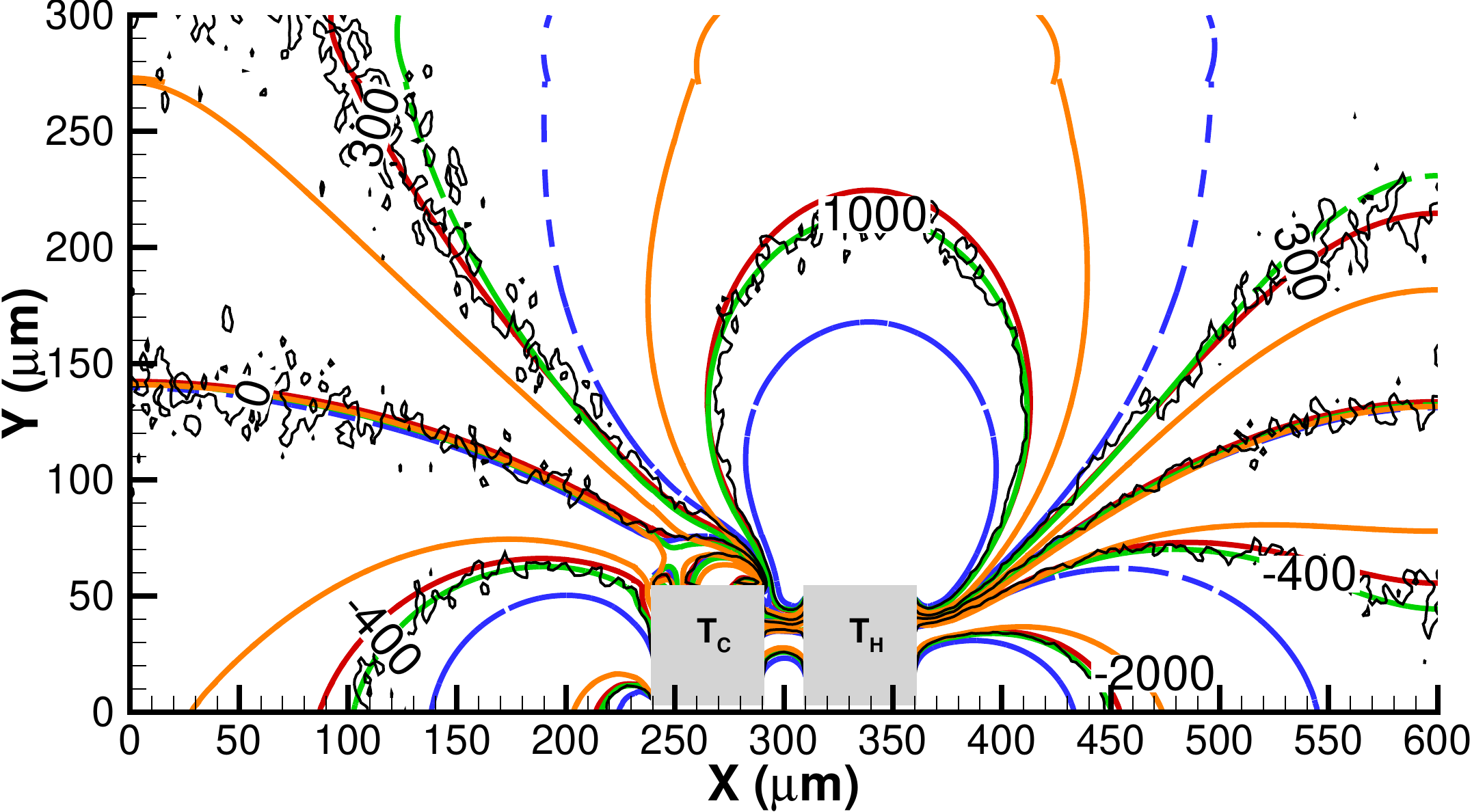}
  \caption{$\Kn=0.3$, $y$-component of heat-flux ($W/m^2$)}
  \label{subfig_mikra_flowfield_966N2_Qy}
\end{subfigure}%
\begin{subfigure}{0.5\textwidth}
  \centering
  \includegraphics[width=0.95\textwidth]{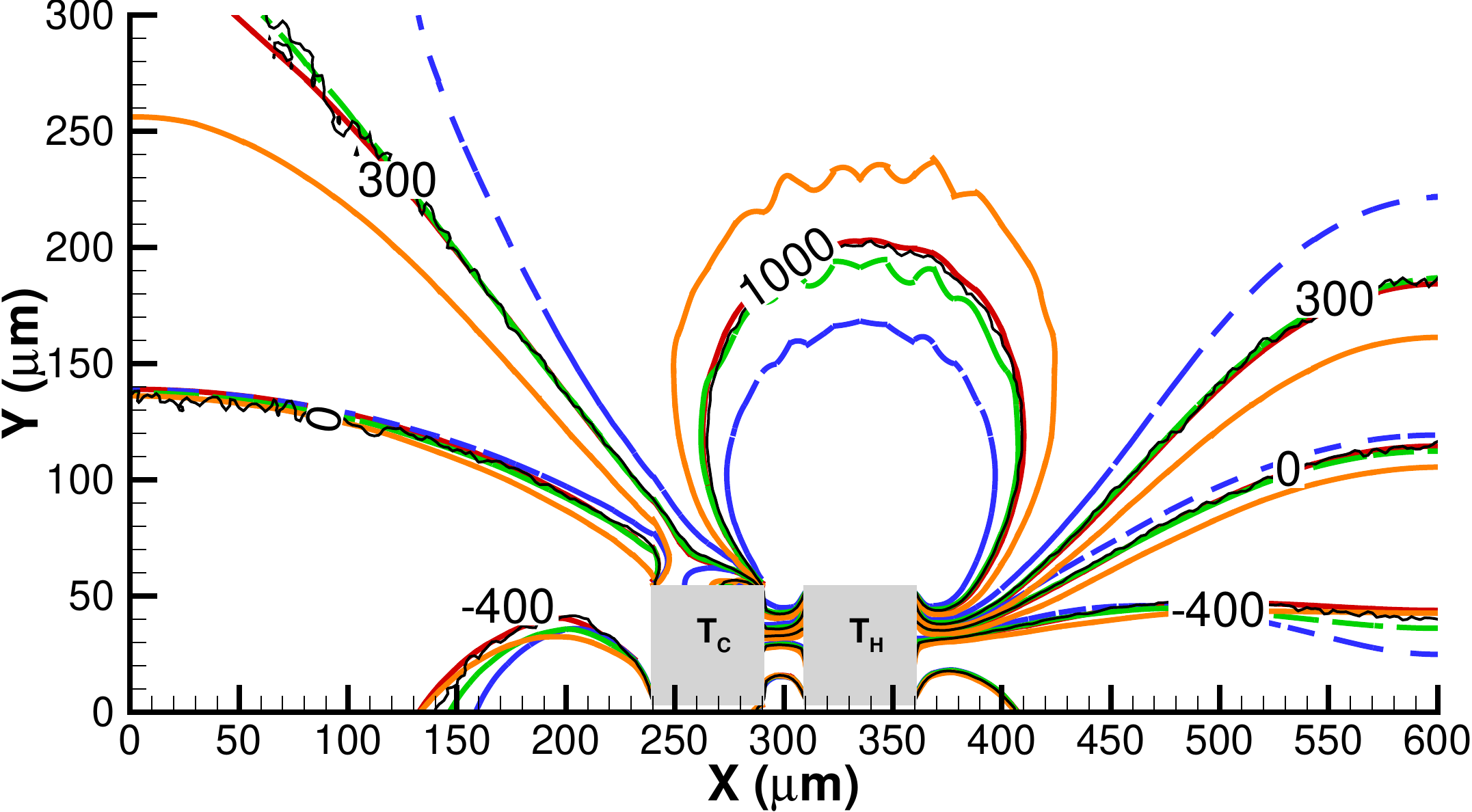}
  \caption{$\Kn=1.85$, $y$-component of heat-flux ($W/m^2$)}
  \label{subfig_mikra_flowfield_155N2_Qy}
\end{subfigure}
\begin{subfigure}{0.5\textwidth}
  \centering
  \includegraphics[width=0.95\textwidth]{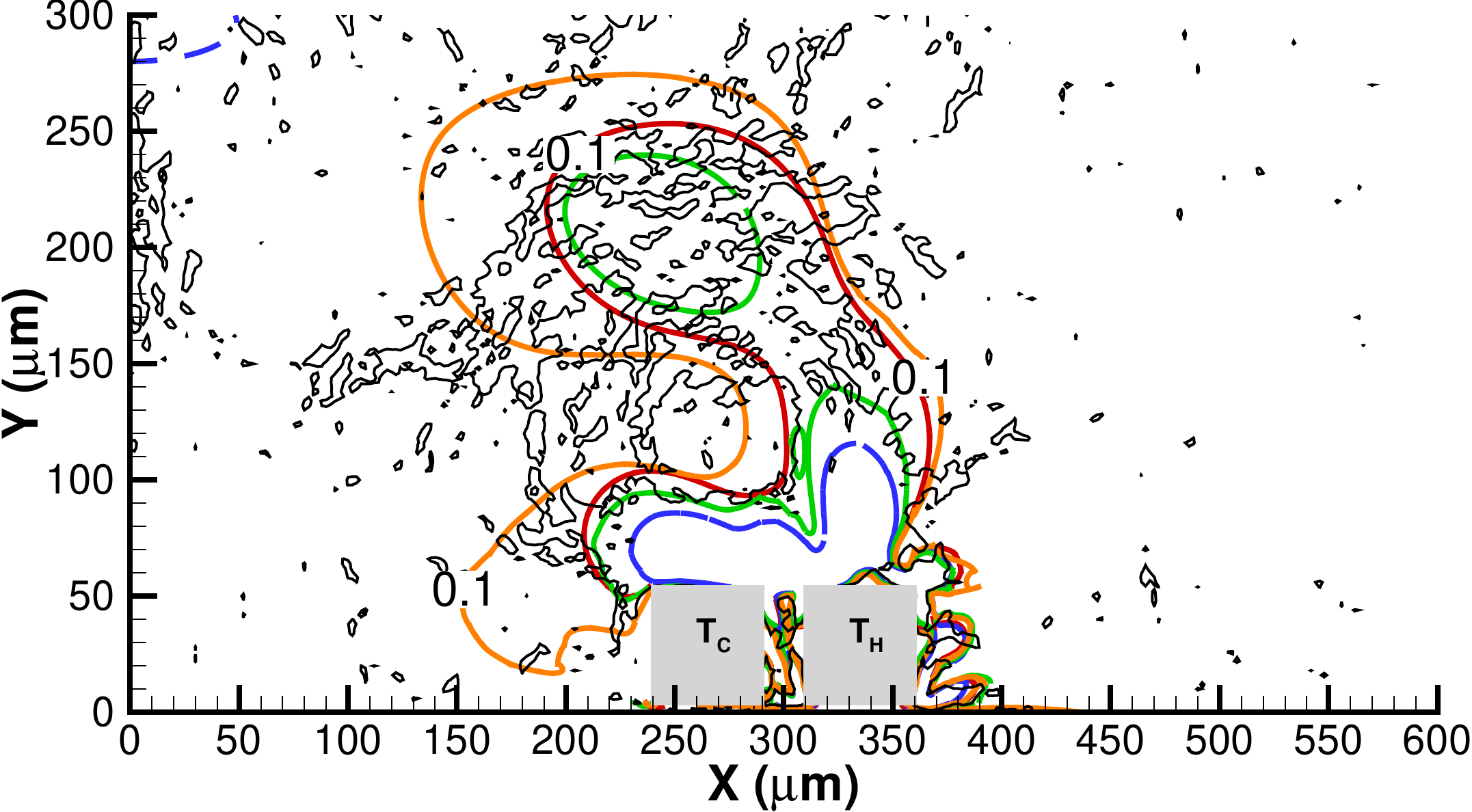}
  \caption{$\Kn=0.3$, Speed ($m/s$)}
  \label{subfig_mikra_flowfield_966N2_Speed}
\end{subfigure}%
\begin{subfigure}{0.5\textwidth}
  \centering
  \includegraphics[width=0.95\textwidth]{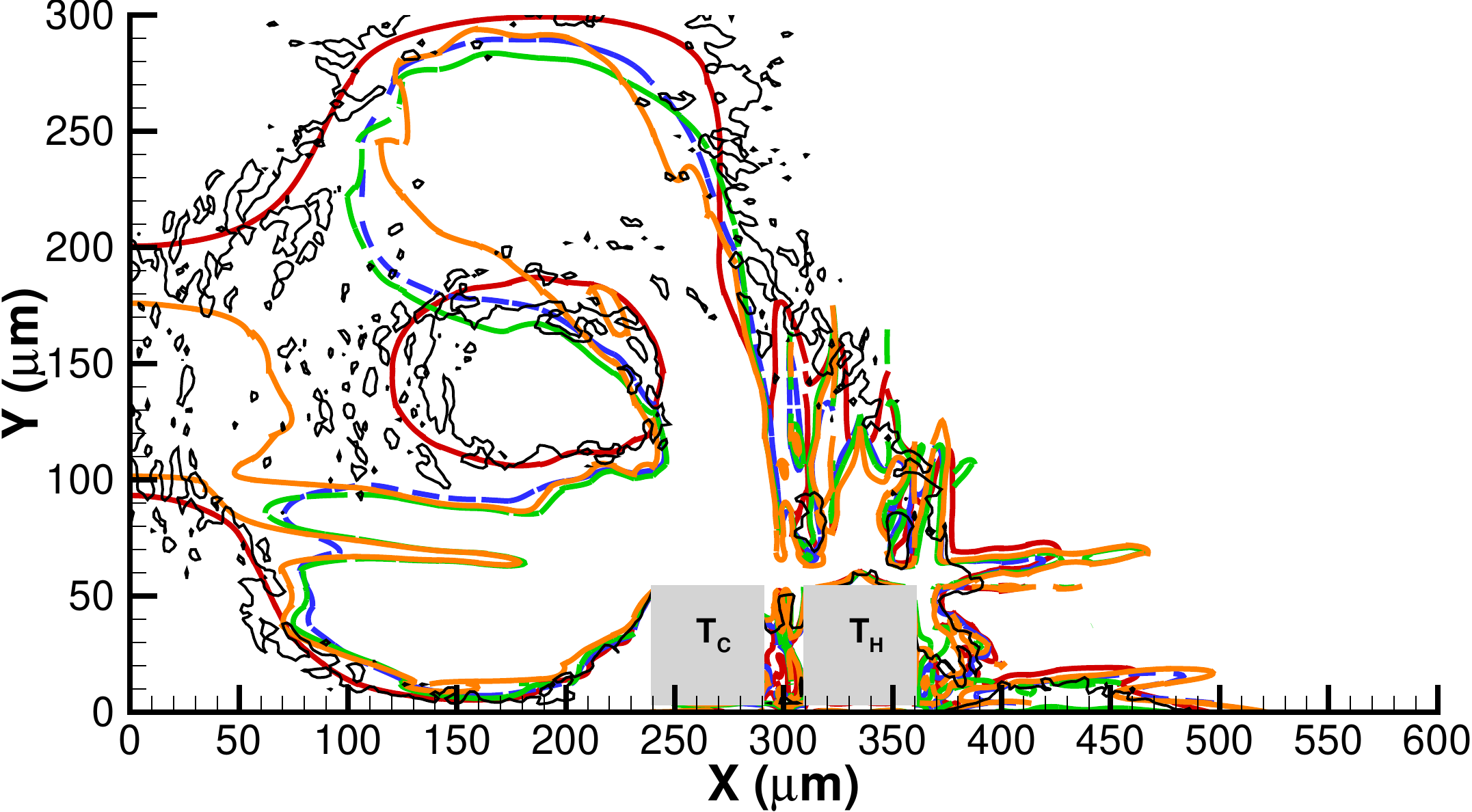}
  \caption{$\Kn=1.85$, Speed ($m/s$)}
  \label{subfig_mikra_flowfield_155N2_Speed}
\end{subfigure}
\caption{Continuation of Fig.~\ref{fig_mikra_flowfield_966N2_nden_T_Pxy}.}
\label{fig_mikra_flowfield_966N2_Qx_Qy_Speed}
\end{figure*}

\begin{figure*}[!ht]
\definecolor{verdigris}{rgb}{0.26, 0.7, 0.68}
\begin{tikzpicture}[scale=1]		
\def\Ho{20};
\def\L{600 / \Ho};  
\def\H{300 / \Ho};  
\def\hG{10 / \Ho};  
\def\hW{4 / \Ho};   
\def\Lh{50 / \Ho};  
\def\Hh{50 / \Ho};  
\def \off{20};
\def \dt{0.8};
\def \pw{10};
\coordinate (c17) at ({\L/2-\hG-\Lh}, {\hW});
\coordinate (c18) at ({\L/2-\hG}, {\hW});
\coordinate (c19) at ({\L/2-\hG}, {\hW+\Hh});
\coordinate (c20) at ({\L/2-\hG-\Lh}, {\hW+\Hh});
\coordinate (c21) at ({\L/2+\hG}, {\hW});
\coordinate (c22) at ({\L/2+\hG+\Lh}, {\hW});
\coordinate (c23) at ({\L/2+\hG+\Lh}, {\hW+\Hh});
\coordinate (c24) at ({\L/2+\hG}, {\hW+\Hh});
\draw[dashed] (c17) -- (c18);
\draw[dashed] (c21) -- (c22);
\draw[dashed] (c17) -- (c20);
\draw[dashed] (c18) -- (c19);
\draw[dashed] (c21) -- (c24);
\draw[dashed] (c22) -- (c23);
\draw[dashed] (c20) -- (c19);
\draw[dashed] (c24) -- (c23);
\draw[pattern=north west lines, pattern color=black!50, line width = 0.1mm, thin, draw=none, fill=blue!60] (c17) -- (c18) -- (c19) -- (c20) -- (c17) node[anchor=south west, color=white] {\large $\mathbf{T_C}$};
\draw[pattern=north west lines, pattern color=black!50, line width = 0.1mm, thin, draw=none, fill=red!60] (c21) -- (c22) -- (c23) -- (c24) -- (c21) node[anchor=south west, color=white] {\large $\mathbf{T_H}$};
\draw[line width=\dt, <->, shorten >=2pt, shorten <=2pt,  black!60] ([yshift=-3*\off]c19) -- ([yshift=-3*\off]c24) node[anchor=south, pos=0.5, yshift=0.1*\off] {$20\mu m$};
\draw[fill=blue!60, draw=black!50] ([xshift=-1.5*\off,yshift=1.5*\off]c19) circle (5pt);
\draw[line width=2, <-, shorten >=2pt, shorten <=2pt,  blue!60] ([xshift=-0.5*\off,yshift=0.5*\off]c19) -- ([xshift=-1.5*\off,yshift=1.5*\off]c19) node[anchor=south west, pos=0.5] {\textcolor{blue}{A}};
\draw[fill=red!60, draw=black!50] ([xshift=1*\off,yshift=1*\off]c19) circle (5pt);
\draw[line width=2, <-, shorten >=2pt, shorten <=2pt,  red!60] ([xshift=0.1*\off,yshift=0.1*\off]c19) -- ([xshift=1.0*\off,yshift=1.0*\off]c19) node[anchor=south east, pos=0.5] {\textcolor{red}{B}};
\draw[dashed, verdigris, line width=1.5] ([xshift=0.4*\off,yshift=0.4*\off]c19) -- ([xshift=-0.4*\off,yshift=0.4*\off]c19) -- ([xshift=-0.4*\off,yshift=-0.4*\off]c19) -- ([xshift=0.4*\off,yshift=-0.4*\off]c19) -- ([xshift=0.4*\off,yshift=0.4*\off]c19) node[anchor=north west, pos=0.5] {\textbf{\textcolor{verdigris}{$dS$}}};
\node[inner sep=0pt] (russell) at ([xshift=6*\off, yshift=-2*\off]c23) {\includegraphics[width=.5\textwidth]{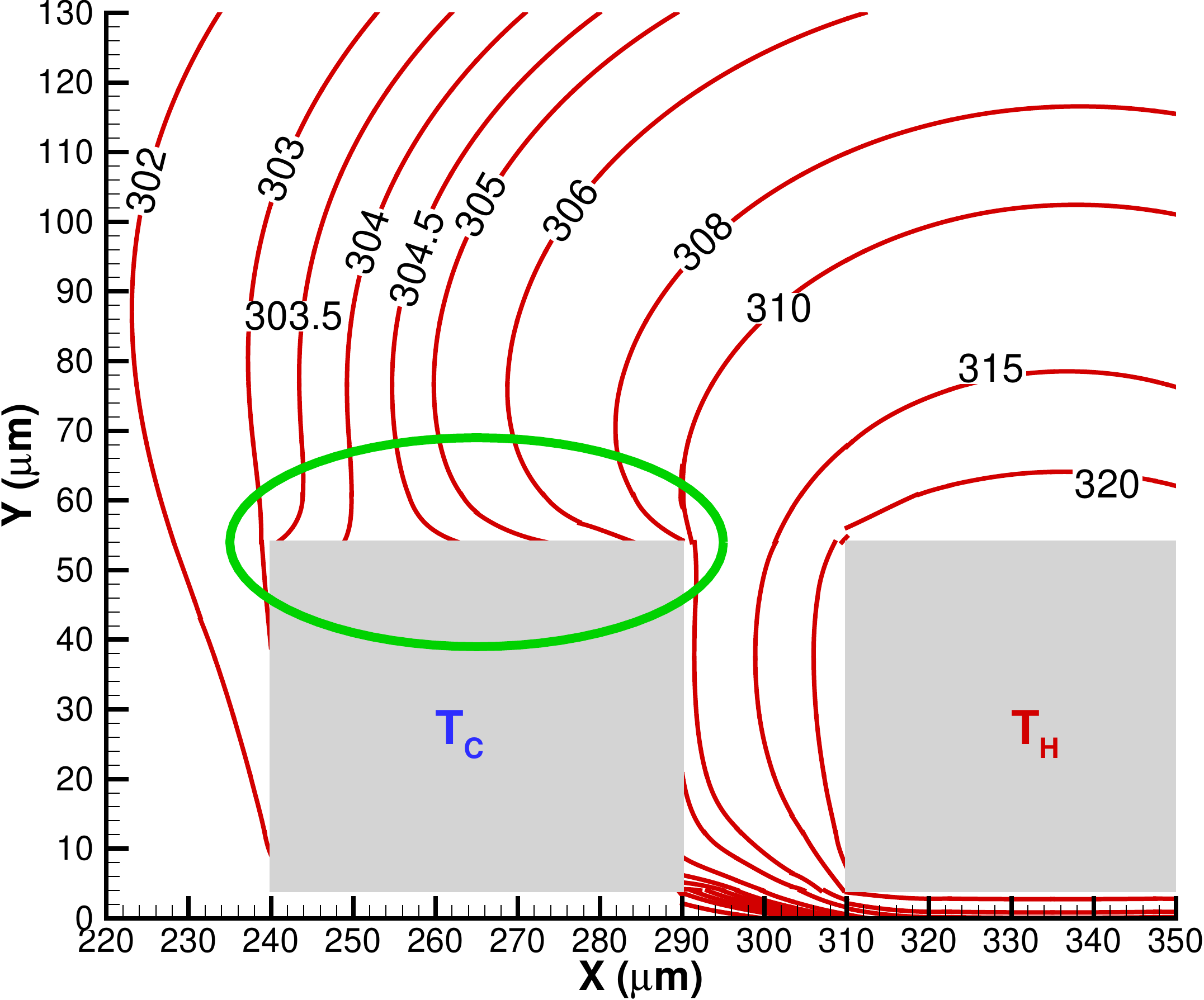}};
\draw[-latex, line width=0.5mm, magenta] ([xshift=-0.5*\off,yshift=-0.5*\off]c17) -- ([xshift=0.5*\off,yshift=-0.5*\off]c17) node[anchor=north] {$x$};
\draw[-latex, line width=0.5mm, magenta] ([xshift=-0.5*\off,yshift=-0.5*\off]c17) -- ([xshift=-0.5*\off,yshift=0.5*\off]c17) node[anchor=south east] {$y$};
\end{tikzpicture}
\caption{Sharp curvature in isotherms near the surface of cold vane at $\Kn=0.3$. This can be interpreted in terms of imbalance of molecules of type $A$ (cold) and type $B$ (hot) at the top-left/top-right ends of the cold vane.}
\label{fig_edgeflow_T}
\end{figure*}

Figures~\ref{fig_mikra_flowfield_966N2_nden_T_Pxy} and \ref{fig_mikra_flowfield_966N2_Qx_Qy_Speed} illustrate the contour plot of various flow properties for the highest pressure case $\Kn=0.3$ (left column) and $\Kn=1.85$ (right column). For each of these plots, the DSMC and DGFS contours have been overlaid, wherein DSMC results have been indicated by thin black lines, and DGFS results have been indicated with thick red lines. Since the flow is strictly driven by temperature gradients, we expect very small deviation in the number density from the equilibrium value of $235.8901\times 10^{21}\,m^{-3}$, as is also evident from Fig.~\ref{subfig_mikra_flowfield_966N2_nden}. In terms of temperature, in Fig.~\ref{subfig_mikra_flowfield_966N2_T}, we observe a rather familiar flow expansion, in the sense that, the hot vane dissipates heat to the surrounding acting as a source, thereby giving rise to a spiral with spiral's origin at the hot vane. Observe the interaction of contour lines (isotherms at $305K$ and $310K$) with the cold vane in the region $(250 \mu m \leq X \leq 300 \mu m,\; 25 \mu m \leq Y \leq 60 \mu m)$. We notice sharply curved isotherms near the top and right sides of the cold vane (see Fig.~\ref{fig_edgeflow_T}). Taking into account the Knudsen number of $0.3$ and the characteristic length scale of system of $20\mu m$, the Knudsen layer should extend few mean free paths from the solid surfaces i.e., $O(\lambda) \approx O(6 \mu m)$. Therefore, one should expect some temperature jump, and therefore non-linearity in the temperature in the near-wall region. More interestingly, we note an inflection in the isotherms at the top surface of the cold vane. This is essentially because the cold vane surface temperature is $304K$, while the free-stream is at $296K$. Hence, near to the heating source, say top-right end of the cold vane, the surface temperature is lower than the temperature of a layer of molecules just above the surface; and far away from the heating source, say top-left end of the cold vane, the surface temperature is higher than the temperature of a layer of molecules just above the surface. Therefore, an inflection in isotherms is expected somewhere between the top-left and top-right corner of the cold vane. 

The origin of Knudsen force can be appreciated as follows. Consider a differential area $dS$ over the cold vane as shown Fig.~\ref{fig_edgeflow_T}. The molecules impinging on the area $dS$ can be thought as made up of two types of molecules: molecules coming from colder point $A$ and molecules coming from hotter point $B$, both separated by few mean free paths. Near to the top right end of the cold vane, nearer to the hot vane, one should expect larger concentration of molecules of type $B$, and smaller concentration of molecules of type $A$. Conversely, near to the top left end of the cold vane, which is (relatively) far away from the hot vane, one should expect a smaller concentration of molecules of type $B$, and larger concentration of molecules of type $A$. Specifically, at the top left end of the cold vane, due to this imbalance of particles hitting the surface area, the momentum transferred to the surface element $dS$ is in the opposite direction to the temperature gradient; however the gas flow is induced in the direction of the temperature gradient \cite{sone2012kinetic,ibrayeva2017numerical}. This overall momentum imbalance contributes to the Knudsen force. 


Figure~\ref{subfig_mikra_flowfield_966N2_Pxy} illustrates the variation of off-diagonal ($xy$) component of stress tensor at $\Kn=0.3$. First, we note the development of four ovals/ellipses originating at the four corners/edges of the hot vane. The effect is more pronounced at the right end (top-right and bottom-right corners) of the hot vane i.e., the length of the semi-major axis is larger for the ellipses on the right. At the top-left corner of the hot vane, in particular, we observe interaction of ovals with the top-right edge of the cold vane (note the distorted shape of the oval/ellipse at the top-left boundaries of the hot vane). Since the Knudsen number is in the \textit{slip/early-transition} regime ($\Kn=0.3$), consider the expression for the stress-tensor, arrived in part by second order Chapman-Enskog expansion\cite{kogan1976stresses}:
\begin{align}
P_{ij} &= p \delta_{ij} + \tau^\one_{ij} + \tau^\two_{ij} + \hdots \nonumber \\ 
\tau^\one_{ij} &= -2 \mu \Big[ \frac{\partial u_i}{\partial x_j} \Big] \nonumber \\
\tau^\two_{ij} &= \underbrace{K_2 \frac{\mu^2}{\rho T} \Big[ \frac{\partial^2 T}{\partial x_i \partial x_j} \Big] + K_3 \frac{\mu^2}{\rho T^2} \Big[ \frac{\partial T}{\partial x_i}\frac{\partial T}{\partial x_j} \Big]}_{\tau^T_{ij}\;:\;\text{Thermal stress tensor}} \nonumber \\ 
&+ K_1 \frac{\mu^2}{\rho} \frac{\partial u_k}{\partial x_k} \Big[\frac{\partial u_i}{\partial x_j}\Big], \quad i,j,k \in \{1,2\} 
\label{eq_stress_Kogan}
\end{align}
where $P_{ij}$, $p$, $u$, $\mu$, $\rho$ are stress tensor, pressure, velocity, dynamic viscosity, and density respectively. $\delta_{ij}$ is the Kronecker delta function, $\tau_{ij}$ is the off-diagonal term of the stress tensor, and $K_i\approx1,\,i=\{1,2,3\}$ are species/molecular-interaction specific constants\cite{kogan1976stresses}. This yields 
\begin{align}
P_{12} &= \tau^\one_{12} + \tau^\two_{12} + \hdots  = P_{xy}\nonumber \\ 
\tau^T_{12} &= K_2 \frac{\mu^2}{\rho T} \Big[ \frac{\partial^2 T}{\partial x_1 \partial x_2} \Big] + K_3 \frac{\mu^2}{\rho T^2} \Big[ \frac{\partial T}{\partial x_1}\frac{\partial T}{\partial x_2} \Big] \nonumber \\
&= K_2 \frac{\mu^2}{\rho T} \Big[ \frac{\partial^2 T}{\partial x \partial y} \Big] + K_3 \frac{\mu^2}{\rho T^2} \Big[ \frac{\partial T}{\partial x}\frac{\partial T}{\partial y} \Big] 
\label{eq_thermalStress_Kogan}
\end{align}

\begin{figure*}[!ht]
\definecolor{verdigris}{rgb}{0.26, 0.7, 0.68}
\begin{tikzpicture}[scale=1]		
\def\Ho{20};
\def\L{600 / \Ho};  
\def\H{300 / \Ho};  
\def\hG{10 / \Ho};  
\def\hW{4 / \Ho};   
\def\Lh{50 / \Ho};  
\def\Hh{50 / \Ho};  
\def \off{20};
\def \dt{0.8};
\def \pw{10};
\coordinate (c17) at ({\L/2-\hG-\Lh}, {\hW});
\coordinate (c18) at ({\L/2-\hG}, {\hW});
\coordinate (c19) at ({\L/2-\hG}, {\hW+\Hh});
\coordinate (c20) at ({\L/2-\hG-\Lh}, {\hW+\Hh});
\coordinate (c21) at ({\L/2+\hG}, {\hW});
\coordinate (c22) at ({\L/2+\hG+\Lh}, {\hW});
\coordinate (c23) at ({\L/2+\hG+\Lh}, {\hW+\Hh});
\coordinate (c24) at ({\L/2+\hG}, {\hW+\Hh});
\draw[dashed] (c17) -- (c18);
\draw[dashed] (c21) -- (c22);
\draw[dashed] (c17) -- (c20);
\draw[dashed] (c18) -- (c19);
\draw[dashed] (c21) -- (c24);
\draw[dashed] (c22) -- (c23);
\draw[dashed] (c20) -- (c19);
\draw[dashed] (c24) -- (c23);
\draw[pattern=north west lines, pattern color=black!50, line width = 0.1mm, thin, draw=none, fill=blue!60] (c17) -- (c18) -- (c19) -- (c20) -- (c17) node[anchor=south west, color=white] {\large $\mathbf{T_C}$};
\draw[pattern=north west lines, pattern color=black!50, line width = 0.1mm, thin, draw=none, fill=red!60] (c21) -- (c22) -- (c23) -- (c24) -- (c21) node[anchor=south west, color=white] {\large $\mathbf{T_H}$};
\draw[fill=blue!60, draw=verdigris] (c19) circle (3pt) node[anchor=south east] {\textcolor{blue!60}{\bf A}};
\draw[fill=red!60, draw=verdigris] (c24) circle (3pt) node[anchor=south west] {\textcolor{red!60}{\bf B}};
\draw[line width=\dt, <->, shorten >=2pt, shorten <=2pt,  black!60] ([yshift=-0.8*\off]c19) -- ([yshift=-0.8*\off]c24) node[anchor=south, pos=0.5, yshift=-1*\off, fill=white, rounded corners=3pt] {$\approx 10K \, \text{difference}$};
\draw[fill=verdigris, draw=verdigris] ([yshift=\off]$ (c24)!0.5!(c19) $) circle (3pt) node[anchor=south east] {\textcolor{verdigris}{\bf C}};
\draw[fill=magenta, draw=magenta] ($ (c24)!0.5!(c19) $) circle (3pt) node[anchor=north, yshift=-0.1*\off] {\textcolor{magenta}{\bf D}};
\draw[line width=\dt, <->, shorten >=5pt, shorten <=5pt,  black!60] ([yshift=\off]$ (c24)!0.5!(c19) $) -- ($ (c24)!0.5!(c19) $) node[anchor=west, pos=0.5, yshift=0.3*\off, xshift=0.15*\off, fill=white, rounded corners=3pt] {$\approx 5K \, \text{difference}$};
\node[inner sep=0pt] (russell) at ([xshift=6*\off, yshift=-2*\off]c23) {\includegraphics[width=.5\textwidth]{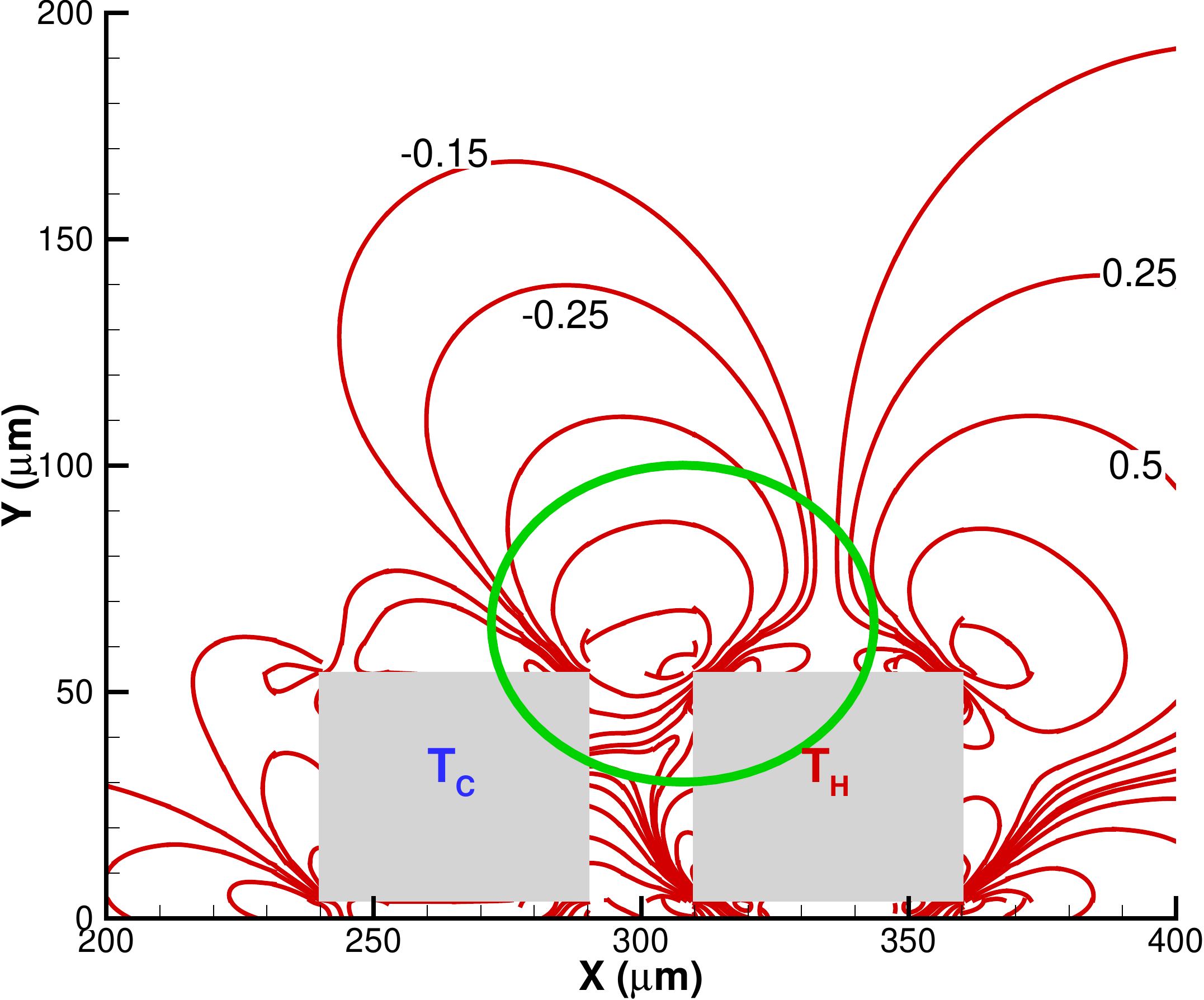}};
\draw[-latex, line width=0.5mm, magenta] ([xshift=-0.5*\off,yshift=-0.5*\off]c17) -- ([xshift=0.5*\off,yshift=-0.5*\off]c17) node[anchor=north] {$x$};
\draw[-latex, line width=0.5mm, magenta] ([xshift=-0.5*\off,yshift=-0.5*\off]c17) -- ([xshift=-0.5*\off,yshift=0.5*\off]c17) node[anchor=south east] {$y$};
\end{tikzpicture}
\caption{$xy$ component of stress tensor at $\Kn=0.3$: origin of oval/ellipses at the edges of the vanes. Note the distorted shape of the ellipse between the top-right corner of cold vane and top-left corner of hot vane. Since the temperature gradient is stronger between point A and B, compared to point C and D, we expect the semi-major axis of the ellipse to be larger than the semi-minor axis, and hence the distorted ellipse/oval -- an observation consistent with Eqs.~(\ref{eq_stress_Kogan}, \ref{eq_thermalStress_Kogan}) since $\partial T/\partial x \gg \partial T/\partial y$.}
\label{fig_edgeflow_Pxy}
\end{figure*}

Let us consider four points in the flow: $A$ (top-right corner of cold vane), $B$ (top-left corner of hot vane), $C$ (third vertex of equilateral triangle $\Delta ABC$ s.t. $\overrightarrow{BC}\times\overrightarrow{CA}/\|\overrightarrow{BC}\times\overrightarrow{CA}\|=\hat{k}$), and $D$ (mid point of $A$ and $B$) as shown in Fig.~\ref{fig_edgeflow_Pxy}. Based on isotherms in Fig.~\ref{fig_edgeflow_T}, it can be inferred that the temperature difference between points $A$ and $B$ is $\approx10K$, whereas the temperature difference between points $C$ and $D$ is $\approx 5K$. Consistent with the Eq.~\ref{eq_thermalStress_Kogan}, theoretically, we expect the thermal stresses (and therefore $P_{xy}$) to be larger between points $A$ and $B$ since $\partial T/\partial x|_{AB} \gg \partial T/\partial y|_{CD}$ (more formally: $\|\nabla T\|_{AB} \gg \|\nabla T\|_{CD}$, $\|\nabla^2 T\|_{AB} \gg \|\nabla^2 T\|_{CD}$). Hence, the distorted ellipse. A more subtle observation is as follows: Why, \textit{precisely}, should an isocontour line of $xy$ component of stress, start from top-left corner of the hot-vane (i.e., point $B$) and end at the top-right corner of cold-vane (i.e., point $A$). What happens to the entire flow field if we introduce roughness on the walls, or smooth the vane corners--few questions that we delegate to a future study.

Next, Figs.~\ref{subfig_mikra_flowfield_966N2_Qx}, and \ref{subfig_mikra_flowfield_966N2_Qy} depict the variation of $x$ and $y$ components of heat flux. We want to reemphasize that DSMC simulations consider the rotational degrees of freedom of $N_2$ into account, whereas DGFS, being in very early stages of research, doesn't. Nevertheless, we observe a fair agreement between DSMC and DGFS. In Fig.~\ref{subfig_mikra_flowfield_966N2_Qx}, in the region $(250 \mu m \leq X \leq 320 \mu m,\; 50 \mu m \leq Y \leq 90 \mu m)$, we again note presence of iso-contour lines between the top-left and top-right corners of the cold and hot vanes. A more subtle observation is as follows: Multiple iso-contours, for instance $-2000\,W/m^2$, $-4000\,W/m^2$, $-6000\,W/m^2$ (the unlabeled contour just below $-4000\,W/m^2$ iso-contour), differing by large magnitudes, start at \textit{approximately} the top-left corner of the hot vane, and end at the top-right corner of the cold vane, resulting in sharply curved isocontours. A partial explanation of such effects appears in Ref.~\onlinecite{sone1997demonstration}, wherein the author attributed the observation to simply \textit{edge effects}, basing the argument on the imbalance of particles of type A (cold) and type B (hot) near to the edges, as was mentioned earlier in the discussion. 

\begin{figure*}[!ht]
\centering
\begin{subfigure}{.5\textwidth}
  \centering
  \includegraphics[width=70mm]{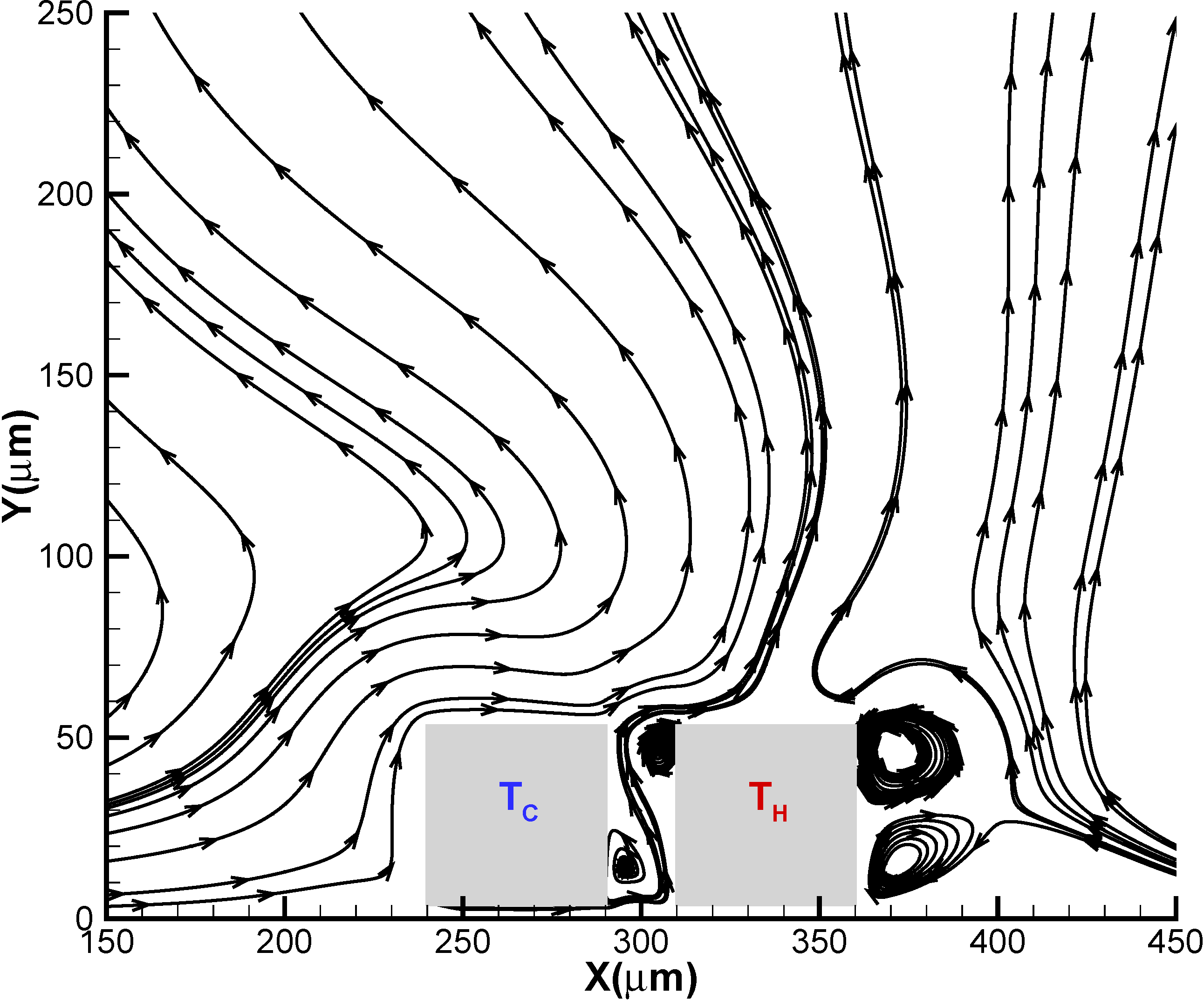}
  \caption{$\Kn=0.3$, Flow streamlines, $t=1.25\times 10^{-3}\,sec$}
  \label{subfig_mikra_speed_evolution_50}
\end{subfigure}%
\begin{subfigure}{.5\textwidth}
  \centering
  \includegraphics[width=70mm]{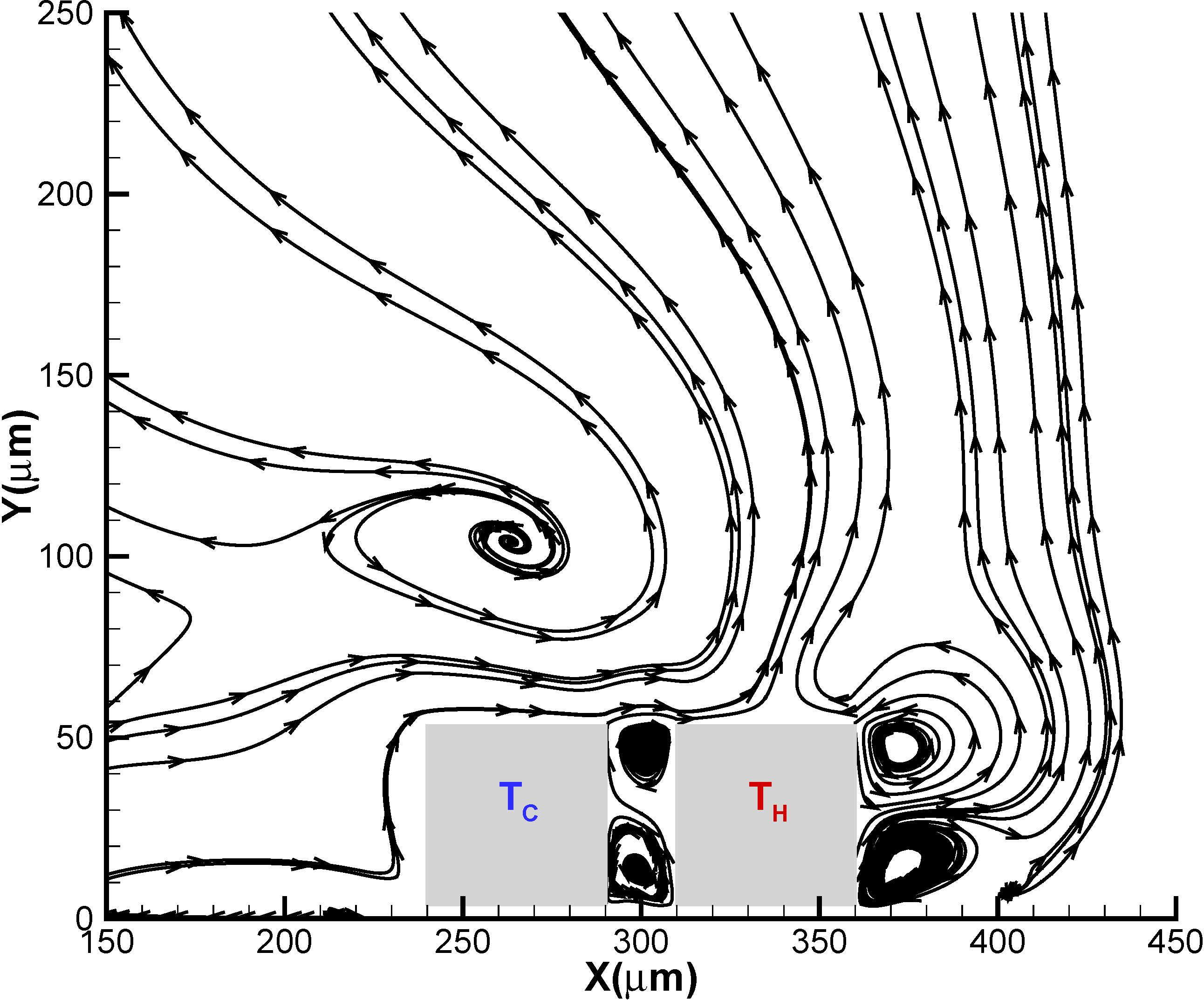}
  \caption{$\Kn=0.3$, Flow streamlines, $t=2.5\times 10^{-3}\,sec$}
\end{subfigure}%

\begin{subfigure}{.5\textwidth}
  \centering
  \includegraphics[width=70mm]{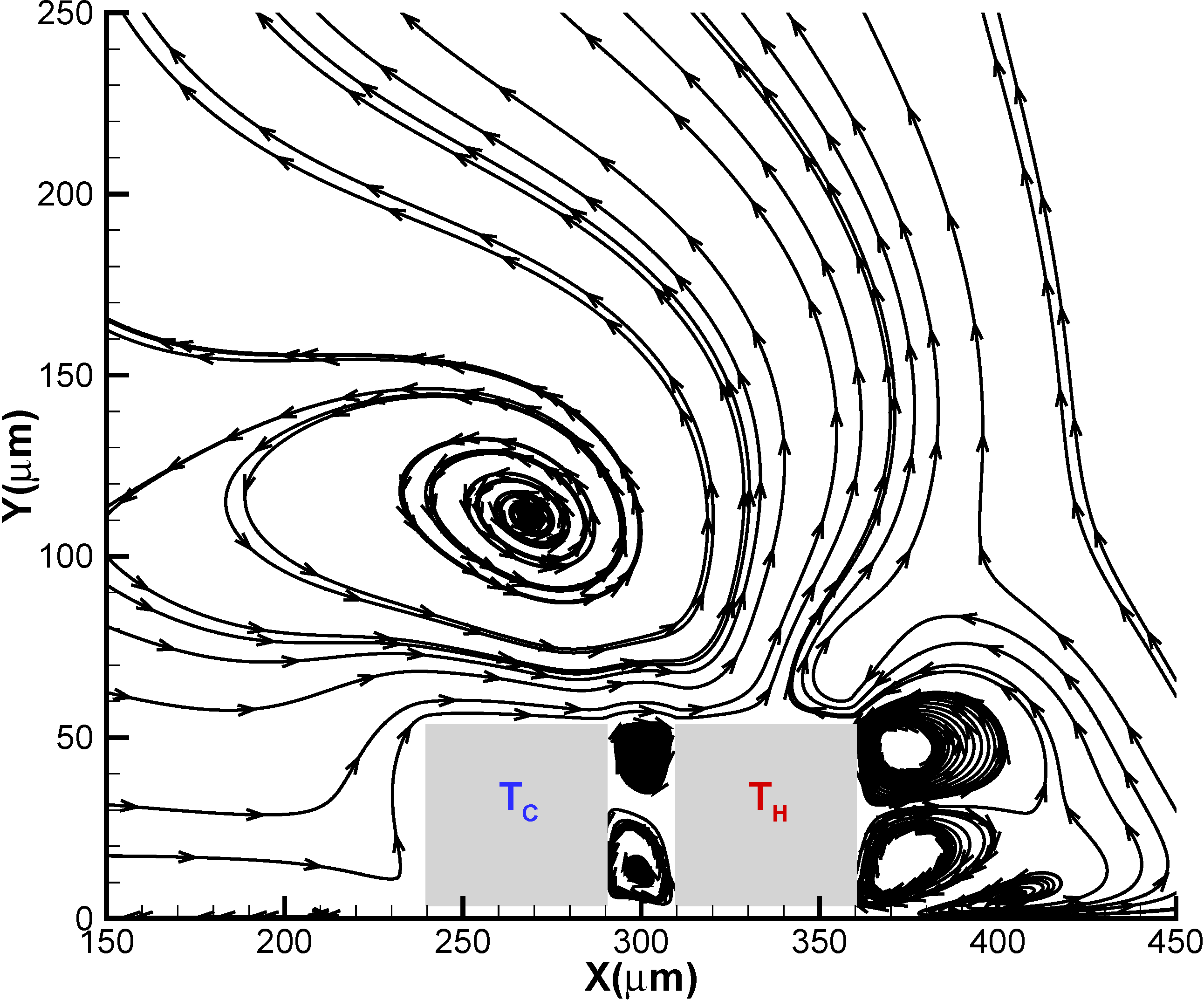}
  \caption{$\Kn=0.3$, Flow streamlines, $t=3.75\times 10^{-3}\,sec$}
\end{subfigure}%
\begin{subfigure}{.5\textwidth}
  \centering
  \includegraphics[width=70mm]{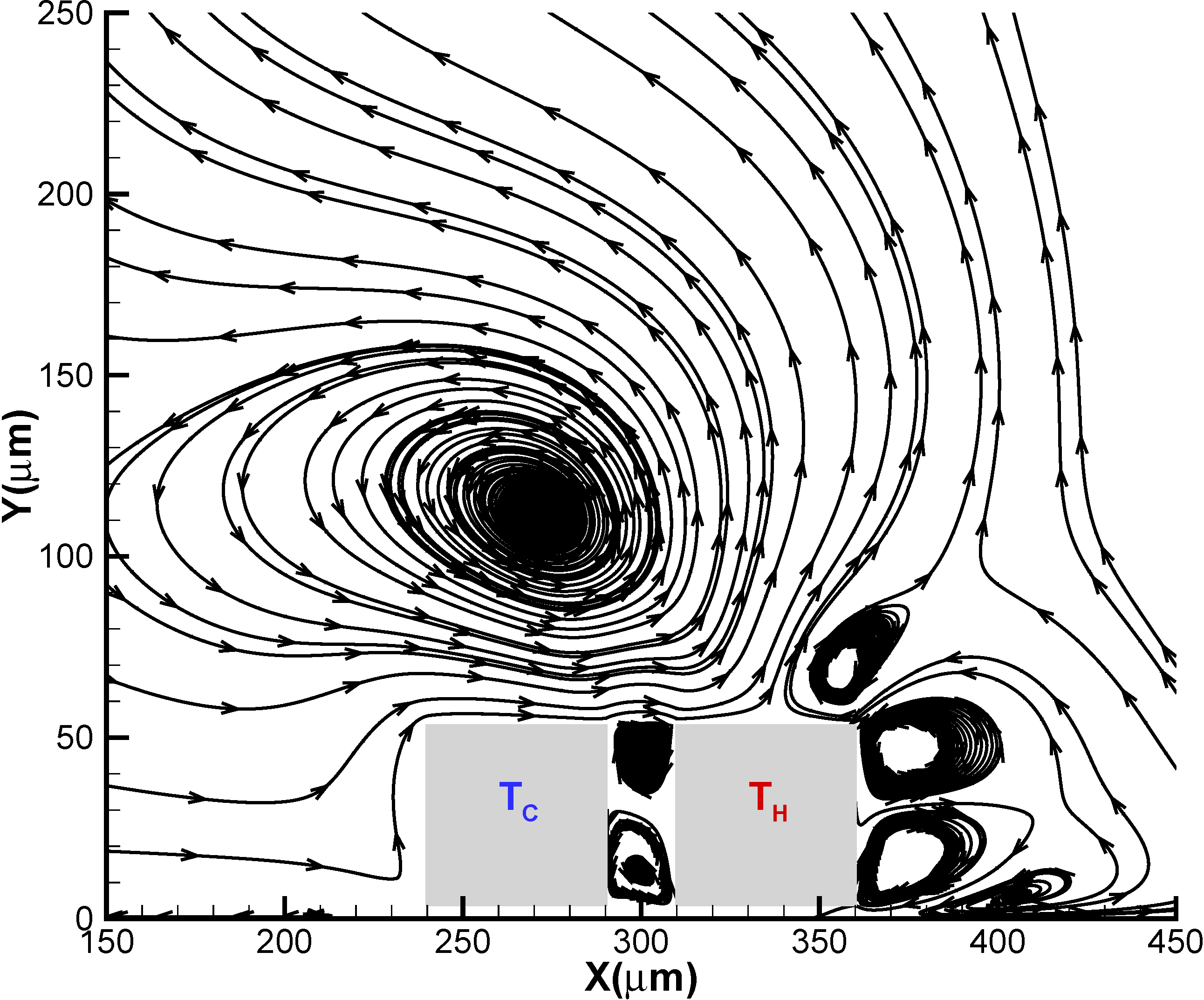}
  \caption{$\Kn=0.3$, Flow streamlines, $t=4.375\times 10^{-3}\,sec$}
\end{subfigure}
\caption{Instantaneous streamlines near the vanes of MIKRA Gen1 device at $\Kn=0.30$ obtained from DGFS using VHS collision model. Observe the vortex formation above the cold vane, and top right corner of hot vane.}
\label{fig_mikra_speed_evolution}
\end{figure*}

Figure~\ref{subfig_mikra_flowfield_966N2_Speed} illustrates the flow speed in the domain. We notice significant statistical fluctuations in DSMC (thin black lines), to an extent that removing DGFS contour lines in red, would make it difficult, if not impossible, to decipher the overall flow structure. A more complete picture of the flow is presented through transient DGFS streamlines in Figs.~\ref{fig_mikra_speed_evolution}. First, we note the streamlines pointing in the upward direction. This is essentially due to the heating of the molecules (and therefore the thermal energy imparted to them) in the lower portions of the domain. In the process, four characteristic vortexes appear at the four corners of the heated vane, relatively early during the course of the simulation, for instance, see Fig.~\ref{subfig_mikra_speed_evolution_50} at $1.25\,ms$. Over the time, secondary vortexes appear in the flow, most notably, a larger vortex at the top of the cold vane, and a smaller vortex near the top-right corner of hot vane. 

\begin{figure*}[!ht]
\centering
\begin{subfigure}{.5\textwidth}
  \centering
  \includegraphics[width=70mm]{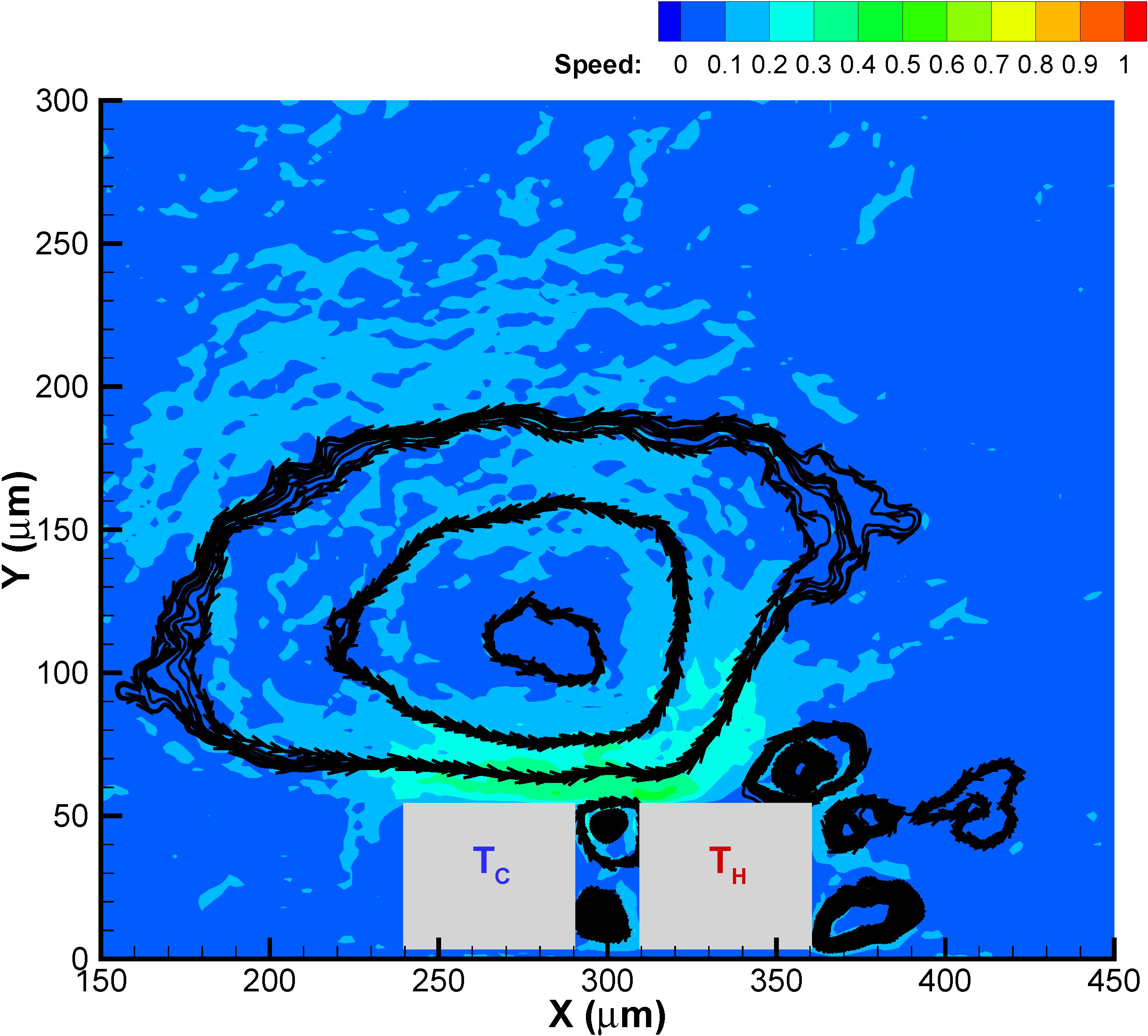}
  \caption{Speed ($m/s$), $\Kn=0.30$, DSMC}
\end{subfigure}%
\begin{subfigure}{.5\textwidth}
  \centering
  \includegraphics[width=70mm]{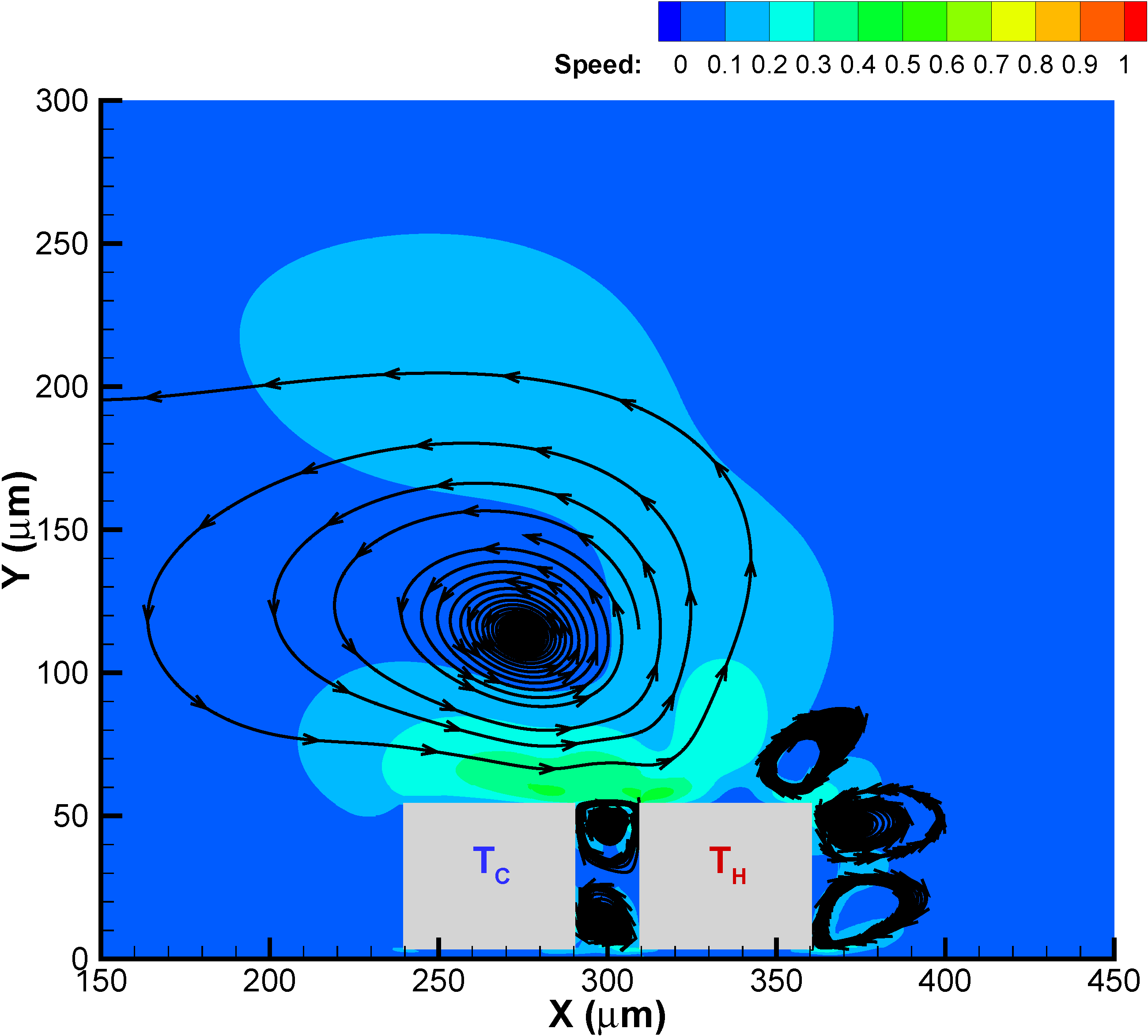}
  \caption{Speed ($m/s$), $\Kn=0.30$, DGFS}
\end{subfigure}
\begin{subfigure}{.5\textwidth}
  \centering
  \includegraphics[width=70mm]{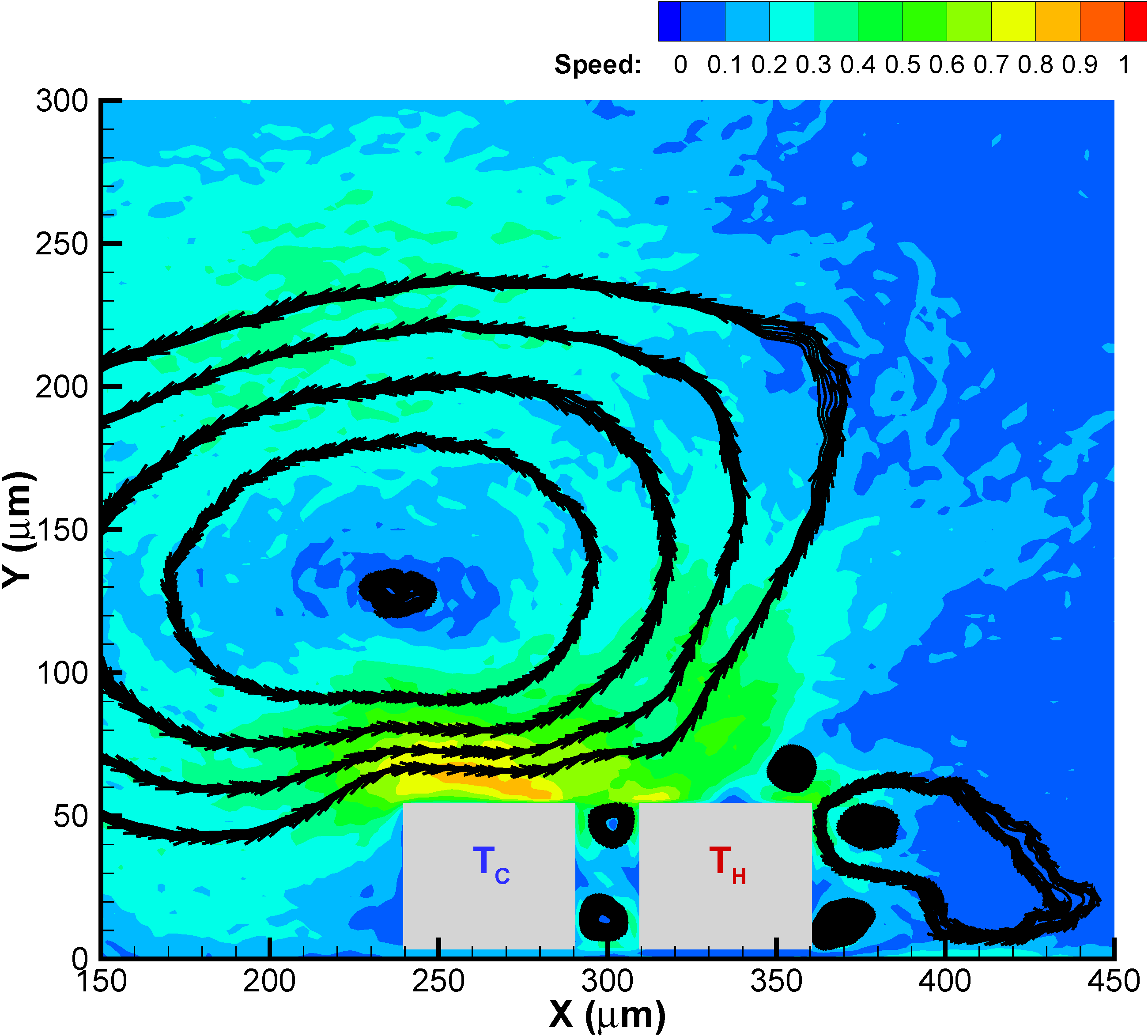}
  \caption{Speed ($m/s$), $\Kn=0.74$, DSMC}
\end{subfigure}%
\begin{subfigure}{.5\textwidth}
  \centering
  \includegraphics[width=70mm]{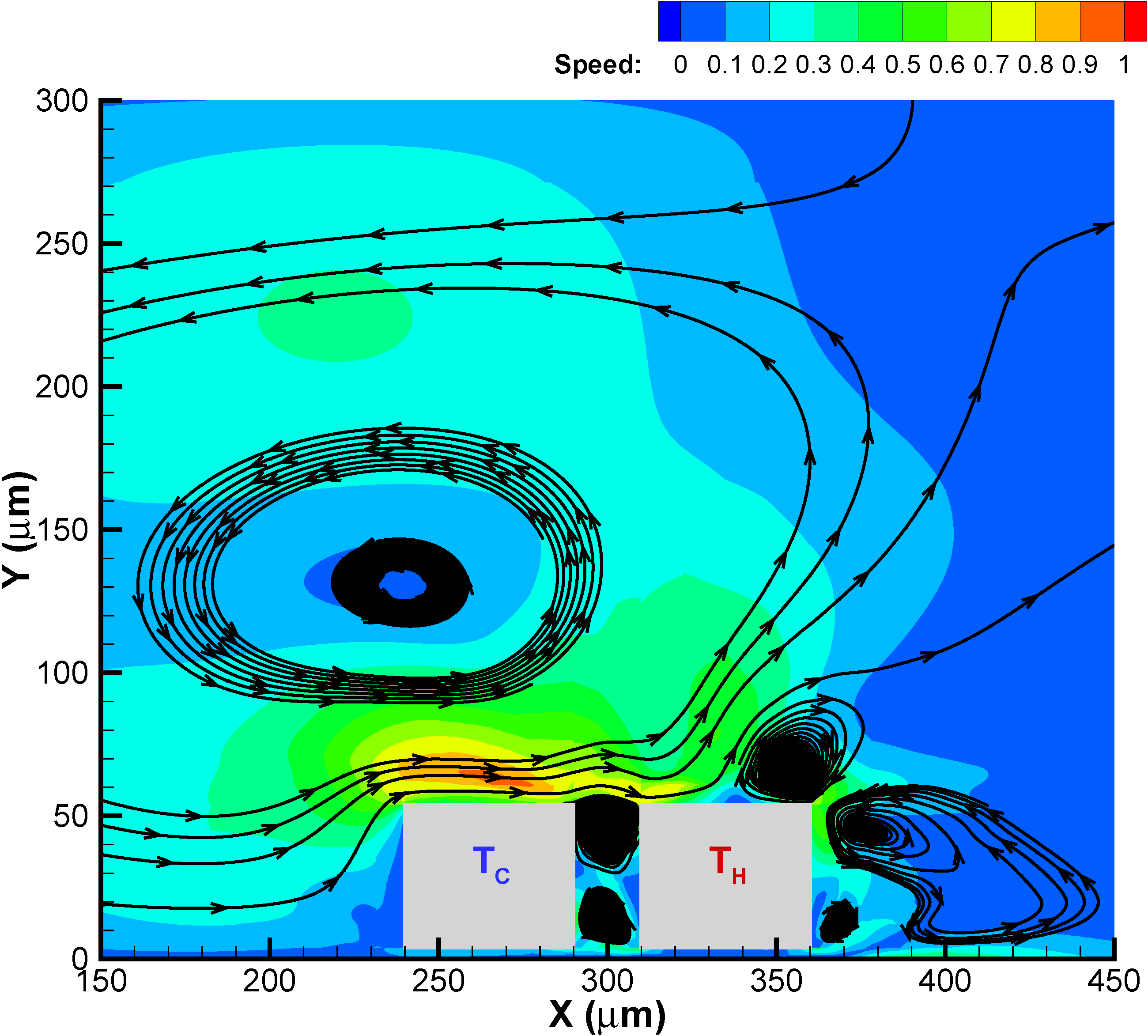}
  \caption{Speed ($m/s$), $\Kn=0.74$, DGFS}
\end{subfigure}
\begin{subfigure}{.5\textwidth}
  \centering
  \includegraphics[width=70mm]{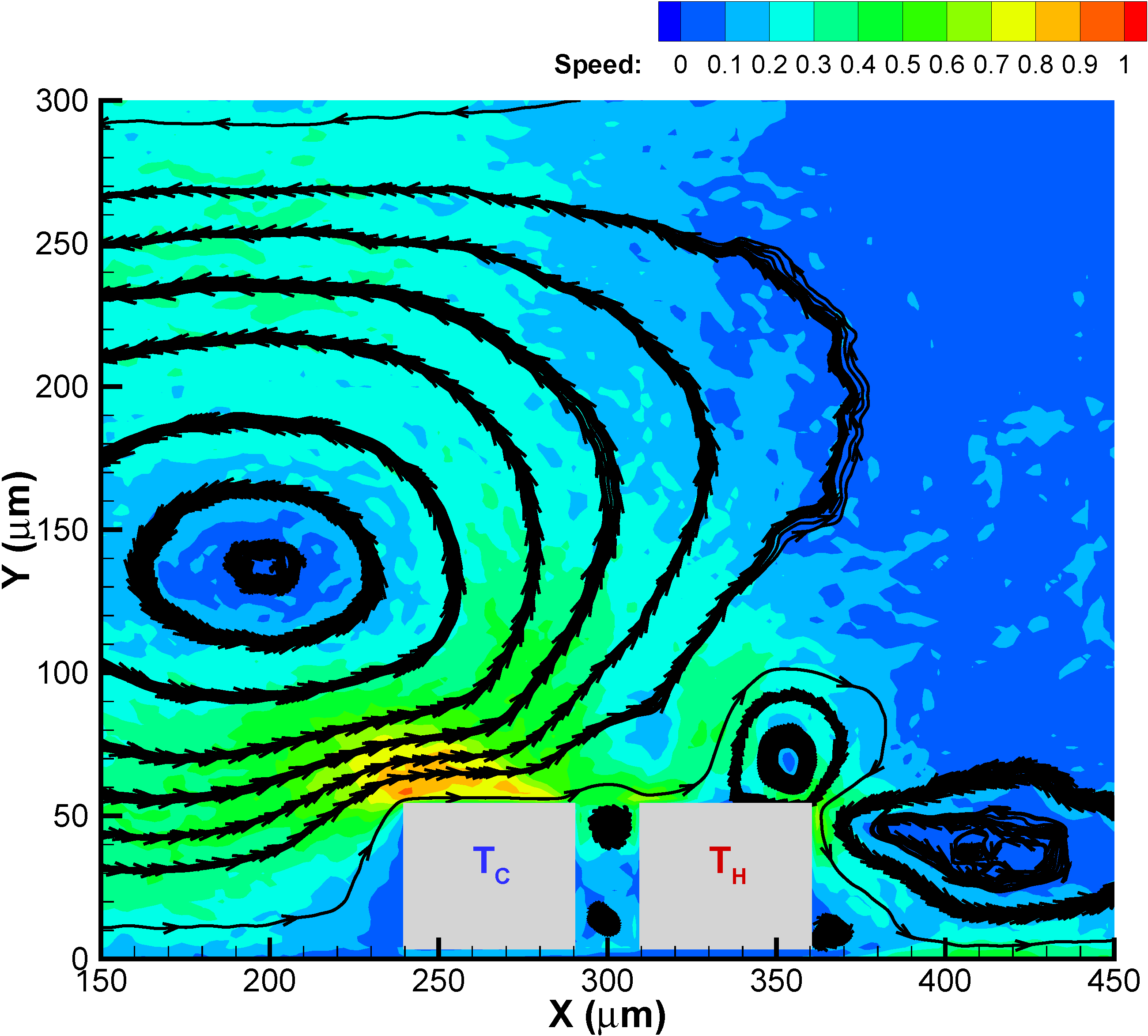}
  \caption{Speed ($m/s$), $\Kn=1.85$, DSMC}
\end{subfigure}%
\begin{subfigure}{.5\textwidth}
  \centering
  \includegraphics[width=70mm]{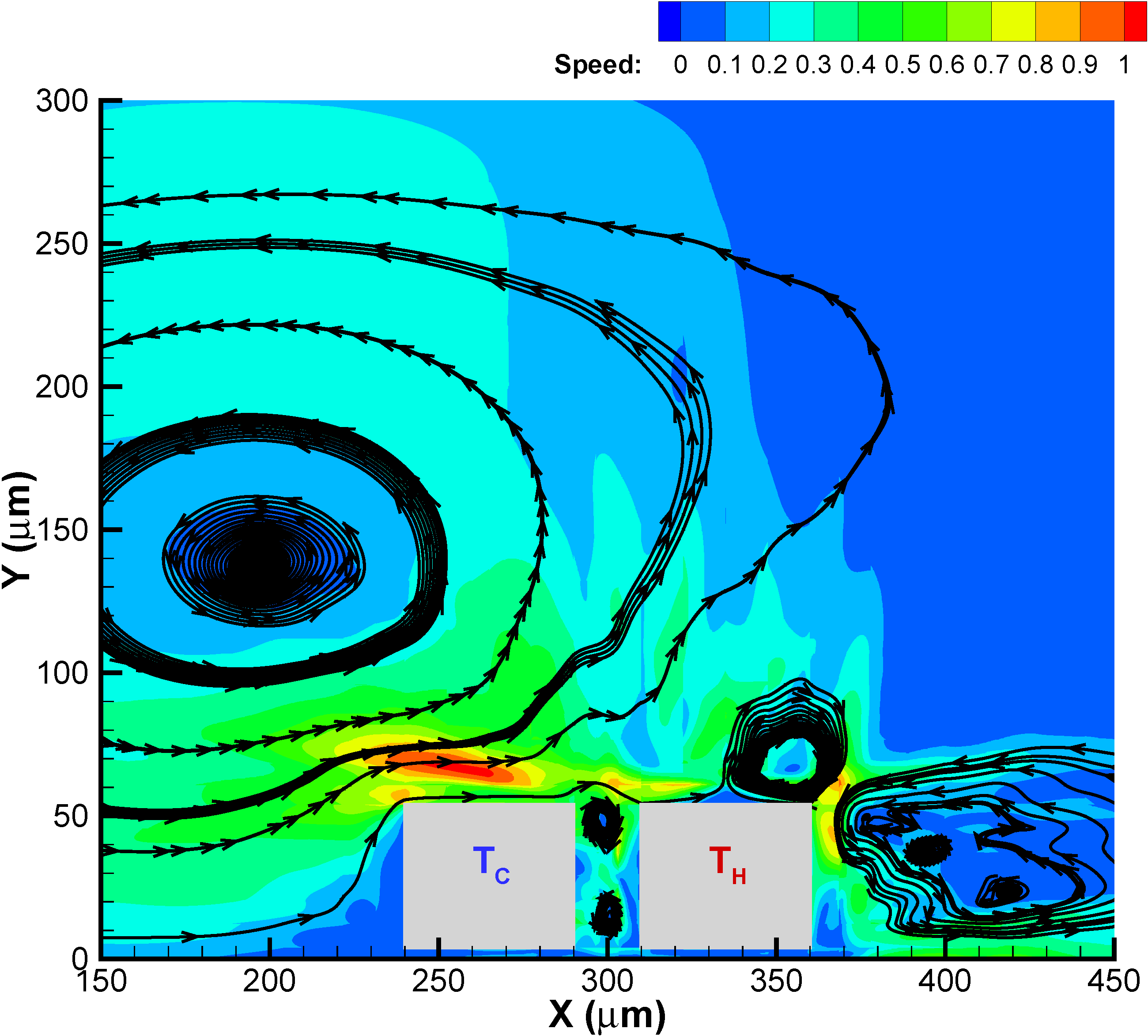}
  \caption{Speed ($m/s$), $\Kn=1.85$, DGFS}
\end{subfigure}
\caption{Variation of flow speed at steady state for MIKRA Gen1 cases obtained from DSMC and DGFS using VHS collision model.}
\label{fig_mikra_speed}
\end{figure*}

Figure~\ref{fig_mikra_speed} shows the \textit{steady state} speed contours at different Knudsen numbers with the corresponding flow streamlines overlaid. With increase in Knudsen number from $\Kn=0.3$ to $\Kn=0.74$, we note sharp increase in flow velocity, approximately by a factor of two. Consequently, the vortexes grow in size. The change in flow speed, however, from $\Kn=0.74$ to $\Kn=1.85$, although appreciable, is relatively mild. 

\begin{figure*}[!ht]
\centering
\begin{subfigure}{.5\textwidth}
  \centering
  \includegraphics[width=75mm]{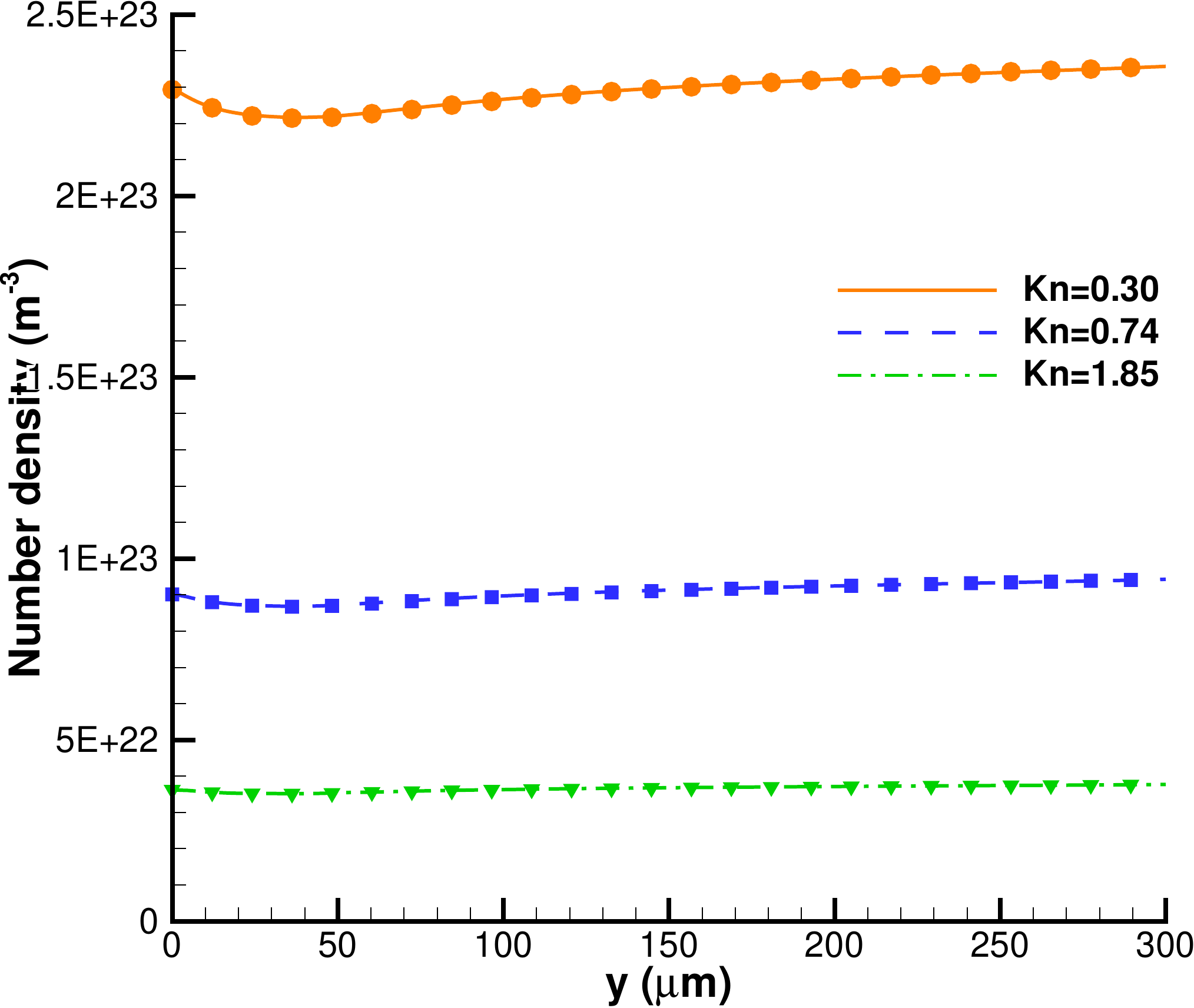}
  \caption{Number density (on vertical centerline)}
  \label{subfig_mikra_plotoverline_nden}
\end{subfigure}%
\begin{subfigure}{.5\textwidth}
  \centering
  \includegraphics[width=75mm]{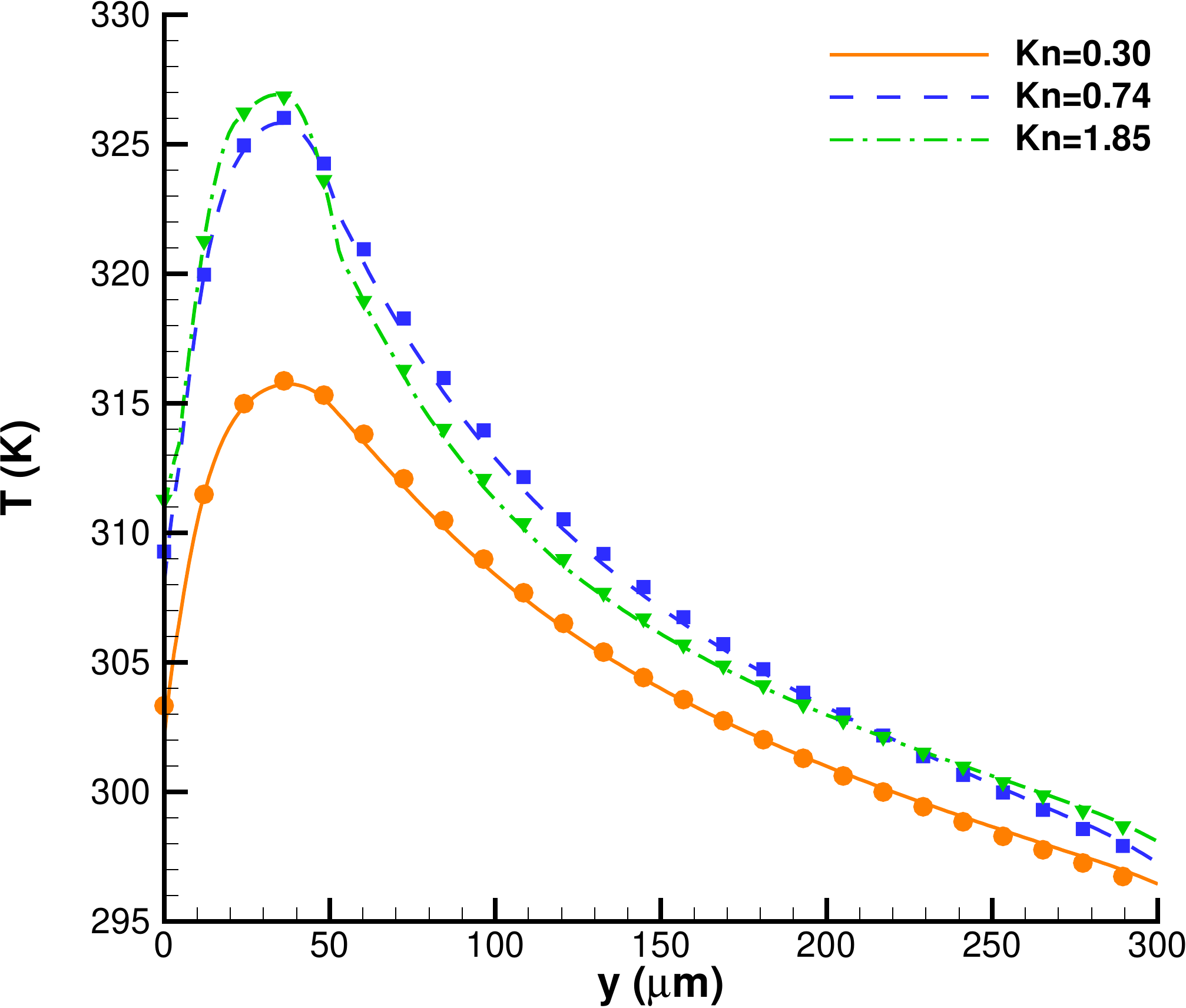}
  \caption{Temperature (on vertical centerline)}
  \label{subfig_mikra_plotoverline_T}
\end{subfigure}
\begin{subfigure}{.5\textwidth}
  \centering
  \includegraphics[width=75mm]{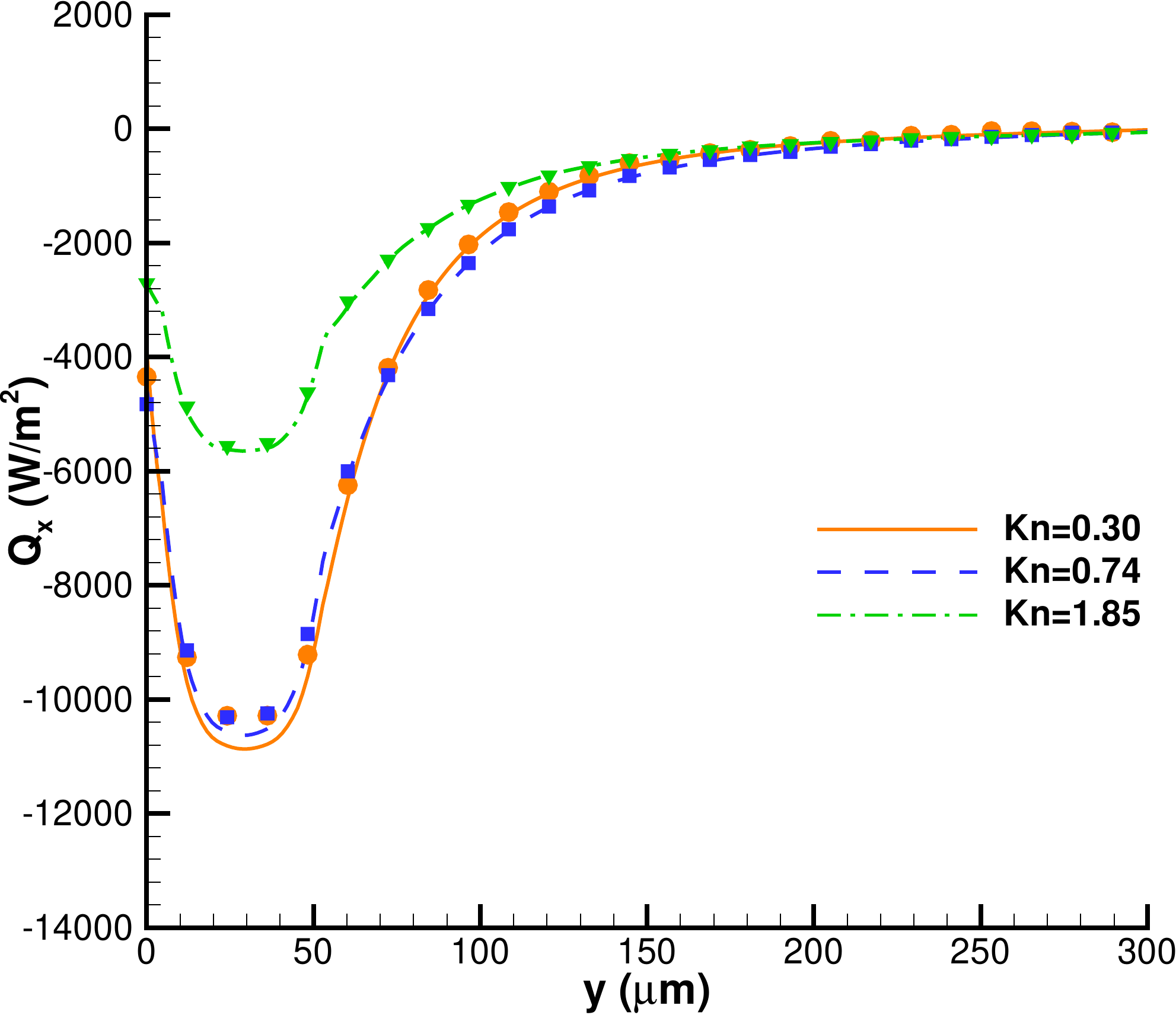}
  \caption{$x$-component of heat-flux (on vertical centerline)}
  \label{subfig_mikra_plotoverline_Qx}
\end{subfigure}%
\begin{subfigure}{.5\textwidth}
  \centering
  \includegraphics[width=75mm]{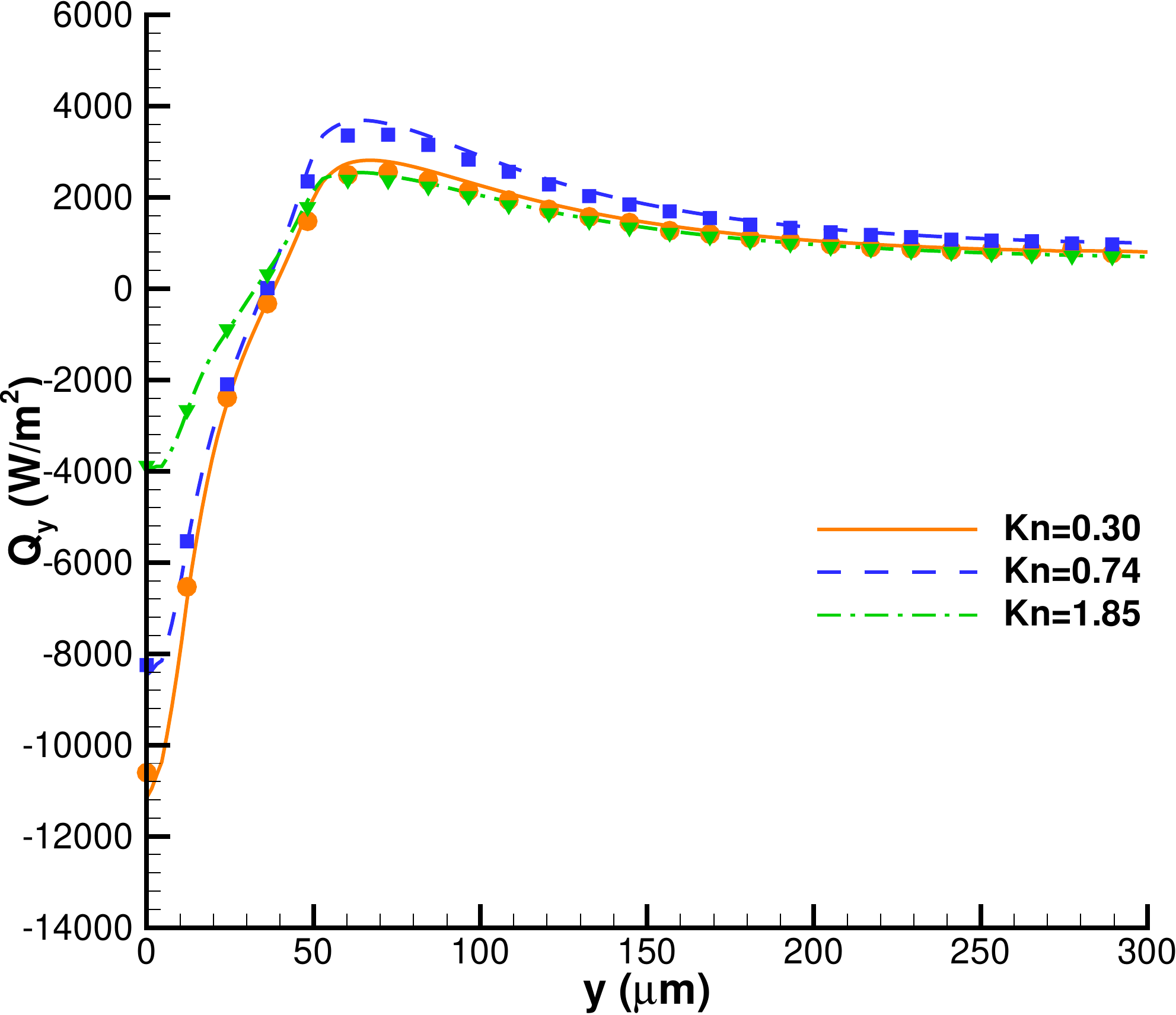}
  \caption{$y$-component of heat-flux (on vertical centerline)}
  \label{subfig_mikra_plotoverline_Qy}
\end{subfigure}
\begin{subfigure}{.5\textwidth}
  \centering
  \includegraphics[width=75mm]{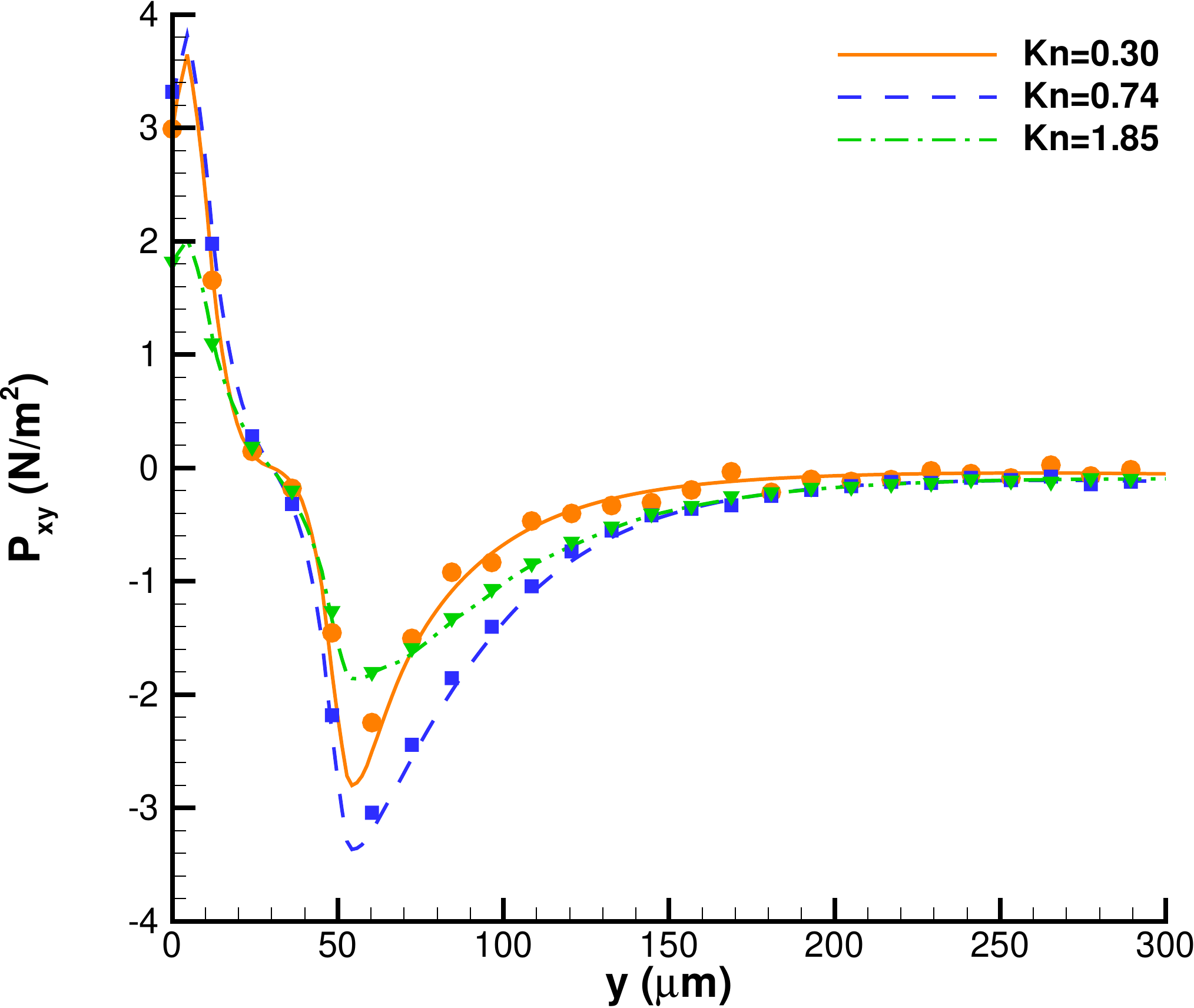}
  \caption{$xy$-component of stress (on vertical centerline)}
  \label{subfig_mikra_plotoverline_Pxy}
\end{subfigure}%
\begin{subfigure}{.5\textwidth}
  \centering
  \includegraphics[width=75mm]{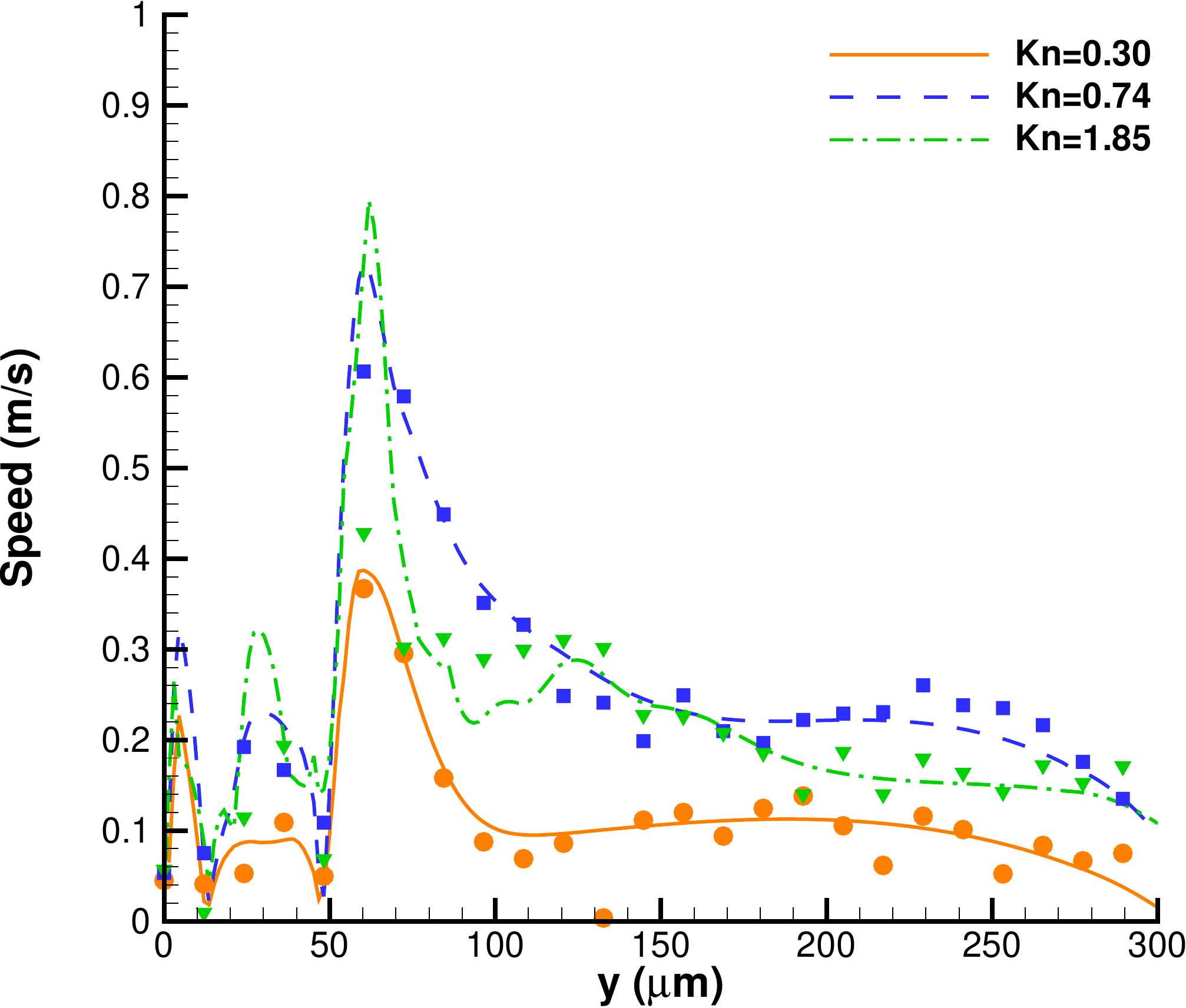}
  \caption{Speed (on vertical centerline)}
  \label{subfig_mikra_plotoverline_Speed}
\end{subfigure}
\caption{Variation of flow properties along the domain vertical centerline ($X=300\mu m$) for MIKRA Gen1 cases obtained from DSMC (symbols) and DGFS (lines) using VHS collision model.}
\label{fig_mikra_plotoverline}
\end{figure*}

\begin{figure*}[!ht]
\centering
\begin{subfigure}{.5\textwidth}
  \centering
  \includegraphics[width=75mm]{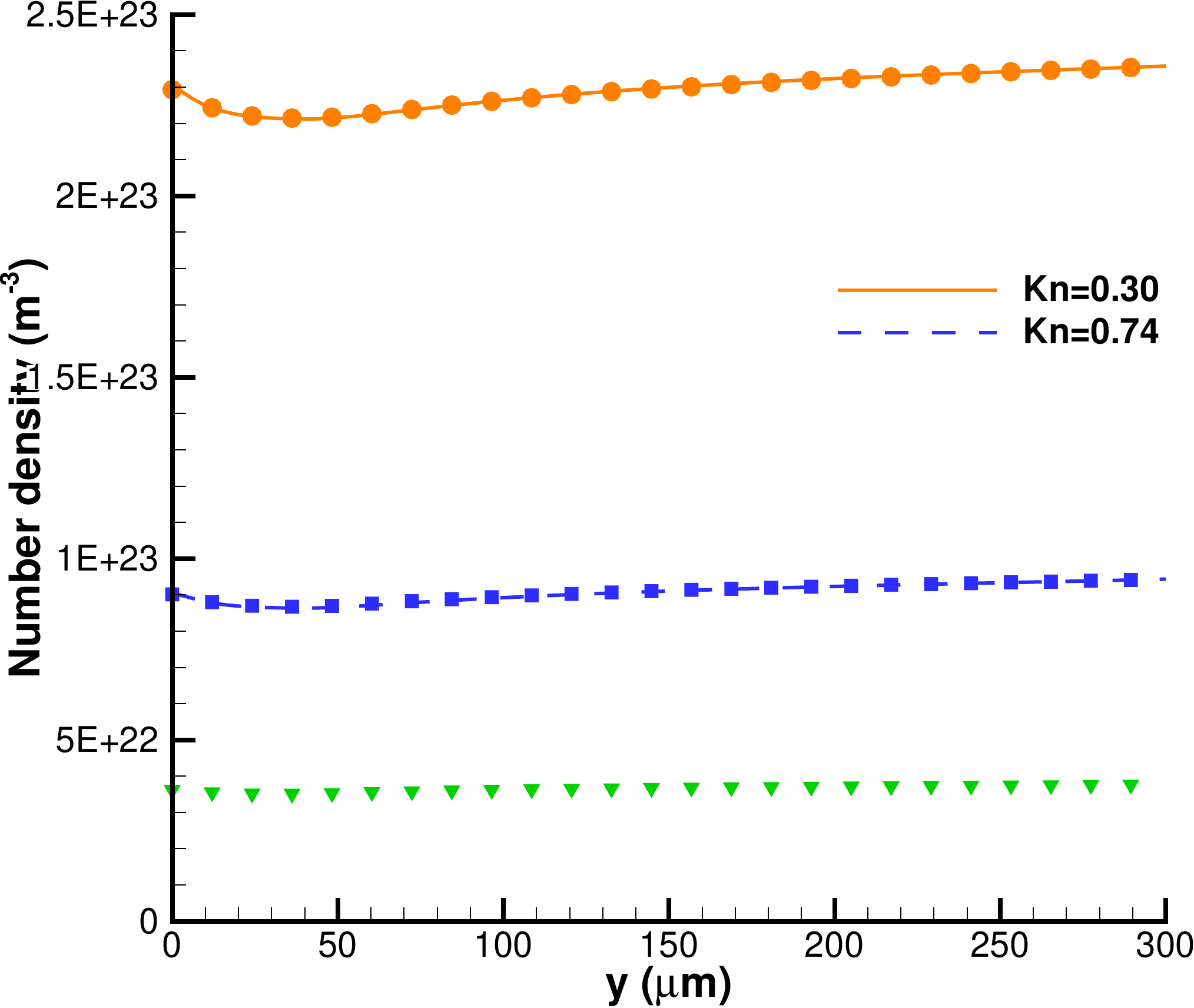}
  \caption{Number density (on vertical centerline)}
  \label{subfig_mikra_plotoverline_nden_BGK}
\end{subfigure}%
\begin{subfigure}{.5\textwidth}
  \centering
  \includegraphics[width=75mm]{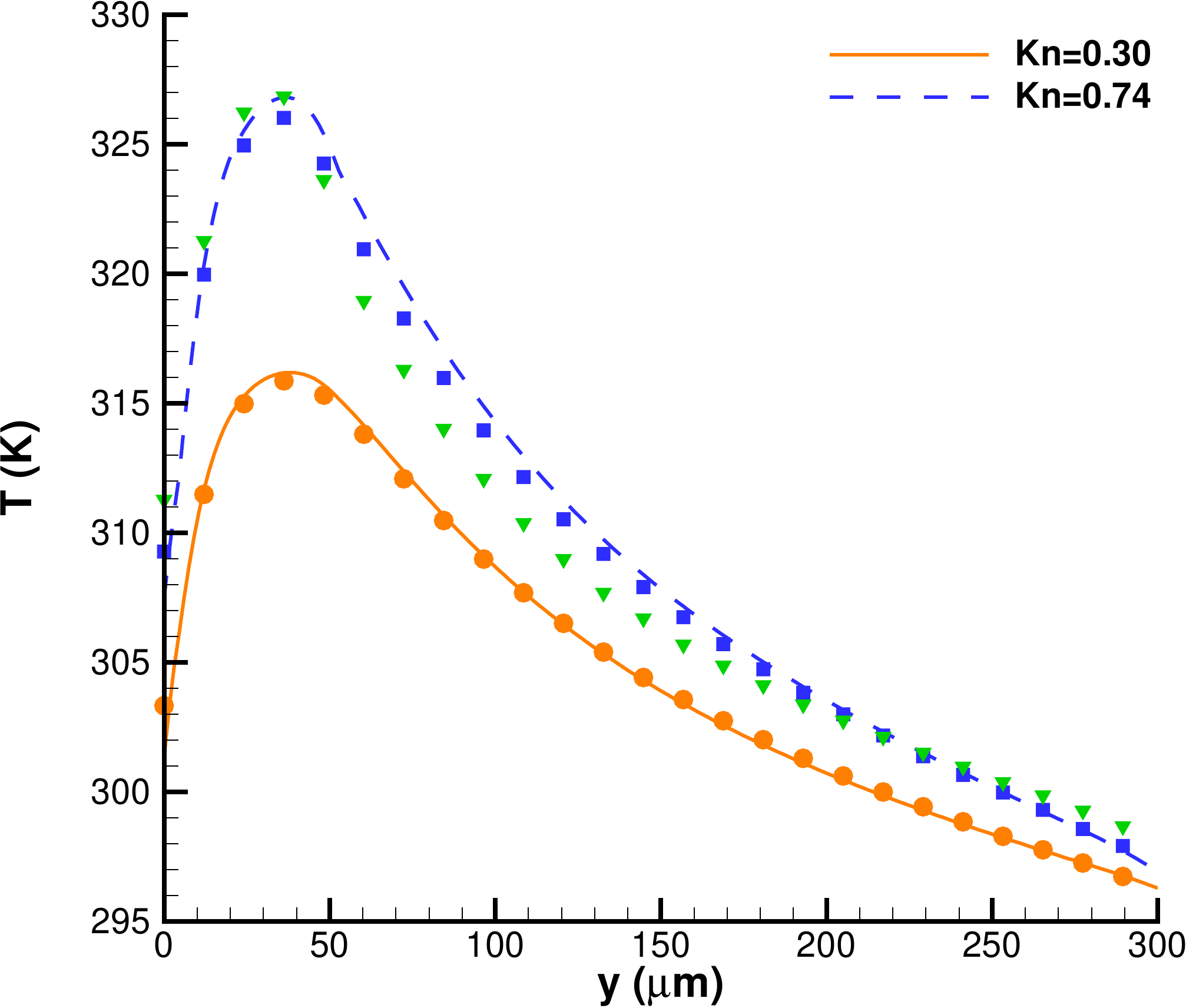}
  \caption{Temperature (on vertical centerline)}
  \label{subfig_mikra_plotoverline_T_BGK}
\end{subfigure}
\begin{subfigure}{.5\textwidth}
  \centering
  \includegraphics[width=75mm]{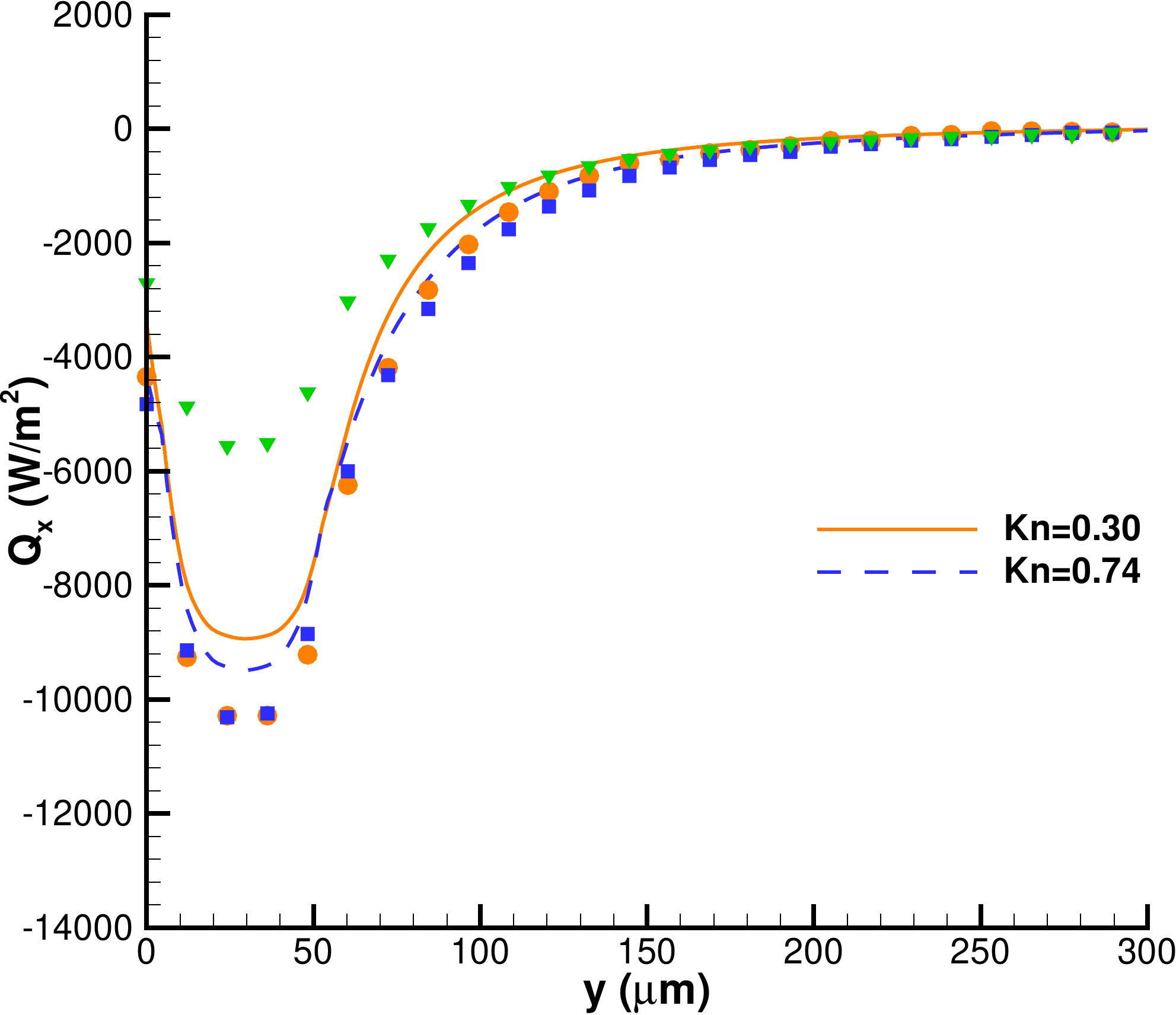}
  \caption{$x$-component of heat-flux (on vertical centerline)}
  \label{subfig_mikra_plotoverline_Qx_BGK}
\end{subfigure}%
\begin{subfigure}{.5\textwidth}
  \centering
  \includegraphics[width=75mm]{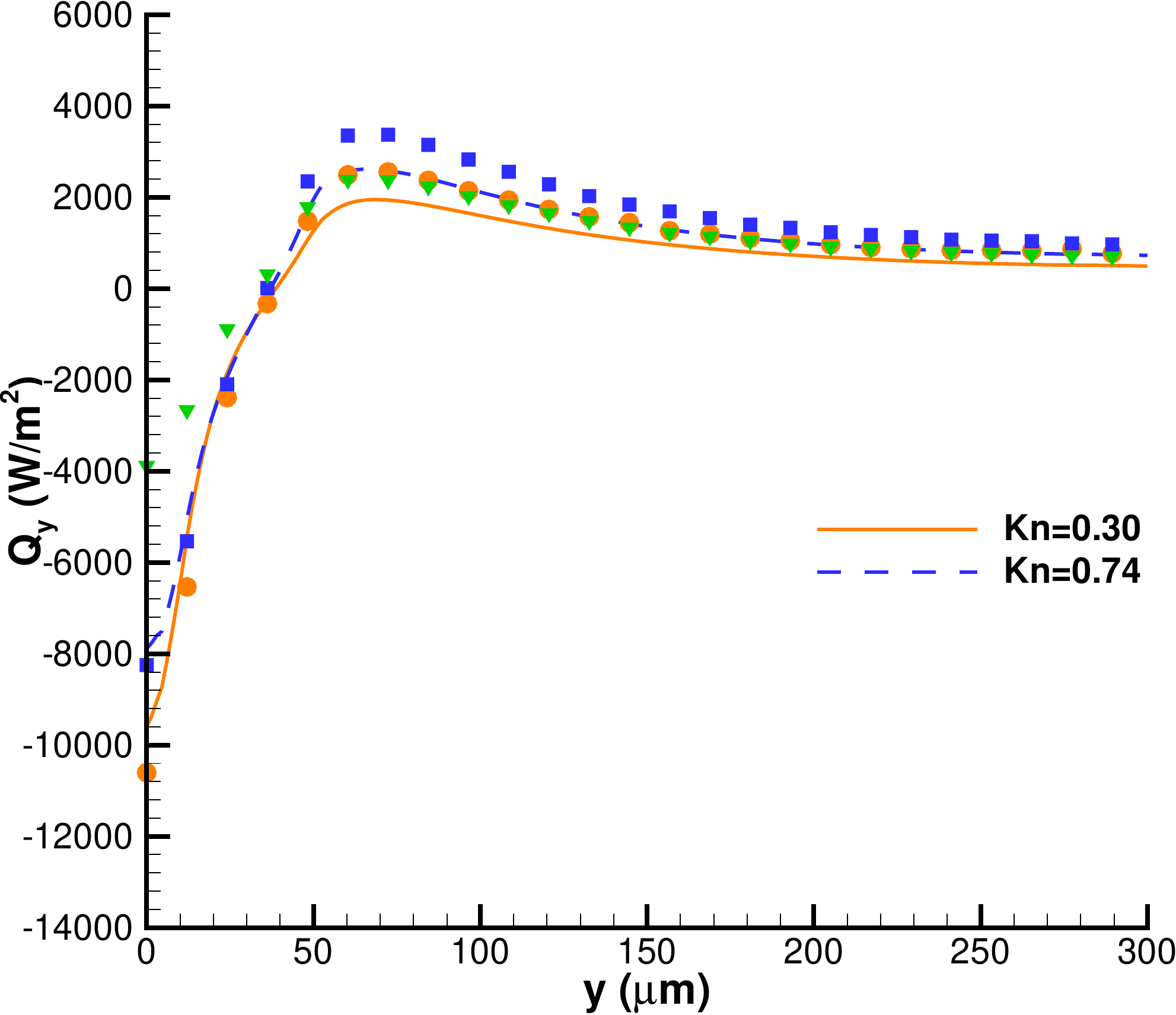}
  \caption{$y$-component of heat-flux (on vertical centerline)}
  \label{subfig_mikra_plotoverline_Qy_BGK}
\end{subfigure}
\begin{subfigure}{.5\textwidth}
  \centering
  \includegraphics[width=75mm]{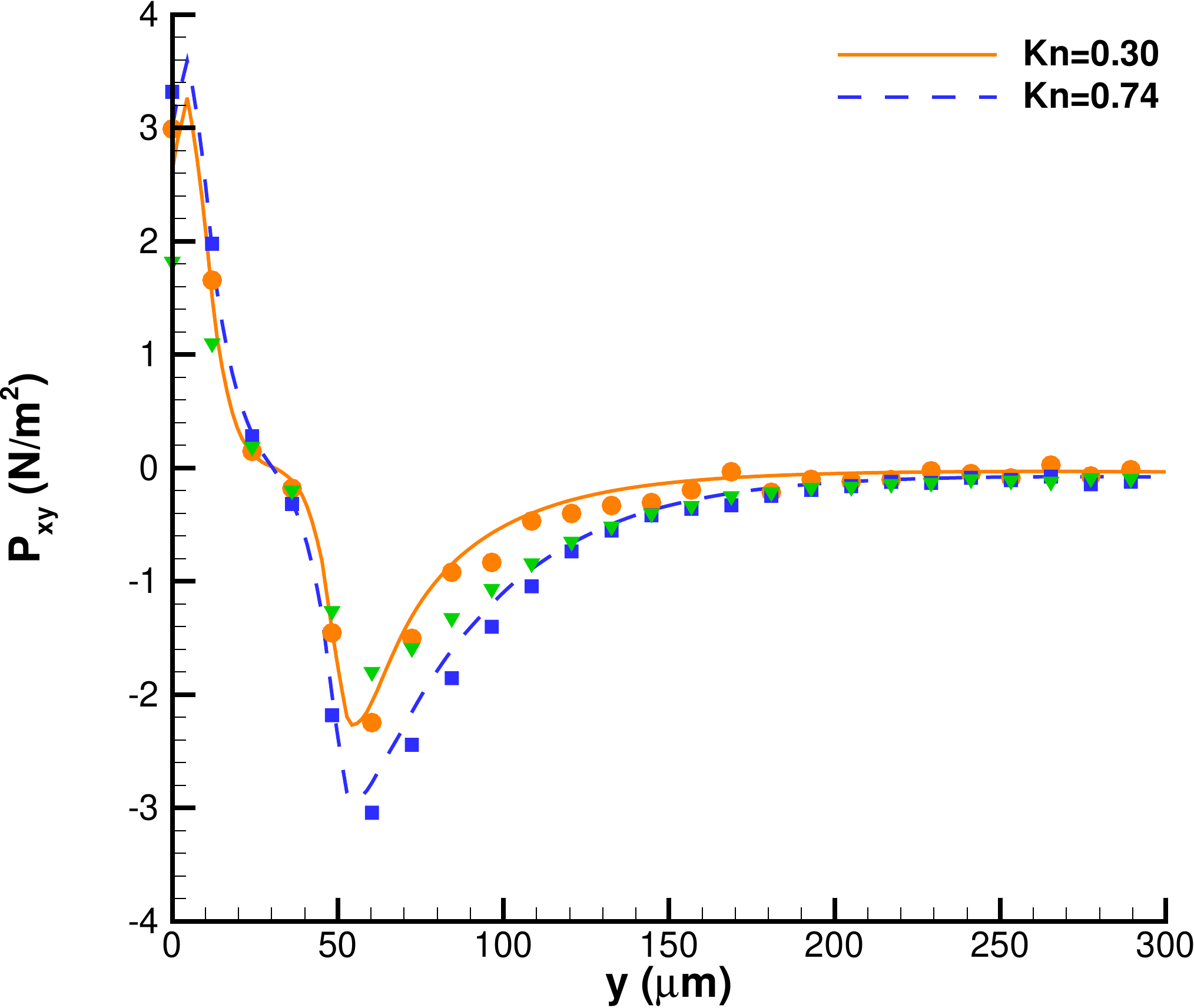}
  \caption{$xy$-component of stress (on vertical centerline)}
  \label{subfig_mikra_plotoverline_Pxy_BGK}
\end{subfigure}%
\begin{subfigure}{.5\textwidth}
  \centering
  \includegraphics[width=75mm]{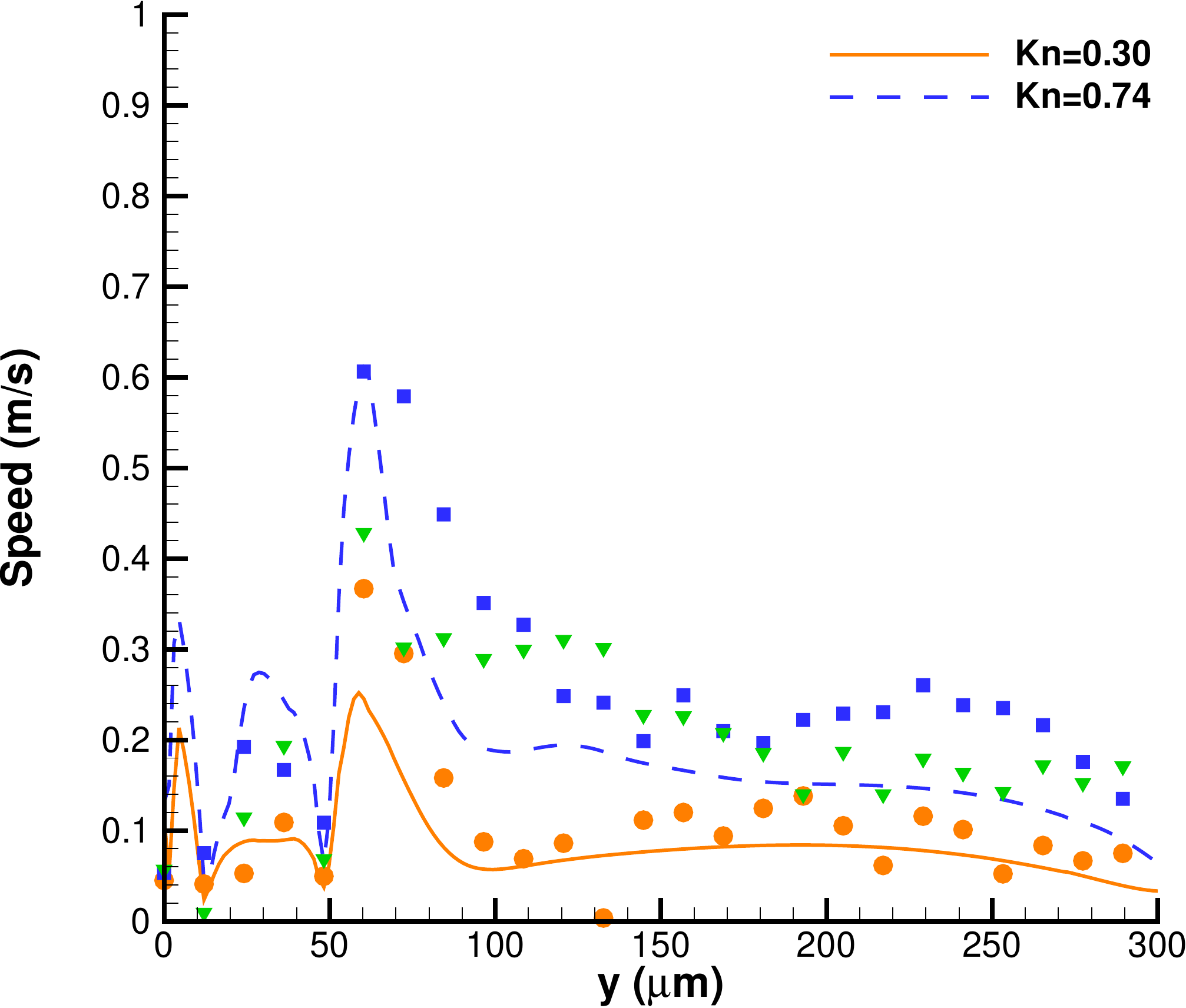}
  \caption{Speed (on vertical centerline)}
  \label{subfig_mikra_plotoverline_Speed_BGK}
\end{subfigure}
\caption{Variation of flow properties along the domain vertical centerline ($X=300\mu m$) for MIKRA Gen1 cases obtained from DSMC (symbols) and BGK (lines).}
\label{fig_mikra_plotoverline_bgk}
\end{figure*}

\begin{figure*}[!ht]
\centering
\begin{subfigure}{.5\textwidth}
  \centering
  \includegraphics[width=75mm]{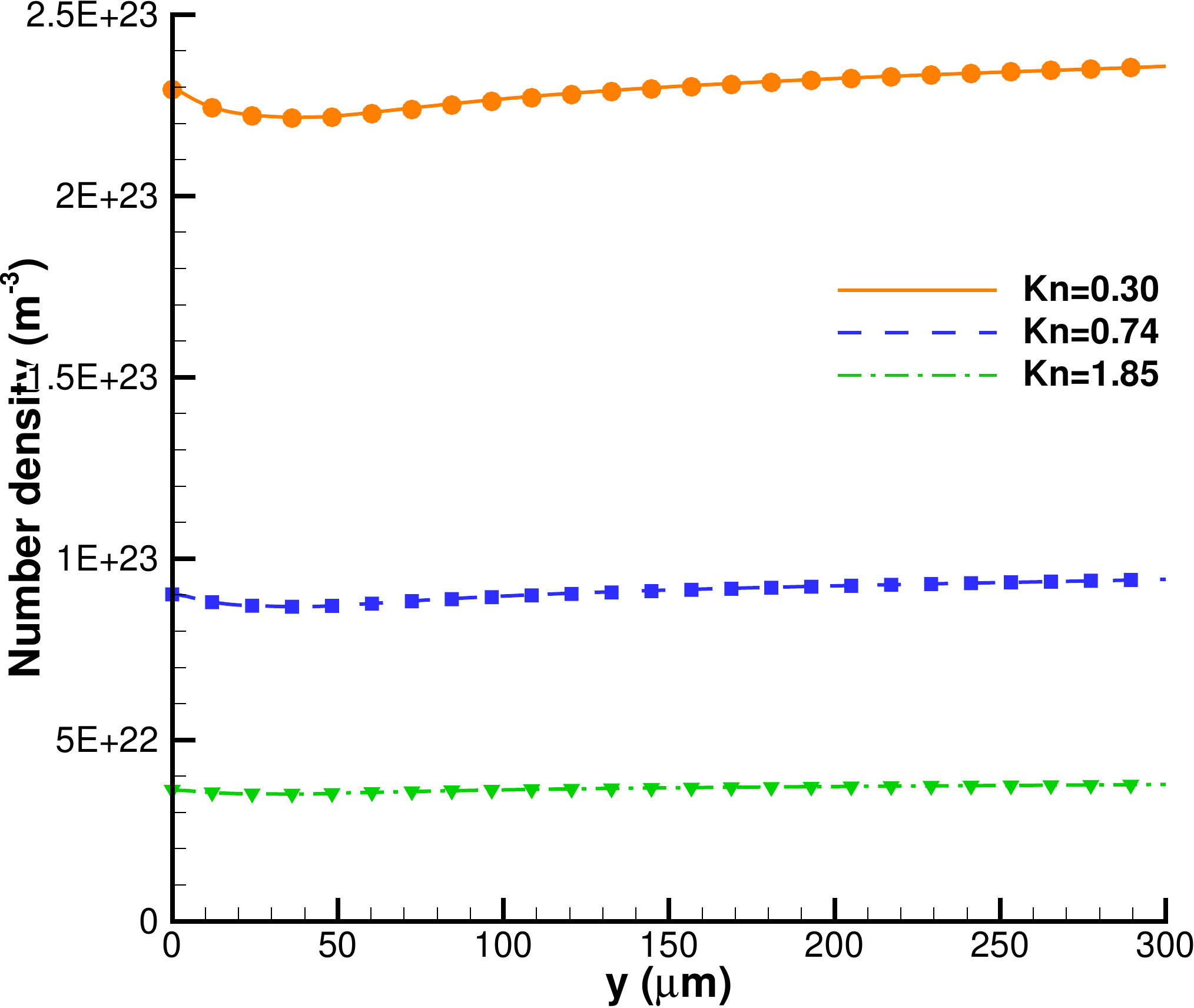}
  \caption{Number density (on vertical centerline)}
  \label{subfig_mikra_plotoverline_nden_ESBGK}
\end{subfigure}%
\begin{subfigure}{.5\textwidth}
  \centering
  \includegraphics[width=75mm]{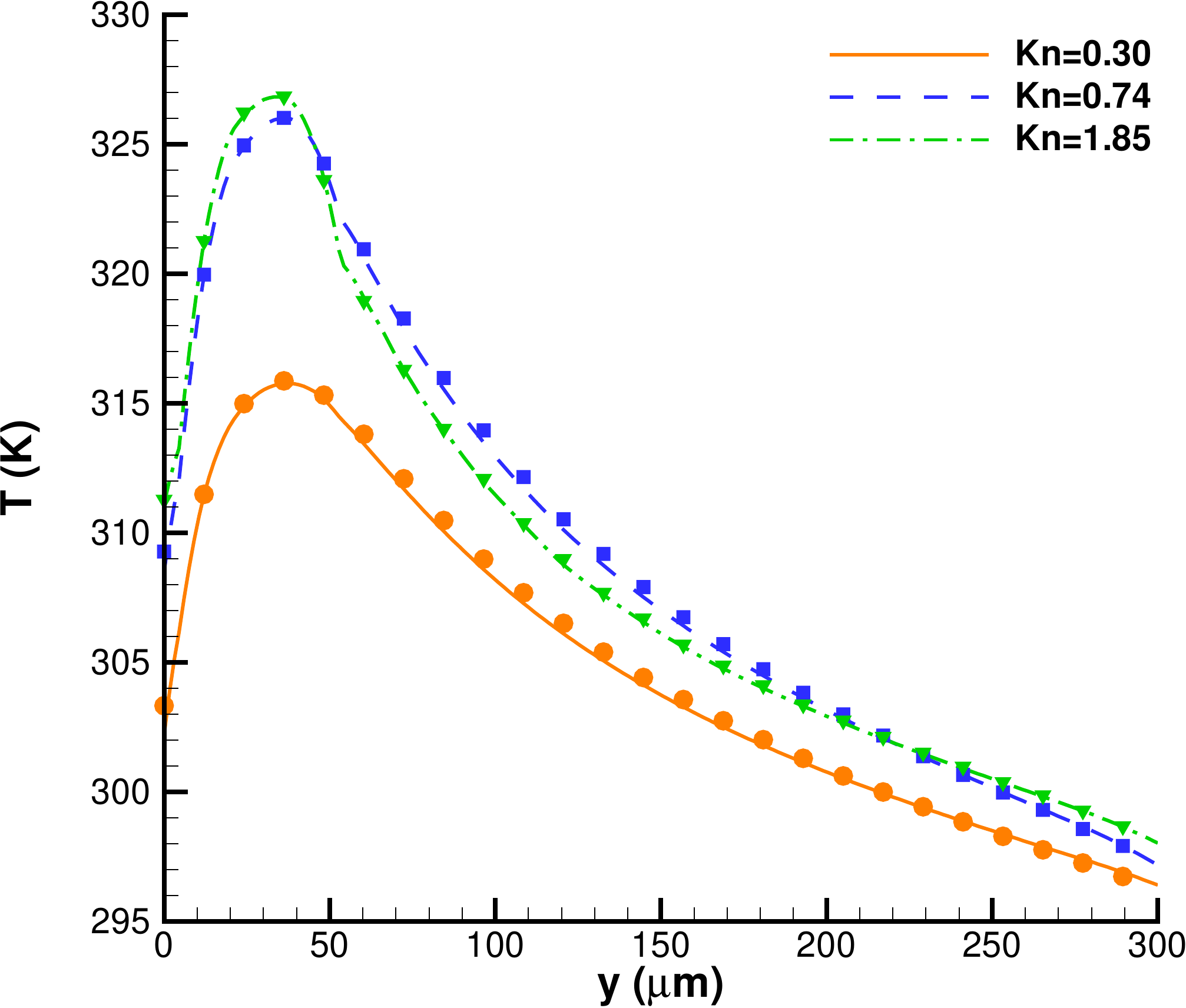}
  \caption{Temperature (on vertical centerline)}
  \label{subfig_mikra_plotoverline_T_ESBGK}
\end{subfigure}
\begin{subfigure}{.5\textwidth}
  \centering
  \includegraphics[width=75mm]{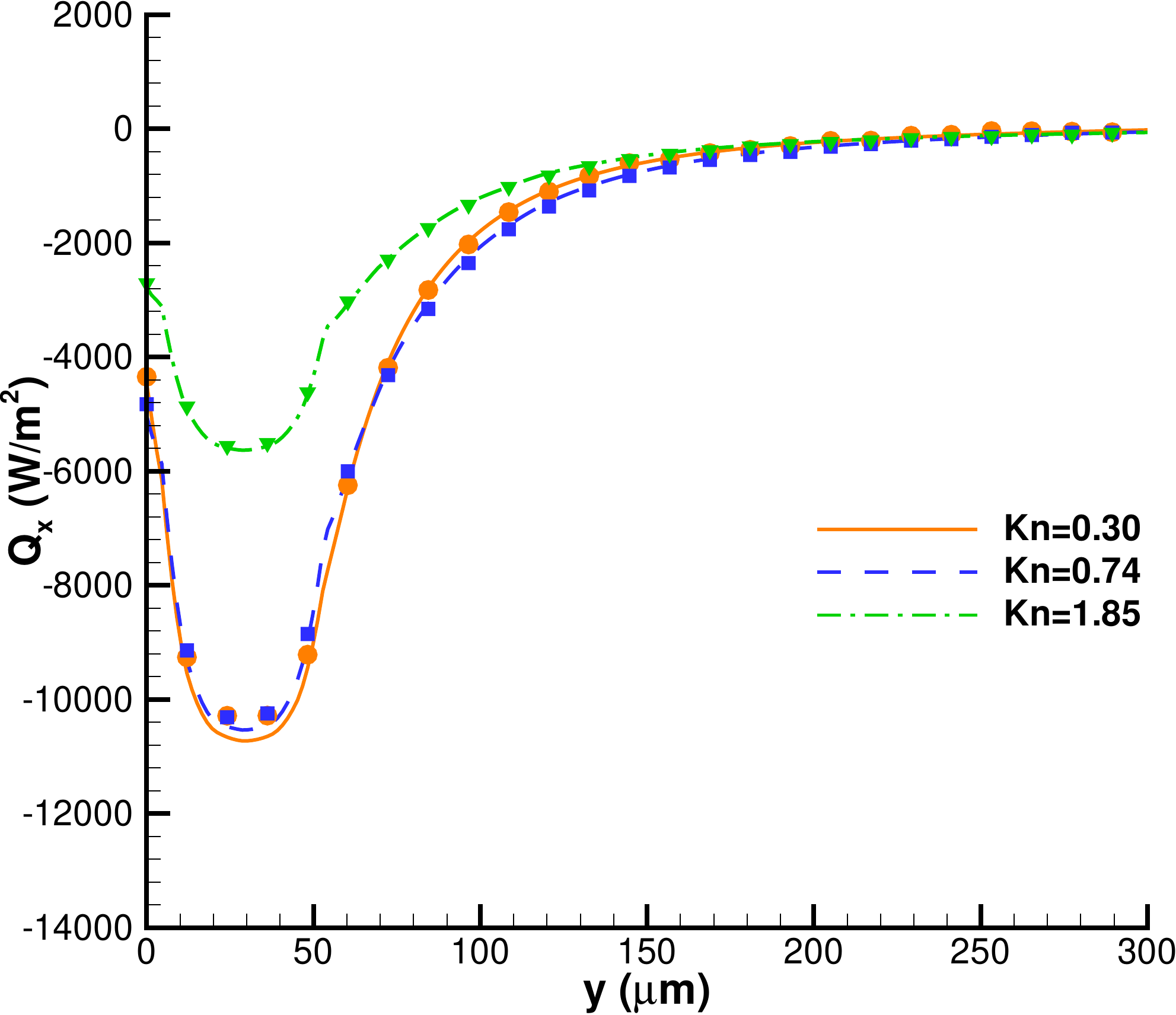}
  \caption{$x$-component of heat-flux (on vertical centerline)}
  \label{subfig_mikra_plotoverline_Qx_ESBGK}
\end{subfigure}%
\begin{subfigure}{.5\textwidth}
  \centering
  \includegraphics[width=75mm]{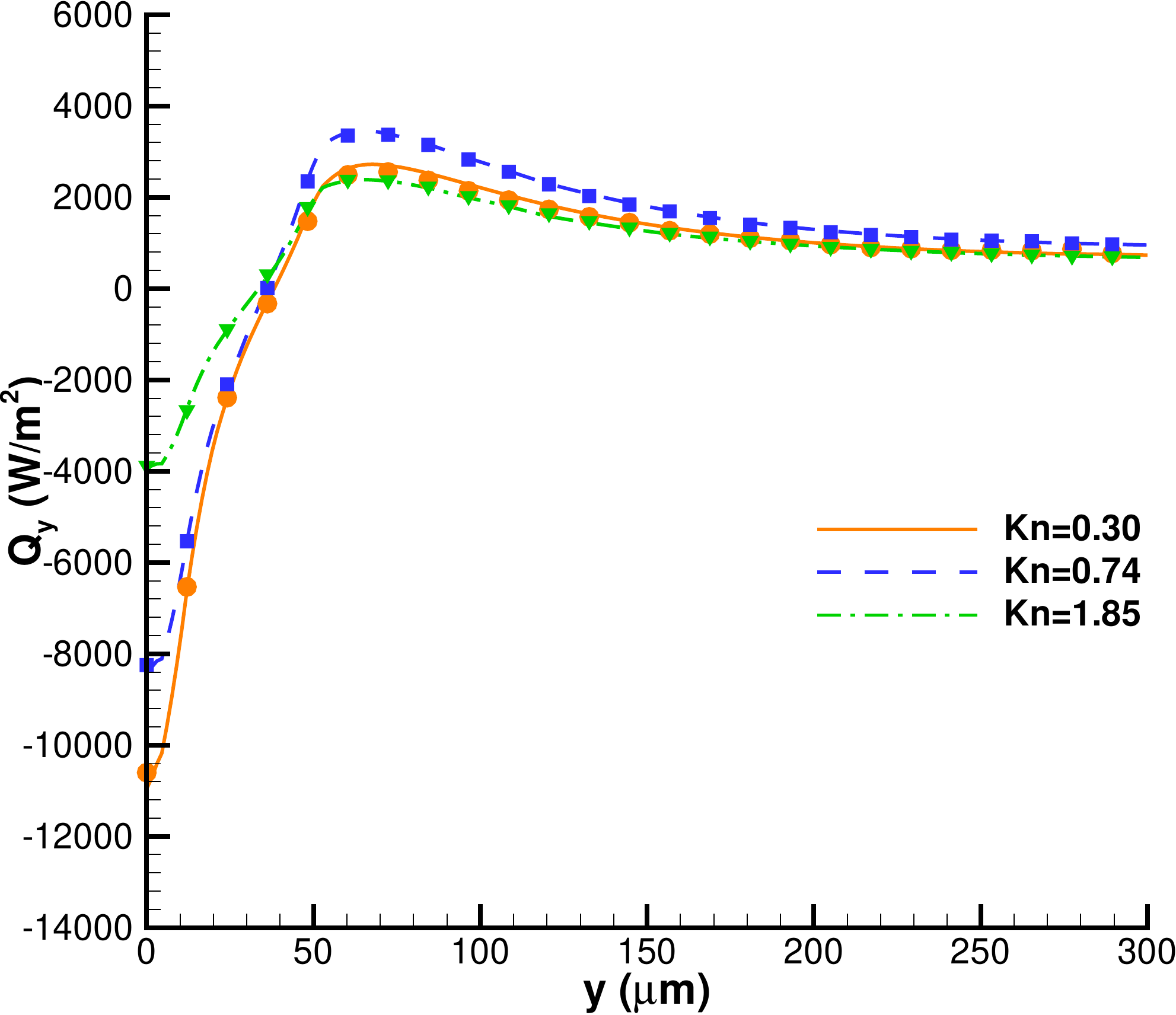}
  \caption{$y$-component of heat-flux (on vertical centerline)}
  \label{subfig_mikra_plotoverline_Qy_ESBGK}
\end{subfigure}
\begin{subfigure}{.5\textwidth}
  \centering
  \includegraphics[width=75mm]{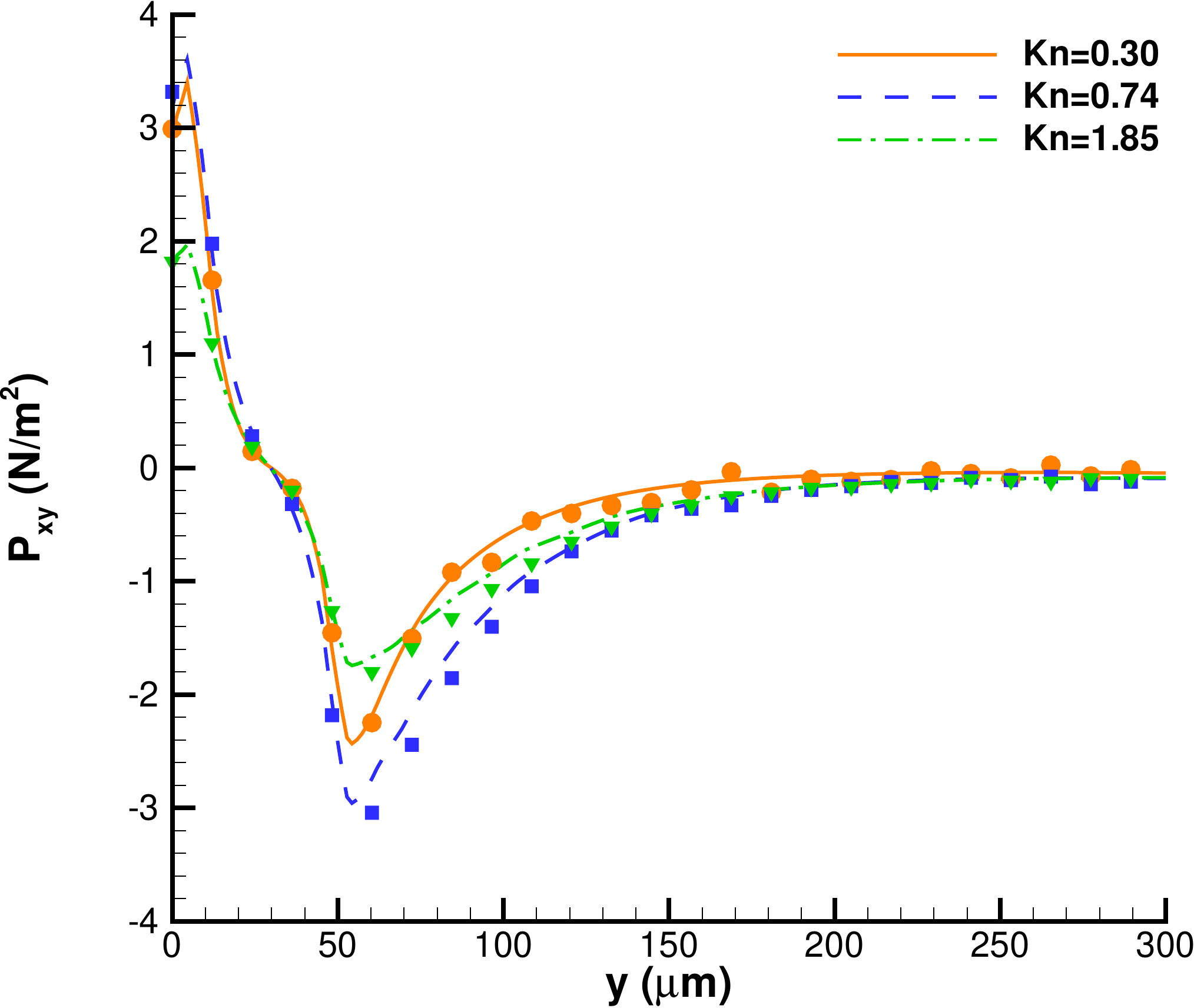}
  \caption{$xy$-component of stress (on vertical centerline)}
  \label{subfig_mikra_plotoverline_Pxy_ESBGK}
\end{subfigure}%
\begin{subfigure}{.5\textwidth}
  \centering
  \includegraphics[width=75mm]{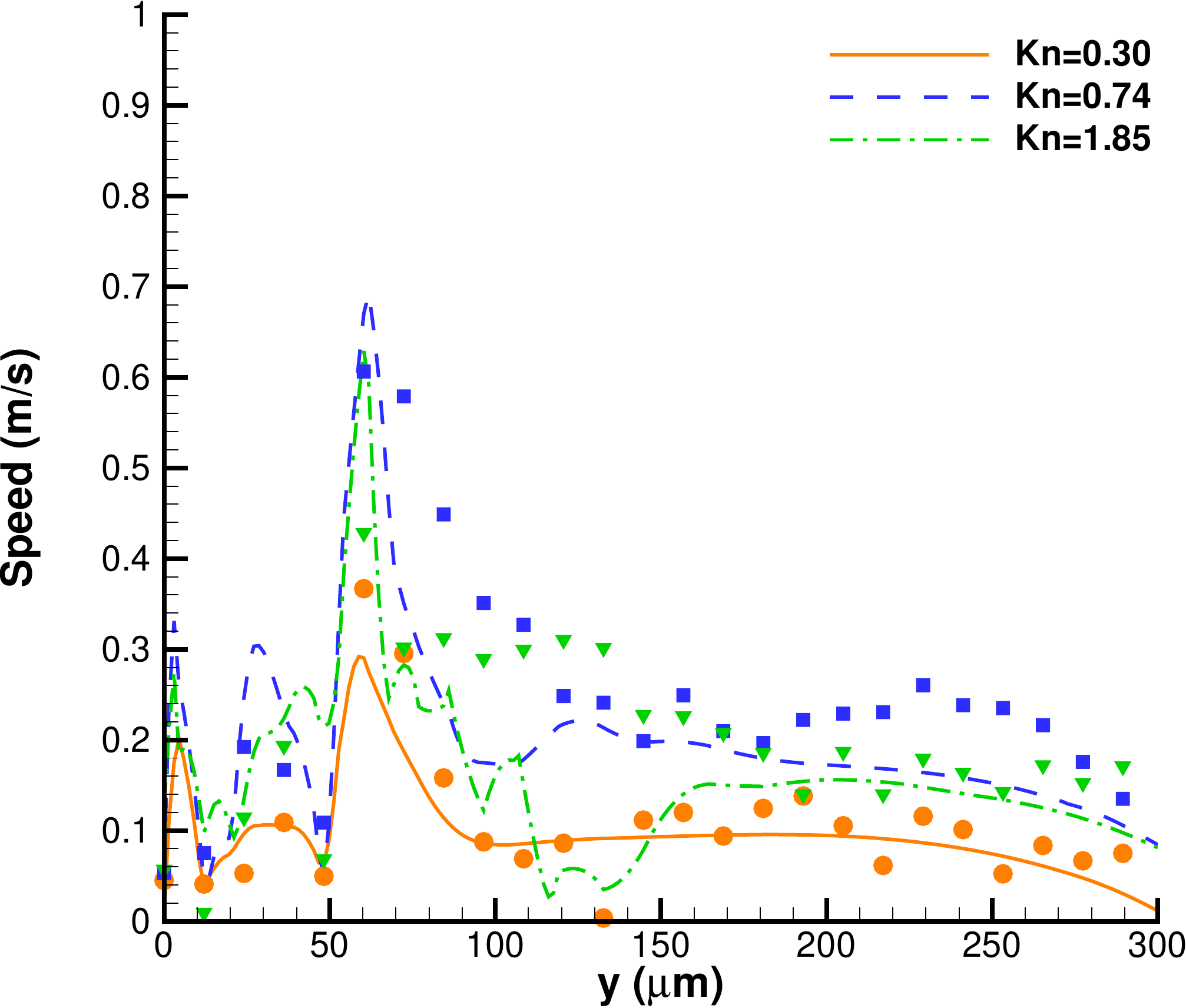}
  \caption{Speed (on vertical centerline)}
  \label{subfig_mikra_plotoverline_Speed_ESBGK}
\end{subfigure}
\caption{Variation of flow properties along the domain vertical centerline ($X=300\mu m$) for MIKRA Gen1 cases obtained from DSMC (symbols) and ESBGK (lines).}
\label{fig_mikra_plotoverline_esbgk}
\end{figure*}

\begin{figure*}[!ht]
\centering
\begin{subfigure}{.5\textwidth}
  \centering
  \includegraphics[width=75mm]{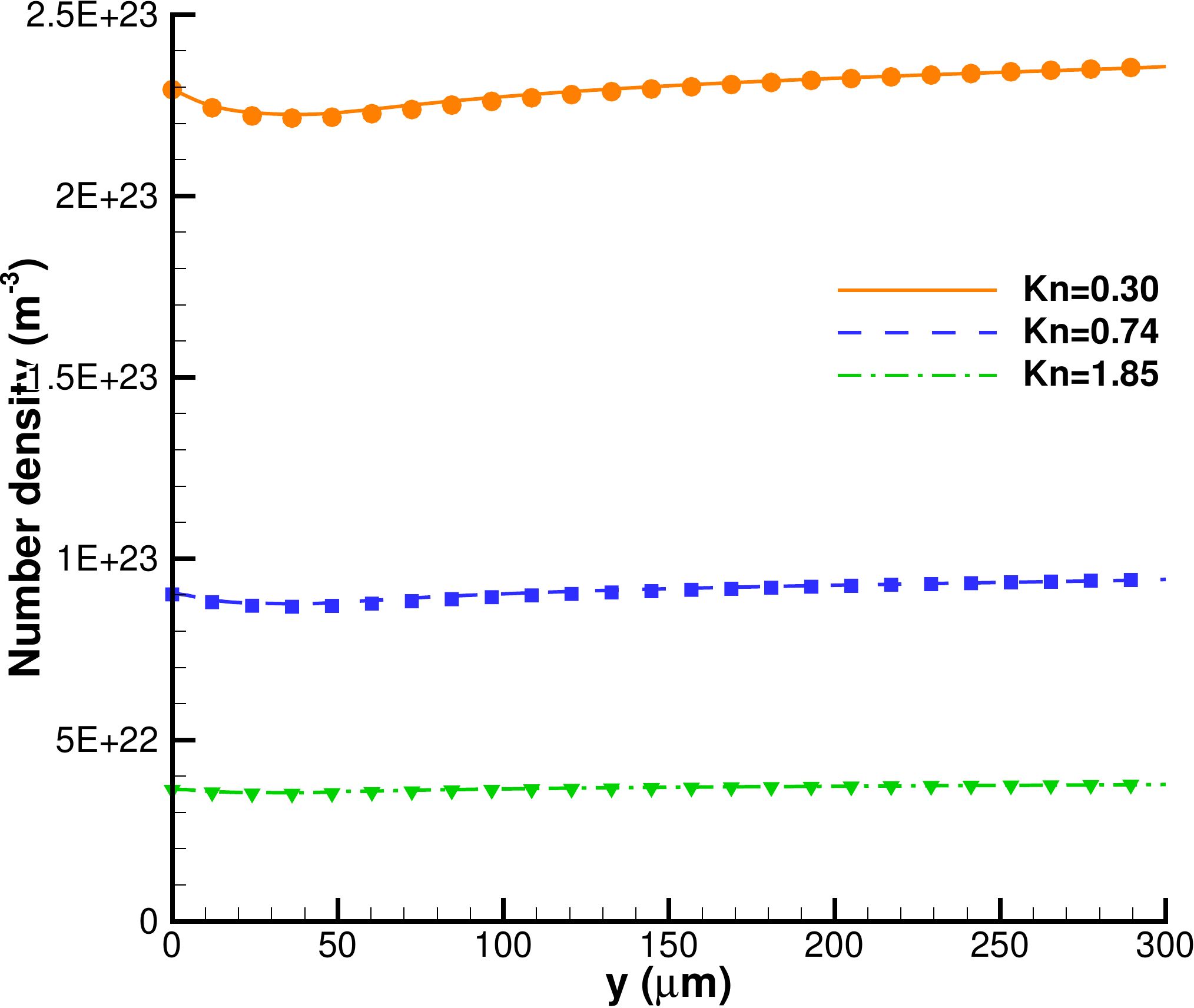}
  \caption{Number density (on vertical centerline)}
  \label{subfig_mikra_plotoverline_nden_Shakhov}
\end{subfigure}%
\begin{subfigure}{.5\textwidth}
  \centering
  \includegraphics[width=75mm]{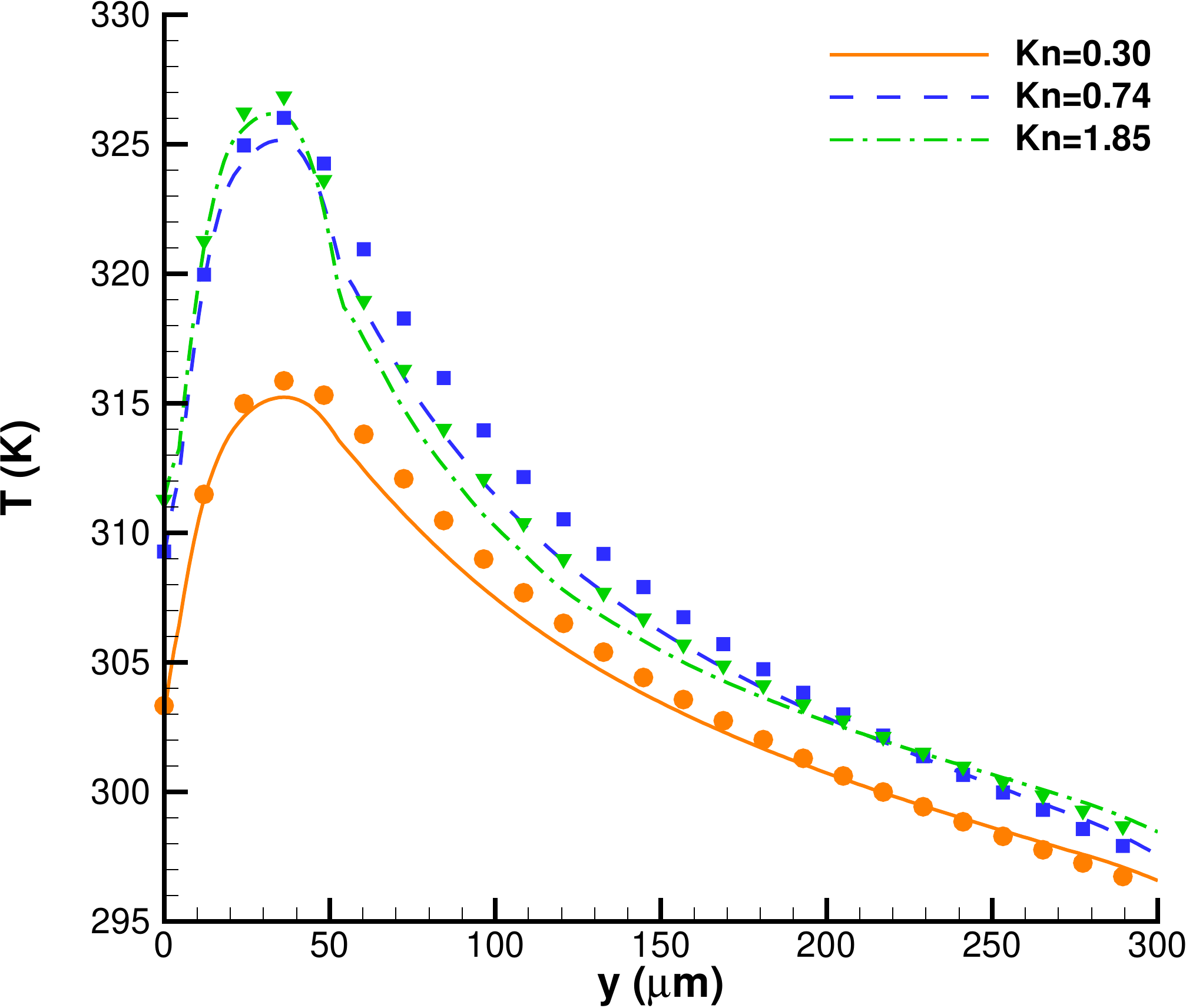}
  \caption{Temperature (on vertical centerline)}
  \label{subfig_mikra_plotoverline_T_Shakhov}
\end{subfigure}
\begin{subfigure}{.5\textwidth}
  \centering
  \includegraphics[width=75mm]{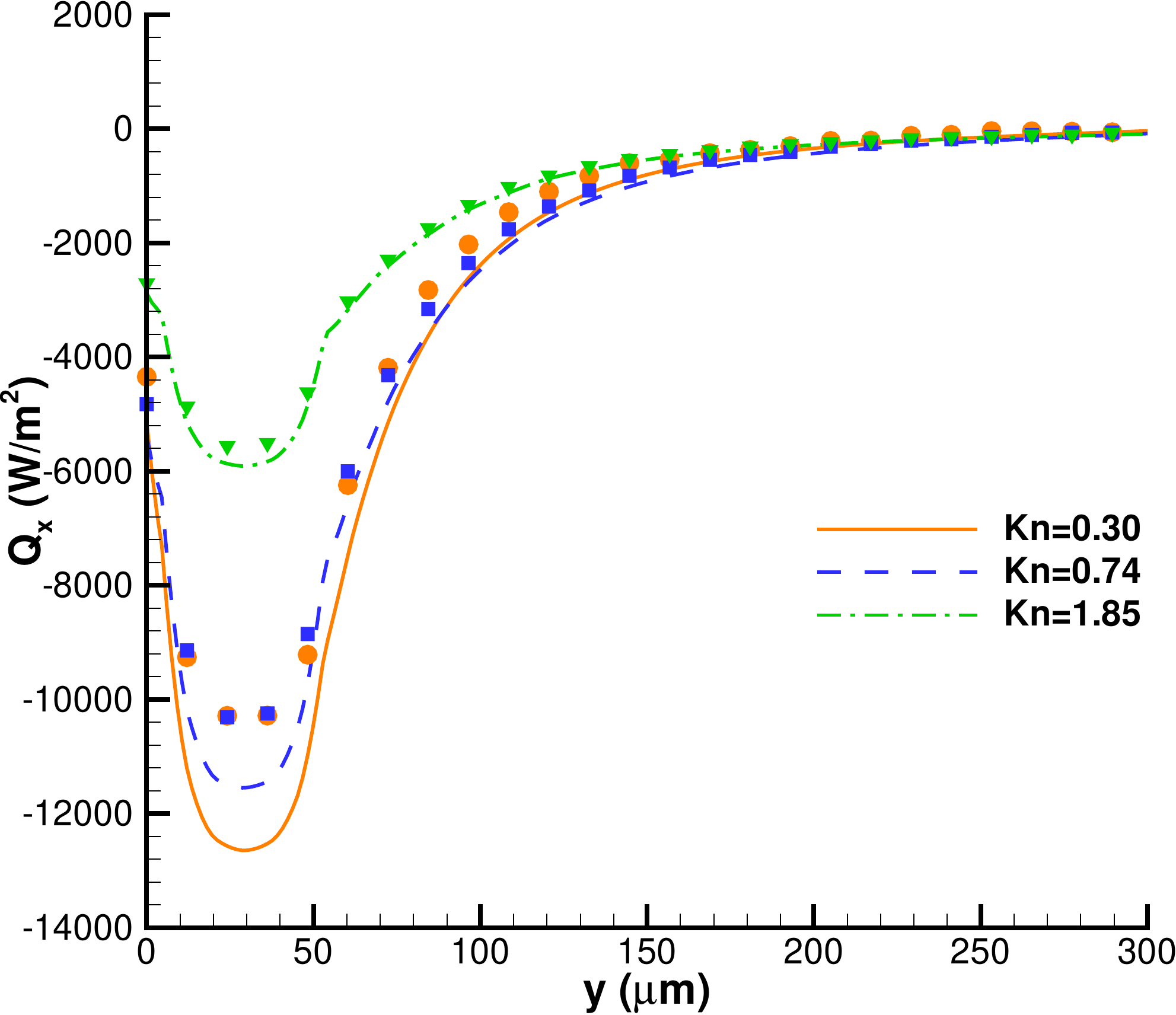}
  \caption{$x$-component of heat-flux (on vertical centerline)}
  \label{subfig_mikra_plotoverline_Qx_Shakhov}
\end{subfigure}%
\begin{subfigure}{.5\textwidth}
  \centering
  \includegraphics[width=75mm]{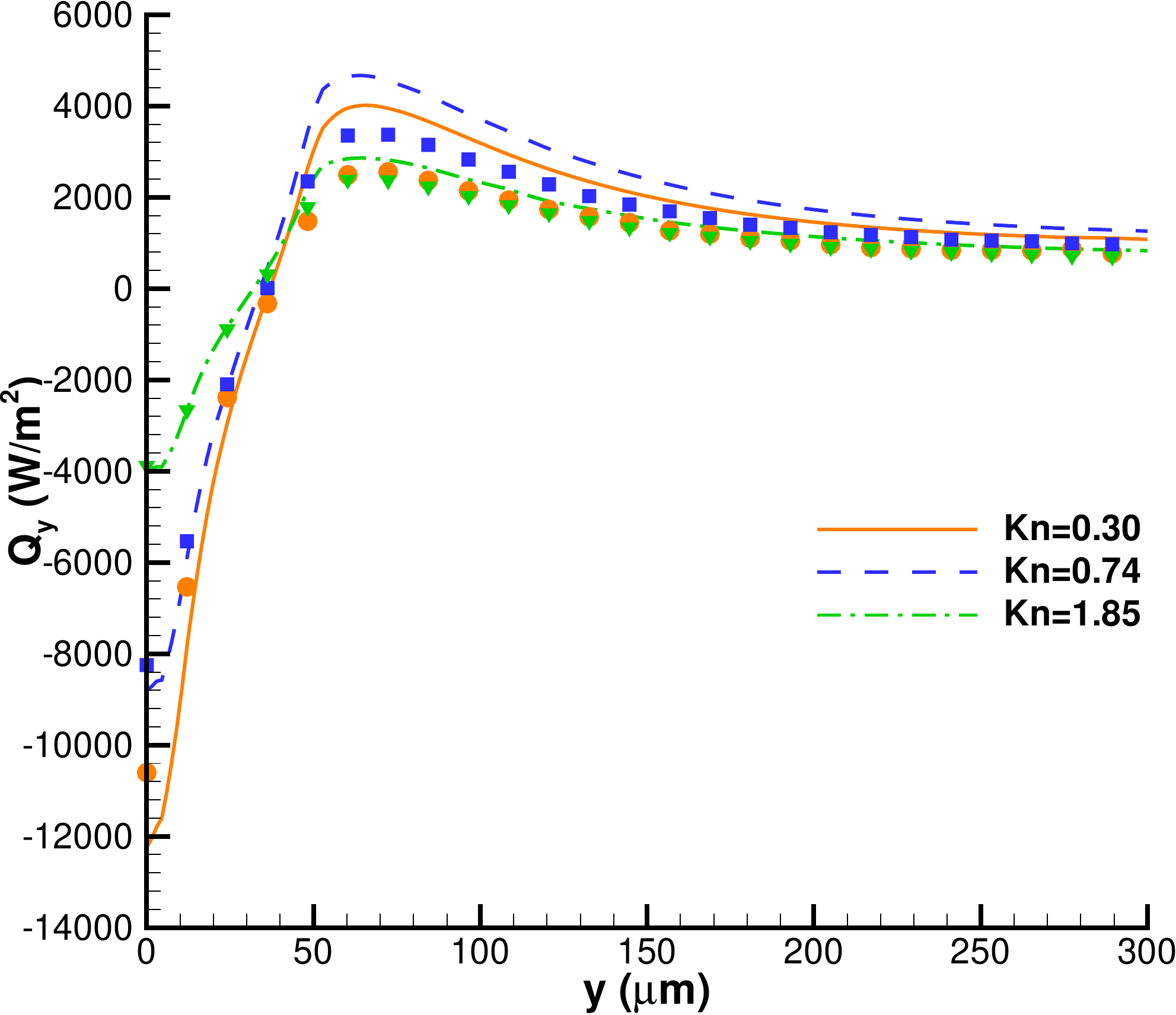}
  \caption{$y$-component of heat-flux (on vertical centerline)}
  \label{subfig_mikra_plotoverline_Qy_Shakhov}
\end{subfigure}
\begin{subfigure}{.5\textwidth}
  \centering
  \includegraphics[width=75mm]{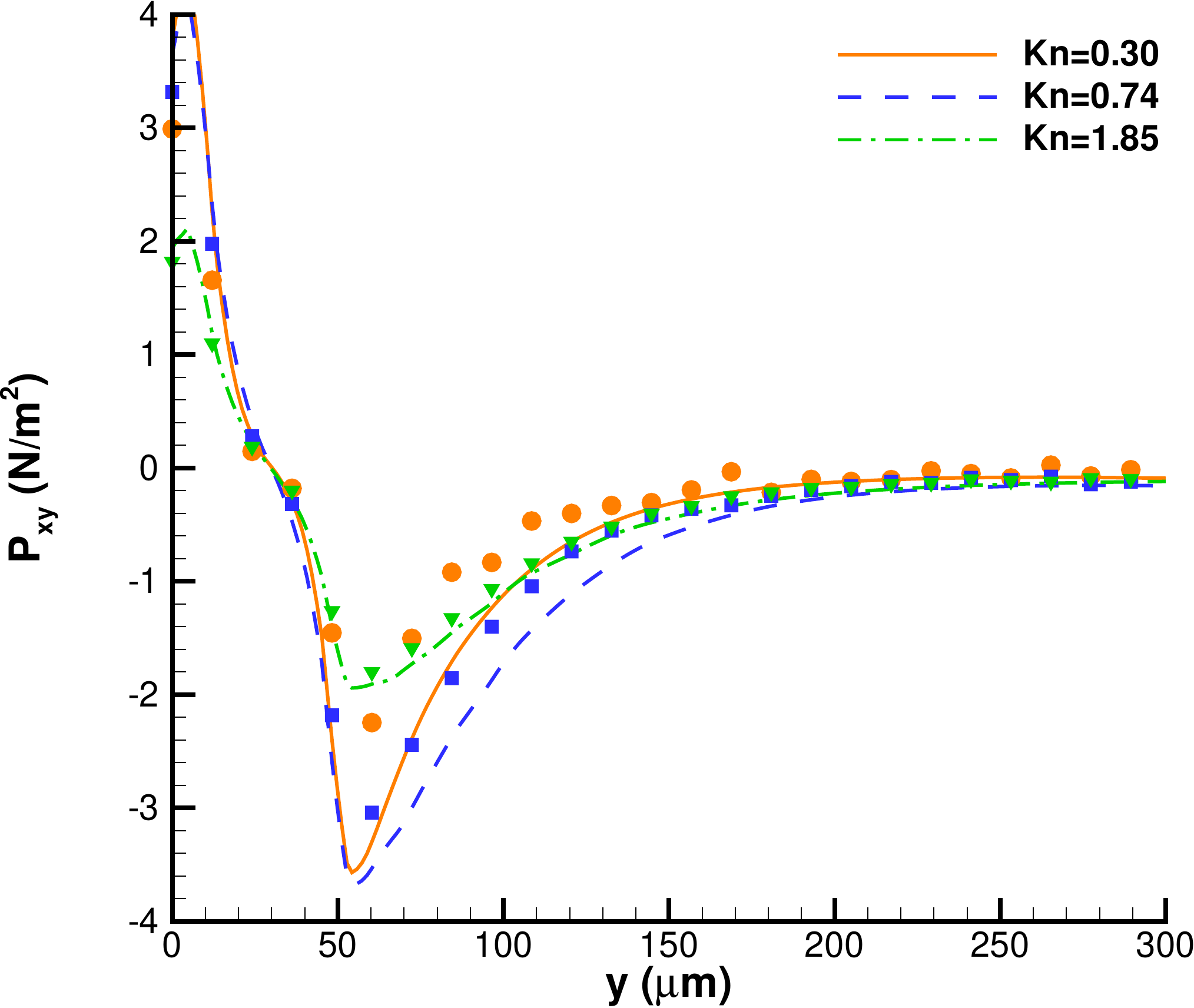}
  \caption{$xy$-component of stress (on vertical centerline)}
  \label{subfig_mikra_plotoverline_Pxy_Shakhov}
\end{subfigure}%
\begin{subfigure}{.5\textwidth}
  \centering
  \includegraphics[width=75mm]{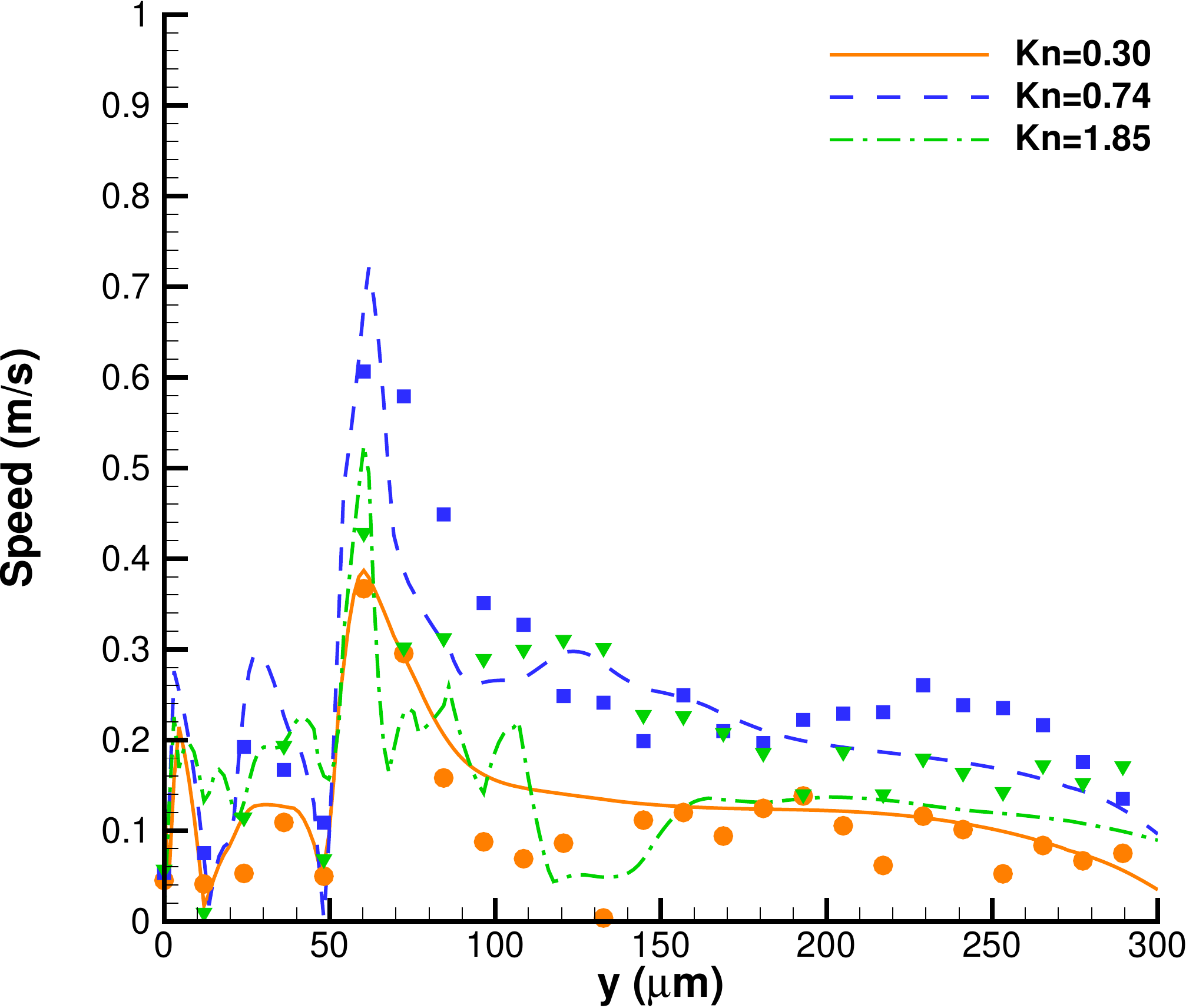}
  \caption{Speed (on vertical centerline)}
  \label{subfig_mikra_plotoverline_Speed_Shakhov}
\end{subfigure}
\caption{Variation of flow properties along the domain vertical centerline ($X=300\mu m$) for MIKRA Gen1 cases obtained from DSMC (symbols) and Shakhov (lines).}
\label{fig_mikra_plotoverline_Shakhov}
\end{figure*}

Finally, we compare the variation of flow properties along the vertical centerline ($x=300\mu m$, $0\leq y \leq 300 \mu m$) in Figs.~\ref{fig_mikra_plotoverline}, \ref{fig_mikra_plotoverline_bgk}, \ref{fig_mikra_plotoverline_esbgk}, and \ref{fig_mikra_plotoverline_Shakhov} for various models. We observe a fair agreement between DSMC and DGFS results ignoring the statistical noise (see Figs.~\ref{fig_mikra_flowfield_966N2_nden_T_Pxy}, \ref{fig_mikra_flowfield_966N2_Qx_Qy_Speed}). In particular, in Fig.~\ref{subfig_mikra_plotoverline_T}, we observe peak temperatures near the \textit{edges} of hot and cold vanes i.e., in the region $x=300\mu m$, $30\leq y \leq 60 \mu m$. Through Figs.~\ref{subfig_mikra_plotoverline_Qx} and \ref{subfig_mikra_plotoverline_Qy}, we infer that the thermal gradients are stronger in the $x$-direction. More notably, we observe the highest thermal-stress in the \textit{edge} region (note the valley in the region $x=300\mu m$, $40\leq y \leq 60 \mu m$). We conjecture the trough of the valley to be shallower if the vane edges ought to be made smoother. A slightly peculiar observation is as follows: the trough of the valley is deeper at $\Kn=0.74$ compared to $\Kn=0.30$, and shallower at $\Kn=1.85$ compared to $\Kn=0.74$. This could be explained as follows: at $\Kn=0.30$ the temperature difference, $T_H-T_C$, is lower than the one correspoding to the $\Kn=0.74$ case and therefore the thermal stress increases in the latter case. For the $\Kn=1.85$ and $\Kn=0.74$ cases, wherein the temperature difference is approximately same, the peak thermal-stress decreases owing to the bimodal nature of the Knudsen forces. 

\section{Multi-species MIKRA}
\label{sec_multispecies_mikra}
In the present section, we carry out the MIKRA simulations for binary mixture consisting of $N_2$ and $H_2O$ using the variable soft sphere model. 

\subsection{Problem Statement}
The flow configuration remains the same as shown in Fig.~\ref{fig_mikraSchematic}. We consider the 2D uniform flow of binary mixture of $N_2$ and $H_2O$. The end goal is to simulate the motion of gas flows in the gap between the two vanes, subject to initial pressure $p_\infty$, hot ($T_H$) and cold ($T_C$) vane temperature as listed in Tab.~\ref{tab_multispecies_mikraCases}, in order to identify the correct circulation, induced low velocity, temperature gradient, Knudsen forces, and heat transfer rate from the vanes. The results are to be obtained from both stochastic (DSMC) and deterministic (DGFS) simulations.

\begin{table}[!ht]
\centering
\begin{ruledtabular}
\begin{tabular}{@{}lc@{}}
Parameter & \multicolumn{1}{c}{Cases} \\ 
 & MSM-01 \\ 
\hline
Pressure: $p$ (Torr) & $1.163$ \\
Total number density: $n$ ($\times 10^{21}\;m^{-3}$) & $37.86091$ \\
Concentration: ($n^\isp/n,\,n^\jsp/n$) & $(0.05,\,0.95)$ \\
Knudsen number\footnotemark[1]: $\Kn$ & $1.85$ \\
Cold vane temperature: $T_C$ ($K$) & $306$ \\
Hot vane temperature: $T_H$ ($K$) & $363$ \\
\hline
\multicolumn{2}{l}{DGFS parameters} \\
Points in velocity mesh: $N^3$ & $32^3$ \\
Points in radial direction\footnotemark[2]: $N_\rho$ & $8$ \\
Points on \textit{full} sphere\footnotemark[2]: $M$ & $12$ \\
Size of velocity mesh\footnotemark[3] & $[-6,\,6]^3$ \\
\end{tabular}
\end{ruledtabular}
\footnotetext[1]{Based on hard-sphere definition using total number density (see Ref.~\onlinecite{Bird})}
\footnotetext[2]{Required in the fast Fourier spectral low-rank decomposition (see Ref.~\onlinecite{jaiswal2018dgfsGPU})}
\footnotetext[3]{Non-dimensional (see Refs.~\onlinecite{JAH19,jaiswal2019dgfsMulti} for details on non-dimensionalization)}
\caption{Numerical parameters for thermo-stress convection in MIKRA Gen1 simulations for DSMC and DGFS using VSS collision model for $N_2$/$H_2O$ binary mixture.}
\label{tab_multispecies_mikraCases}
\end{table}

\begin{table}[!ht]
\centering
\begin{ruledtabular}
\begin{tabular}{@{}lcc@{}}
& $N_2$ & $H_2O$ \\
\cline{2-3}
Mass: $m$ ($kg$) & $46.5 \times 10^{-27}$ & $29.9 \times 10^{-27}$ \\
Viscosity index\footnotemark[1]: $\omega_{i},\,$ ($-$) & $0.74$ & $1.00$ \\ 
Scattering index: $\alpha_{i},\,$ ($-$) & $1.36$ & $1.00$ \\
Ref. diameter: $d_{\mathrm{ref},i}$ ($m$) & $4.07 \times 10^{-10}$ & $5.78 \times 10^{-10}$ \\
Ref. temperature: $T_{\mathrm{ref},i}$ ($K$) & 273 & 273 \\ 
\end{tabular}
\footnotetext[1]{For cross-collision (see Refs.~\onlinecite{Bird,jaiswal2019dgfsMulti}): $\Psi_{ij}=0.5(\Psi_{i}+\Psi_{j}),\,i\neq j,\,\text{where}\,\Psi=\{\omega,\,\alpha,\,d_{\mathrm{ref}},\,T_{\mathrm{ref}}\}$}
\end{ruledtabular}
\caption{$N_2$ and $H_2O$ gas VSS parameters used in MIKRA Gen1 DSMC and DGFS simulations.}
\label{tab_multispecies_props}
\end{table}

\subsection{Numerical details}
\label{subsec_multispecies_numdetails}
The multi-species simulations are carried out for flows in transition regime. The specific differences between stochastic (DSMC) and deterministic (DGFS) modelling is described next.

\begin{itemize}
\item \textbf{DSMC}: SPARTA\cite{gallis2014direct} has been employed for carrying out DSMC simulations in the present work. The geometric parameters remain the same as described in section~\ref{subsec_numdetails}. A minimum of 300 DSMC simulator particles per cell is used in conjunction with the no-time collision (NTC) algorithm and VSS scattering model. The simulations are first run for 200,000 unsteady steps, and subsequently another 5,000,000 steady steps wherein the flow sampling is performed. Similar to the previous single-species MIKRA case, the DSMC domain is discretized into $300 \times 150$ cells, resulting in a uniform cell size of $2\,\mu m$, with 285 particles of $H_2O$ and 15 particles of $N_2$, per cell on average during initialization. A time step of $10^{-9}$ sec is used during \textit{move} step of DSMC algorithm throughout the course of simulation. $N_2$ and $H_2O$ are used as the working gases in simulations. The properties of the working gas is given in Tab.~\ref{tab_multispecies_props}. Note that for $N_2$, we consider $\zeta_R=2$ rotational degrees of freedom,  rotational relaxation $Z_R=0.2$, $\zeta_V=2$ vibrational degrees of freedom, vibrational relaxation $Z_V=1.90114 \times 10^{-5}$, and vibrational temperature $T_v=3371\,K$; and for $H_2O$, we consider $\zeta_R=3$ rotational degrees of freedom,  rotational relaxation $Z_R=0.2$, $\zeta_V=3$ vibrational degrees of freedom, vibrational relaxation $Z_V=1.90114 \times 10^{-5}$, and vibrational temperature $T_v=5261\,K$.

\item \textbf{DGFS}: We use the DGFS implementation described in Ref.~\onlinecite{jaiswal2019dgfsMulti}. The geometrical parameters remain the same as described in section~\ref{subsec_numdetails}. Multi-species case specific DGFS parameters have been provided in Tab.~\ref{tab_multispecies_mikraCases}. Note that, we employ $N_2$ and $H_2O$ as the working gas in simulations. $N_2$ is diatomic, and $H_2O$ is triatomic, however, DGFS, as of now, is applicable for monoatomic gases only. Since the working temperature range is low, we anticipate the effects of vibrational degrees of freedom to be negligible. 
\end{itemize}

\subsection{Results and Discussion}
\label{subsec_multispecies_mikra_results}

\subsubsection{Flow pattern}
Figures~\ref{fig_mikramulti_flowfield_155N2_nden_T} and \ref{fig_mikramulti_flowfield_155N2_Pxy_Speed} illustrate the contour plot of various flow properties for the MSM-01 case in transition regime, wherein the $N_2$ and $H_2O$ are in $0.05:0.95$ concentration ratio. Similar to the single species case, for each of these plots, the DSMC and DGFS contours have been overlaid, wherein DSMC results have been indicated by thin black lines, and DGFS results have been indicated with thick red lines. Since the flow is strictly driven by temperature gradients, we expect very small deviation in the number density from the equilibrium values of $35.9678645\times 10^{21}\,m^{-3}$ for $H_2O$ and $1.8930455\times 10^{21}\,m^{-3}$ for $N_2$, as is also evident from Figs.~\ref{subfig_mikramulti_flowfield_155N2_nden_N2} and \ref{subfig_mikramulti_flowfield_155N2_nden_H2O}. In terms of temperature, in Figs.~\ref{subfig_mikramulti_flowfield_155N2_T_N2} and \ref{subfig_mikramulti_flowfield_155N2_T_H2O}, we again observe a rather familiar flow expansion, in the sense that, the hot vane dissipates heat to the surrounding acting as a source, thereby giving rise to a spiral with spiral's origin at the hot vane. From the fundamental mass/momentum conservation principles, one can infer that, in the presence of temperature gradients, the heavier species, here $N_2$, moves slower and the lighter species, here $H_2O$, moves faster giving rise to the well-known thermal diffusion. This explain why the isotherms for $H_2O$ spread farther apart compared to the those of $N_2$.

\begin{figure*}[!ht]
\centering
\begin{subfigure}[t]{0.68\textwidth}
  \centering
  \includegraphics[width=0.68\textwidth]{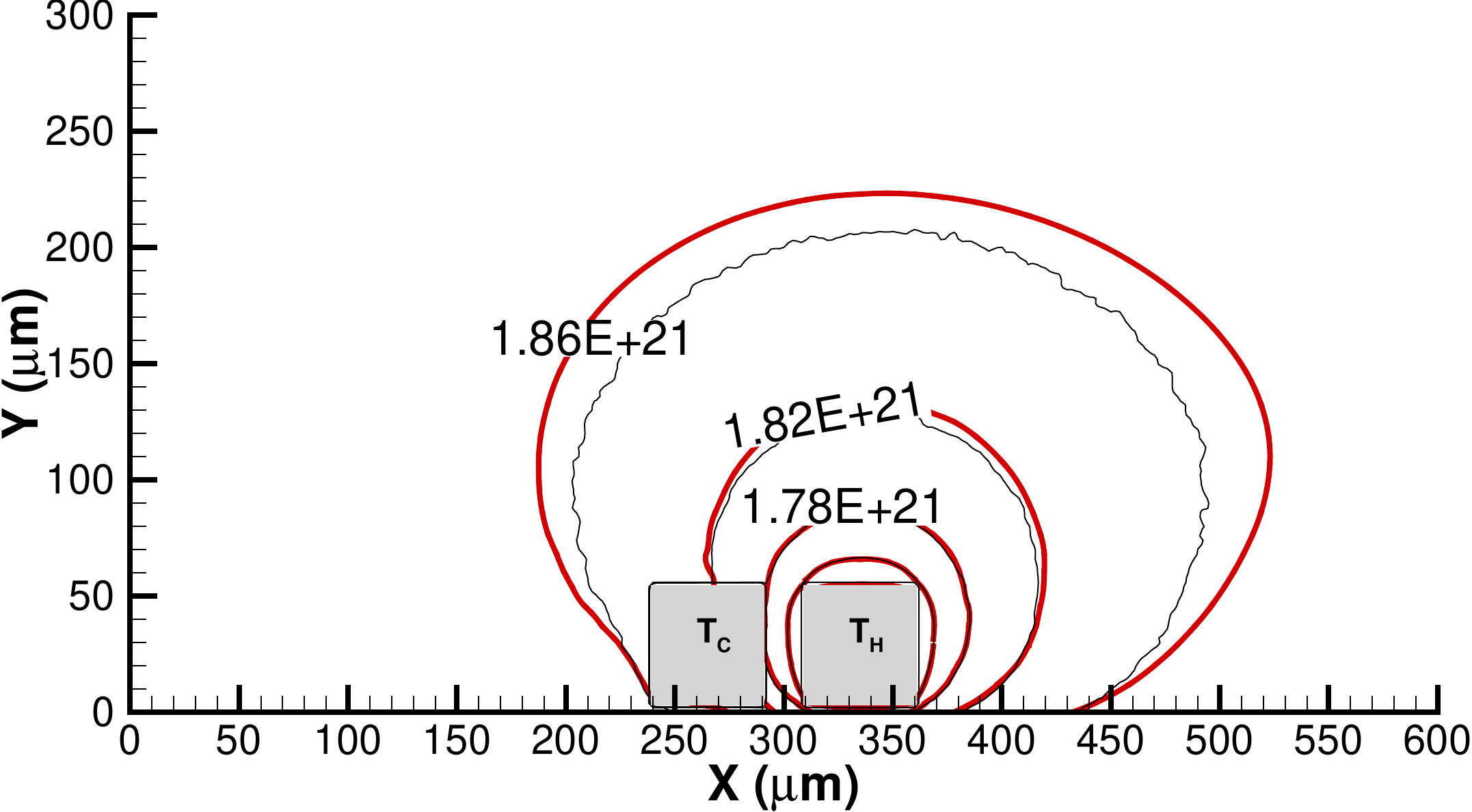}
  \caption{$N_2$, Number density ($m^{-3}$)}
  \label{subfig_mikramulti_flowfield_155N2_nden_N2}
\end{subfigure}%

\begin{subfigure}[t]{0.68\textwidth}
  \centering
  \includegraphics[width=0.68\textwidth]{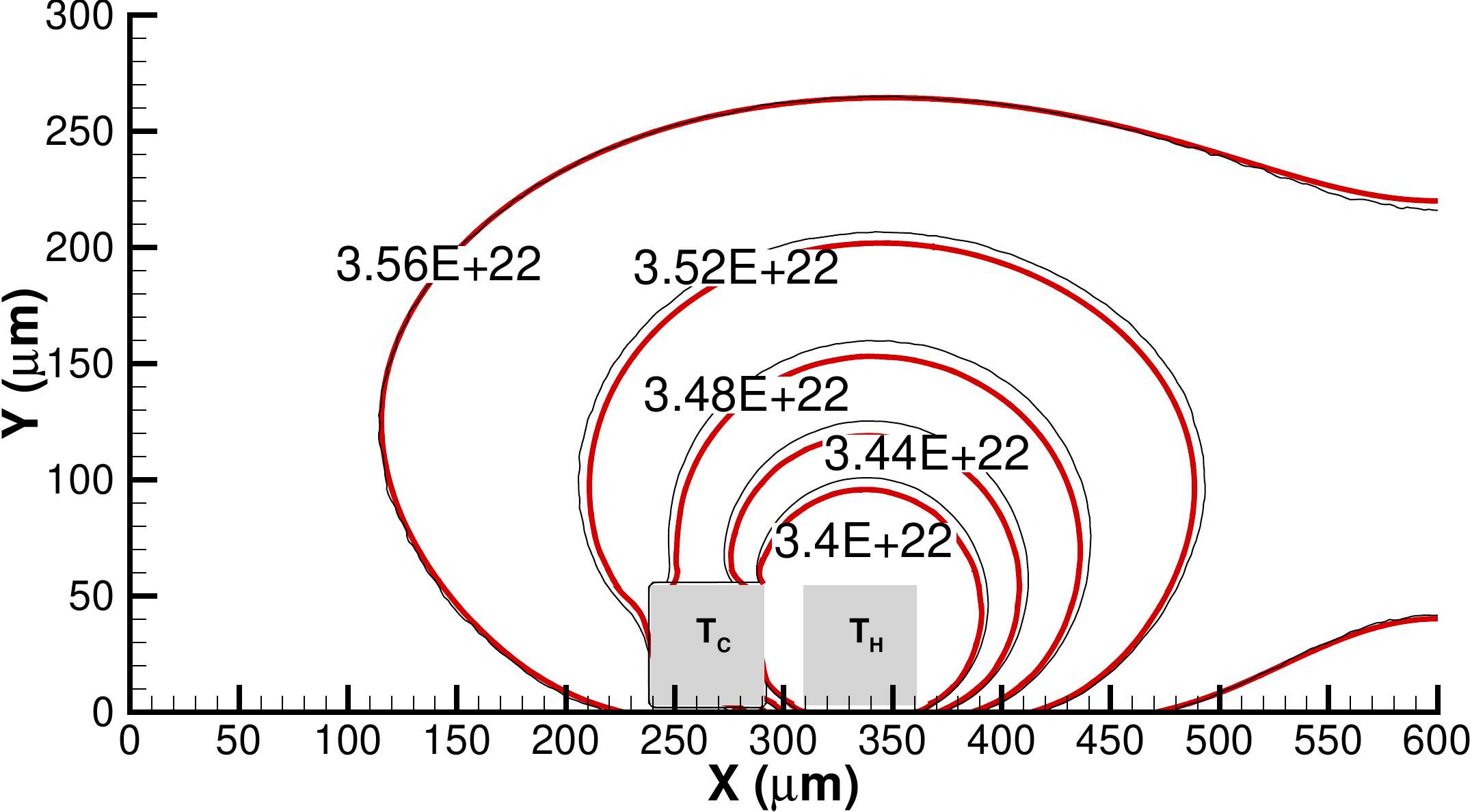}
  \caption{$H_2O$, Number density ($m^{-3}$)}
  \label{subfig_mikramulti_flowfield_155N2_nden_H2O}
\end{subfigure}

\begin{subfigure}[t]{0.68\textwidth}
  \centering
  \includegraphics[width=0.68\textwidth]{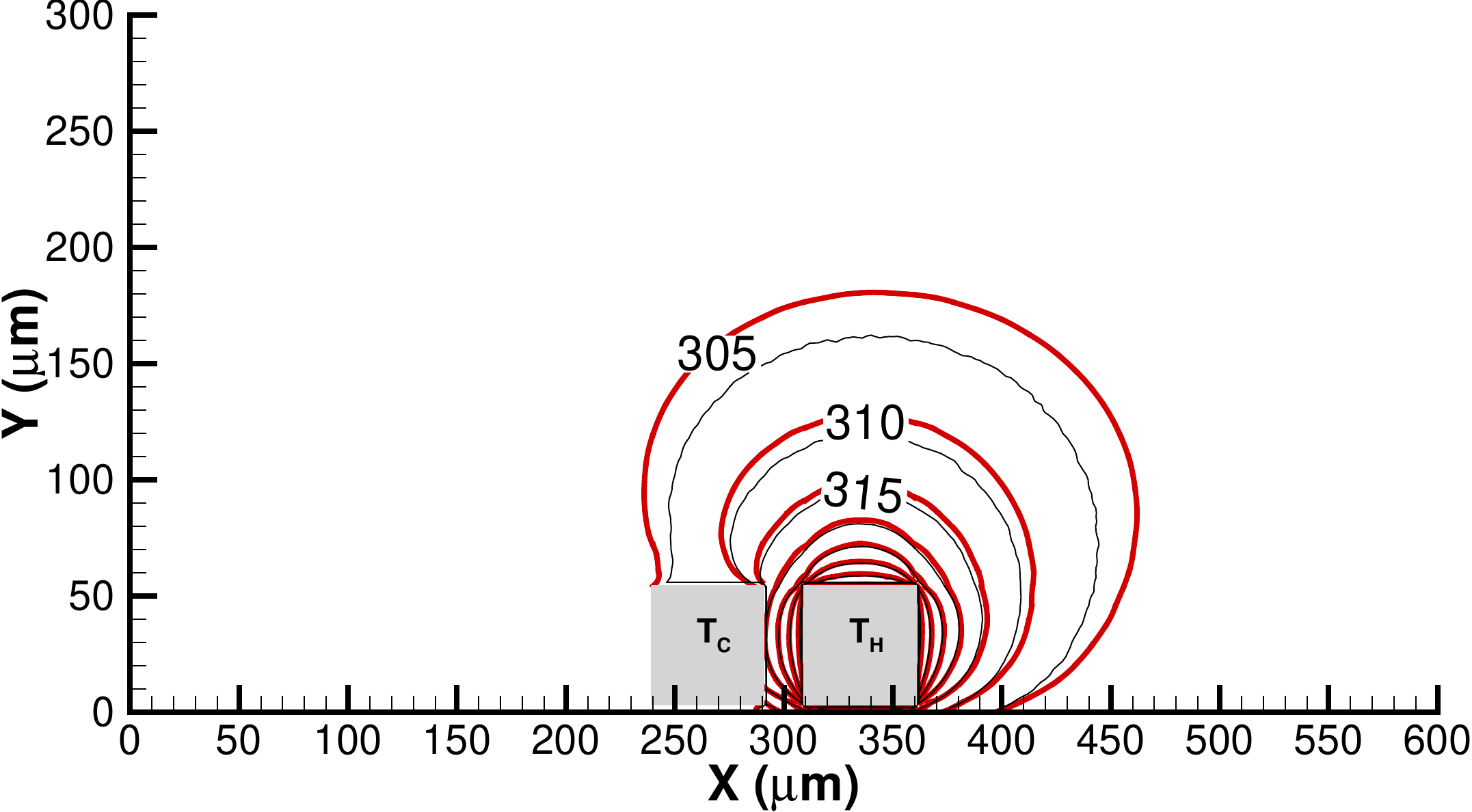}
  \caption{$N_2$, Temperature ($K$)}
  \label{subfig_mikramulti_flowfield_155N2_T_N2}
\end{subfigure}%

\begin{subfigure}[t]{0.68\textwidth}
  \centering
  \includegraphics[width=0.68\textwidth]{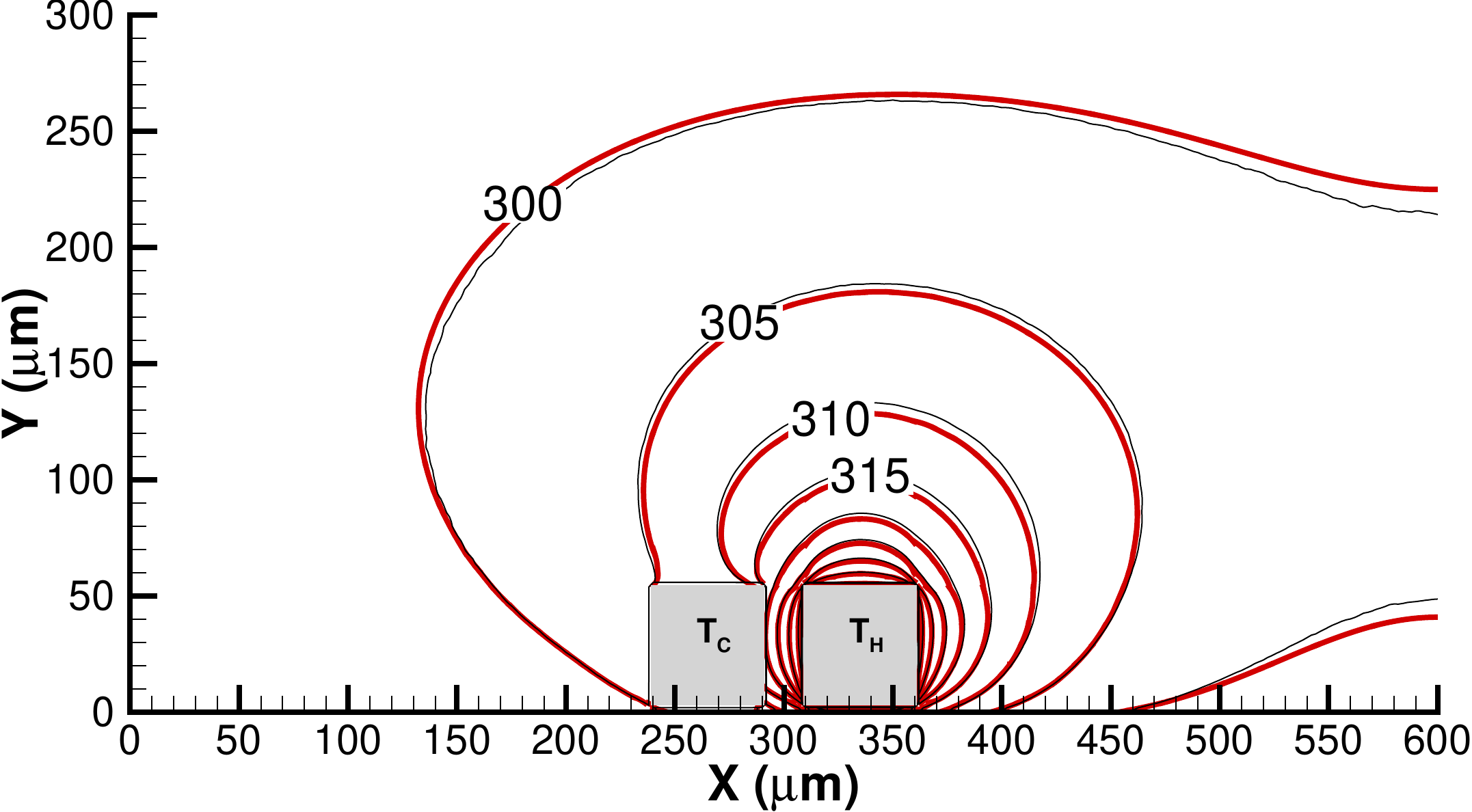}
  \caption{$H_2O$, Temperature ($K$)}
  \label{subfig_mikramulti_flowfield_155N2_T_H2O}
\end{subfigure}
\caption{Variation of flow properties along the domain for multi-species MIKRA Gen1 case (MSM-01: $\Kn=1.85$) obtained with DSMC (thin black lines) and DGFS (thick red lines) using VSS collision model. We want to reemphasize that DSMC simulations consider the rotational degrees of freedom of $N_2$ and $H_2O$ into account, whereas DGFS, being in very early stages of research, doesn't; and therefore we expect some differences between DSMC and DGFS results.}
\label{fig_mikramulti_flowfield_155N2_nden_T}
\end{figure*}

Figures~\ref{subfig_mikramulti_flowfield_155N2_Pxy_N2} and \ref{subfig_mikramulti_flowfield_155N2_Pxy_H2O} illustrate the variation of off-diagonal ($xy$) component of stress tensor. Again, we observe the development of four ovals/ellipses originating at the four corners/edges of the hot vane, wherein the effects are more pronounced at the right end (top-right and bottom-right corners) of the hot vane. The stress is higher for $H_2O$ compared to $N_2$. Figures~\ref{subfig_mikramulti_flowfield_155N2_Speed_N2} and \ref{subfig_mikramulti_flowfield_155N2_Speed_H2O} illustrate the flow speed in the domain. We notice significant statistical fluctuations in DSMC (thin black lines) contour lines for $N_2$ due to lower number of DSMC simulator particles. In particular, we observe that DGFS results/contours are insusceptible to the concentration of the individual species, thereby opening the possibility of its application for simulating flows involving species in trace concentrations. 

\begin{figure*}[!ht]
\centering
\begin{subfigure}[t]{0.7\textwidth}
  \centering
  \includegraphics[width=0.7\textwidth]{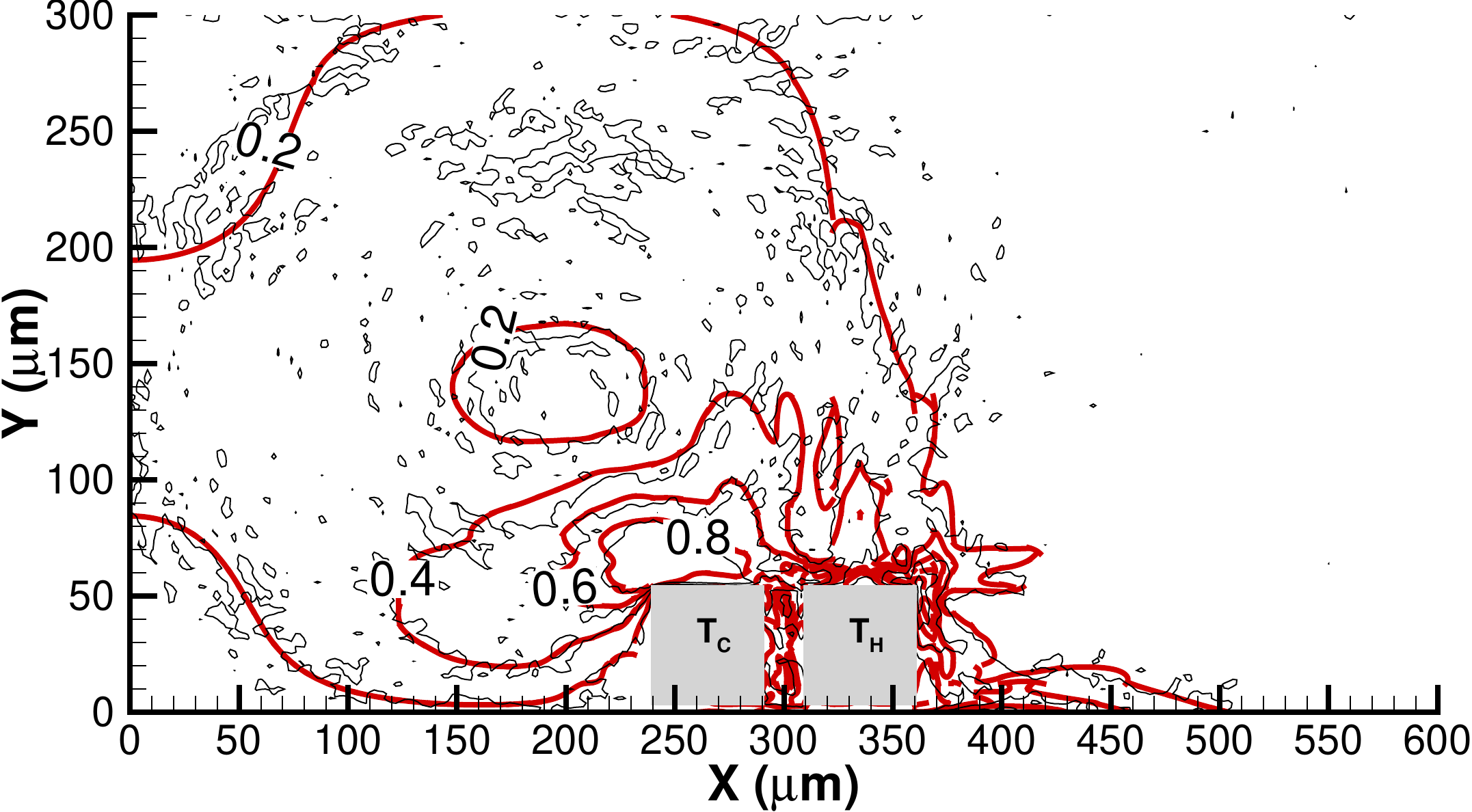}
  \caption{$N_2$, Speed ($m/s$)}
  \label{subfig_mikramulti_flowfield_155N2_Speed_N2}
\end{subfigure}%

\begin{subfigure}[t]{0.7\textwidth}
  \centering
  \includegraphics[width=0.7\textwidth]{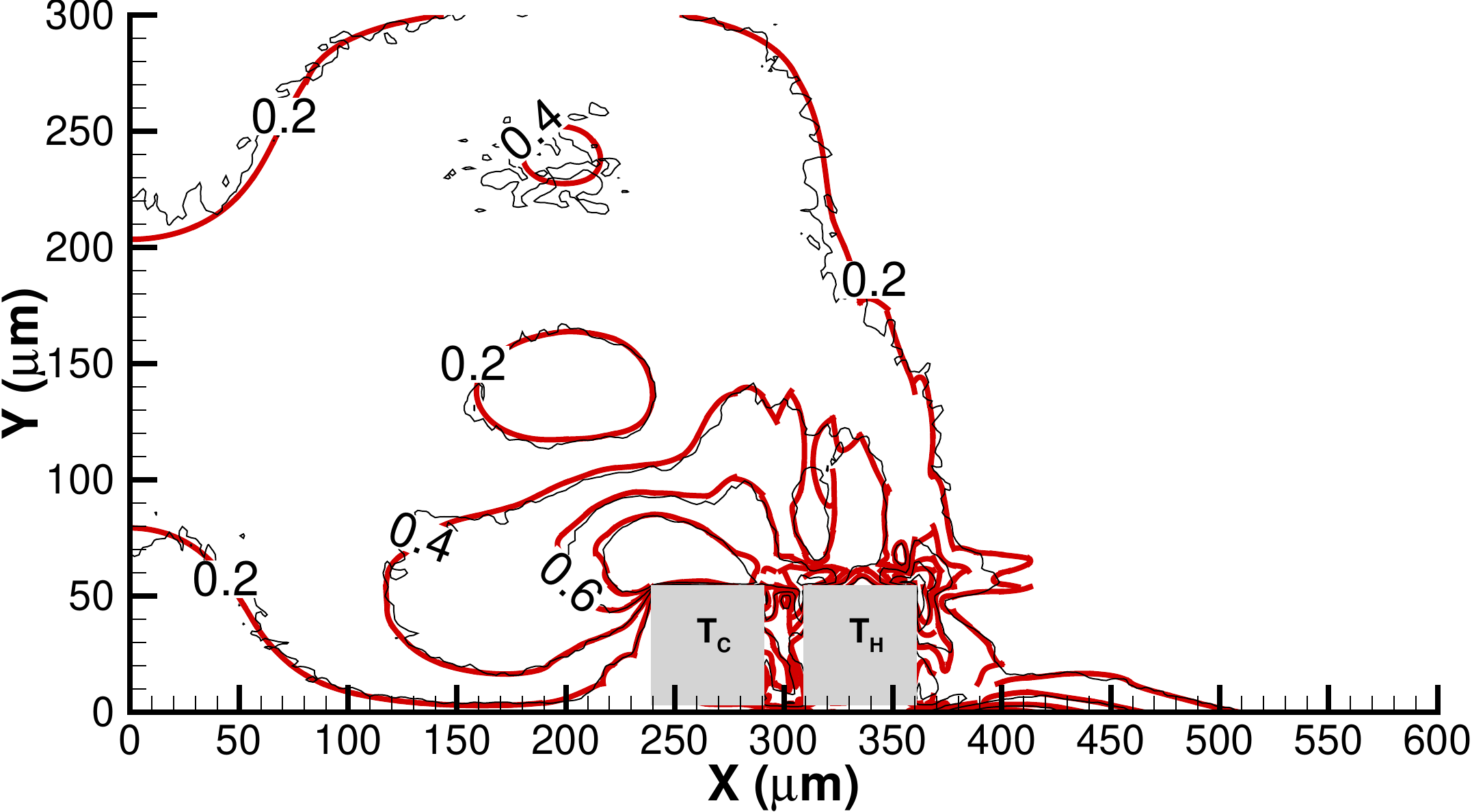}
  \caption{$H_2O$, Speed ($m/s$)}
  \label{subfig_mikramulti_flowfield_155N2_Speed_H2O}
\end{subfigure}

\begin{subfigure}[t]{0.7\textwidth}
  \centering
  \includegraphics[width=0.7\textwidth]{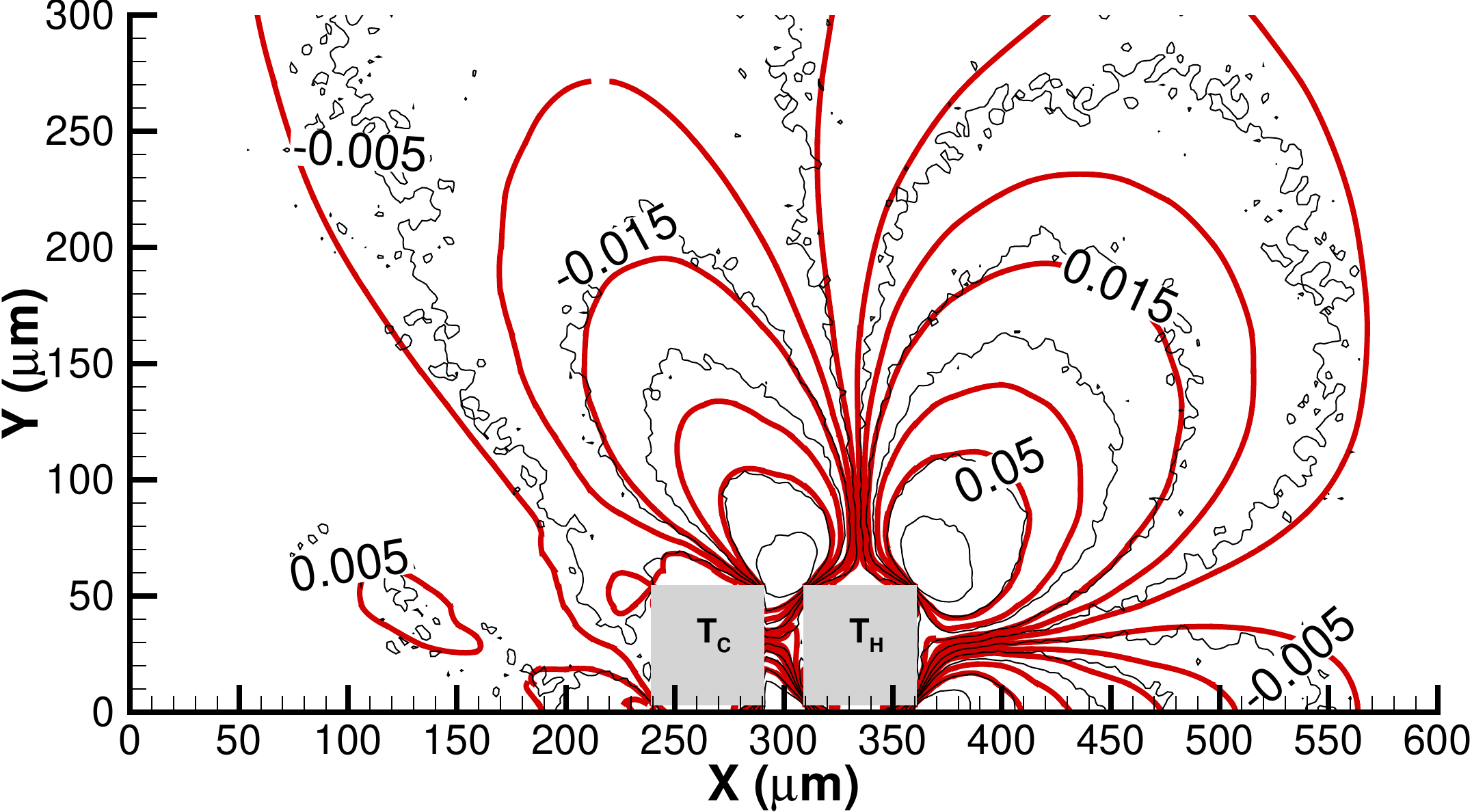}
  \caption{$N_2$, $xy$-component of stress ($N/m^{2}$)}
  \label{subfig_mikramulti_flowfield_155N2_Pxy_N2}
\end{subfigure}%

\begin{subfigure}[t]{0.7\textwidth}
  \centering
  \includegraphics[width=0.7\textwidth]{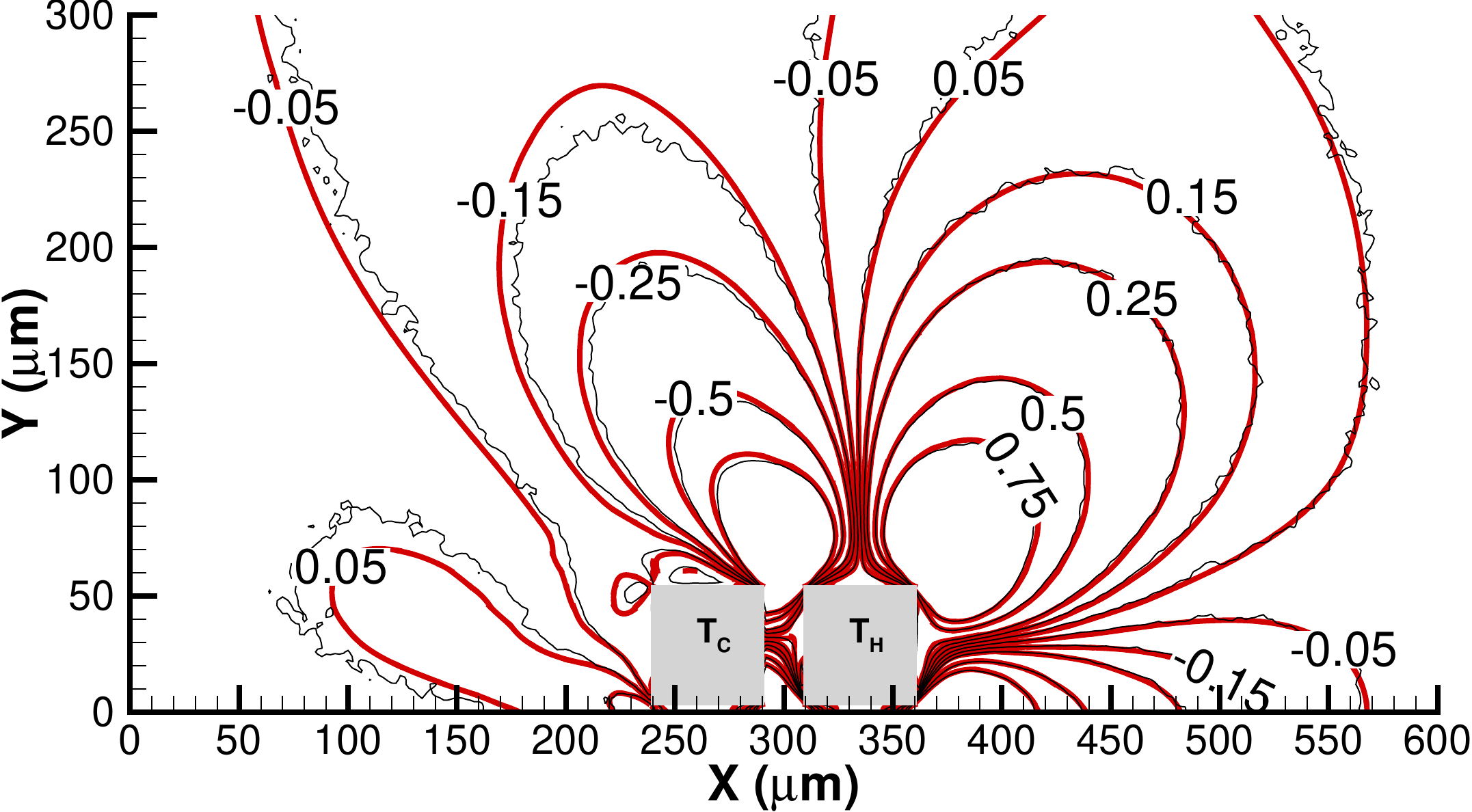}
  \caption{$H_2O$, $xy$-component of stress ($N/m^{2}$)}
  \label{subfig_mikramulti_flowfield_155N2_Pxy_H2O}
\end{subfigure}

\caption{Continuation of Fig.~\ref{fig_mikramulti_flowfield_155N2_nden_T}.}
\label{fig_mikramulti_flowfield_155N2_Pxy_Speed}
\end{figure*}

\begin{figure*}[!ht]
\centering
\begin{subfigure}{.5\textwidth}
  \centering
  \includegraphics[width=80mm]{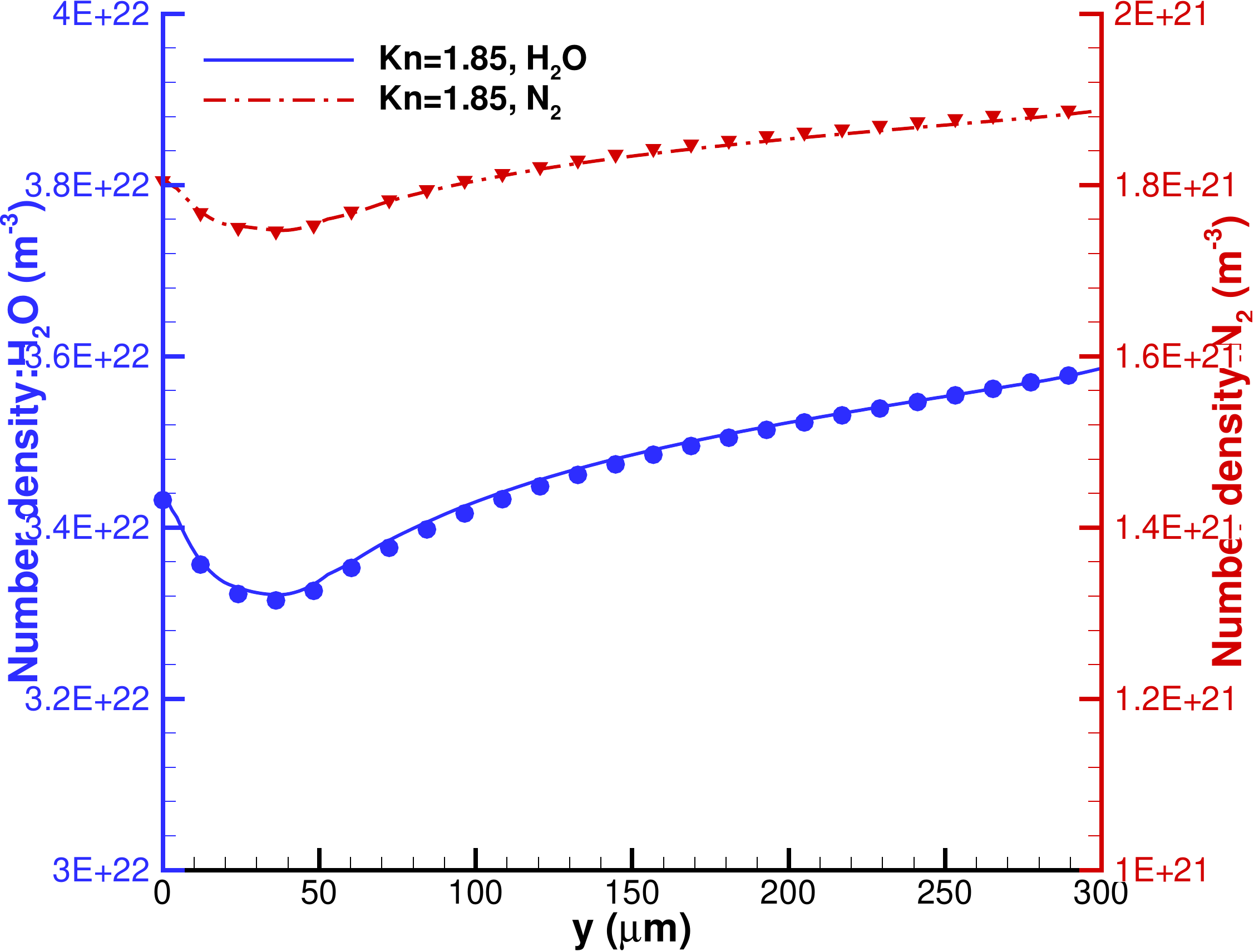}
  \caption{Number density (on vertical centerline)}
  \label{subfig_mikramulti_plotoverline_nden}
\end{subfigure}%
\begin{subfigure}{.5\textwidth}
  \centering
  \includegraphics[width=80mm]{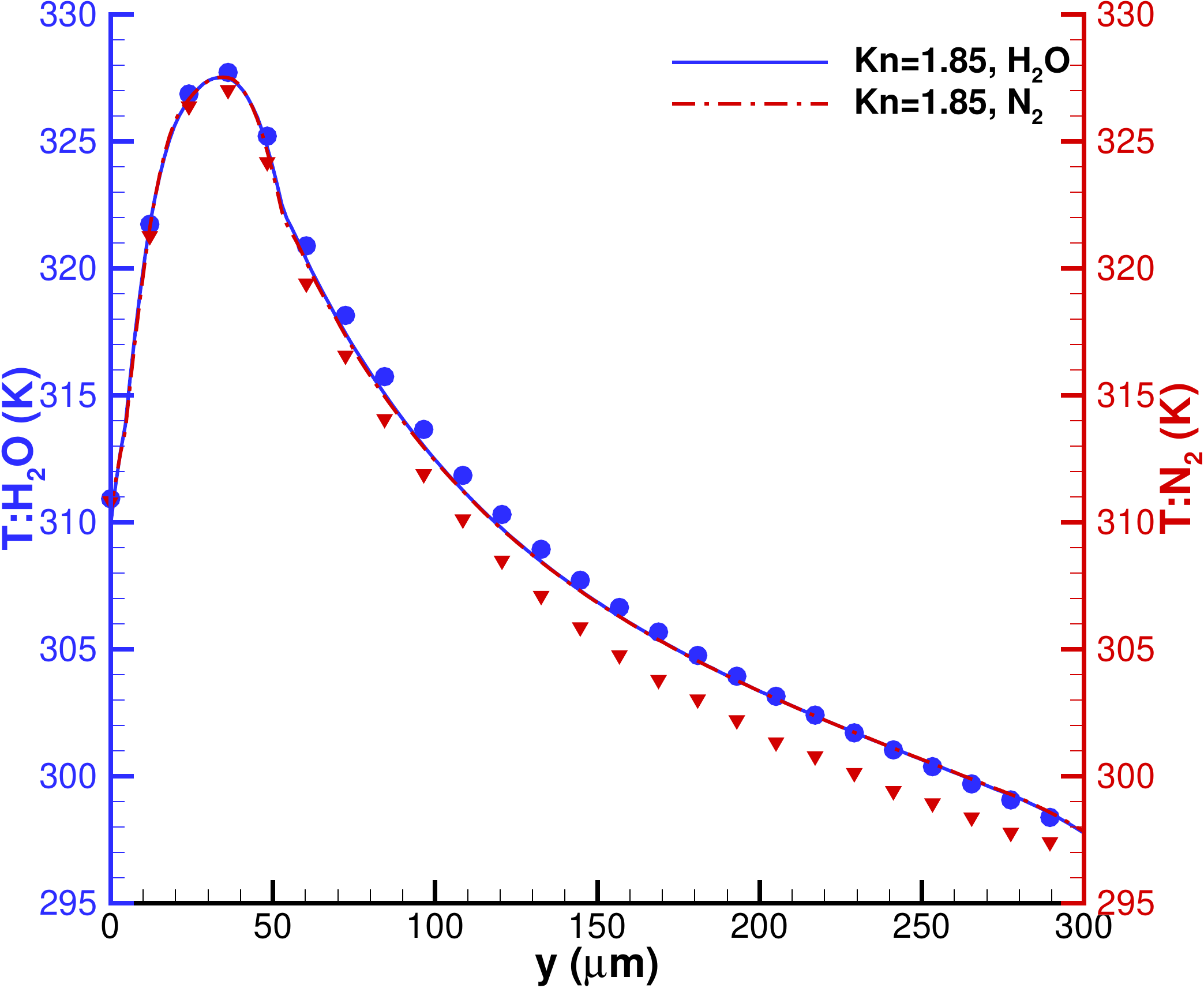}
  \caption{Temperature (on vertical centerline)}
  \label{subfig_mikramulti_plotoverline_T}
\end{subfigure}
\begin{subfigure}{.5\textwidth}
  \centering
  \includegraphics[width=80mm]{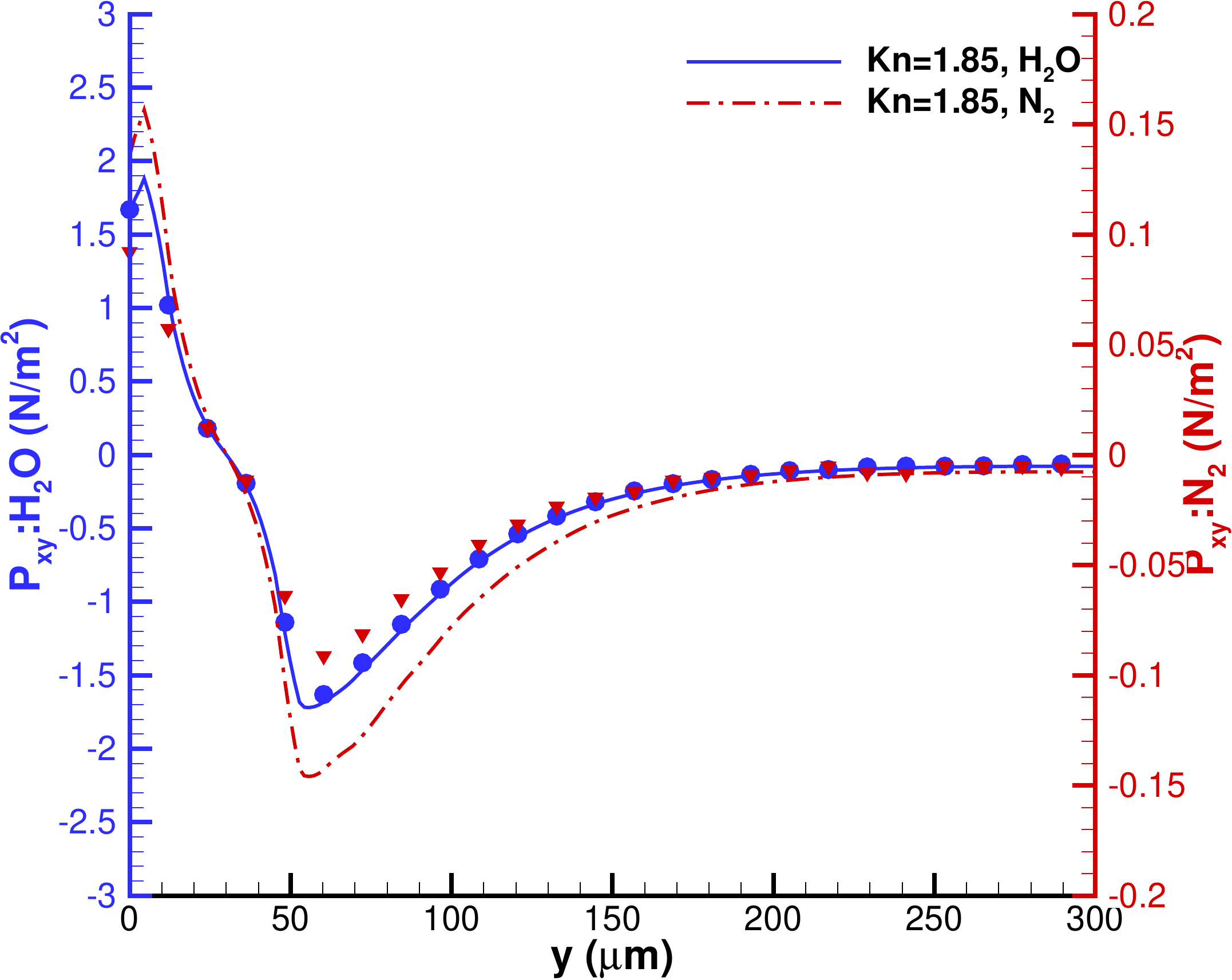}
  \caption{$xy$-component of stress (on vertical centerline)}
  \label{subfig_mikramulti_plotoverline_Pxy}
\end{subfigure}%
\begin{subfigure}{.5\textwidth}
  \centering
  \includegraphics[width=80mm]{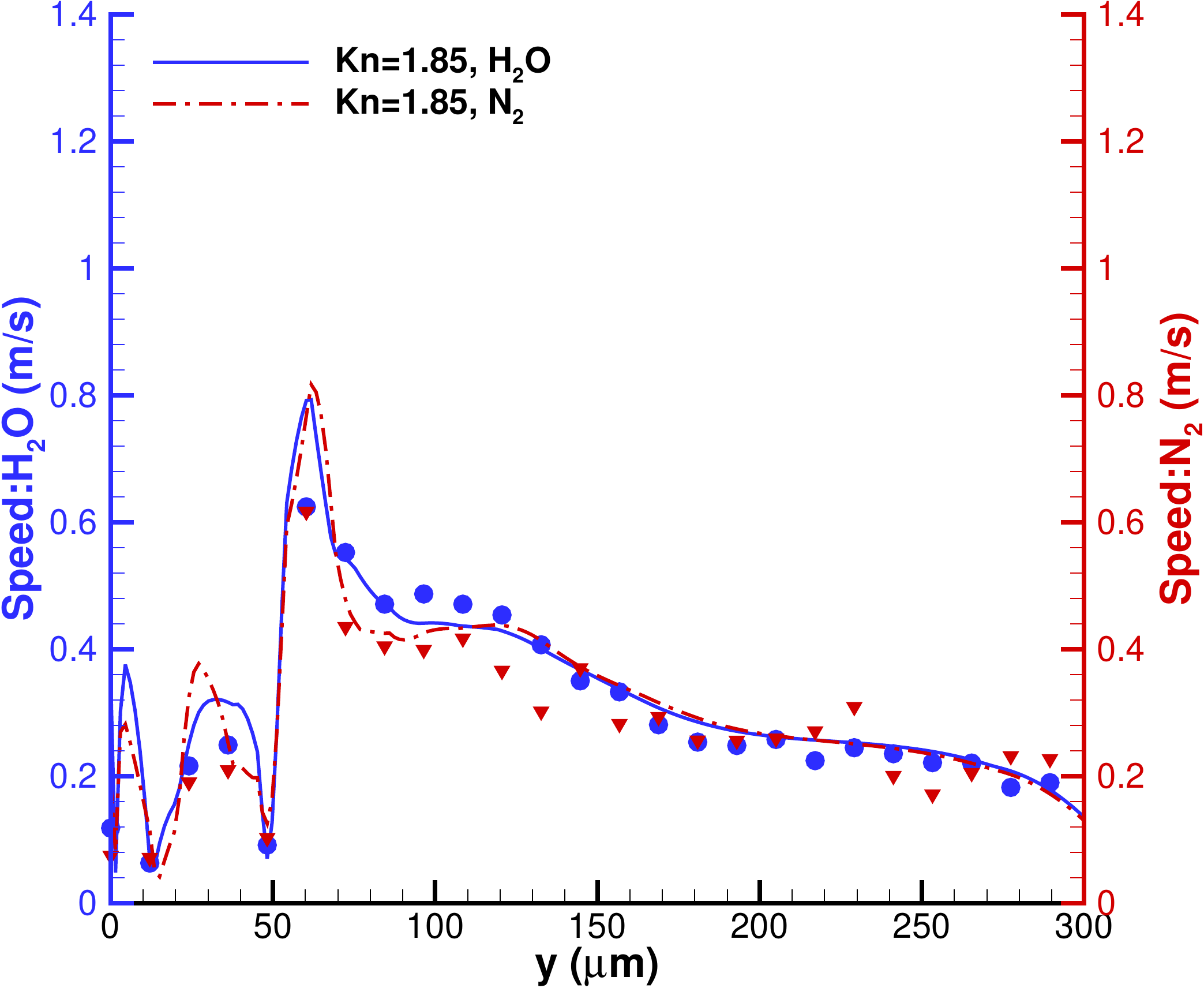}
  \caption{Speed (on vertical centerline)}
  \label{subfig_mikramulti_plotoverline_Speed}
\end{subfigure}
\caption{Variation of flow properties along the domain vertical centerline ($X=300\mu m$) for multi-species MIKRA Gen1 case obtained with DSMC (symbols) and DGFS (lines) using VSS collision model.}
\label{fig_multispecies_mikra_plotoverline}
\end{figure*}

Finally, we compare the variation of inidividual species flow properties along the vertical centerline ($x=300\mu m$, $0\leq y \leq 300 \mu m$). We observe a fair agreement between DSMC and DGFS results ignoring the statistical noise for the bulk properties. In Fig.~\ref{subfig_mikra_plotoverline_T}, we note peak temperatures near the \textit{edges} of hot and cold vanes i.e., in the region $x=300\mu m$, $30\leq y \leq 60 \mu m$. More specifically, the temperature is higher for $H_2O$ compared to $N_2$, an observation consistent with fundamental conservation principles. One can infer that the magnitude of the thermal gradients are stronger in the $x$-direction. Notably, in Fig.~\ref{subfig_mikramulti_plotoverline_Pxy}, we observe the highest thermal-stress in the \textit{edge} region (note the valley in the region $x=300\mu m$, $40\leq y \leq 60 \mu m$). Finally, consistent with aforementioned observations, we observe higher velocity at $x=300\mu m$, $y \approx 54 \mu m$ --- the location of the top edges of the two vanes.


\section{Conclusions}
\label{sec_conclusions}
We have presented an application of the recently introduced deterministic discontinuous Galerkin fast spectral (DGFS) method for assessing the flow phenomenon in the thermo-stress convection enabled microscale device MIKRA --- a compact low-power pressure sensor. We carried out MIKRA simulations in slip-to-transition regime gas flows at different Knudsen numbers. The single-species cases are run with variable hard sphere scattering model. We conclude that the results obtained with DGFS and DSMC are inextricable ignoring the statistical noise. The DSMC provides verification benchmark for the solution of Boltzmann equation with real gas effects. The overall DGFS method is simple from mathematical and implementation perspective; highly accurate in both physical and velocity spaces as well as time; robust, i.e. applicable for general geometry and spatial mesh; exhibits nearly linear parallel scaling; and directly applies to general collision kernels, for instance, Bird's variable hard/soft sphere models, needed for high fidelity modelling. DGFS presents a viable alternative for simulation of highly information rich thermo-stress convection processes at microscale.

\begin{acknowledgments}
SJ and JH's research is partially supported by NSF grant DMS-1620250 and NSF CAREER grant DMS-1654152.
\end{acknowledgments}

\section*{References}
\bibliography{boltzmannInPractice}

\begin{thebibliography}{81}%
\makeatletter
\providecommand \@ifxundefined [1]{%
 \@ifx{#1\undefined}
}%
\providecommand \@ifnum [1]{%
 \ifnum #1\expandafter \@firstoftwo
 \else \expandafter \@secondoftwo
 \fi
}%
\providecommand \@ifx [1]{%
 \ifx #1\expandafter \@firstoftwo
 \else \expandafter \@secondoftwo
 \fi
}%
\providecommand \natexlab [1]{#1}%
\providecommand \enquote  [1]{``#1''}%
\providecommand \bibnamefont  [1]{#1}%
\providecommand \bibfnamefont [1]{#1}%
\providecommand \citenamefont [1]{#1}%
\providecommand \href@noop [0]{\@secondoftwo}%
\providecommand \href [0]{\begingroup \@sanitize@url \@href}%
\providecommand \@href[1]{\@@startlink{#1}\@@href}%
\providecommand \@@href[1]{\endgroup#1\@@endlink}%
\providecommand \@sanitize@url [0]{\catcode `\\12\catcode `\$12\catcode
  `\&12\catcode `\#12\catcode `\^12\catcode `\_12\catcode `\%12\relax}%
\providecommand \@@startlink[1]{}%
\providecommand \@@endlink[0]{}%
\providecommand \url  [0]{\begingroup\@sanitize@url \@url }%
\providecommand \@url [1]{\endgroup\@href {#1}{\urlprefix }}%
\providecommand \urlprefix  [0]{URL }%
\providecommand \Eprint [0]{\href }%
\providecommand \doibase [0]{http://dx.doi.org/}%
\providecommand \selectlanguage [0]{\@gobble}%
\providecommand \bibinfo  [0]{\@secondoftwo}%
\providecommand \bibfield  [0]{\@secondoftwo}%
\providecommand \translation [1]{[#1]}%
\providecommand \BibitemOpen [0]{}%
\providecommand \bibitemStop [0]{}%
\providecommand \bibitemNoStop [0]{.\EOS\space}%
\providecommand \EOS [0]{\spacefactor3000\relax}%
\providecommand \BibitemShut  [1]{\csname bibitem#1\endcsname}%
\let\auto@bib@innerbib\@empty
\bibitem [{\citenamefont {Ho}\ and\ \citenamefont {Tai}(1998)}]{ho1998micro}%
  \BibitemOpen
  \bibfield  {author} {\bibinfo {author} {\bibfnamefont {C.-M.}\ \bibnamefont
  {Ho}}\ and\ \bibinfo {author} {\bibfnamefont {Y.-C.}\ \bibnamefont {Tai}},\
  }\bibfield  {title} {\enquote {\bibinfo {title}
  {Micro-electro-mechanical-systems ({MEMS}) and fluid flows},}\ }\href@noop {}
  {\bibfield  {journal} {\bibinfo  {journal} {Annual review of fluid
  mechanics}\ }\textbf {\bibinfo {volume} {30}},\ \bibinfo {pages} {579--612}
  (\bibinfo {year} {1998})}\BibitemShut {NoStop}%
\bibitem [{\citenamefont {Kogan}, \citenamefont {Galkin},\ and\ \citenamefont
  {Fridlender}(1976)}]{kogan1976stresses}%
  \BibitemOpen
  \bibfield  {author} {\bibinfo {author} {\bibfnamefont {M.}~\bibnamefont
  {Kogan}}, \bibinfo {author} {\bibfnamefont {V.}~\bibnamefont {Galkin}}, \
  and\ \bibinfo {author} {\bibfnamefont {O.}~\bibnamefont {Fridlender}},\
  }\bibfield  {title} {\enquote {\bibinfo {title} {Stresses produced in gases
  by temperature and concentration inhomogeneities. new types of free
  convection},}\ }\href@noop {} {\bibfield  {journal} {\bibinfo  {journal}
  {Usp. Fiz. Nauk}\ }\textbf {\bibinfo {volume} {119}},\ \bibinfo {pages}
  {111--125} (\bibinfo {year} {1976})}\BibitemShut {NoStop}%
\bibitem [{\citenamefont {Strongrich}\ \emph {et~al.}(2017)\citenamefont
  {Strongrich}, \citenamefont {Pikus}, \citenamefont {Sebasti{\~a}o},\ and\
  \citenamefont {Alexeenko}}]{strongrich2017microscale}%
  \BibitemOpen
  \bibfield  {author} {\bibinfo {author} {\bibfnamefont {A.}~\bibnamefont
  {Strongrich}}, \bibinfo {author} {\bibfnamefont {A.}~\bibnamefont {Pikus}},
  \bibinfo {author} {\bibfnamefont {I.~B.}\ \bibnamefont {Sebasti{\~a}o}}, \
  and\ \bibinfo {author} {\bibfnamefont {A.}~\bibnamefont {Alexeenko}},\
  }\bibfield  {title} {\enquote {\bibinfo {title} {Microscale in-plane knudsen
  radiometric actuator: Design, characterization, and performance modeling},}\
  }\href@noop {} {\bibfield  {journal} {\bibinfo  {journal} {Journal of
  Microelectromechanical Systems}\ }\textbf {\bibinfo {volume} {26}},\ \bibinfo
  {pages} {528--538} (\bibinfo {year} {2017})}\BibitemShut {NoStop}%
\bibitem [{\citenamefont {Knudsen}(1910)}]{knudsen1910thermischer}%
  \BibitemOpen
  \bibfield  {author} {\bibinfo {author} {\bibfnamefont {M.}~\bibnamefont
  {Knudsen}},\ }\bibfield  {title} {\enquote {\bibinfo {title} {Thermischer
  molekulardruck der gase in r{\"o}hren},}\ }\href@noop {} {\bibfield
  {journal} {\bibinfo  {journal} {Annalen der Physik}\ }\textbf {\bibinfo
  {volume} {338}},\ \bibinfo {pages} {1435--1448} (\bibinfo {year}
  {1910})}\BibitemShut {NoStop}%
\bibitem [{\citenamefont {Karniadakis}, \citenamefont {Beskok},\ and\
  \citenamefont {Aluru}(2006)}]{karniadakis2006microflows}%
  \BibitemOpen
  \bibfield  {author} {\bibinfo {author} {\bibfnamefont {G.}~\bibnamefont
  {Karniadakis}}, \bibinfo {author} {\bibfnamefont {A.}~\bibnamefont {Beskok}},
  \ and\ \bibinfo {author} {\bibfnamefont {N.}~\bibnamefont {Aluru}},\
  }\href@noop {} {\emph {\bibinfo {title} {Microflows and nanoflows:
  fundamentals and simulation}}},\ Vol.~\bibinfo {volume} {29}\ (\bibinfo
  {publisher} {Springer Science \& Business Media},\ \bibinfo {year}
  {2006})\BibitemShut {NoStop}%
\bibitem [{\citenamefont {Knudsen}(1950)}]{knudsen1950kinetic}%
  \BibitemOpen
  \bibfield  {author} {\bibinfo {author} {\bibfnamefont {M.}~\bibnamefont
  {Knudsen}},\ }\href@noop {} {\emph {\bibinfo {title} {The kinetic theory of
  gases: some modern aspects}}}\ (\bibinfo  {publisher} {Methuen},\ \bibinfo
  {year} {1950})\BibitemShut {NoStop}%
\bibitem [{\citenamefont {Crookes}\ \emph {et~al.}(1874)\citenamefont {Crookes}
  \emph {et~al.}}]{crookes1874xv}%
  \BibitemOpen
  \bibfield  {author} {\bibinfo {author} {\bibfnamefont {W.}~\bibnamefont
  {Crookes}} \emph {et~al.},\ }\bibfield  {title} {\enquote {\bibinfo {title}
  {On attraction and repulsion resulting from radiation},}\ }\href@noop {}
  {\bibfield  {journal} {\bibinfo  {journal} {Philosophical transactions of the
  Royal society of London}\ }\textbf {\bibinfo {volume} {164}},\ \bibinfo
  {pages} {501--527} (\bibinfo {year} {1874})}\BibitemShut {NoStop}%
\bibitem [{\citenamefont {Ketsdever}\ \emph {et~al.}(2012)\citenamefont
  {Ketsdever}, \citenamefont {Gimelshein}, \citenamefont {Gimelshein},\ and\
  \citenamefont {Selden}}]{ketsdever2012radiometric}%
  \BibitemOpen
  \bibfield  {author} {\bibinfo {author} {\bibfnamefont {A.}~\bibnamefont
  {Ketsdever}}, \bibinfo {author} {\bibfnamefont {N.}~\bibnamefont
  {Gimelshein}}, \bibinfo {author} {\bibfnamefont {S.}~\bibnamefont
  {Gimelshein}}, \ and\ \bibinfo {author} {\bibfnamefont {N.}~\bibnamefont
  {Selden}},\ }\bibfield  {title} {\enquote {\bibinfo {title} {Radiometric
  phenomena: From the 19th to the 21st century},}\ }\href@noop {} {\bibfield
  {journal} {\bibinfo  {journal} {Vacuum}\ }\textbf {\bibinfo {volume} {86}},\
  \bibinfo {pages} {1644--1662} (\bibinfo {year} {2012})}\BibitemShut {NoStop}%
\bibitem [{\citenamefont {Maxwell}(1879)}]{maxwell1879vii}%
  \BibitemOpen
  \bibfield  {author} {\bibinfo {author} {\bibfnamefont {J.~C.}\ \bibnamefont
  {Maxwell}},\ }\bibfield  {title} {\enquote {\bibinfo {title} {On stresses in
  rarified gases arising from inequalities of temperature},}\ }\href@noop {}
  {\bibfield  {journal} {\bibinfo  {journal} {Philosophical Transactions of the
  royal society of London}\ }\textbf {\bibinfo {volume} {170}},\ \bibinfo
  {pages} {231--256} (\bibinfo {year} {1879})}\BibitemShut {NoStop}%
\bibitem [{\citenamefont {Chapman}\ and\ \citenamefont
  {Cowling}(1970)}]{chapman1970mathematical}%
  \BibitemOpen
  \bibfield  {author} {\bibinfo {author} {\bibfnamefont {S.}~\bibnamefont
  {Chapman}}\ and\ \bibinfo {author} {\bibfnamefont {T.}~\bibnamefont
  {Cowling}},\ }\bibfield  {title} {\enquote {\bibinfo {title} {The
  mathematical theory of non-uniform gases: An account of the kinetic theory of
  viscosity, thermal conduction and diffusion in gases. cambridge mathematical
  library},}\ }\href@noop {} {\bibfield  {journal} {\bibinfo  {journal}
  {Cambridge University Press}\ }\textbf {\bibinfo {volume} {1}},\ \bibinfo
  {pages} {27--52} (\bibinfo {year} {1970})}\BibitemShut {NoStop}%
\bibitem [{\citenamefont {Bird}(1994)}]{Bird}%
  \BibitemOpen
  \bibfield  {author} {\bibinfo {author} {\bibfnamefont {G.~A.}\ \bibnamefont
  {Bird}},\ }\href@noop {} {\emph {\bibinfo {title} {Molecular {G}as {D}ynamics
  and the {D}irect {S}imulation of {G}as {F}lows}}}\ (\bibinfo  {publisher}
  {Clarendon Press, Oxford},\ \bibinfo {year} {1994})\BibitemShut {NoStop}%
\bibitem [{\citenamefont {Jaiswal}, \citenamefont {Alexeenko},\ and\
  \citenamefont {Hu}(2019{\natexlab{a}})}]{jaiswal2019dgfsMulti}%
  \BibitemOpen
  \bibfield  {author} {\bibinfo {author} {\bibfnamefont {S.}~\bibnamefont
  {Jaiswal}}, \bibinfo {author} {\bibfnamefont {A.~A.}\ \bibnamefont
  {Alexeenko}}, \ and\ \bibinfo {author} {\bibfnamefont {J.}~\bibnamefont
  {Hu}},\ }\bibfield  {title} {\enquote {\bibinfo {title} {A discontinuous
  galerkin fast spectral method for the multi-species full boltzmann
  equation},}\ }\href@noop {} {\bibfield  {journal} {\bibinfo  {journal} {arXiv
  preprint arXiv:1903.03056}\ } (\bibinfo {year}
  {2019}{\natexlab{a}})}\BibitemShut {NoStop}%
\bibitem [{\citenamefont {Pavlov}(2019)}]{pavlov2019diffusion}%
  \BibitemOpen
  \bibfield  {author} {\bibinfo {author} {\bibfnamefont {A.~V.}\ \bibnamefont
  {Pavlov}},\ }\bibfield  {title} {\enquote {\bibinfo {title} {Diffusion and
  thermodiffusion of atmospheric neutral gases: A review},}\ }\href {\doibase
  10.1007/s10712-019-09522-2} {\bibfield  {journal} {\bibinfo  {journal}
  {Surveys in Geophysics}\ }\textbf {\bibinfo {volume} {40}},\ \bibinfo {pages}
  {247--276} (\bibinfo {year} {2019})}\BibitemShut {NoStop}%
\bibitem [{\citenamefont {Sone}(2012)}]{sone2012kinetic}%
  \BibitemOpen
  \bibfield  {author} {\bibinfo {author} {\bibfnamefont {Y.}~\bibnamefont
  {Sone}},\ }\href@noop {} {\emph {\bibinfo {title} {Kinetic theory and fluid
  dynamics}}}\ (\bibinfo  {publisher} {Springer Science \& Business Media},\
  \bibinfo {year} {2012})\BibitemShut {NoStop}%
\bibitem [{\citenamefont {Kennard}\ \emph {et~al.}(1938)\citenamefont {Kennard}
  \emph {et~al.}}]{kennard1938kinetic}%
  \BibitemOpen
  \bibfield  {author} {\bibinfo {author} {\bibfnamefont {E.~H.}\ \bibnamefont
  {Kennard}} \emph {et~al.},\ }\href@noop {} {\emph {\bibinfo {title} {Kinetic
  theory of gases, with an introduction to statistical mechanics}}}\ (\bibinfo
  {publisher} {McGraw-Hill, 1938.},\ \bibinfo {year} {1938})\BibitemShut
  {NoStop}%
\bibitem [{\citenamefont {Sone}(1966)}]{sone1966thermal}%
  \BibitemOpen
  \bibfield  {author} {\bibinfo {author} {\bibfnamefont {Y.}~\bibnamefont
  {Sone}},\ }\bibfield  {title} {\enquote {\bibinfo {title} {Thermal creep in
  rarefied gas},}\ }\href@noop {} {\bibfield  {journal} {\bibinfo  {journal}
  {Journal of the Physical Society of Japan}\ }\textbf {\bibinfo {volume}
  {21}},\ \bibinfo {pages} {1836--1837} (\bibinfo {year} {1966})}\BibitemShut
  {NoStop}%
\bibitem [{\citenamefont {Sharipov}\ and\ \citenamefont
  {Seleznev}(1998)}]{sharipov1998data}%
  \BibitemOpen
  \bibfield  {author} {\bibinfo {author} {\bibfnamefont {F.}~\bibnamefont
  {Sharipov}}\ and\ \bibinfo {author} {\bibfnamefont {V.}~\bibnamefont
  {Seleznev}},\ }\bibfield  {title} {\enquote {\bibinfo {title} {Data on
  internal rarefied gas flows},}\ }\href@noop {} {\bibfield  {journal}
  {\bibinfo  {journal} {Journal of Physical and Chemical Reference Data}\
  }\textbf {\bibinfo {volume} {27}},\ \bibinfo {pages} {657--706} (\bibinfo
  {year} {1998})}\BibitemShut {NoStop}%
\bibitem [{\citenamefont {Aoki}, \citenamefont {Sone},\ and\ \citenamefont
  {Waniguchi}(1998)}]{aoki1998rarefied}%
  \BibitemOpen
  \bibfield  {author} {\bibinfo {author} {\bibfnamefont {K.}~\bibnamefont
  {Aoki}}, \bibinfo {author} {\bibfnamefont {Y.}~\bibnamefont {Sone}}, \ and\
  \bibinfo {author} {\bibfnamefont {Y.}~\bibnamefont {Waniguchi}},\ }\bibfield
  {title} {\enquote {\bibinfo {title} {A rarefied gas flow induced by a
  temperature field: Numerical analysis of the flow between two coaxial
  elliptic cylinders with different uniform temperatures},}\ }\href@noop {}
  {\bibfield  {journal} {\bibinfo  {journal} {Computers \& Mathematics with
  Applications}\ }\textbf {\bibinfo {volume} {35}},\ \bibinfo {pages} {15--28}
  (\bibinfo {year} {1998})}\BibitemShut {NoStop}%
\bibitem [{\citenamefont {Sone}(1972)}]{sone1972flow}%
  \BibitemOpen
  \bibfield  {author} {\bibinfo {author} {\bibfnamefont {Y.}~\bibnamefont
  {Sone}},\ }\bibfield  {title} {\enquote {\bibinfo {title} {Flow induced by
  thermal stress in rarefied gas},}\ }\href@noop {} {\bibfield  {journal}
  {\bibinfo  {journal} {The Physics of Fluids}\ }\textbf {\bibinfo {volume}
  {15}},\ \bibinfo {pages} {1418--1423} (\bibinfo {year} {1972})}\BibitemShut
  {NoStop}%
\bibitem [{\citenamefont {Sone}\ and\ \citenamefont
  {Yoshimoto}(1997)}]{sone1997demonstration}%
  \BibitemOpen
  \bibfield  {author} {\bibinfo {author} {\bibfnamefont {Y.}~\bibnamefont
  {Sone}}\ and\ \bibinfo {author} {\bibfnamefont {M.}~\bibnamefont
  {Yoshimoto}},\ }\bibfield  {title} {\enquote {\bibinfo {title} {Demonstration
  of a rarefied gas flow induced near the edge of a uniformly heated plate},}\
  }\href@noop {} {\bibfield  {journal} {\bibinfo  {journal} {Physics of
  Fluids}\ }\textbf {\bibinfo {volume} {9}},\ \bibinfo {pages} {3530--3534}
  (\bibinfo {year} {1997})}\BibitemShut {NoStop}%
\bibitem [{\citenamefont {Selden}\ \emph
  {et~al.}(2009{\natexlab{a}})\citenamefont {Selden}, \citenamefont {Ngalande},
  \citenamefont {Gimelshein}, \citenamefont {Muntz}, \citenamefont
  {Alexeenko},\ and\ \citenamefont {Ketsdever}}]{selden2009area}%
  \BibitemOpen
  \bibfield  {author} {\bibinfo {author} {\bibfnamefont {N.}~\bibnamefont
  {Selden}}, \bibinfo {author} {\bibfnamefont {C.}~\bibnamefont {Ngalande}},
  \bibinfo {author} {\bibfnamefont {S.}~\bibnamefont {Gimelshein}}, \bibinfo
  {author} {\bibfnamefont {E.}~\bibnamefont {Muntz}}, \bibinfo {author}
  {\bibfnamefont {A.}~\bibnamefont {Alexeenko}}, \ and\ \bibinfo {author}
  {\bibfnamefont {A.}~\bibnamefont {Ketsdever}},\ }\bibfield  {title} {\enquote
  {\bibinfo {title} {Area and edge effects in radiometric forces},}\
  }\href@noop {} {\bibfield  {journal} {\bibinfo  {journal} {Physical Review
  E}\ }\textbf {\bibinfo {volume} {79}},\ \bibinfo {pages} {041201} (\bibinfo
  {year} {2009}{\natexlab{a}})}\BibitemShut {NoStop}%
\bibitem [{\citenamefont {Fowee}\ \emph {et~al.}(2016)\citenamefont {Fowee},
  \citenamefont {Ibrayeva}, \citenamefont {Strongrich},\ and\ \citenamefont
  {Alexeenko}}]{fowee2016experimental}%
  \BibitemOpen
  \bibfield  {author} {\bibinfo {author} {\bibfnamefont {K.}~\bibnamefont
  {Fowee}}, \bibinfo {author} {\bibfnamefont {A.}~\bibnamefont {Ibrayeva}},
  \bibinfo {author} {\bibfnamefont {A.}~\bibnamefont {Strongrich}}, \ and\
  \bibinfo {author} {\bibfnamefont {A.}~\bibnamefont {Alexeenko}},\ }\bibfield
  {title} {\enquote {\bibinfo {title} {Experimental measurements and numerical
  modeling of a thermostress convection-based actuator},}\ }in\ \href@noop {}
  {\emph {\bibinfo {booktitle} {AIP Conference Proceedings}}},\ Vol.\ \bibinfo
  {volume} {1786}\ (\bibinfo {organization} {AIP Publishing},\ \bibinfo {year}
  {2016})\ p.\ \bibinfo {pages} {200004}\BibitemShut {NoStop}%
\bibitem [{\citenamefont {Ibrayeva}(2017)}]{ibrayeva2017numerical}%
  \BibitemOpen
  \bibfield  {author} {\bibinfo {author} {\bibfnamefont {A.}~\bibnamefont
  {Ibrayeva}},\ }\emph {\bibinfo {title} {Numerical Modeling of Thermal Edge
  Flow}},\ \href@noop {} {Master's thesis},\ \bibinfo  {school} {Purdue
  University} (\bibinfo {year} {2017})\BibitemShut {NoStop}%
\bibitem [{\citenamefont {Passian}\ \emph {et~al.}(2002)\citenamefont
  {Passian}, \citenamefont {Wig}, \citenamefont {Meriaudeau}, \citenamefont
  {Ferrell},\ and\ \citenamefont {Thundat}}]{passian2002knudsen}%
  \BibitemOpen
  \bibfield  {author} {\bibinfo {author} {\bibfnamefont {A.}~\bibnamefont
  {Passian}}, \bibinfo {author} {\bibfnamefont {A.}~\bibnamefont {Wig}},
  \bibinfo {author} {\bibfnamefont {F.}~\bibnamefont {Meriaudeau}}, \bibinfo
  {author} {\bibfnamefont {T.}~\bibnamefont {Ferrell}}, \ and\ \bibinfo
  {author} {\bibfnamefont {T.}~\bibnamefont {Thundat}},\ }\bibfield  {title}
  {\enquote {\bibinfo {title} {Knudsen forces on microcantilevers},}\
  }\href@noop {} {\bibfield  {journal} {\bibinfo  {journal} {Journal of applied
  physics}\ }\textbf {\bibinfo {volume} {92}},\ \bibinfo {pages} {6326--6333}
  (\bibinfo {year} {2002})}\BibitemShut {NoStop}%
\bibitem [{\citenamefont {Passian}\ \emph {et~al.}(2003)\citenamefont
  {Passian}, \citenamefont {Warmack}, \citenamefont {Ferrell},\ and\
  \citenamefont {Thundat}}]{passian2003thermal}%
  \BibitemOpen
  \bibfield  {author} {\bibinfo {author} {\bibfnamefont {A.}~\bibnamefont
  {Passian}}, \bibinfo {author} {\bibfnamefont {R.}~\bibnamefont {Warmack}},
  \bibinfo {author} {\bibfnamefont {T.}~\bibnamefont {Ferrell}}, \ and\
  \bibinfo {author} {\bibfnamefont {T.}~\bibnamefont {Thundat}},\ }\bibfield
  {title} {\enquote {\bibinfo {title} {Thermal transpiration at the microscale:
  a crookes cantilever},}\ }\href@noop {} {\bibfield  {journal} {\bibinfo
  {journal} {Physical review letters}\ }\textbf {\bibinfo {volume} {90}},\
  \bibinfo {pages} {124503} (\bibinfo {year} {2003})}\BibitemShut {NoStop}%
\bibitem [{\citenamefont {Foroutan}\ \emph {et~al.}(2014)\citenamefont
  {Foroutan}, \citenamefont {Majumdar}, \citenamefont {Mahdavipour},
  \citenamefont {Ward},\ and\ \citenamefont
  {Paprotny}}]{foroutan2014levitation}%
  \BibitemOpen
  \bibfield  {author} {\bibinfo {author} {\bibfnamefont {V.}~\bibnamefont
  {Foroutan}}, \bibinfo {author} {\bibfnamefont {R.}~\bibnamefont {Majumdar}},
  \bibinfo {author} {\bibfnamefont {O.}~\bibnamefont {Mahdavipour}}, \bibinfo
  {author} {\bibfnamefont {S.}~\bibnamefont {Ward}}, \ and\ \bibinfo {author}
  {\bibfnamefont {I.}~\bibnamefont {Paprotny}},\ }\bibfield  {title} {\enquote
  {\bibinfo {title} {Levitation of untethered stress engineered microflyers
  using thermophoretic (knudsen) force},}\ }in\ \href@noop {} {\emph {\bibinfo
  {booktitle} {Technical Digest of the Hilton Head Workshop}}}\ (\bibinfo
  {year} {2014})\ pp.\ \bibinfo {pages} {105--106}\BibitemShut {NoStop}%
\bibitem [{\citenamefont {Nallapu}, \citenamefont {Tallapragada},\ and\
  \citenamefont {Thangavelautham}(2017)}]{nallapu2017radiometric}%
  \BibitemOpen
  \bibfield  {author} {\bibinfo {author} {\bibfnamefont {R.~T.}\ \bibnamefont
  {Nallapu}}, \bibinfo {author} {\bibfnamefont {A.}~\bibnamefont
  {Tallapragada}}, \ and\ \bibinfo {author} {\bibfnamefont {J.}~\bibnamefont
  {Thangavelautham}},\ }\bibfield  {title} {\enquote {\bibinfo {title}
  {Radiometric actuators for spacecraft attitude control},}\ }\href@noop {}
  {\bibfield  {journal} {\bibinfo  {journal} {arXiv preprint arXiv:1701.07545}\
  } (\bibinfo {year} {2017})}\BibitemShut {NoStop}%
\bibitem [{\citenamefont {Cornella}\ \emph {et~al.}(2012)\citenamefont
  {Cornella}, \citenamefont {Ketsdever}, \citenamefont {Gimelshein},\ and\
  \citenamefont {Gimelshein}}]{cornella2012analysis}%
  \BibitemOpen
  \bibfield  {author} {\bibinfo {author} {\bibfnamefont {B.~M.}\ \bibnamefont
  {Cornella}}, \bibinfo {author} {\bibfnamefont {A.~D.}\ \bibnamefont
  {Ketsdever}}, \bibinfo {author} {\bibfnamefont {N.~E.}\ \bibnamefont
  {Gimelshein}}, \ and\ \bibinfo {author} {\bibfnamefont {S.~F.}\ \bibnamefont
  {Gimelshein}},\ }\bibfield  {title} {\enquote {\bibinfo {title} {Analysis of
  multivane radiometer arrays in high-altitude propulsion},}\ }\href@noop {}
  {\bibfield  {journal} {\bibinfo  {journal} {Journal of Propulsion and Power}\
  }\textbf {\bibinfo {volume} {28}},\ \bibinfo {pages} {831--839} (\bibinfo
  {year} {2012})}\BibitemShut {NoStop}%
\bibitem [{\citenamefont {Grad}(1949)}]{grad1949}%
  \BibitemOpen
  \bibfield  {author} {\bibinfo {author} {\bibfnamefont {H.}~\bibnamefont
  {Grad}},\ }\bibfield  {title} {\enquote {\bibinfo {title} {On the kinetic
  theory of rarefied gases},}\ }\href@noop {} {\bibfield  {journal} {\bibinfo
  {journal} {Communications on pure and applied mathematics}\ }\textbf
  {\bibinfo {volume} {2}},\ \bibinfo {pages} {331--407} (\bibinfo {year}
  {1949})}\BibitemShut {NoStop}%
\bibitem [{\citenamefont {Bird}(1963)}]{bird1963approach}%
  \BibitemOpen
  \bibfield  {author} {\bibinfo {author} {\bibfnamefont {G.}~\bibnamefont
  {Bird}},\ }\bibfield  {title} {\enquote {\bibinfo {title} {Approach to
  translational equilibrium in a rigid sphere gas},}\ }\href@noop {} {\bibfield
   {journal} {\bibinfo  {journal} {The Physics of Fluids}\ }\textbf {\bibinfo
  {volume} {6}},\ \bibinfo {pages} {1518--1519} (\bibinfo {year}
  {1963})}\BibitemShut {NoStop}%
\bibitem [{\citenamefont {Maitland}\ and\ \citenamefont
  {Smith}(1972)}]{maitland1972critical}%
  \BibitemOpen
  \bibfield  {author} {\bibinfo {author} {\bibfnamefont {G.~C.}\ \bibnamefont
  {Maitland}}\ and\ \bibinfo {author} {\bibfnamefont {E.~B.}\ \bibnamefont
  {Smith}},\ }\bibfield  {title} {\enquote {\bibinfo {title} {Critical
  reassessment of viscosities of 11 common gases},}\ }\href@noop {} {\bibfield
  {journal} {\bibinfo  {journal} {Journal of Chemical and Engineering Data}\
  }\textbf {\bibinfo {volume} {17}},\ \bibinfo {pages} {150--156} (\bibinfo
  {year} {1972})}\BibitemShut {NoStop}%
\bibitem [{\citenamefont {Koura}, \citenamefont {Matsumoto},\ and\
  \citenamefont {Shimada}(1991)}]{koura1991test}%
  \BibitemOpen
  \bibfield  {author} {\bibinfo {author} {\bibfnamefont {K.}~\bibnamefont
  {Koura}}, \bibinfo {author} {\bibfnamefont {H.}~\bibnamefont {Matsumoto}}, \
  and\ \bibinfo {author} {\bibfnamefont {T.}~\bibnamefont {Shimada}},\
  }\bibfield  {title} {\enquote {\bibinfo {title} {A test of equivalence of the
  variable-hard-sphere and inverse-power-law models in the direct-simulation
  monte carlo method},}\ }\href@noop {} {\bibfield  {journal} {\bibinfo
  {journal} {Physics of Fluids A: Fluid Dynamics}\ }\textbf {\bibinfo {volume}
  {3}},\ \bibinfo {pages} {1835--1837} (\bibinfo {year} {1991})}\BibitemShut
  {NoStop}%
\bibitem [{\citenamefont {Koura}\ and\ \citenamefont
  {Matsumoto}(1991)}]{koura1991variable}%
  \BibitemOpen
  \bibfield  {author} {\bibinfo {author} {\bibfnamefont {K.}~\bibnamefont
  {Koura}}\ and\ \bibinfo {author} {\bibfnamefont {H.}~\bibnamefont
  {Matsumoto}},\ }\bibfield  {title} {\enquote {\bibinfo {title} {Variable soft
  sphere molecular model for inverse-power-law or {L}ennard-{J}ones
  potential},}\ }\href@noop {} {\bibfield  {journal} {\bibinfo  {journal}
  {Physics of Fluids A: Fluid Dynamics}\ }\textbf {\bibinfo {volume} {3}},\
  \bibinfo {pages} {2459--2465} (\bibinfo {year} {1991})}\BibitemShut {NoStop}%
\bibitem [{\citenamefont {Kersch}, \citenamefont {Morokoff},\ and\
  \citenamefont {Werner}(1994)}]{kersch1994selfconsistent}%
  \BibitemOpen
  \bibfield  {author} {\bibinfo {author} {\bibfnamefont {A.}~\bibnamefont
  {Kersch}}, \bibinfo {author} {\bibfnamefont {W.}~\bibnamefont {Morokoff}}, \
  and\ \bibinfo {author} {\bibfnamefont {C.}~\bibnamefont {Werner}},\
  }\bibfield  {title} {\enquote {\bibinfo {title} {Selfconsistent simulation of
  sputter deposition with the monte carlo method},}\ }\href@noop {} {\bibfield
  {journal} {\bibinfo  {journal} {Journal of applied physics}\ }\textbf
  {\bibinfo {volume} {75}},\ \bibinfo {pages} {2278--2285} (\bibinfo {year}
  {1994})}\BibitemShut {NoStop}%
\bibitem [{\citenamefont {Fan}(2002)}]{fan2002generalized}%
  \BibitemOpen
  \bibfield  {author} {\bibinfo {author} {\bibfnamefont {J.}~\bibnamefont
  {Fan}},\ }\bibfield  {title} {\enquote {\bibinfo {title} {A generalized
  soft-sphere model for monte carlo simulation},}\ }\href@noop {} {\bibfield
  {journal} {\bibinfo  {journal} {Physics of Fluids}\ }\textbf {\bibinfo
  {volume} {14}},\ \bibinfo {pages} {4399--4405} (\bibinfo {year}
  {2002})}\BibitemShut {NoStop}%
\bibitem [{\citenamefont {Jaiswal}, \citenamefont {Sebasti\~ao},\ and\
  \citenamefont {Alexeenko}(2018)}]{jaiswal2018dsmc}%
  \BibitemOpen
  \bibfield  {author} {\bibinfo {author} {\bibfnamefont {S.}~\bibnamefont
  {Jaiswal}}, \bibinfo {author} {\bibfnamefont {I.}~\bibnamefont
  {Sebasti\~ao}}, \ and\ \bibinfo {author} {\bibfnamefont {A.~A.}\ \bibnamefont
  {Alexeenko}},\ }\bibfield  {title} {\enquote {\bibinfo {title} {{DSMC-SPARTA}
  {I}mplementation of {M-1} {S}cattering {M}odel},}\ }in\ \href@noop {} {\emph
  {\bibinfo {booktitle} {Proceedings of 31st Rarefied Gas Dynamics
  Symposium}}}\ (\bibinfo  {publisher} {AIP},\ \bibinfo {year} {2018})\
  \bibinfo {note} {to appear:
  \href{http://goo.gl/N7qFao}{http://goo.gl/N7qFao}}\BibitemShut {NoStop}%
\bibitem [{\citenamefont {Weaver}(2015)}]{weaver2015assessment}%
  \BibitemOpen
  \bibfield  {author} {\bibinfo {author} {\bibfnamefont {A.~B.}\ \bibnamefont
  {Weaver}},\ }\emph {\bibinfo {title} {Assessment of high-fidelity collision
  models in the direct simulation {M}onte {C}arlo method}},\ \href@noop {}
  {Ph.D. thesis},\ \bibinfo  {school} {Purdue University, West Lafayette}
  (\bibinfo {year} {2015})\BibitemShut {NoStop}%
\bibitem [{\citenamefont {Jaiswal}, \citenamefont {Alexeenko},\ and\
  \citenamefont {Hu}(2019{\natexlab{b}})}]{JAH19}%
  \BibitemOpen
  \bibfield  {author} {\bibinfo {author} {\bibfnamefont {S.}~\bibnamefont
  {Jaiswal}}, \bibinfo {author} {\bibfnamefont {A.}~\bibnamefont {Alexeenko}},
  \ and\ \bibinfo {author} {\bibfnamefont {J.}~\bibnamefont {Hu}},\ }\bibfield
  {title} {\enquote {\bibinfo {title} {A discontinuous {G}alerkin fast spectral
  method for the full {B}oltzmann equation with general collision kernels},}\
  }\href@noop {} {\bibfield  {journal} {\bibinfo  {journal} {Journal of
  Computational Physics}\ }\textbf {\bibinfo {volume} {378}},\ \bibinfo {pages}
  {178--208} (\bibinfo {year} {2019}{\natexlab{b}})}\BibitemShut {NoStop}%
\bibitem [{\citenamefont {Jaiswal}, \citenamefont {Alexeenko},\ and\
  \citenamefont {Hu}(2018)}]{jaiswal2018dgfsGPU}%
  \BibitemOpen
  \bibfield  {author} {\bibinfo {author} {\bibfnamefont {S.}~\bibnamefont
  {Jaiswal}}, \bibinfo {author} {\bibfnamefont {A.~A.}\ \bibnamefont
  {Alexeenko}}, \ and\ \bibinfo {author} {\bibfnamefont {J.}~\bibnamefont
  {Hu}},\ }\bibfield  {title} {\enquote {\bibinfo {title} {{F}ast
  {D}eterministic solution of the full {B}oltzmann equation on {G}raphics
  {P}rocessing {U}nits},}\ }in\ \href@noop {} {\emph {\bibinfo {booktitle}
  {Proceedings of 31st Rarefied Gas Dynamics Symposium}}}\ (\bibinfo
  {organization} {AIP},\ \bibinfo {year} {2018})\ \bibinfo {note} {to appear:
  \href{http://goo.gl/x4A7sy}{http://goo.gl/x4A7sy}}\BibitemShut {NoStop}%
\bibitem [{\citenamefont {Jaiswal}\ \emph {et~al.}(2019)\citenamefont
  {Jaiswal}, \citenamefont {Hu}, \citenamefont {Brillon},\ and\ \citenamefont
  {Alexeenko}}]{jaiswal2019dgfsMultiSpeciesGPU}%
  \BibitemOpen
  \bibfield  {author} {\bibinfo {author} {\bibfnamefont {S.}~\bibnamefont
  {Jaiswal}}, \bibinfo {author} {\bibfnamefont {J.}~\bibnamefont {Hu}},
  \bibinfo {author} {\bibfnamefont {J.~K.}\ \bibnamefont {Brillon}}, \ and\
  \bibinfo {author} {\bibfnamefont {A.~A.}\ \bibnamefont {Alexeenko}},\
  }\bibfield  {title} {\enquote {\bibinfo {title} {A discontinuous {G}alerkin
  fast spectral method for multi-species full {B}oltzmann on streaming
  multi-processors},}\ }in\ \href {https://goo.gl/PCr3AG} {\emph {\bibinfo
  {booktitle} {Proceedings of Platform for Advanced Scientific Computing
  (PASC'19)}}}\ (\bibinfo  {publisher} {ACM},\ \bibinfo {address} {Zurich,
  Switzerland},\ \bibinfo {year} {2019})\ \bibinfo {note} {{A}ccepted:
  \href{https://goo.gl/PCr3AG}{https://goo.gl/PCr3AG}}\BibitemShut {NoStop}%
\bibitem [{\citenamefont {Loyalka}(1977)}]{loyalka1977knudsen}%
  \BibitemOpen
  \bibfield  {author} {\bibinfo {author} {\bibfnamefont {S.}~\bibnamefont
  {Loyalka}},\ }\bibfield  {title} {\enquote {\bibinfo {title} {Knudsen forces
  in vacuum microbalance},}\ }\href@noop {} {\bibfield  {journal} {\bibinfo
  {journal} {The Journal of Chemical Physics}\ }\textbf {\bibinfo {volume}
  {66}},\ \bibinfo {pages} {4935--4940} (\bibinfo {year} {1977})}\BibitemShut
  {NoStop}%
\bibitem [{\citenamefont {Selden}\ \emph
  {et~al.}(2009{\natexlab{b}})\citenamefont {Selden}, \citenamefont {Ngalande},
  \citenamefont {Gimelshein}, \citenamefont {Gimelshein},\ and\ \citenamefont
  {Ketsdever}}]{selden2009origins}%
  \BibitemOpen
  \bibfield  {author} {\bibinfo {author} {\bibfnamefont {N.}~\bibnamefont
  {Selden}}, \bibinfo {author} {\bibfnamefont {C.}~\bibnamefont {Ngalande}},
  \bibinfo {author} {\bibfnamefont {N.}~\bibnamefont {Gimelshein}}, \bibinfo
  {author} {\bibfnamefont {S.}~\bibnamefont {Gimelshein}}, \ and\ \bibinfo
  {author} {\bibfnamefont {A.}~\bibnamefont {Ketsdever}},\ }\bibfield  {title}
  {\enquote {\bibinfo {title} {Origins of radiometric forces on a circular vane
  with a temperature gradient},}\ }\href@noop {} {\bibfield  {journal}
  {\bibinfo  {journal} {Journal of Fluid Mechanics}\ }\textbf {\bibinfo
  {volume} {634}},\ \bibinfo {pages} {419--431} (\bibinfo {year}
  {2009}{\natexlab{b}})}\BibitemShut {NoStop}%
\bibitem [{\citenamefont {Fierro}\ and\ \citenamefont
  {Garcia}(1981)}]{fierro1981gas}%
  \BibitemOpen
  \bibfield  {author} {\bibinfo {author} {\bibfnamefont {J.~G.}\ \bibnamefont
  {Fierro}}\ and\ \bibinfo {author} {\bibfnamefont {A.~A.}\ \bibnamefont
  {Garcia}},\ }\bibfield  {title} {\enquote {\bibinfo {title} {Gas dynamics at
  low pressures in a vacuum microbalance},}\ }\href@noop {} {\bibfield
  {journal} {\bibinfo  {journal} {Vacuum}\ }\textbf {\bibinfo {volume} {31}},\
  \bibinfo {pages} {79--84} (\bibinfo {year} {1981})}\BibitemShut {NoStop}%
\bibitem [{\citenamefont {Alexeenko}\ \emph
  {et~al.}(2006{\natexlab{a}})\citenamefont {Alexeenko}, \citenamefont {Muntz},
  \citenamefont {Gallis},\ and\ \citenamefont
  {Torczynski}}]{alexeenko2006comparison}%
  \BibitemOpen
  \bibfield  {author} {\bibinfo {author} {\bibfnamefont {A.}~\bibnamefont
  {Alexeenko}}, \bibinfo {author} {\bibfnamefont {E.~P.}\ \bibnamefont
  {Muntz}}, \bibinfo {author} {\bibfnamefont {M.}~\bibnamefont {Gallis}}, \
  and\ \bibinfo {author} {\bibfnamefont {J.}~\bibnamefont {Torczynski}},\
  }\bibfield  {title} {\enquote {\bibinfo {title} {Comparison of kinetic models
  for gas damping of moving microbeams},}\ }in\ \href@noop {} {\emph {\bibinfo
  {booktitle} {36th AIAA Fluid Dynamics Conference and Exhibit}}}\ (\bibinfo
  {year} {2006})\ p.\ \bibinfo {pages} {3715}\BibitemShut {NoStop}%
\bibitem [{\citenamefont {Zhu}\ and\ \citenamefont {Ye}(2010)}]{zhu2010origin}%
  \BibitemOpen
  \bibfield  {author} {\bibinfo {author} {\bibfnamefont {T.}~\bibnamefont
  {Zhu}}\ and\ \bibinfo {author} {\bibfnamefont {W.}~\bibnamefont {Ye}},\
  }\bibfield  {title} {\enquote {\bibinfo {title} {Origin of knudsen forces on
  heated microbeams},}\ }\href@noop {} {\bibfield  {journal} {\bibinfo
  {journal} {Physical Review E}\ }\textbf {\bibinfo {volume} {82}},\ \bibinfo
  {pages} {036308} (\bibinfo {year} {2010})}\BibitemShut {NoStop}%
\bibitem [{\citenamefont {Nabeth}, \citenamefont {Chigullapalli},\ and\
  \citenamefont {Alexeenko}(2011)}]{nabeth2011quantifying}%
  \BibitemOpen
  \bibfield  {author} {\bibinfo {author} {\bibfnamefont {J.}~\bibnamefont
  {Nabeth}}, \bibinfo {author} {\bibfnamefont {S.}~\bibnamefont
  {Chigullapalli}}, \ and\ \bibinfo {author} {\bibfnamefont {A.~A.}\
  \bibnamefont {Alexeenko}},\ }\bibfield  {title} {\enquote {\bibinfo {title}
  {Quantifying the knudsen force on heated microbeams: A compact model and
  direct comparison with measurements},}\ }\href@noop {} {\bibfield  {journal}
  {\bibinfo  {journal} {Physical Review E}\ }\textbf {\bibinfo {volume} {83}},\
  \bibinfo {pages} {066306} (\bibinfo {year} {2011})}\BibitemShut {NoStop}%
\bibitem [{\citenamefont {Anikin}(2011)}]{anikin2011numerical}%
  \BibitemOpen
  \bibfield  {author} {\bibinfo {author} {\bibfnamefont {Y.~A.}\ \bibnamefont
  {Anikin}},\ }\bibfield  {title} {\enquote {\bibinfo {title} {Numerical study
  of radiometric forces via the direct solution of the boltzmann kinetic
  equation},}\ }\href@noop {} {\bibfield  {journal} {\bibinfo  {journal}
  {Computational Mathematics and Mathematical Physics}\ }\textbf {\bibinfo
  {volume} {51}},\ \bibinfo {pages} {1251--1266} (\bibinfo {year}
  {2011})}\BibitemShut {NoStop}%
\bibitem [{\citenamefont
  {Tcheremissine}(1998)}]{tcheremissine1998conservative}%
  \BibitemOpen
  \bibfield  {author} {\bibinfo {author} {\bibfnamefont {F.}~\bibnamefont
  {Tcheremissine}},\ }\bibfield  {title} {\enquote {\bibinfo {title}
  {Conservative evaluation of boltzmann collision integral in discrete
  ordinates approximation},}\ }\href@noop {} {\bibfield  {journal} {\bibinfo
  {journal} {Computers \& Mathematics with Applications}\ }\textbf {\bibinfo
  {volume} {35}},\ \bibinfo {pages} {215--221} (\bibinfo {year}
  {1998})}\BibitemShut {NoStop}%
\bibitem [{\citenamefont {Lotfian}\ and\ \citenamefont
  {Roohi}(2019)}]{lotfian2019radiometric}%
  \BibitemOpen
  \bibfield  {author} {\bibinfo {author} {\bibfnamefont {A.}~\bibnamefont
  {Lotfian}}\ and\ \bibinfo {author} {\bibfnamefont {E.}~\bibnamefont
  {Roohi}},\ }\bibfield  {title} {\enquote {\bibinfo {title} {Radiometric flow
  in periodically patterned channels: fluid physics and improved
  configurations},}\ }\href@noop {} {\bibfield  {journal} {\bibinfo  {journal}
  {Journal of Fluid Mechanics}\ }\textbf {\bibinfo {volume} {860}},\ \bibinfo
  {pages} {544--576} (\bibinfo {year} {2019})}\BibitemShut {NoStop}%
\bibitem [{\citenamefont {Gimelshein}\ \emph {et~al.}(2011)\citenamefont
  {Gimelshein}, \citenamefont {Gimelshein}, \citenamefont {Ketsdever},\ and\
  \citenamefont {Selden}}]{gimelshein2011impact}%
  \BibitemOpen
  \bibfield  {author} {\bibinfo {author} {\bibfnamefont {N.}~\bibnamefont
  {Gimelshein}}, \bibinfo {author} {\bibfnamefont {S.}~\bibnamefont
  {Gimelshein}}, \bibinfo {author} {\bibfnamefont {A.}~\bibnamefont
  {Ketsdever}}, \ and\ \bibinfo {author} {\bibfnamefont {N.}~\bibnamefont
  {Selden}},\ }\bibfield  {title} {\enquote {\bibinfo {title} {Impact of vane
  size and separation on radiometric forces for microactuation},}\ }\href@noop
  {} {\bibfield  {journal} {\bibinfo  {journal} {Journal of Applied Physics}\
  }\textbf {\bibinfo {volume} {109}},\ \bibinfo {pages} {074506} (\bibinfo
  {year} {2011})}\BibitemShut {NoStop}%
\bibitem [{\citenamefont {Strongrich}\ \emph {et~al.}(2014)\citenamefont
  {Strongrich}, \citenamefont {O'Neill}, \citenamefont {Cofer},\ and\
  \citenamefont {Alexeenko}}]{strongrich2014experimental}%
  \BibitemOpen
  \bibfield  {author} {\bibinfo {author} {\bibfnamefont {A.~D.}\ \bibnamefont
  {Strongrich}}, \bibinfo {author} {\bibfnamefont {W.~J.}\ \bibnamefont
  {O'Neill}}, \bibinfo {author} {\bibfnamefont {A.~G.}\ \bibnamefont {Cofer}},
  \ and\ \bibinfo {author} {\bibfnamefont {A.~A.}\ \bibnamefont {Alexeenko}},\
  }\bibfield  {title} {\enquote {\bibinfo {title} {Experimental measurements
  and numerical simulations of the knudsen force on a non-uniformly heated
  beam},}\ }\href@noop {} {\bibfield  {journal} {\bibinfo  {journal} {Vacuum}\
  }\textbf {\bibinfo {volume} {109}},\ \bibinfo {pages} {405--416} (\bibinfo
  {year} {2014})}\BibitemShut {NoStop}%
\bibitem [{\citenamefont {Alexeenko}\ and\ \citenamefont
  {Strongrich}(2016)}]{alexeenko2016microelectromechanical}%
  \BibitemOpen
  \bibfield  {author} {\bibinfo {author} {\bibfnamefont {A.}~\bibnamefont
  {Alexeenko}}\ and\ \bibinfo {author} {\bibfnamefont {A.}~\bibnamefont
  {Strongrich}},\ }\href@noop {} {\enquote {\bibinfo {title}
  {Microelectromechanical gas sensor based on knudsen thermal force},}\ }
  (\bibinfo {year} {2016}),\ \bibinfo {note} {{US} Patent App.
  15/183,259}\BibitemShut {NoStop}%
\bibitem [{\citenamefont {Strongrich}\ and\ \citenamefont
  {Alexeenko}(2015)}]{strongrich2015microstructure}%
  \BibitemOpen
  \bibfield  {author} {\bibinfo {author} {\bibfnamefont {A.}~\bibnamefont
  {Strongrich}}\ and\ \bibinfo {author} {\bibfnamefont {A.}~\bibnamefont
  {Alexeenko}},\ }\bibfield  {title} {\enquote {\bibinfo {title}
  {Microstructure actuation and gas sensing by the knudsen thermal force},}\
  }\href@noop {} {\bibfield  {journal} {\bibinfo  {journal} {Applied Physics
  Letters}\ }\textbf {\bibinfo {volume} {107}},\ \bibinfo {pages} {193508}
  (\bibinfo {year} {2015})}\BibitemShut {NoStop}%
\bibitem [{\citenamefont {Pikus}\ \emph {et~al.}(2019)\citenamefont {Pikus},
  \citenamefont {Sebasti{\~a}o}, \citenamefont {Strongrich},\ and\
  \citenamefont {Alexeenko}}]{pikus2019characterization}%
  \BibitemOpen
  \bibfield  {author} {\bibinfo {author} {\bibfnamefont {A.}~\bibnamefont
  {Pikus}}, \bibinfo {author} {\bibfnamefont {I.~B.}\ \bibnamefont
  {Sebasti{\~a}o}}, \bibinfo {author} {\bibfnamefont {A.}~\bibnamefont
  {Strongrich}}, \ and\ \bibinfo {author} {\bibfnamefont {A.}~\bibnamefont
  {Alexeenko}},\ }\bibfield  {title} {\enquote {\bibinfo {title}
  {Characterization of a knudsen force based vacuum sensor for {N2H2O} gas
  mixtures},}\ }\href@noop {} {\bibfield  {journal} {\bibinfo  {journal}
  {Vacuum}\ }\textbf {\bibinfo {volume} {161}},\ \bibinfo {pages} {130--137}
  (\bibinfo {year} {2019})}\BibitemShut {NoStop}%
\bibitem [{\citenamefont {McCormack}(1973)}]{mccormack1973construction}%
  \BibitemOpen
  \bibfield  {author} {\bibinfo {author} {\bibfnamefont {F.~J.}\ \bibnamefont
  {McCormack}},\ }\bibfield  {title} {\enquote {\bibinfo {title} {Construction
  of linearized kinetic models for gaseous mixtures and molecular gases},}\
  }\href@noop {} {\bibfield  {journal} {\bibinfo  {journal} {Phys. Fluids}\
  }\textbf {\bibinfo {volume} {16}},\ \bibinfo {pages} {2095--2105} (\bibinfo
  {year} {1973})}\BibitemShut {NoStop}%
\bibitem [{\citenamefont {Luo}\ and\ \citenamefont
  {Girimaji}(2003)}]{luo2003theory}%
  \BibitemOpen
  \bibfield  {author} {\bibinfo {author} {\bibfnamefont {L.-S.}\ \bibnamefont
  {Luo}}\ and\ \bibinfo {author} {\bibfnamefont {S.~S.}\ \bibnamefont
  {Girimaji}},\ }\bibfield  {title} {\enquote {\bibinfo {title} {Theory of the
  lattice boltzmann method: two-fluid model for binary mixtures},}\ }\href@noop
  {} {\bibfield  {journal} {\bibinfo  {journal} {Physical Review E}\ }\textbf
  {\bibinfo {volume} {67}},\ \bibinfo {pages} {036302} (\bibinfo {year}
  {2003})}\BibitemShut {NoStop}%
\bibitem [{\citenamefont {Bhatnagar}, \citenamefont {Gross},\ and\
  \citenamefont {Krook}(1954)}]{bhatnagar1954model}%
  \BibitemOpen
  \bibfield  {author} {\bibinfo {author} {\bibfnamefont {P.~L.}\ \bibnamefont
  {Bhatnagar}}, \bibinfo {author} {\bibfnamefont {E.~P.}\ \bibnamefont
  {Gross}}, \ and\ \bibinfo {author} {\bibfnamefont {M.}~\bibnamefont
  {Krook}},\ }\bibfield  {title} {\enquote {\bibinfo {title} {A model for
  collision processes in gases. i. small amplitude processes in charged and
  neutral one-component systems},}\ }\href@noop {} {\bibfield  {journal}
  {\bibinfo  {journal} {Physical review}\ }\textbf {\bibinfo {volume} {94}},\
  \bibinfo {pages} {511} (\bibinfo {year} {1954})}\BibitemShut {NoStop}%
\bibitem [{\citenamefont {Sirovich}(1962)}]{sirovich1962kinetic}%
  \BibitemOpen
  \bibfield  {author} {\bibinfo {author} {\bibfnamefont {L.}~\bibnamefont
  {Sirovich}},\ }\bibfield  {title} {\enquote {\bibinfo {title} {Kinetic
  modeling of gas mixtures},}\ }\href@noop {} {\bibfield  {journal} {\bibinfo
  {journal} {Phys. Fluids}\ }\textbf {\bibinfo {volume} {5}},\ \bibinfo {pages}
  {908--918} (\bibinfo {year} {1962})}\BibitemShut {NoStop}%
\bibitem [{\citenamefont {Andries}, \citenamefont {Aoki},\ and\ \citenamefont
  {Perthame}(2002)}]{AAP02}%
  \BibitemOpen
  \bibfield  {author} {\bibinfo {author} {\bibfnamefont {P.}~\bibnamefont
  {Andries}}, \bibinfo {author} {\bibfnamefont {K.}~\bibnamefont {Aoki}}, \
  and\ \bibinfo {author} {\bibfnamefont {B.}~\bibnamefont {Perthame}},\
  }\bibfield  {title} {\enquote {\bibinfo {title} {A consistent {BGK}-type
  model for gas mixtures},}\ }\href@noop {} {\bibfield  {journal} {\bibinfo
  {journal} {J. Stat. Phys.}\ }\textbf {\bibinfo {volume} {106}},\ \bibinfo
  {pages} {993--1018} (\bibinfo {year} {2002})}\BibitemShut {NoStop}%
\bibitem [{\citenamefont {Haack}, \citenamefont {Hauck},\ and\ \citenamefont
  {Murillo}(2017)}]{haack2017conservative}%
  \BibitemOpen
  \bibfield  {author} {\bibinfo {author} {\bibfnamefont {J.~R.}\ \bibnamefont
  {Haack}}, \bibinfo {author} {\bibfnamefont {C.~D.}\ \bibnamefont {Hauck}}, \
  and\ \bibinfo {author} {\bibfnamefont {M.~S.}\ \bibnamefont {Murillo}},\
  }\bibfield  {title} {\enquote {\bibinfo {title} {A conservative, entropic
  multispecies bgk model},}\ }\href@noop {} {\bibfield  {journal} {\bibinfo
  {journal} {Journal of Statistical Physics}\ }\textbf {\bibinfo {volume}
  {168}},\ \bibinfo {pages} {826--856} (\bibinfo {year} {2017})}\BibitemShut
  {NoStop}%
\bibitem [{\citenamefont {Bobylev}\ \emph {et~al.}(2018)\citenamefont
  {Bobylev}, \citenamefont {Bisi}, \citenamefont {Groppi}, \citenamefont
  {Spiga},\ and\ \citenamefont {Potapenko}}]{bobylev2018general}%
  \BibitemOpen
  \bibfield  {author} {\bibinfo {author} {\bibfnamefont {A.~V.}\ \bibnamefont
  {Bobylev}}, \bibinfo {author} {\bibfnamefont {M.}~\bibnamefont {Bisi}},
  \bibinfo {author} {\bibfnamefont {M.}~\bibnamefont {Groppi}}, \bibinfo
  {author} {\bibfnamefont {G.}~\bibnamefont {Spiga}}, \ and\ \bibinfo {author}
  {\bibfnamefont {I.~F.}\ \bibnamefont {Potapenko}},\ }\bibfield  {title}
  {\enquote {\bibinfo {title} {A general consistent bgk model for gas
  mixtures.}}\ }\href@noop {} {\bibfield  {journal} {\bibinfo  {journal}
  {Kinetic \& Related Models}\ }\textbf {\bibinfo {volume} {11}} (\bibinfo
  {year} {2018})}\BibitemShut {NoStop}%
\bibitem [{\citenamefont {Holway~Jr}(1966)}]{holway1966new}%
  \BibitemOpen
  \bibfield  {author} {\bibinfo {author} {\bibfnamefont {L.~H.}\ \bibnamefont
  {Holway~Jr}},\ }\bibfield  {title} {\enquote {\bibinfo {title} {New
  statistical models for kinetic theory: methods of construction},}\
  }\href@noop {} {\bibfield  {journal} {\bibinfo  {journal} {The physics of
  fluids}\ }\textbf {\bibinfo {volume} {9}},\ \bibinfo {pages} {1658--1673}
  (\bibinfo {year} {1966})}\BibitemShut {NoStop}%
\bibitem [{\citenamefont {Brull}(2015)}]{brull2015ellipsoidal}%
  \BibitemOpen
  \bibfield  {author} {\bibinfo {author} {\bibfnamefont {S.}~\bibnamefont
  {Brull}},\ }\bibfield  {title} {\enquote {\bibinfo {title} {An ellipsoidal
  statistical model for gas mixtures},}\ }\href@noop {} {\bibfield  {journal}
  {\bibinfo  {journal} {Communications in Mathematical Sciences}\ }\textbf
  {\bibinfo {volume} {8}},\ \bibinfo {pages} {1--13} (\bibinfo {year}
  {2015})}\BibitemShut {NoStop}%
\bibitem [{\citenamefont {Shakhov}(1968)}]{shakhov1968generalization}%
  \BibitemOpen
  \bibfield  {author} {\bibinfo {author} {\bibfnamefont {E.}~\bibnamefont
  {Shakhov}},\ }\bibfield  {title} {\enquote {\bibinfo {title} {Generalization
  of the krook kinetic relaxation equation},}\ }\href@noop {} {\bibfield
  {journal} {\bibinfo  {journal} {Fluid Dynamics}\ }\textbf {\bibinfo {volume}
  {3}},\ \bibinfo {pages} {95--96} (\bibinfo {year} {1968})}\BibitemShut
  {NoStop}%
\bibitem [{\citenamefont {Xu}\ and\ \citenamefont
  {Huang}(2010)}]{xu2010unified}%
  \BibitemOpen
  \bibfield  {author} {\bibinfo {author} {\bibfnamefont {K.}~\bibnamefont
  {Xu}}\ and\ \bibinfo {author} {\bibfnamefont {J.-C.}\ \bibnamefont {Huang}},\
  }\bibfield  {title} {\enquote {\bibinfo {title} {A unified gas-kinetic scheme
  for continuum and rarefied flows},}\ }\href@noop {} {\bibfield  {journal}
  {\bibinfo  {journal} {Journal of Computational Physics}\ }\textbf {\bibinfo
  {volume} {229}},\ \bibinfo {pages} {7747--7764} (\bibinfo {year}
  {2010})}\BibitemShut {NoStop}%
\bibitem [{\citenamefont {Guo}, \citenamefont {Xu},\ and\ \citenamefont
  {Wang}(2013)}]{guo2013discrete}%
  \BibitemOpen
  \bibfield  {author} {\bibinfo {author} {\bibfnamefont {Z.}~\bibnamefont
  {Guo}}, \bibinfo {author} {\bibfnamefont {K.}~\bibnamefont {Xu}}, \ and\
  \bibinfo {author} {\bibfnamefont {R.}~\bibnamefont {Wang}},\ }\bibfield
  {title} {\enquote {\bibinfo {title} {Discrete unified gas kinetic scheme for
  all knudsen number flows: Low-speed isothermal case},}\ }\href@noop {}
  {\bibfield  {journal} {\bibinfo  {journal} {Physical Review E}\ }\textbf
  {\bibinfo {volume} {88}},\ \bibinfo {pages} {033305} (\bibinfo {year}
  {2013})}\BibitemShut {NoStop}%
\bibitem [{\citenamefont {Alexeenko}, \citenamefont {Cofer},\ and\
  \citenamefont {Heister}(2017)}]{alexeenko2017microelectronic}%
  \BibitemOpen
  \bibfield  {author} {\bibinfo {author} {\bibfnamefont {A.}~\bibnamefont
  {Alexeenko}}, \bibinfo {author} {\bibfnamefont {A.~G.}\ \bibnamefont
  {Cofer}}, \ and\ \bibinfo {author} {\bibfnamefont {S.~D.}\ \bibnamefont
  {Heister}},\ }\href@noop {} {\enquote {\bibinfo {title} {Microelectronic
  thermal valve},}\ } (\bibinfo {year} {2017}),\ \bibinfo {note} {{US} Patent
  App. 15/370,633}\BibitemShut {NoStop}%
\bibitem [{\citenamefont {Cercignani}(1988)}]{Cercignani}%
  \BibitemOpen
  \bibfield  {author} {\bibinfo {author} {\bibfnamefont {C.}~\bibnamefont
  {Cercignani}},\ }\href@noop {} {\emph {\bibinfo {title} {The {B}oltzmann
  {E}quation and {I}ts {A}pplications}}}\ (\bibinfo  {publisher}
  {Springer-Verlag, New York},\ \bibinfo {year} {1988})\BibitemShut {NoStop}%
\bibitem [{\citenamefont {Harris}(2004)}]{Harris}%
  \BibitemOpen
  \bibfield  {author} {\bibinfo {author} {\bibfnamefont {S.}~\bibnamefont
  {Harris}},\ }\href@noop {} {\emph {\bibinfo {title} {An Introduction to the
  Theory of the Boltzmann Equation}}}\ (\bibinfo  {publisher} {Dover
  Publications},\ \bibinfo {year} {2004})\BibitemShut {NoStop}%
\bibitem [{\citenamefont {Gamba}\ \emph {et~al.}(2017)\citenamefont {Gamba},
  \citenamefont {Haack}, \citenamefont {Hauck},\ and\ \citenamefont
  {Hu}}]{GHHH17}%
  \BibitemOpen
  \bibfield  {author} {\bibinfo {author} {\bibfnamefont {I.}~\bibnamefont
  {Gamba}}, \bibinfo {author} {\bibfnamefont {J.}~\bibnamefont {Haack}},
  \bibinfo {author} {\bibfnamefont {C.}~\bibnamefont {Hauck}}, \ and\ \bibinfo
  {author} {\bibfnamefont {J.}~\bibnamefont {Hu}},\ }\bibfield  {title}
  {\enquote {\bibinfo {title} {A fast spectral method for the {B}oltzmann
  collision operator with general collision kernels},}\ }\href@noop {}
  {\bibfield  {journal} {\bibinfo  {journal} {SIAM J. Sci. Comput.}\ }\textbf
  {\bibinfo {volume} {39}},\ \bibinfo {pages} {B658--B674} (\bibinfo {year}
  {2017})}\BibitemShut {NoStop}%
\bibitem [{\citenamefont {Mieussens}\ and\ \citenamefont
  {Struchtrup}(2004)}]{mieussens2004numerical}%
  \BibitemOpen
  \bibfield  {author} {\bibinfo {author} {\bibfnamefont {L.}~\bibnamefont
  {Mieussens}}\ and\ \bibinfo {author} {\bibfnamefont {H.}~\bibnamefont
  {Struchtrup}},\ }\bibfield  {title} {\enquote {\bibinfo {title} {Numerical
  comparison of bhatnagar--gross--krook models with proper prandtl number},}\
  }\href@noop {} {\bibfield  {journal} {\bibinfo  {journal} {Physics of
  Fluids}\ }\textbf {\bibinfo {volume} {16}},\ \bibinfo {pages} {2797--2813}
  (\bibinfo {year} {2004})}\BibitemShut {NoStop}%
\bibitem [{\citenamefont {Mieussens}(2000)}]{mieussens2000discrete}%
  \BibitemOpen
  \bibfield  {author} {\bibinfo {author} {\bibfnamefont {L.}~\bibnamefont
  {Mieussens}},\ }\bibfield  {title} {\enquote {\bibinfo {title}
  {Discrete-velocity models and numerical schemes for the boltzmann-bgk
  equation in plane and axisymmetric geometries},}\ }\href@noop {} {\bibfield
  {journal} {\bibinfo  {journal} {Journal of Computational Physics}\ }\textbf
  {\bibinfo {volume} {162}},\ \bibinfo {pages} {429--466} (\bibinfo {year}
  {2000})}\BibitemShut {NoStop}%
\bibitem [{\citenamefont {Gallis}\ \emph {et~al.}(2014)\citenamefont {Gallis},
  \citenamefont {Torczynski}, \citenamefont {Plimpton}, \citenamefont {Rader},
  \citenamefont {Koehler},\ and\ \citenamefont {Fan}}]{gallis2014direct}%
  \BibitemOpen
  \bibfield  {author} {\bibinfo {author} {\bibfnamefont {M.~A.}\ \bibnamefont
  {Gallis}}, \bibinfo {author} {\bibfnamefont {J.~R.}\ \bibnamefont
  {Torczynski}}, \bibinfo {author} {\bibfnamefont {S.~J.}\ \bibnamefont
  {Plimpton}}, \bibinfo {author} {\bibfnamefont {D.~J.}\ \bibnamefont {Rader}},
  \bibinfo {author} {\bibfnamefont {T.}~\bibnamefont {Koehler}}, \ and\
  \bibinfo {author} {\bibfnamefont {J.}~\bibnamefont {Fan}},\ }\bibfield
  {title} {\enquote {\bibinfo {title} {Direct simulation monte carlo: The quest
  for speed},}\ }in\ \href@noop {} {\emph {\bibinfo {booktitle} {AIP Conference
  Proceedings}}},\ Vol.\ \bibinfo {volume} {1628}\ (\bibinfo {organization}
  {AIP},\ \bibinfo {year} {2014})\ pp.\ \bibinfo {pages} {27--36}\BibitemShut
  {NoStop}%
\bibitem [{\citenamefont {Gallis}\ \emph {et~al.}(2017)\citenamefont {Gallis},
  \citenamefont {Bitter}, \citenamefont {Koehler}, \citenamefont {Torczynski},
  \citenamefont {Plimpton},\ and\ \citenamefont
  {Papadakis}}]{gallis2017molecular}%
  \BibitemOpen
  \bibfield  {author} {\bibinfo {author} {\bibfnamefont {M.~A.}\ \bibnamefont
  {Gallis}}, \bibinfo {author} {\bibfnamefont {N.~P.}\ \bibnamefont {Bitter}},
  \bibinfo {author} {\bibfnamefont {T.~P.}\ \bibnamefont {Koehler}}, \bibinfo
  {author} {\bibfnamefont {J.~R.}\ \bibnamefont {Torczynski}}, \bibinfo
  {author} {\bibfnamefont {S.~J.}\ \bibnamefont {Plimpton}}, \ and\ \bibinfo
  {author} {\bibfnamefont {G.}~\bibnamefont {Papadakis}},\ }\bibfield  {title}
  {\enquote {\bibinfo {title} {Molecular-level simulations of turbulence and
  its decay},}\ }\href@noop {} {\bibfield  {journal} {\bibinfo  {journal}
  {Physical Review Letters}\ }\textbf {\bibinfo {volume} {118}},\ \bibinfo
  {pages} {064501} (\bibinfo {year} {2017})}\BibitemShut {NoStop}%
\bibitem [{\citenamefont {Gallis}\ \emph {et~al.}(2016)\citenamefont {Gallis},
  \citenamefont {Koehler}, \citenamefont {Torczynski},\ and\ \citenamefont
  {Plimpton}}]{gallis2016direct}%
  \BibitemOpen
  \bibfield  {author} {\bibinfo {author} {\bibfnamefont {M.~A.}\ \bibnamefont
  {Gallis}}, \bibinfo {author} {\bibfnamefont {T.~P.}\ \bibnamefont {Koehler}},
  \bibinfo {author} {\bibfnamefont {J.~R.}\ \bibnamefont {Torczynski}}, \ and\
  \bibinfo {author} {\bibfnamefont {S.~J.}\ \bibnamefont {Plimpton}},\
  }\bibfield  {title} {\enquote {\bibinfo {title} {Direct simulation monte
  carlo investigation of the rayleigh-taylor instability},}\ }\href@noop {}
  {\bibfield  {journal} {\bibinfo  {journal} {Physical Review Fluids}\ }\textbf
  {\bibinfo {volume} {1}},\ \bibinfo {pages} {043403} (\bibinfo {year}
  {2016})}\BibitemShut {NoStop}%
\bibitem [{\citenamefont {Sebastiao}, \citenamefont {Qiao},\ and\ \citenamefont
  {Alexeenko}(2018)}]{sebastiao2018direct}%
  \BibitemOpen
  \bibfield  {author} {\bibinfo {author} {\bibfnamefont {I.~B.}\ \bibnamefont
  {Sebastiao}}, \bibinfo {author} {\bibfnamefont {L.}~\bibnamefont {Qiao}}, \
  and\ \bibinfo {author} {\bibfnamefont {A.~A.}\ \bibnamefont {Alexeenko}},\
  }\bibfield  {title} {\enquote {\bibinfo {title} {Direct {S}imulation {M}onte
  {C}arlo {M}odeling of {H}2-{O}2 deflagration waves},}\ }\href@noop {}
  {\bibfield  {journal} {\bibinfo  {journal} {Combustion and Flame}\ }\textbf
  {\bibinfo {volume} {198}},\ \bibinfo {pages} {40--53} (\bibinfo {year}
  {2018})}\BibitemShut {NoStop}%
\bibitem [{\citenamefont {Jaiswal}\ \emph {et~al.}(2018)\citenamefont
  {Jaiswal}, \citenamefont {Sebasti\~ao}, \citenamefont {Strongrich},\ and\
  \citenamefont {Alexeenko}}]{jaiswal2018femta}%
  \BibitemOpen
  \bibfield  {author} {\bibinfo {author} {\bibfnamefont {S.}~\bibnamefont
  {Jaiswal}}, \bibinfo {author} {\bibfnamefont {I.}~\bibnamefont
  {Sebasti\~ao}}, \bibinfo {author} {\bibfnamefont {A.}~\bibnamefont
  {Strongrich}}, \ and\ \bibinfo {author} {\bibfnamefont {A.~A.}\ \bibnamefont
  {Alexeenko}},\ }\bibfield  {title} {\enquote {\bibinfo {title} {{FEMTA}
  {M}icropropulsion {S}ystem {C}haracterization by {DSMC}},}\ }in\ \href@noop
  {} {\emph {\bibinfo {booktitle} {Proceedings of 31st Rarefied Gas Dynamics
  Symposium (RGD-31)}}}\ (\bibinfo  {publisher} {American Institute of Physics
  (AIP)},\ \bibinfo {address} {Glasgow, UK},\ \bibinfo {year} {2018})\ \bibinfo
  {note} {to appear:
  \href{http://goo.gl/LMLg8Y}{http://goo.gl/LMLg8Y}}\BibitemShut {NoStop}%
\bibitem [{\citenamefont {Alexeenko}\ \emph
  {et~al.}(2006{\natexlab{b}})\citenamefont {Alexeenko}, \citenamefont
  {Gimelshein}, \citenamefont {Muntz},\ and\ \citenamefont
  {Ketsdever}}]{alexeenko2006kinetic}%
  \BibitemOpen
  \bibfield  {author} {\bibinfo {author} {\bibfnamefont {A.~A.}\ \bibnamefont
  {Alexeenko}}, \bibinfo {author} {\bibfnamefont {S.~F.}\ \bibnamefont
  {Gimelshein}}, \bibinfo {author} {\bibfnamefont {E.~P.}\ \bibnamefont
  {Muntz}}, \ and\ \bibinfo {author} {\bibfnamefont {A.~D.}\ \bibnamefont
  {Ketsdever}},\ }\bibfield  {title} {\enquote {\bibinfo {title} {Kinetic
  modeling of temperature driven flows in short microchannels},}\ }\href@noop
  {} {\bibfield  {journal} {\bibinfo  {journal} {International Journal of
  Thermal Sciences}\ }\textbf {\bibinfo {volume} {45}},\ \bibinfo {pages}
  {1045--1051} (\bibinfo {year} {2006}{\natexlab{b}})}\BibitemShut {NoStop}%
\bibitem [{\citenamefont {Biswas}, \citenamefont {Devine},\ and\ \citenamefont
  {Flaherty}(1994)}]{biswas1994parallel}%
  \BibitemOpen
  \bibfield  {author} {\bibinfo {author} {\bibfnamefont {R.}~\bibnamefont
  {Biswas}}, \bibinfo {author} {\bibfnamefont {K.~D.}\ \bibnamefont {Devine}},
  \ and\ \bibinfo {author} {\bibfnamefont {J.~E.}\ \bibnamefont {Flaherty}},\
  }\bibfield  {title} {\enquote {\bibinfo {title} {Parallel, adaptive finite
  element methods for conservation laws},}\ }\href@noop {} {\bibfield
  {journal} {\bibinfo  {journal} {Applied Numerical Mathematics}\ }\textbf
  {\bibinfo {volume} {14}},\ \bibinfo {pages} {255--283} (\bibinfo {year}
  {1994})}\BibitemShut {NoStop}%
\bibitem [{\citenamefont {Cockburn}, \citenamefont {Karniadakis},\ and\
  \citenamefont {Shu}(2000)}]{cockburn2000development}%
  \BibitemOpen
  \bibfield  {author} {\bibinfo {author} {\bibfnamefont {B.}~\bibnamefont
  {Cockburn}}, \bibinfo {author} {\bibfnamefont {G.~E.}\ \bibnamefont
  {Karniadakis}}, \ and\ \bibinfo {author} {\bibfnamefont {C.-W.}\ \bibnamefont
  {Shu}},\ }\bibfield  {title} {\enquote {\bibinfo {title} {The development of
  discontinuous galerkin methods},}\ }in\ \href@noop {} {\emph {\bibinfo
  {booktitle} {Discontinuous Galerkin Methods}}}\ (\bibinfo  {publisher}
  {Springer},\ \bibinfo {year} {2000})\ pp.\ \bibinfo {pages}
  {3--50}\BibitemShut {NoStop}%
\bibitem [{\citenamefont {Bakhtiari}\ \emph {et~al.}(2016)\citenamefont
  {Bakhtiari}, \citenamefont {Malhotra}, \citenamefont {Raoofy}, \citenamefont
  {Mehl}, \citenamefont {Bungartz},\ and\ \citenamefont
  {Biros}}]{bakhtiari2016parallel}%
  \BibitemOpen
  \bibfield  {author} {\bibinfo {author} {\bibfnamefont {A.}~\bibnamefont
  {Bakhtiari}}, \bibinfo {author} {\bibfnamefont {D.}~\bibnamefont {Malhotra}},
  \bibinfo {author} {\bibfnamefont {A.}~\bibnamefont {Raoofy}}, \bibinfo
  {author} {\bibfnamefont {M.}~\bibnamefont {Mehl}}, \bibinfo {author}
  {\bibfnamefont {H.-J.}\ \bibnamefont {Bungartz}}, \ and\ \bibinfo {author}
  {\bibfnamefont {G.}~\bibnamefont {Biros}},\ }\bibfield  {title} {\enquote
  {\bibinfo {title} {A parallel arbitrary-order accurate amr algorithm for the
  scalar advection-diffusion equation},}\ }in\ \href
  {http://dl.acm.org/citation.cfm?id=3014904.3014963} {\emph {\bibinfo
  {booktitle} {Proceedings of the International Conference for High Performance
  Computing, Networking, Storage and Analysis}}},\ \bibinfo {series and number}
  {SC '16}\ (\bibinfo  {publisher} {IEEE Press},\ \bibinfo {address}
  {Piscataway, NJ, USA},\ \bibinfo {year} {2016})\ pp.\ \bibinfo {pages}
  {44:1--44:12}\BibitemShut {NoStop}%
\end{thebibliography}%

\end{document}